\documentclass[10pt]{iopart}
\usepackage{epsfig}
\usepackage{iopams}
\usepackage{graphicx}

\renewcommand{\epsfig}[1]{}

\newcommand{\be}{\begin{equation}}
\newcommand{\ee}{\end{equation}}
\newcommand{\bea}{\begin{eqnarray}}
\newcommand{\eea}{\end{eqnarray}}
\newcommand{\s}{\sigma}
\newcommand{\eps}{\epsilon}
\newcommand{\la}{\langle}
\newcommand{\ra}{\rangle}
\newcommand{\CC}{{\cal C}}
\newcommand{\HH}{{\cal H}}
\newcommand{\RR}{{\cal R}}
\newcommand{\PP}{{\cal P}}
\newcommand{\WW}{{\cal W}}
\newcommand{\FF}{{\cal F}}
\newcommand{\Tc}{T_{\rm c}}
\newcommand{\Te}{T_{\rm eff}}
\newcommand{\Tg}{T_{\rm g}}
\newcommand{\TK}{T_{\rm K}}
\newcommand{\Trsb}{T_{\rm RSB}}
\newcommand{\eq}[1]{(\ref{#1})}
\newcommand{\eqq}[2]{ (\ref{#1}),(\ref{#2})}
\newcommand{\eqqq}[3]{ (\ref{#1}),(\ref{#2}),(\ref{#3})}
\newcommand{\eg}{{\it e.g.}}
\newcommand{\ie}{{\it i.e.}}

\catcode`@=11 
\renewcommand\tableofcontents{%
  \section*{\contentsname}%
  \@starttoc{toc}%
}
\catcode`@=12

\begin{document}

\topical[Violation of FDT in glassy systems] {Violation of the
  fluctuation-dissipation theorem in glassy systems: basic notions and
  the numerical evidence}

\author{A Crisanti\dag and F Ritort\ddag}
\address{
  \dag\ Dipartimento di Fisica, Universit\`a di Roma ``La
  Sapienza'',  INFM Sezione di Roma I and SMC, 
  P.le Aldo Moro 2, 00185 Roma, Italy
}
\address{
  \ddag\ Departament de F\'{\i}sica Fonamental, Facultat de
  F\'{\i}sica, Universitat de Barcelona, Diagonal 647, 08028 Barcelona,
  Spain
}

\ead{
  \mailto{andrea.crisanti@phys.uniroma1.it},
  \mailto{ritort@ffn.ub.es}
}

\begin{abstract}
This review reports on the research done during the past years on
violations of the fluctuation-dissipation theorem (FDT) in glassy
systems. It is focused on the existence of a
quasi-fluctuation-dissipation theorem (QFDT) in glassy systems and the
currently supporting knowledge gained from numerical simulation
studies. It covers a broad range of non-stationary aging and
stationary driven systems such as structural-glasses, spin-glasses,
coarsening systems, ferromagnetic models at criticality, trap models,
models with entropy barriers, kinetically constrained models, sheared
systems and granular media.  The review is divided into four main
parts: 1) An introductory section explaining basic notions related to
the existence of the FDT in equilibrium and its possible extension to
the glassy regime (QFDT), 2) A description of the basic analytical
tools and results derived in the framework of some exactly solvable
models, 3) A detailed report of the current evidence in favour of the
QFDT and 4) A brief digression on the experimental evidence in its
favour.  This review is intended for inexpert readers who want to
learn about the basic notions and concepts related to the existence of
the QFDT as well as for the more expert readers who may be interested
in more specific results.
\end{abstract}

\pacs{64.40.-i, 64.60.Cn, 75.10.Nr}

\maketitle

\tableofcontents

\section{Introduction}
\label{intro}

The search for a general theory of non-equilibrium processes has been
a primary goal in modern statistical physics. Despite of many efforts
in this direction we have a limited understanding of the basic
principles behind non-equilibrium theories. Compared with ensemble
equilibrium theory, a general principle
like the equal probability Boltzmann principle (that forms the basis of
equilibrium statistical mechanics and provides a statistical foundation of
thermodynamics) is still lacking.  During the last century, the field of non-equilibrium
phenomena has grown in two directions: 1) by developing new statistical
models as an inspiring source of fruitful new concepts and ideas 2) by
establishing partial links among different, apparently
disconnected, non-equilibrium phenomena. 

Although much progress has been made in the first direction, the
second one remains less unexplored.  While a general principle
governing non-equilibrium systems probably does not exist, substantial progress
could be done following the second route in the search for
basic principles governing a restricted category or class of systems.
The applications of such
basic principles may
be very important because {\em a priori} many different
systems can fall into the same category. Hence the interest
in the research on the existence of such restricted formulations.

During the past years it has become increasingly clear that {\em glassy
systems} may constitute one of these large categories where their
physical behavior can be rationalized within a restricted formulation.
Glassy systems are rather common in nature and many systems such as
structural glasses, spin glasses, disordered and granular materials or
proteins present what is called {\it glassy behaviors}. This means a
dramatic slowing down of relaxational processes when some control
parameters are varied. A typical signature of glassy behaviour is a {\it
power-law} or {\it stretched exponential} behavior of correlation
functions, as opposed to exponential decay.  As the characteristic
relaxation time may change by several orders of magnitude it can easily
exceed the observation time. As a consequence the system {\it ages}: the
observed static and dynamic properties depend on the {\it age} of the
system defined as the time since the system was prepared (also called
{\it waiting time}). For this reason this residual very slow
non-equilibrium phenomena is commonly known as {\it aging}.

Aging systems include a large variety of materials. In fact, nearly all
physical systems, within an appropriate set of conditions and observed
during a specific time-window, display glassy properties. The origin of
glassy behaviour, however, can vary from system to system.  The most
important class of glassy systems (which include window glasses) are
glass forming liquids where glassy behaviour is due to the appearance,
as some external parameter is changed, of a long-lived complex pattern
of interacting bonds between their microscopic constituents which
strongly inhibits relaxation toward equilibrium. Aging follows from the
very-slow motion of such a complex pattern of interacting bonds which
induces a slow change of the atomic structure of the liquid. For this
reason glass-forming liquids are usually called {\it structural
glasses}. Our current understanding of the slow glassy relaxation
dynamics is greatly limited by the lack of a general non-equilibrium
theory that accounts for these phenomena.

Glasses can be generated by the fast cooling of a liquid.  Upon
cooling from high temperatures down to the melting transition
temperature $T_{\rm M}$, sometimes crystallization does not occur and the
liquid continues its way down in temperature beyond $T_{\rm M}$ by following
a line (called supercooled liquid line) which is the continuation of
the liquid line. As the liquid line is thermodynamically stable only
above $T_{\rm M}$, the supercooled liquid line is metastable with locally
equilibrated properties, so its lifetime can be extremely large.
As cooling proceeds it is observed that the
supercooled liquid falls out of equilibrium (i.e. departs from the
supercooled liquid line), below a temperature $T^*(r)$ which depends on
the cooling rate $r$.  The state reached below $T^*(r)$ is called a
glass and the corresponding relaxational regime is indistinctly termed
as {\em aging} or {\em glassy}.  For small values of $r$ a sharp transition is
observed at $T^*(r)$, usually referred as structural arrest, where the
heat capacity jumps down, indicating the freezing
of degrees of freedom.
 Contrarily to the supercooled state, the
glass state is of non-equilibrium nature and $T^*(r)$ is observed to
decrease with $r$. As $T^*(r)$ depends on the cooling rate, no
equilibrium phase transition occurs at that temperature.
This means that the liquid will eventually equilibrate back to the supercooled
state. The equilibration process may take an extremely long time (even
for temperatures only a few degrees below $T^*$) being inaccessible from
any practical point of view. Under some conditions the equilibration
time can be larger than the age of the universe!. In these conditions the glass
state is the only {\it observable} state.

Long equilibration times imply that the
glass state is characterized by very low energy
dissipation rates, also called entropy production. This may give the
false impression that the glass is in a stationary state.
For instance,  a piece of silica glass at room temperature looks pretty 
stable, indeed its optical, electrical and mechanical properties 
appear constant in time. However a more
careful examination reveals that the physical properties are constant
only if observed on time-scales much smaller than the time elapsed since
the glass was prepared or formed. Beyond that timescale, the physical
properties change revealing that the glass is aging.

Although aging was identified long time ago in the study polymers 
\cite{Struik78} it has received a renewed interest in connection with 
the study of spin glasses.  Measurements of the magnetization in spin
glasses have shown that aging is a general property of the
low-temperature spin-glass phase.  There are several types of spin glass
materials, the most common ones are metallic spin glasses. These are
random diluted magnetic systems where glassy behaviour arises from the
disordered pattern of exchange interactions, rather than being
self-generated as in structural glasses. Indeed random dilution
generates exchange interactions with random competing signs, the system
is then frustrated since a finite fraction of bonds cannot
be satisfied.  Aging is consequence of the slow
evolution of the pattern of satisfied bonds which becomes strongly
inhibited as the temperature is lowered.

Another class of systems with glassy properties are {\it driven
systems} which, under certain conditions, reach a stationary state
characterized by non-Gibbsian probability distributions.
After applying a time-dependent perturbation of frequency $\omega>
1/t_{\rm eq}$, upon an initially equilibrated system of relaxation
time $t_{\rm eq}$, a new stationary state is reached
which for many aspects is similar to the aging state of the
relaxational system of age $\sim 1/\omega$.

Another important aspect of glassy systems that has received
considerable attention for a long time \cite{Tool46} is the idea about
the existence of an effective temperature (sometimes also called fictive
temperature) describing the non-equilibrium properties of the glassy
state. During the last decades it has emerged that a possible way to
rationalize the existence of an effective temperature is by measuring
violations of the fluctuation-dissipation theorem (FDT). In
glassy systems, a new modified relation between correlations and
responses that goes under the name of quasi FDT (QFDT) provides a
description of the dynamics in the glassy state by quantifying the
violations of the FDT. In this new theorem the effective temperature
plays the role of the temperature of the bath. Related to the concept of
the effective temperature is the idea of the existence of a heat flow
from the glass to the thermal bath put in contact with the system. As
the glass has an effective temperature larger than that of the bath, the
heat flows from the glass to the bath. However, the energy dissipation
rate from the glass toward the bath is extremely low (hardly
measurable) and, in general, this flow can be also understood in terms
of an effective very low thermal conductivity.  The reader should be aware that
using QFDT is only one among other possible ways of introducing an
effective temperature for the description of the glassy state. In
general other definitions which use a generalization of different
equilibrium relations to the non-equilibrium regime are possible. This
arises the problem of the equivalence of all possible definitions. We
shall not discuss this point in this review and we will stick to the
QFDT definition of an effective temperature.

This review will concentrate on the existence of a QFDT, its physical
meaning, in what conditions it can emerge and the numerical evidence
reported in favour of its validity. This is a rapidly growing area of
research which is attracting new condensed matter and statistical
physicists. We will report here the most important results obtained
until the summer of the year 2002. Although we have tried to cover
most of the published work some contributions may have been
overlooked. We apologize in advance to those colleagues. Although some
of the results here reviewed are currently well understood many others
still lack a full comprehension so it is not exaggerated to say that
some of the ideas and suggestions described in this review could be
modified in the future to adapt to the forthcoming theoretical,
numerical or experimental evidence.  Most of the results here reported
deal with relaxational aging systems (as compared to driven systems)
since these are the ones that have mostly attracted the attention of
the researchers in the field. However, future developments in this
exciting area of research might compensate this original unbalance as
driven systems appear more amenable to experimental research than
aging systems.  Moreover, in this review we shall only consider the
FDT in its classical version.  Although most of the the ideas can be
extended to the quantum regime, to our knowledge there are neither
numerical or experimental works challenging FDT violations in the
quantum aging regime. Therefore we shall not address them, the
interested reader is referred to a recent review \cite{Cugliandolo02}.

Many textbooks and article reviews can be useful to complement the
contents of this review. Basic reviews on the glass transition since mid
80's until now can be found in
\cite{Jackle86,Angell88,EdiAngNag96,Benedetti96}. Other accounts dealing
with aspects of the glass transition include: thermodynamic theories of
the glass transition \cite{Wolynes97,Marzio02}, mode-coupling theory
(MCT) \cite{GotSjo92,Got99} and numerical simulations \cite{Kob99}.
For spin glasses a good selection of review articles can be found in
\cite{Young98}. A clear discussion of mode-coupling approximations in
the context of disordered systems can be found in
\cite{BouCugKurMez96}. A recent discussion of several aspects of
concerning FDT violations can be found also in a recent review on
kinetically constrained models \cite{RitSol02}.  Finally, a throughout
compendium of analytical methods for glassy dynamics has been recently
collected in \cite{Cugliandolo02} and a review of granular systems in
\cite{Bouchaud02}. Proceeding articles covering several aspects of
glasses and spin-glasses can be found in \cite{Sitges96,Trieste99} and
for kinetically constrained models in \cite{Barcelona02}.

The contents of this review have been written with two kinds of
readers in mind: inexpert and expert. Those inexpert readers who want
to understand the most basic ideas as well as the interest of
investigating FDT violations must read Sections
\ref{basic},\ref{ME},\ref{EXTME}. These sections have been written at
an introductory level, so expert readers who know about the subject
may directly start reading from Section \ref{thermo}. However, a
careful reading of section \ref{EXTME} is recommended to those readers
who want to have a more physically appealing description of the
possible origin of FDT violations.  Section \ref{thermo} deals with
some of the thermodynamic consequences of FDT violations. Sections
\ref{esm},\ref{qfdt:numerics} constitute the core of the
review. Section \ref{esm} describes our knowledge of FDT violations
gathered from several exactly solvable models where many aspects of
their non-equilibrium behavior can be understood by analytical
means. Section \ref{qfdt:numerics} covers all evidence collected in
the past years in favour of the existence of a QFDT in glassy
systems. Many of the model systems described in this section
correspond to realistic as well as model systems for which analytical
solutions are hardly known. The expert reader who wants to grasp the
state of the art concerning these questions will be mainly interested
on these two sections. Finally, a brief account of some experimental
results on FDT violations is described in Section \ref{exp}. Section
\ref{conclusions} presents some conclusions.

\section{Basic definitions and concepts}
\label{basic}

In this section we recall some concepts of equilibrium 
theory which will be needed later for the description the 
glassy state.

\subsection{The microcanonical and canonical ensembles}
\label{basic:micro}

The foundations of equilibrium statistical mechanics rely on 
the maximum entropy postulate and the Boltzmann's
equal probability hypothesis.
An introduction to the basic postulates can be found in the
classical books on statistical mechanics, 
rather excellent are that by Ma \cite{Ma85} and Callen \cite{Callen85}.  
Good discussions also comes from the Information Theory, see for example
the book of Beck and Schloegel \cite{BecSch93}. 

In what follows we shall denote with $\CC$ a generic system
configuration in the phase space. The phase space can be either
continuous or discrete depending on the particular system. For
example, for a system of $N$ particles $\CC$ are the positions and
momenta in a continuous $6N$-dimensional space, while for a system of
$N$ $1/2$-spins $\CC$ is a point in a a discrete $N$ dimensional space
with $2^N$ points.  The system evolves in time following a dynamical
rule which generally speaking is a rule that for each configuration
$\CC$ associates a new configuration $\CC'$.  The set of
configurations which can be visited given a dynamical rule, defines
the region of motion in the phase space. Let $\Gamma$ be the volume of
the region of motion allowed by the invariant quantities. The basic
assumption of statistical mechanics asserts that the entropy is the
logarithm of $\Gamma$.  This makes the entropy computable without
having to solve the dynamics.
If we assume that the system is described by an energy function $E(\CC)$ 
then the motion is confined to a region in phase space of constant energy.
The calculation of entropy is then reduced to 
\begin{equation}
  \label{eq:ent-micro}
  S(E) =\ln \sum_{\CC}\delta\bigl(E-E(\CC)\bigr)
\end{equation}
where in the case of continuous variables the sum must be read as an
integral. This
equation defines the {\it microcanonical ensemble}.
Since all allowed states are included $S(E)$ is clearly a maximum over
all possible regions of constant energy $E$ into which the phase space can
be divided. 

Let us consider now an observable $A(\CC)$ which we will
assume  to be neither a constant of motion or a univocal function of the energy
$E$. We also assume that $A(\CC)$ is extensive, {\it i.e.}, it is proportional
to the system size (volume or the number of constituents).
We can then divide the phase space according to the value of $A(\CC)$
and, defining the degeneration $\Omega(E,A)$ of the partition as the total 
number of configurations $\CC$ of
energy $E$ and observable value $A$:
\begin{equation}
  \label{basic:micro:1}
  \Omega(E,A)=\sum_{\CC}\delta\bigl(E-E(\CC)\bigr)\,
                              \delta\bigl(A-A(\CC)\bigr),
\end{equation}
introduce the entropy 
in analogy with (\ref{eq:ent-micro}):
\begin{equation}
  \label{basic:micro:5}
  S(E,A)=\ln \Omega(E,A).
\end{equation}
Using the integral representation of delta functions 
$\Omega(E,A)$ can be rewritten as 
\begin{eqnarray}
\label{basic:micro:2}
\fl 
\Omega(E,A) = \int_{-\infty}^\infty\, \frac{d\alpha_1\,d\alpha_2}{(2\pi)^2}\,
       \exp{\bigl(i\alpha_1 E+i\alpha_2 A\bigr)} \,
       \sum_{\CC}\,\exp{\bigl[-i\alpha_1 E(C)-i\alpha_2 A(C)\bigr]}
\nonumber\\
\lo = \int_{-\infty}^\infty\,\frac{d\alpha_1\,d\alpha_2}{(2\pi)^2}\,
 \exp{\bigl[ {\cal S}(E,A,\alpha_1,\alpha_2)\bigr]}
\end{eqnarray}
where the function  ${\cal S}(E,A,\alpha_1,\alpha_2)$ is given by,
\begin{equation}
  \label{basic:micro:3}
  {\cal S}(E,A,\alpha_1,\alpha_2)= i\alpha_1 E + i\alpha_2 A +
                     \ln{\cal Z}(\alpha_1,\alpha_2)
\end{equation}
and ${\cal Z}(\alpha_1,\alpha_2)$ is the {\it partition function} 
given by,
\begin{equation}
  {\cal Z}(\alpha_1,\alpha_2)=\sum_{\CC}
  \exp{\bigl[-i\alpha_1 E(\CC)-i\alpha_2 A(\CC)\bigr]}
  \label{basic:micro:4}
\end{equation}
Since both energy and the observable are extensive quantities the sum in
(\ref{basic:micro:4}) is dominated, in the limit of large system size,
by the largest contribution, and ${\cal Z}$ is exponentially large in
the system size.  The function ${\cal S}(E,A,\alpha_1,\alpha_2)$ is then
an extensive quantity and, in that limit, the integrations can be done
selecting the dominant contribution using the saddle point method:
\begin{equation} 
\label{basic:micro:6a}
  \frac{\partial {\cal S}(E,A,\alpha_1,\alpha_2)}
       {\partial \alpha_1} =
  \frac{\partial {\cal S}(E,A,\alpha_1,\alpha_2)}
       {\partial \alpha_2}=0
\end{equation}
which leads to the saddle point equations
\begin{eqnarray}
  \label{basic:micro:6}
  \lo E = \frac{1}{{\cal Z}(\alpha_1,\alpha_2)}\,\sum_{\CC}\, E(\CC)\,
  \exp{\bigl[-i\alpha_1 E(\CC)-i\alpha_2 A(\CC)\bigr]}
        \equiv \langle E\rangle
	\\
  \label{basic:micro:7}
  \lo A = \frac{1}{{\cal Z}(\alpha_1,\alpha_2)}\,\sum_{\CC}\, A(\CC)\,
  \exp{\bigl[-i\alpha_1 E(\CC)-i\alpha_2 A(\CC)\bigr]}
        \equiv \langle A\rangle
\end{eqnarray}
Reality of $\Omega(E,A)$ implies that the solution 
$\alpha_1^*(E,A)$, $\alpha_2^*(E,A)$ of the saddle point equations must
be pure imaginary: 
$\alpha_1^*=-i\beta$, $\alpha_2^*=-i\mu$ with $\beta$ and $\mu$ real numbers.
The entropy \eq{basic:micro:5} is given by the value
of ${\cal S}(E,A,\alpha_1,\alpha_2)$ evaluated at the saddle point:
\begin{equation}
  \label{basic:micro:9}
  S(E,A) = {\cal S}(E,A,-i\beta, -i\mu)
         = \beta E+\mu A+ \ln {\cal Z}(\beta,\mu).
\end{equation}
We can now ask the following question. What is best choice of the value 
of $A(\CC)$ for which $S$ attains its maximum?
Stationarity of ${\cal S}$ with respect to $\alpha_1$ and $\alpha_2$ 
at the saddle point implies
\begin{eqnarray}
  \label{basic:micro:11}
  \frac{\partial S(E,A)}{\partial E} &=& \beta
  \\
  \label{basic:micro:12}
  \frac{\partial S(E,A)}{\partial A} &=& \mu
\end{eqnarray}
so that the maximum entropy assumption requires $\mu=0$, {\it i.e.},
the entropy $S(E,A)$ must be stationary with respect to variations of $A$.
For any energy $E$ the best choice of $A$ is then:
\begin{equation}
  \label{basic:micro:16}
  A = \langle A\rangle 
   = \frac{1}{{\cal Z}(\beta)}\,\sum_{\CC}\, A(\CC)\,
  \exp{\bigl[-\beta E(\CC)\bigr]}
\end{equation}
where
\begin{equation}
  {\cal Z}(\beta) = \sum_{\CC}\,\exp{\bigl[-\beta E(\CC)\bigr]}
                  = \exp{\bigl[-\beta F(\beta)\bigr]}
\label{basic:micro:15}
\end{equation}
The value of the entropy is 
\begin{equation}
  S(E) = \beta E+\ln{\cal Z}(\beta)
  \label{basic:micro:14}
\end{equation}
and is independent on $A$ as required from stationarity.
Finally from (\ref{basic:micro:11}) it follows that $\beta^{-1}$ 
can be identified with the temperature $T$ of equilibrium thermodynamics,
while insertion of  (\ref{basic:micro:15}) into (\ref{basic:micro:14})
yields the thermodynamic relation $F(\beta)=E(T)-TS(E)$ identifying 
$F(\beta)$ with the Helmholtz free energy. 
Equations (\ref{basic:micro:11}), (\ref{basic:micro:16}), 
(\ref{basic:micro:15}) and (\ref{basic:micro:14}) define 
the {\it canonical ensemble}. 

At difference with the microcanonical ensemble the measure of the 
canonical ensemble is not restricted on states of constant energy.
All possible states $\CC$ enter but with a weight proportional to 
$\exp[-\beta E(\CC)]$. For a given temperature, however, only 
states with energy $E=\langle E\rangle$ given by (\ref{basic:micro:16}) for 
$A(\CC)= E(\CC)$ [See (\ref{basic:micro:6})] significantly
contribute to the measure.
The temperature can be seen as a {\it Lagrange multiplier} used to fix the
value of the energy.
Vice versa each value of $E$ in equilibrium selects
a temperature $T$ through \eq{basic:micro:11}.

\subsection{Einstein fluctuation theory}
\label{basic:fluct}

A key contribution in the development of equilibrium statistical
mechanics is the statistical theory of fluctuations developed by
Einstein \cite{Callen85}. 
In the previous section we have seen that in equilibrium the
value of any observable $A(\CC)$ is given by (\ref{basic:micro:16}),
which for the purpose of this section will be denoted by $A_{\rm eq}$.
The equilibrium value corresponds to the most probable value
of $A(\CC)$, {\it i.e.}, 
to the value of $A(\CC)$ which has the overwhelming  probability to be seen
in equilibrium. The same considerations apply to the energy $E(\CC)$ 
in the canonical ensemble.

We may then ask what is the probability of observing a value of $A(\CC)$ 
different from the equilibrium value.  
This probability is simply
proportional to the number of configurations with $A(\CC) = A$
which from (\ref{basic:micro:5}) is
\begin{equation}
  P(\delta A)\propto \Omega(E_{\rm eq},A=A_{\rm eq}+\delta A)
             \propto \exp{\bigl[ S(E_{\rm eq},A_{\rm eq}+\delta A) \bigr]}
\label{basic:fluct:10}
\end{equation}
where following our notation $E_{\rm eq}$ is equilibrium energy.
For small value of the fluctuations $\delta A$ the exponent can be expanded
and using stationarity of the entropy with respect to variation of 
$A$ we get,
\begin{equation}
  S(E_{\rm eq},A) = S(E_{\rm eq},A_{\rm eq}) +
              \frac{\bigl(\delta A\bigr )^2}{2} 
	      \left(\frac{\partial^2 S}{\partial A^2}
                          \right)_{A=A_{\rm eq}}\,
          +O \bigl(\delta A\bigr)^3
  \label{basic:fluct:11}
\end{equation}
If $A_{\rm eq}$ is a maximum, then a necessary condition is
\begin{equation} 
  \left(\frac{\partial^2 S}{\partial A^2}
  \right)_{A=A_{\rm eq}}= -\frac{1}{T \chi_A} < 0.
 \label{basic:fluct:12}
\end{equation}
 From (\ref{basic:fluct:10}) we finally obtain,
\begin{equation}
  P(\delta A) = \frac{1}{\sqrt{2\pi T\chi_A}} 
     \exp{\left[-\frac{(\delta A)^2}{2T\chi_A} \right]}
  \label{basic:fluct:9}
\end{equation}
As $\chi_A$ is an extensive quantity only subextensive fluctuations 
$\delta A\sim \sqrt{V}$, where $V$ is the system size, 
have finite probability in equilibrium. This justifies the most probable 
character of the equilibrium value $A_{\rm eq}$.

The quantity $\chi_A$ is called {\it susceptibility}, and from 
(\ref{basic:fluct:9}) is related to fluctuations of $A$ through
\begin{equation}
  T\chi_A = \langle A^2\rangle - \langle A\rangle^2.
  \label{basic:fluct:8}
\end{equation}
This relation is the simplest form of the static
Fluctuation-Dissipation Theorem (FDT) which
relates the magnitude of thermal fluctuations with the response of the
system to a (small) perturbation. 

Suppose we add a constant perturbation $-\eps A(\CC)$ to the energy $E(\CC)$. 
Then in the new equilibrium value $\la A\ra_{\eps}$ of $A$ is
[See (\ref{basic:micro:16})]
\begin{equation}
  \langle A\rangle_{\eps}=\frac{
  \sum_{\CC}\,A(\CC)\,\exp{\bigl[-\beta E(\CC) + \beta \eps A(\CC)\bigr]}
  }
  {\sum_{\CC}\,\exp{\bigl[-\beta E(\CC)+\beta \eps A(\CC)\bigr]}
  }.
  \label{basic:fluct:21}
\end{equation}
The susceptibility $\chi_A$ is defined as the variation of  
$\langle A\rangle$ induced by a small perturbation
\begin{equation}
  \chi_{A} = \left.\frac{\partial \langle A\rangle_{\eps}}
                        {\partial \eps}
                   \right|_{\eps=0}
  \label{basic:fluct:23}
\end{equation}
and hence measures the response of the system to the perturbation.
Inserting (\ref{basic:fluct:21}) into (\ref{basic:fluct:23}) a
straightforward calculation leads to (\ref{basic:fluct:8}).

The FDT formula (\ref{basic:fluct:8}) is non-trivial result, since it
relates different physical processes:
the susceptibility describes an extensive $O(V)$ variation of the observable 
$A$ while the r.h.s. of (\ref{basic:fluct:8})
describes subextensive $O(\sqrt{V})$ thermal fluctuations.
This fact is at the basis of Onsager regression principle discussed in
the next section.

\subsection{The Onsager regression principle: a simple derivation of the
fluctuation-dissipation theorem (FDT)}
\label{basic:onsager}

Onsager proposed \cite{Onsager31a,Onsager31b}
a simple derivation of FDT for time-dependent perturbations.
The derivation bypasses the more cumbersome analytical developments 
using linear response theory formalism, the
Fokker-Planck equation or the generalized master-equation
approach. 

Onsager derivation 
is based on the following {\it regression principle}: If a
system initially in an equilibrium state $1$ is driven by an external
perturbation to a different equilibrium state $2$, 
then the evolution of the system from state $1$ toward state $2$
in the presence of the perturbation can be treated as a spontaneous
equilibrium fluctuation (in the presence of the perturbation) from the
(now) non-equilibrium state $1$ to the (now) equilibrium state $2$.

Suppose that the system is initially in equilibrium with a thermal 
bath at temperature 
$T$, then the probability distribution of system 
configuration $\CC$  in state $1$ is given by the canonical ensemble 
(\ref{basic:micro:16}):
\begin{equation}
  P_0(\CC) = \frac
     {\exp{\bigl[-\beta E(\CC)\bigr]}}
     {\sum_{\CC}\, \exp{\bigl[-\beta E(\CC)\bigr]}}
\label{basic:onsager:1}
\end{equation}
The subscript '$0$' indicates that the system in unperturbed.

\begin{figure}
  \centering
  \includegraphics[scale=0.35]{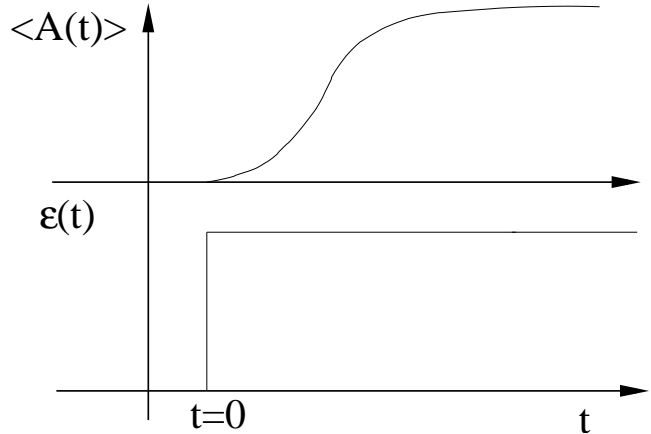}
  \caption{Perturbation $\eps(t)$ and typical evolution curve for 
    $\langle A(t)\rangle_{\epsilon}$.
  }
  \label{fig:onsager:1}
\end{figure}

At time $t=0$ a constant perturbation coupled to the observable
$B(\CC)$ is applied to the system changing its energy into
\begin{equation}
  E_{\epsilon}(\CC)=E(\CC)-\eps(t) B(\CC), \qquad 
  \label{basic:onsager:2}
\end{equation}
where $\eps(t)=\eps$ if $t> 0$, and zero otherwise.
The effect of the perturbation can be monitored by looking at the 
evolution of the expectation value $\langle A(t)\rangle_{\epsilon}$ 
of an observable $A(\CC)$,
not necessarily equal to $B(\CC)$, from the equilibrium value 
in state $1$
$\langle A(t=0)\rangle_{\epsilon}=\langle A\rangle_0$ 
toward the new equilibrium  value in state $2$.
The shape of $\eps(t)$ and a typical evolution
of $\langle A(t)\rangle_{\epsilon}$ are shown in Figure \ref{fig:onsager:1}. 

The expectation value of $\langle A(t)\rangle_{\epsilon}$ 
is given by the average over all possible dynamical
paths originating from initial configurations weighted with the 
probability distribution (\ref{basic:onsager:1}):
\begin{equation}
  \langle A(t)\rangle_{\epsilon} = \sum_{\CC,\CC_0}\,
                 A(\CC)\,P_{\eps}(\CC,t|\CC_0,0)\,P_0(\CC_0)
\label{basic:onsager:3}
\end{equation}
where $P_{\epsilon}(\CC,t|\CC_0,0)$ is the conditional probability 
for the evolution from the configuration $\CC_0$ at time $t=0$ 
to the configuration $\CC$ at time $t$.  
If $\epsilon=0$ the expectation value becomes time independent
since the initial state is in equilibrium and $P_{0}(\CC,t|\CC_0,0)$
describes spontaneous equilibrium fluctuations. 

The Onsager's regression principle asserts that
the conditional probabilities after having applied the perturbation
are equal to those of spontaneous equilibrium fluctuations
in state $2$. Hence since the state $2$ 
is still described by the canonical ensemble (\ref{basic:onsager:1}), 
but with the energy (\ref{basic:onsager:2}) now including the 
perturbation term, then 
\begin{equation}
  P_{\epsilon}(\CC,t|\CC_0,0) = P_{0}(\CC,t|\CC_0,0)\,
      \exp{\bigl\{\beta\epsilon [B(\CC)-B(\CC_0] \bigr\} }.
\label{basic:onsager:5}
\end{equation}
where the r.h.s of this equations is just the product of the
spontaneous equilibrium fluctuations conditional probabilities
$P_{0}(\CC,t|\CC_0,0)$ in state $1$ corrected by the presence of the
perturbation term $\epsilon B(\CC)$~\footnote{The
relation~\eq{basic:onsager:5} is only valid for $\CC_0\neq
\CC$. Indeed, the normalization condition of conditional probabilities
$\sum_{\CC}P_{\eps}(\CC,t|\CC_0,0)=1$ implies that
$P_{\epsilon}(\CC,t|\CC,0)\neq P_{0}(\CC,t|\CC,0)$ so relation
\eq{basic:onsager:5} does not hold for $\CC_0= \CC$. The difference
between both probabilities $P_{\epsilon}(\CC,t|\CC,0),
P_{0}(\CC,t|\CC,0)$ does not matter as the transition $\CC\to \CC$ does not contribute to the response function in \eq{basic:onsager:6}}.  Inserting
(\ref{basic:onsager:5}) into (\ref{basic:onsager:3}) and expanding the
exponential up to the linear order we get,
\begin{eqnarray}
  \langle A(t)\rangle_{\epsilon}-\langle A \rangle_0 &=& 
  \beta \epsilon\sum_{\CC,\CC_0}\,A(\CC)\,\bigl[B(\CC)-B(\CC_0)\bigr]\,
               P_0(\CC_0)
\nonumber\\
 &=& \beta\epsilon\bigl[\langle A(t)B(t)\rangle_0 - 
                        \langle A(t)B(0)\rangle_0 \bigr].
  \label{basic:onsager:6}
\end{eqnarray}
If we define the correlation function and time-dependent susceptibility as
\begin{eqnarray}
  C_{A,B}(t,s) &=&  \langle A(t)B(s)\rangle_0 \\     
  \chi_{A,B}(t) &=& \lim_{\epsilon\to 0}
              \frac{\langle A(t)\rangle_{\epsilon} -\langle A\rangle_0}
                   {\epsilon}
\end{eqnarray}
then from (\ref{basic:onsager:6}) we get the integrated form of the 
FDT relation
\begin{equation}
  \chi_{A,B}(t) = \beta\bigl[C_{A,B}(t,t)-C_{A,B}(t,0) \bigr]
\label{basic:onsager:8}
\end{equation}
The static form of the FDT (\ref{basic:fluct:8}) is easily obtained 
from (\ref{basic:onsager:8}) by taking $A=B$ and the limit $t\to\infty$. 
In this case $\chi_{A,A}(t)\to \chi_A$ [cfr. (\ref{basic:fluct:23})]
and $C_{AA}(t,t)=\langle A^2\rangle_{0}$ while 
$C_{AA}(t,0)\to \langle A\rangle_{0}^2$ as
correlations factorize for infinitely separated times.

Sometimes the FDT relation is written in a differential form by
considering the two-times response or retarded Green function:
\begin{equation}
  R_{A,B}(t,s) = \frac{\delta \langle A(t)\rangle}{\delta \epsilon(s)},
  \qquad t>s
\label{basic:onsager:9}
\end{equation}
which gives the response to an impulsive
perturbation $\eps(s)$ acting at time $s$. Causality imposes that 
the response function $R_{A,B}(t,s)$ is zero for $t<s$: perturbations cannot
propagate backward in time.
The susceptibility $\chi_{A,B}(t)$ is the integral of the response
function $R_{A,B}(t,s)$ then, using \eq{basic:onsager:8}:
\begin{eqnarray}
  \int_{0}^{t}R_{A,B}(t,s)\,ds &=& \chi_{A,B}(t) 
\nonumber\\
                              &=& \beta\bigl[C_{A,B}(0)-C_{A,B}(t)\bigr]
\nonumber\\
     &=&\beta\int_{0}^{t}\frac{\partial}{\partial s}\, C_{A,B}(t,s)\,ds.
\label{basic:onsager:10}
\end{eqnarray}
The last equality, and the arbitrarity of time $t$, implies that 
\begin{equation}
  R_{A,B}(t,s) = \beta \frac{\partial}{\partial s}\, C_{A,B}(t,s)\theta(t-s)
  \label{basic:onsager:11}
\end{equation}
This is the differential form of the FDT relation.

\section{The Master Equation Approach}
\label{ME}

In this section we introduce the Master equation for the dynamical
evolution of a generic system and show how the FDT arises within this
approach.  Besides the previous derivation in Section \ref{basic:onsager},
many other derivations of the FDT exist. We give a few collection of
references where these are presented.  Derivations can be classified in
two families: deterministic or stochastic. Deterministic approaches are
linear response theory \cite{Kubo}, operator formalism for master
equations \cite{Risken96} and quantum statistical mechanics
\cite{Parisi98}. Stochastic approaches are the Langevin and Fokker-Planck
equations \cite{ZinnJustin96}. Here we present a stochastic derivation which
is convenient for the purpose of the present review.

\subsection{The Master Equation (ME)}
\label{subsec:ME}
Any dynamical law describing the evolution of a system is
a rule which for each system configuration $\CC$ associates a new 
configuration $\CC'$. The time is just a label for bookkeeping 
the sequence so generated. Therefore to simplify the presentation and the 
notation we shall consider the time as an integer variable giving the
equivalent expressions for the continuous limit when needed.
This picture has also the advantage of being more closely related
to numerical simulations since all numerical methods use discrete
time schemes.

The dynamics can be encoded into the {\it conditional} or 
{\it transition} probability
$W(\CC,t|\CC',t-1)$ of going from
configuration $\CC'$ at time $t-1$ to configuration $\CC$ at time
$t$. 
Indeed if $P({\cal C},t)$ denotes the probability that the
system at time $t$ is in the configuration ${\cal C}$, then
from the Bayes theorem it follows that
\begin{equation}
  P(\CC,t)=\sum_{\CC'}\,W(\CC,t|\CC',t-1)\, P(\CC',t-1).
\label{eq1a}
\end{equation}
and the  $W(\CC,t|\CC',t-1)$ together with the initial condition 
$P(\CC,0)$ fully define the dynamical evolution of the system in the 
phase space. Equation (\ref{eq1a}) is an identity valid for all processes and
is the first of a hierarchy of equations for joint probabilities. Only if the
process is Markovian, {\it i.e.}, only if the conditional probability is 
determined entirely by the knowledge of the most recent past, then the
hierarchy can be closed.
Equation (\ref{eq1a}) and probability conservation at all times
($\sum_{\CC'}P(\CC',t)=1$) implies that $W(\CC,t|\CC',t-1)$ must satisfy the 
normalization condition
\begin{equation}
  \sum_{\CC'}\,W(\CC',t|\CC,t-1) = 1\qquad 
  \mbox{for all}\  \CC\ \mbox{and}\  t.
\label{eq2a}
\end{equation}
In the continuous time limit (\ref{eq1a}) is not well defined. To have 
an expression valid in this limit one then considers the variation
of $P(\CC,t)$ between two successive times which, using the normalization 
condition (\ref{eq2a}), reads
\begin{eqnarray}
\fl  P(\CC,t+1)-P(\CC,t) = 
\nonumber\\
\lo\phantom{} \sum_{\CC'}\,W(\CC,t+1|\CC',t)\, P(\CC',t)-
      \sum_{\CC'}\,W(\CC',t+1|\CC,t)\, P(\CC,t).
\end{eqnarray}
Dividing both sides of this equality for the time increment $\Delta t$ and 
and taking it to zero we get the {\it Master Equation} (ME)
\begin{equation}
\frac{\partial P(\CC,t)}{\partial t} =
      \sum_{\CC'}\,W(\CC|\CC';t)\, P(\CC',t)-
      \sum_{\CC'}\,W(\CC'|\CC;t)\, P(\CC,t)
\label{eq3a}
\end{equation}
where $W(\CC'|\CC;t)=\lim_{\Delta t \to 0} W(\CC,t+\Delta t|\CC',t)/\Delta t$
is called {\it transition rate} and gives the transition probability per 
unit of time. Solving the ME is often
an extremely difficult and unfordable task, even in the Markovian case.

Loosely speaking the transition probability $W(\CC,t|\CC',t-1)$ can be 
seen as a transition rate for a unit time interval ($\Delta t =1$), thus 
in what follows we shall not make distinction between transition 
probabilities and transition rates and shall call them generically 
transition rates using for both the notation $W(\CC'|\CC;t)$.
Which one is the appropriate will be clear from the context.

Transition rates depend on the specific dynamical rules
and hence by the Hamiltonian and eventual constraints (holonomic or
non-holonomic). 
Let the system under consideration be described by an Hamiltonian
which can depend on time
through a set of time-dependent external parameters
$\lambda^j_t$.
For instance, $\lambda_t$ may denote an time-dependent
external pressure applied to a liquid or a time varying electric or
magnetic field applied in a dielectric or a magnetic medium.  
We shall denote the set of these parameters by the vector $\blambda_t$
and the Hamiltonian by $\HH_{\blambda_t}(\CC)$ to indicate the
time dependence through $\blambda_t$.
Accordingly we shall also denote $W(\CC|\CC',t)$ by 
$W_{\blambda_t}(\CC|\CC')$.  If
the Hamiltonian is time-independent either $\blambda_t$ 
or just the subindex $t$ will be dropped, depending on the context.

Regardless of their form the transition rates must satisfy
the following requirements:

\begin{itemize}

\item{Non-negativeness and normalization.} The $W_{\blambda_t}(\CC|\CC';t)$ 
are probabilities so they must be non-negative and satisfy the normalization
condition (\ref{eq2a}). 

\item{Ergodicity.} Transition rates must be such that starting from any
configuration any other configuration of a finite system can be visited
in a finite time. For continuous variables 
the condition is stated by considering an arbitrary finite phase space 
region around a given point ('neighborhood').

\item{Detailed balance.} If the $\blambda$ are time-independent the 
equilibrium  distribution  $P_{\blambda}^{\rm eq}(\CC)$ 
is the stationary solution of the ME (\ref{eq3a}). 
A sufficient condition for this is that the transition rates be 
time-independent and 
\begin{equation} 
  \frac{W_{\blambda}(\CC'|\CC)}{W_{\blambda}(\CC|\CC')}=
  \frac{P_{\blambda}^{\rm eq}(\CC')}{P_{\blambda}^{\rm eq}(\CC)}
  \label{eq4a}
\end{equation}
This condition receives the name of {\it detailed balance}. 
Different equilibrium ensembles are thus encoded into the
different forms of transition rates.
The Perron-Frobenius theorem assures \cite{Gantmacher59} that this
condition, together with non-negativeness, normalization and 
ergodicity guarantees that the equilibrium distribution in a finite 
system is reached in a finite time.

\item{Causality.} This is an important assumption for the time-dependent 
transition rates and means that future is only determined by the past,
{\it i.e.}, a perturbation applied at a
given time can only propagate forward in time and not backward.
Consequence of this is that the 
transition rates must depend only on the values of 
$\blambda_t$ taken at the {\it lowest} time $t$. 

\end{itemize} 

For arbitrary time-dependent $\blambda_t$ the stationary solution of ME 
will in  general also depend on time. In this case, however, the previous 
conditions, and in particular the detailed balance condition, 
are not enough for determining the stationary state which in general 
will not be Gibbsian,  {\it i.e.},  not described by the
Boltzmann-Gibbs distribution. 
Only when $\blambda_t$ can be treated as small perturbations
some predictions can be obtained from the
linear response theory.

\subsection{Correlations, responses and the FDT}
\label{ME:corr_rep}

Consider two arbitrary observables $A(\CC),B(\CC)$ which for
simplicity are assumed time-independent. 
These can be either local or global quantities defined over a microscopic or macroscopic
region respectively. 
For instance, in a liquid an observable could be the local density at a 
point or the total mass of a given macroscopic region.
In systems with
discrete variables such as magnetic systems, it can be a spin of a given
magnetic atom or the magnetization of a macroscopic part of the system.

The two-times correlation function $C_{A,B}(t,s)$ between $A(t)$ and $B(s)$ is
defined as
the average $\langle A(t) B(s)\rangle$ over all possible
dynamical paths from time $0$ to time $t$ and all possible initial conditions
weighted by the probability distribution $P(\CC,0)$,
\begin{equation}
  C_{A,B}(t,s) = \langle A(t)B(s)\rangle
               = \sum_{\CC,\CC'}\, A(\CC')\, P(\CC',t|\CC,s)\,
                                   B(\CC)\,P(\CC,s)
\label{eq5a}
\end{equation}
where $P(\CC',t|\CC,s)$ is the conditional probability to evolve
from $\CC$ at time $s$ to $\CC'$ at later time $t$. 
Unless otherwise stated, in what follows we shall adopt the convention that $t\ge s$. 

To simplify the notation we switch to discrete (integer) 
time variable so that a
dynamical path from time $0$ to time $t$ is given by a sequence of
$t$ configurations $\lbrace \CC_0, \CC_1,\ldots,\CC_t \rbrace$. 
along which $\blambda$ takes the sequence of values
$\lbrace \blambda_0, \blambda_1, \ldots,\blambda_t \rbrace$. 
In this case, using (\ref{eq1a}), the correlation can be easily
rewritten in term of transition rates as
\begin{equation}
  C_{A,B}(t,s) = \sum_{\CC_s,\ldots,\CC_{t}}\,A(\CC_t)
   \left[ \prod_{k=s}^{t-1}\,W_{\blambda_k}(\CC_{k+1}|\CC_{k})
   \right]\, B(\CC_s)\, P(\CC_s,s)
\label{eq6a}
\end{equation}
where we have used the short-hand notation
$W_{\blambda_k}(\CC_{k+1}|\CC_k) \equiv W_{\blambda_k}(\CC_{k+1}|\CC_k;k)$.

In equilibrium
correlations satisfy the time-translational invariance (TTI)
property: $C_{A,B}(t,s)=C_{A,B}(t-s)$. Indeed in this case 
$P(\CC_s,s)$ is replaced by $P^{\rm eq}(\CC_s)$ and the transition
rates satisfy the detailed balance condition (\ref{eq4a}):
\begin{equation}
  W_{\blambda}(\CC_{k+1}|\CC_{k}) = W_{\blambda}(\CC_{k}|\CC_{k+1})\,
                        \frac{P_{\blambda}^{\rm eq}(\CC_{k+1})}
                             {P_{\blambda}^{\rm eq}(\CC_k)}.
\label{eq7a}
\end{equation}
Inserting this relation into (\ref{eq6a})
%
%
the factors $P^{\rm eq}(\CC_k)$ in the numerator and denominator of the
product cancel one by one and,  exchanging the indexes $t\leftrightarrow s$,
we obtain
\begin{eqnarray}
\fl C_{A,B}(t,s) = 
     \sum_{\CC_s,\ldots,\CC_t}\, A(\CC_t)\, P_{\blambda}^{\rm eq}(\CC_t)\,
             \left[\prod_{k=s}^{t-1}W_{\blambda}(\CC_{k+1}|\CC_{k})\right]\, 
             B(\CC_s) 
  \nonumber\\
  \lo= (t \leftrightarrow s)
  \nonumber \\
  \lo= \sum_{\CC_t,\ldots,\CC_s}\, B(\CC_t)\, 
             \left[\prod_{k=t}^{s-1}W_{\blambda}(\CC_{k}|\CC_{k+1})\right]\,
          A(\CC_s)\,P_{\blambda}^{\rm eq}(\CC_s)
\nonumber\\
   \lo= \langle B(t)A(s)\rangle 
\nonumber\\
   \lo= C_{B,A}(t,s).
\label{eq9a}
\end{eqnarray}
which implies that 
$C_{A,B}(t,s)=C_{A,B}(t-s)$. 

The correlation function $C_{A,B}(t,s)$ is a measure of how the 
system loses memory of its previous past history and hence
decays for large time separations. To measure how a 
system responds to external perturbations one introduces the
response functions.
Similar to correlations, responses also
tend to decay with time because the effect of the perturbation is
progressively forgotten in a thermal environment. However there is 
an important difference between correlations and responses: causality. 
While two observables can be correlated forward or backward in time, 
a perturbation cannot propagate backward in time and the response of the 
system for times before the perturbation is applied must be zero.
Nevertheless despite this difference response and correlations can be 
treated on equal footing by employing a supersymmetric formalism. 
The interested reader can find more details, {\it e.g.} in 
the classical book by Zinn-Justin \cite{ZinnJustin96}.

To study the response of the system to an external perturbation we
assume that at time $s$ an impulsive perturbation of small 
intensity $\epsilon$ is applied to the observable $B(\CC)$ and 
measure the variation of the average value of an observable $A(\CC)$ 
at later times.
The response function $R_{A,B}(t,s)$ is defined in the limit of vanishing
perturbation strength as 
\begin{equation}
  R_{A,B}(t,s) = \lim_{\epsilon\to 0} 
   \frac{\langle A(t)\rangle_{\epsilon_s} - \langle A(t)\rangle_{0}}
	{\epsilon},
   \qquad t > s
\label{eq11a}
\end{equation}
with
\begin{eqnarray}
  \langle A(t)\rangle_{\epsilon_s}=
  \sum_{\CC}\,A(\CC)\,P_{\epsilon_s}(\CC,t), 
  \label{eq12aa}\\
  \langle A(t)\rangle_{0}=\sum_{\CC}\,A(\CC)\,P_0(\CC,t).
  \label{eq12a}
\end{eqnarray}
where $P_{\epsilon_s}(\CC,t)$ and $P_{0}(\CC,t)$ are the probability
that the system is in the configuration $\CC$ at time $t>s$ 
in the perturbed and unperturbed case, respectively.
If the system is described by the unperturbed Hamiltonian  $\HH_0(\CC)$,
then in presence of the perturbation the Hamiltonian becomes:
\begin{equation}
  \HH_{\epsilon_s}(\CC) = \HH_0(\CC) - \delta_{t,s}\epsilon B(\CC)     .
  \label{eq10a}
\end{equation}
By using (\ref{eq1a}) and the fact that the Hamiltonians only differ at
time $s$ when the impulse is applied, the probabilities
$P_{\epsilon_s}(\CC,t)$ and $P_{0}(\CC,t)$ can be written as:
\begin{eqnarray}
\fl
  P_{\epsilon_s}(\CC_t,t) = \sum_{\CC_s,\cdots,\CC_{t-1}}\,
   \left[\prod_{k=s+1}^{t-1}W_0(\CC_{k+1}|\CC_{k})\right]\,
                      W_{\epsilon}(\CC_{s+1}|C_{s})\, P_0(\CC_{s},s)
\label{eq13a}\\
\fl
  P_{0}(\CC_t,t) = \sum_{\CC_s,\cdots,\CC_{t-1}}\,
   \left[\prod_{k=s+1}^{t-1}W_0(\CC_{k+1}|\CC_{k})\right]\,
                      W_{0}(\CC_{s+1}|C_{s})\, P_0(\CC_{s},s)
\label{eq13aa} 
\end{eqnarray}
where $W_{0,\epsilon}$ denotes the transition rates 
in the unperturbed/perturbed case, and 
we have used the short-hand notation
$W(\CC_{k+1}|\CC_k) \equiv W(\CC_{k+1}|\CC_k;k)$.

Because we are interested in the $\epsilon\to 0$ limit the transition
rates in the perturbed and unperturbed case can be related by
expanding the detailed balance condition (\ref{eq4a}) for the perturbed state
around $\epsilon=0$ up to the first order in $\epsilon$,
\begin{eqnarray}
\fl \frac{W_{\epsilon}(\CC'|\CC)}
         {W_{\epsilon}(\CC|\CC')} =
    \frac{P_{\epsilon}^{\rm eq}(\CC')}
         {P_{\epsilon}^{\rm eq}(\CC)} \nonumber\\
\lo= \frac{W_0(\CC'|\CC)}
         {W_0(\CC|\CC')}\,
	 \left\{
	         1 + \epsilon\, 
            \left.
               \frac{\partial}{\partial \epsilon}
                \ln \Bigl[{P_{\epsilon}^{\rm eq}(\CC')} /
                         {P_{\epsilon}^{\rm eq}(\CC)}\Bigr]
             \right|_{\epsilon=0} 
	       + {\cal O}(\epsilon^2)
	 \right\}
\label{eq14aa}
\end{eqnarray}
where $P_{\epsilon}^{\rm eq}(\CC)$ is the equilibrium probability 
distribution in the perturbed state \footnote[1]{ An observation concerning \eq{eq14aa} and its
relation with the Onsager postulate \eq{basic:onsager:5} is
important. Ideally, in order to demonstrate FDT, one would like to have
a relation similar to \eq{basic:onsager:5} relating the unperturbed and
the perturbed rates rather than a relation between the forward and
backward rates as given in \eq{eq14aa}. However, such a relation does
not exist as it depends upon the type of dynamics through the particular
form of the transition rules. For instance, by considering rates of the
type $W_{\blambda}(\CC'|\CC)\propto P_{\blambda}^{\rm eq}(\CC')$,
i.e. depending only upon the final configuration, one finds that
\eq{basic:onsager:5} automatically holds. However such rates cannot be
used to derive the FDT \eqq{basic:onsager:11}{eq22a} as they lead to the
trivial identity $0=0$ because any impulse does not affect dynamics at
later times $R(t,s)=0$ and there are no time correlations. In other
words, the Onsager postulate extended to the non-equilibrium regime
generates violation terms different to those obtained in
the present approach (for instance the term $R^{(1)}$ in the r.h.s of
\eq{eq21a}).  }
. In general
we can write,
\begin{equation} 
  \log[P^{\rm eq}_{\epsilon}(\CC)] = \log[P^{\rm eq}_0(\CC)]
                                      + \phi_0 -\phi_{\epsilon} 
				      + \beta \epsilon B(\CC)
\label{eq10aa}
\end{equation}
where $\phi$ denotes the corresponding thermodynamic potential. For
instance, in the canonical ensemble it corresponds to minus the
Helhmoltz free energy $F$ while in the grandcanonical ensemble it
corresponds to the grandcanonical potential given by the pressure times
the volume.

Using (\ref{eq14aa}) we finally obtain to the leading order in $\epsilon$, 
\begin{eqnarray}
W_{\epsilon}(\CC'|\CC) - W_{0}(\CC'|\CC) &=
    \left[\frac{W_{\epsilon}(\CC|\CC')}{W_{0}(\CC|\CC')} - 1\right]\,
     W_{0}(\CC'|\CC) 
\nonumber\\
&+\epsilon\, 
            \left.
               \frac{\partial}{\partial \epsilon}
                \ln \Bigl[{P_{\epsilon}^{\rm eq}(\CC')} /
                         {P_{\epsilon}^{\rm eq}(\CC)}\Bigr]
             \right|_{\epsilon=0}  \, W_{0}(\CC'|\CC) 
\label{eq15a}
\end{eqnarray}

The physical meaning of the two terms appearing in this expression is
different.  The first term, absent in the Onsager's postulate
\eq{basic:onsager:5}, accounts for the variation due to the change in
the transition rates and does not directly depend upon the particular
form of the equilibrium distribution.  The second term depends directly
on the equilibrium distribution. This distinction is important since
they give different contributions to the response.  For the response
function (\ref{eq11a}) to be well defined the difference $\la
A(t)\ra_{\epsilon}-\la A(t)\ra_{0}$ must be at least linear in
$\epsilon$. This requirement is at the roots of the applicability of
linear response theory and implies that both terms in (\ref{eq15a}) must
be at least linear in $\epsilon$. Concerning the first term, it is
required that the difference $W_{\epsilon}(\CC|\CC') / W_{0}(\CC|\CC') -
1$ to be linear in $\epsilon$, so the transition rates must change
linearly with $\epsilon$ when the system is perturbed. This means that,
during the dynamics, and after applying the perturbation, one cannot
switch arbitrarily from one class of dynamics to another class of
dynamics in a random fashion. To better illustrate what this means let
us consider a Monte Carlo stochastic dynamics. The are different
possible algorithms or transition rules which fulfill detailed balance
(e.g. Metropolis, heat-bath, Glauber). One could imagine of switching
randomly from one algorithm to another one while doing the
dynamics. Nothing forbids this quite {\em artificial} choice. But, when
measuring the response function, it is required that the same time
sequence of algorithms must be used for the unperturbed and perturbed
dynamical evolutions.  Usually, the same algorithm is chosen in a given
simulation so one does not care about this subtlety. For the second
term, we require that $({\partial}/{\partial \epsilon}) \ln
[{P_{\epsilon}^{\rm eq}(\CC')} / {P_{\epsilon}^{\rm eq}(\CC)}]
|_{\epsilon=0} $ be finite. Inspection of \eq{eq10aa} reveals that this
holds if the observable $B(C)$ does not jump discontinuously when the
perturbation is switched on. This condition requires that the system is
not at a first order transition point.  This situation is encountered,
for instance, by perturbing the Ising model at zero temperature with a
uniform magnetic field.

Inserting (\ref{eq15a}) into (\ref{eq11a}) the response function decomposes 
into two parts,
\be R_{A,B}(t,s)=R^{(1)}_{A,B}(t,s)+R^{(2)}_{A,B}(t,s)\label{eq16a}\ee
where,
\begin{eqnarray}
\fl  R^{(1)}_{A,B}(t,s) = \lim_{\epsilon\to 0}\frac{1}{\epsilon}
  \sum_{\CC_s,\ldots,\CC_t} A(\CC_t)\, 
    \left[\prod_{k=s+1}^{t-1}W_0(\CC_{k+1}|\CC_k)\right]
\nonumber\\
\times
    \left[\frac{W_{\epsilon}(\CC_{s}|\CC_{s+1})}
               {W_{0}(\CC_{s}|\CC_{s+1})} - 1
     \right]\,
     W_{0}(\CC_{s+1}|\CC_{s}) \, P_0(\CC_s,s)
\label{eq17a}\\
\fl  R^{(2)}_{A,B}(t,s) = 
  \sum_{\CC_s,\ldots,\CC_t} A(\CC_t)\, 
    \left[\prod_{k=s+1}^{t-1}W_0(\CC_{k+1}|\CC_k)\right]
\nonumber\\
\times
    \left.
          \frac{\partial}{\partial \epsilon}
           \ln \Bigl[{P_{\epsilon}^{\rm eq}(\CC')} /
                     {P_{\epsilon}^{\rm eq}(\CC)}\Bigr]
            \right|_{\epsilon=0} 
     \,
     W_{0}(\CC_{s+1}|\CC_{s}) \, P_0(\CC_s,s)
\label{eq18a}
\eea
Inserting \eq{eq10aa} for the second term we get
\begin{eqnarray}
 R^{(2)}_{A,B}(t,s) =& \left\la A(t)
                      \left. 
		        \frac{\partial \ln P_{\epsilon}^{\rm eq}(\CC_{s+1})}
			                     {\partial\epsilon} 
                      \right|_{\epsilon=0}
                      \right\ra_0
\nonumber\\
		     & \phantom{xxxxxxx} -
		      \left\la A(t)
                      \left. 
		        \frac{\partial \ln P_{\epsilon}^{\rm eq}(\CC_{s})}
			                     {\partial\epsilon} 
                      \right|_{\epsilon=0}
                      \right\ra_0
\label{eq22a}
\end{eqnarray}
which using (\ref{eq10aa}) becomes
\begin{equation}
 R^{(2)}_{A,B}(t,s)= \beta \la A(t)B(s+1)\ra_0 -
                     \beta \la A(t)B(s)\ra_0
\label{eq23ad}
\end{equation}
In the limit of continuous time $R^{(2)}$ in (\ref{eq23ad}) must be replaced by
\begin{equation}
  R^{(2)}_{A,B}(t,s)= \beta\frac{\partial}{\partial s}\la A(t)B(s)\ra_0
\label{eq23a}
\end{equation} 
The first term $R^{(1)}$ cannot be expressed in a simple form. Only in
equilibrium it is possible to show that it vanishes.  To prove it
requires the following steps: first use the identity
\begin{equation}
  \prod_{k=s+1}^{t-1}W_0(\CC_{k+1}|\CC_{k}) = 
  \left[
  \prod_{k=s+1}^{t-1} W_0(\CC_{k}|\CC_{k+1})
  \right]\,
  \frac{P_0^{\rm eq}(\CC_{t})}{P_0^{\rm eq}(\CC_{s+1})}
\label{eq19a}
\ee
and then the normalization \eq{eq2a}.
Collecting all terms we finally obtain for the response function
\begin{eqnarray}
  R_{A,B}(t,s) &= R^{(1)}_{A,B}(t,s) + 
 \beta \frac{\partial}{\partial s}\,\la A(t)B(s)\ra_0
\label{eq21a}
\end{eqnarray}

In equilibrium, besides $R^{(1)}=0$, the correlation function satisfy TTI, so 
from (\ref{eq21a}) we get the equilibrium FDT
\begin{eqnarray}
R_{A,B}^{\rm eq}(t-s) &= \beta \frac{\partial}{\partial s} 
                                C_{A,B}^{\rm eq}(t-s)
                      &= - \beta \frac{\partial}{\partial t} 
                                C_{A,B}^{\rm eq}(t-s), \qquad t>s
\label{FDT}
\end{eqnarray}
The term $R^{(1)}$ may also vanish in the non-equilibrium state
if the first term in (\ref{eq15a}) vanishes faster than linearly with $\epsilon$ for
$\epsilon\to 0$. In this case  (\ref{eq21a}) reduces to the
usual FDT relation (\ref{basic:onsager:11}).
\be
  R_{A,B}(t,s) = \beta\, \theta(t-s)\,\frac{\partial}{\partial s} C_{A,B}(t,s)
\label{FDT2}
\ee

Therefore, the lowest time $s$ has 
a special role in the relation (\ref{eq21a}) between correlations and
responses. This is not a surprise since the relation has been
obtained by assuming causality which privileges the lowest time. In
equilibrium, the role of the lowest time disappears because the system
is TTI.

\subsection{The Component Master Equation}
\label{gen:EXTME}

Let us now divide the phase space into different non-overlapping subsets 
$\RR$ that can be called regions, phases, components or domains. 
In what follows, if not stated
otherwise, we shall use the term {\it component}.
Later in Section \ref{micro:EXTME} we will see that the reduction of the phase space by a suitable 
partitioning can be relevant for the study of the non-equilibrium
regime in glassy systems and in particular for the understanding
of the FDT relations.
For the moment, however, we do not attach any physical meaning to such 
a partitioning, and assume it to be completely arbitrary  postponing the 
identification of a suitable partitioning scheme for glassy systems
to Section \ref{is}.

For each partition of the phase space the probability 
$\PP(\RR,t)$ that system be in the component $\RR$ at time $t$ is given by,
\begin{equation}
  \PP(\RR,t) = \sum_{\CC\in\RR} P(\CC,t),
  \label{eq1b}
\end{equation}
\begin{equation}
  \sum_{\RR}\PP(\RR,t)=1
  \label{eq2b}
\end{equation}
where the normalization condition (\ref{eq2b}) follows from 
normalization of $P(\CC,t)$ and the non-overlapping assumption on $\RR$.
The probability distribution $\PP(\RR,t)$ obeys the master equation
obtained by projecting the (microscopic) master equation (\ref{eq3a}) 
over the component space. The Markovian character of the 
dynamics is preserved under projection. 
Summing (\ref{eq3a}) 
over all configurations $\CC$ belonging to a given component 
$\RR$ we get the {\it Component} Master Equation:
\begin{equation}
  \frac{\partial \PP(\RR,t)}{\partial t} = 
    \sum_{\RR'} \WW(\RR|\RR';t)\, \PP(\RR',t) -
    \sum_{\RR'} \WW(\RR'|\RR;t)\, \PP(\RR,t)
\label{eq3b}
\end{equation}
where $\WW(\RR'|\RR;t)$ are the component transition rates which 
in terms of the original transition rates $W(\CC'|\CC;t)$ read, 
\begin{equation}
  \WW(\RR'|\RR;t) = 
              \frac{\sum_{\CC'\in\RR',\CC\in\RR}W(\CC'|\CC,t)P(\CC,t)}
                   {\PP(\RR,t)}.
\label{eq4b}
\ee
The component transition rates satisfies the same normalization conditions 
as $W$. For example in the case of discrete time we have 
[Cfr. (\ref{eq2a})], 
\begin{equation} 
  \sum_{\RR'}\WW(\RR',t|\RR,t-1) = 1\qquad
  \mbox{for all}\ \RR\ \mbox{and}\ t.
\label{eq5b}
\end{equation}
However there is an important difference between the (microscopic) 
master equation \eq{eq3a} and component master equation \eq{eq3b}: 
the transition rates (\ref{eq4b}) are time-dependent even 
in for time-independent Hamiltonians since are computed with the
run-time configuration probability distribution function $P(\CC,t)$.
Direct consequence of this is that, while the properties of
non-negativeness, normalization, ergodicity and causality  
discussed in section \ref{subsec:ME} for $W$ 
do apply to $\WW$, the
detailed balance is not valid anymore in the component space.
Indeed from (\ref{eq4b}) it follows that
\begin{equation}
  \frac{\WW(\RR'|\RR;t)}{\WW(\RR|\RR';t)}=
          \frac{\sum_{\CC\in\RR,\CC'\in\RR'} W(\CC'|\CC)\,P(\CC,t)}
               {\sum_{\CC\in\RR,\CC'\in\RR'} W(\CC|\CC')\,P(\CC',t)}
	       \times
	       \frac{\PP(\RR',t)}{\PP(\RR,t)}
	       \label{eq6b}
\end{equation}
The detailed balance condition is recovered, however, at equilibrium
where $\WW$, accordintg to \eq{eq4b}, becomes time-independent:
\begin{equation} 
  \WW^{\rm eq}(\RR'|\RR) = 
        \frac{\sum_{\CC\in\RR,\CC'\in\RR'}W(\CC'|\CC)\,P^{\rm eq}(\CC)}
	     {\PP^{\rm eq}(\RR)}
\label{eq8b}
\end{equation}
where $\PP^{\rm eq}(\RR)$ denotes the equilibrium probability
distribution function in the component space associated 
through (\ref{eq1b}) to the
phase space equilibrium probability distribution 
function $\PP^{\rm eq}(\CC)$.
Using the detailed balance condition (\ref{eq4a}) we obtain,
\begin{equation}
  \frac{\WW^{\rm eq}(\RR'|\RR)}{\WW^{\rm eq}(\RR|\RR')}=
  \frac{\PP^{\rm eq}(\RR')}{\PP^{\rm eq}(\RR)}
\label{eq9b}
\end{equation}
which is the detailed balance condition in the component space.

\noindent

In conclusion, in the component space the rates $\WW$ satisfy the same
set of conditions as the rates $W$ except for the detailed balance
condition which in the component space only holds at equilibrium. In
general, rates $\WW$ do not satisfy detailed balance so the FDT
derived in the previous sections for the microscopic master equation
will not be valid in the component space. They are,
however, valid in equilibrium so, for example, after an appropriate
redefinition of correlations and responses the equilibrium
fluctuation-dissipation theorem \eq{FDT} still holds in the component
space.

\section{FDT extensions to the non-equilibrium regime}
\label{EXTME}

In this section we discuss how to extend the previous ideas to the
non-equilibrium glassy regime. After a brief introduction of aging
(intended for the non-specialist) we discuss how to derive a free-energy
master equation by introducing the configurational entropy and the
notion of the effective temperature. We then discuss how to extend the
FDT beyond equilibrium by introducing the fluctuation-dissipation ratio
and the quasi-FDT. This requires the notion of neutral
observables. Finally, a possible partitioning scheme is presented.

\subsection{An intermezzo on aging}
\label{aging}

In this section we discuss one of the main signatures of non-equilibrium
regime of glassy systems, i.e. the existence of aging and the quantifying of FDT
violations through a modified FDT in terms of a set of effective temperatures. 
This discussion is intended for the inexpert reader who
wants to have a glance on the key aspects of glassy systems before
entering into the more detailed exposition. Therefore, the level of this
discussion is highly introductory and the expert reader can move
directly into the next section.

In relaxational glassy systems the two basic properties of correlations
and response, time-translational invariance (TTI)\eq{eq9a} and the FDT
\eq{FDT}, do not hold anymore. In particular, TTI is observed to be
violated in the following way: both correlations and responses decay
slower as the system gets older, i.e. the system is aging. This fact
stems from several experimental observations in polymers and
deformable materials \cite{Struik78}, structural
glasses \cite{MilMacP97} and spin glasses \cite{Young98}.  In driven
systems the situation appears less complicated as TTI holds and only the
FDT is violated. Some of the aspects described below carry over also to
driven systems, however for sake of clarity we will stick in what
follows to the case of aging systems.

Experimentally aging is manifest through the measurement of the so
called integrated response function (IRF) or time-dependent
susceptibility described in Section \ref{IRF:FD}. In aging systems the
weak long term memory makes response functions $R(t,s)$ hardly
measurable as they asymptotically decay to zero, instead it is easier to
measure the cumulative response or time-dependent susceptibility
$\chi(t,s)=\int_{s}^tR(t,t')dt'$ (defined after perturbing the system at
the waiting time $s$ -often denoted as $t_w$-) which in general are
finite: in dielectric measurements of glasses the IRF corresponds to the
polarizability of the sample after cutting or applying an electric
field; in magnetic systems it corresponds to the thermoremanent or
zero-field cooled susceptibility; in
mechanical systems aging is observed by measuring the deformation of the
sample after applying a tensile load. All these measurements reveal that
$\chi(t,s)$ is well approximated by the sum of two contributions,
\be
\chi(t,s)=\chi_{\rm st}(t-s)+\chi_{\rm ag}(t,s)
\label{aging1}
\ee
where the first contribution $\chi_{\rm st}(t-s)$ stands for an
stationary part that asymptotically decays to a finite value or plateau
and $\chi_{\rm ag}(t,s)$ is the aging part that is well approximated by
the scaling relation,
\be
\chi_{\rm ag}(t,s)=\widehat{\chi}\Bigl(\frac{t}{s} \Bigr)
\label{aging2}
\ee
This scaling behavior, that is obtained within many solvable models, is
known as full aging or simple $t/s$ scaling~\footnote{Coarsening system
also deviate from the simple form (\ref{aging2}) in favour of $\chi_{\rm
ag}(t,s)= s^{-a}\widehat{\chi}(t/s)$ with $a\ge 0$.}. However,
deviations from full aging have been reported in many experiments
suggesting that this simple scaling behavior does not fully account for
the experimental data.  In particular, spin glass measurements clearly
favour a subaging scenario where $\hat{\chi}$ has as scaling argument
the variable $\frac{t}{s^{\delta}}$ with $\delta<1$ (we note, however,
that experiments report values around 0.95, i.e. very close to 1).  The
physical origin of these deviations is still unknown.  For both
correlation and response functions similar decompositions as in
(\ref{aging1}), (\ref{aging2}) are expected to be valid but replacing
$\chi$ by $C$ or $R$ (however, for the aging part of the response there
is an additional factor $1/t$ multiplying $R_{\rm
ag}(t,s)$) .  Correlations and responses are difficult to experimentally
access. In theoretical or numerical simulation studies, the calculation
of correlation functions is always preferred. Typical curves for the
susceptibilities or correlations are depicted in Figure
\ref{aging:fig1}.
\begin{figure}
  \centering
  \includegraphics[scale=0.45]{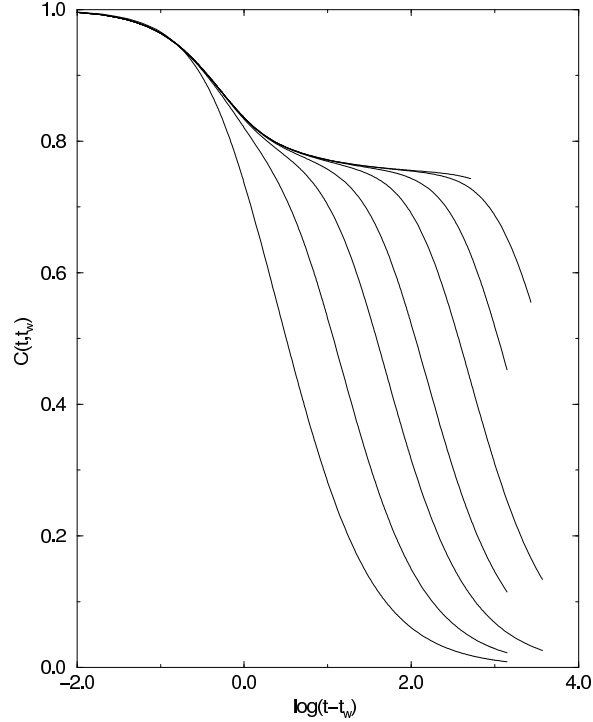}
  \caption{Typical shape of a two-time correlation function $C(t,t_w)$
    (the same would be valid for the integrated susceptibility
    $\chi(t,t_w)$), plotted as a function of $t-t_w$ (in log-scale) with
    system age $t_w$ increasing from left to right. The decomposition
    \eq{aging1} shows the two separate time sectors: in the first
    fast part of the relaxation $C(t,t_w)\simeq C_{\rm st}(t-t_w)$ is
    independent of $s$ and obeys TTI while the
    slow second part $C_{\rm ag}(t-t_w)$ decays from the plateau in a
    timescale growing with $t_w$. This Figure has been taken from a lattice
    gas coarsening model \protect\cite{PadRit97} with properly normalized
    correlations at equal times $C(t,t)=1$.}  
  \label{aging:fig1}
\end{figure}
Within the full aging scenario (\ref{aging2}) it is usually shown that, 
in the asymptotic limit
where both $s$ and $t$ are large, the FDT \eq{FDT} is violated
depending on the ratio $(t-s)/s$. If $(t-s)/s\ll 1$, $C(t,s)\simeq  C_{\rm
st}(t-s)$ and FDT holds,
\be
\frac{\partial C_{\rm st}(t-s)}{\partial s}=TR_{\rm st}(t-s).
\label{aging3}
\ee
However, in the other case $(t-s)/s \ge 1$, then $C(t,s)\simeq C_{\rm
ag}(t,s)$ and the FDT \eq{FDT} is violated according to the new relation,
\be
\frac{\partial C_{\rm ag}(t,s)}{\partial s}=\Te(s)R_{\rm ag}(t,s)
\label{aging4}
\ee
where $\Te(s)$ is a new parameter that enters into the new
relation playing the role of an effective temperature. The particular
way in which the FDT is violated suggests calling the new relation
\eq{aging4} a quasi fluctuation-dissipation theorem (QFDT). Although
other possible terms have been used to refer to the modified FDT, here
we will adhere to the term QFDT, as this is the one originally
introduced by Horner \cite{Horner84} that expresses the idea of partial
equilibration among a subset of degrees of freedom.

Aging carries associated the decomposition of time into time sectors. 
In the previous example of full aging \eq{aging2} there are two time
sectors depending on the ratio $(t-s)/s$ as described in
\eqq{aging3}{aging4}. However, as we already said, deviations from the
full aging behavior are expected to be present in general. In
those cases, the stationary result \eq{aging3} for the short-time sector
still holds but \eq{aging4} is replaced by the more general relation,
\be
\frac{\partial C_{\rm ag}(t,s)}{\partial s}=\Te(C_{\rm ag}(t,s))R_{\rm ag}(t,s)
\label{aging5}
\ee
where the new effective temperature $\Te(C_{\rm ag}(t,s)) \equiv
\Te(C(t,s))$, since $C\equiv C_{\rm ag}$ when $(t-s)/s\sim
O(1)$, and depends on both times only through the value of $C$.  In this
case, the aging part develops time sectors defined as those
values of $t,s$ where $t/s\sim O(1)$ \cite{CugKur95b,FraMez94}. Each
sector is then labeled by the value of the correlation function $C(t,s)$
and many effective temperatures arise in the description of the
non-equilibrium regime, the QFDT \eq{aging5} quantifying FDT
violations within each sector. 
Glassy systems are often classified into
three different groups according to the dependence of $T_{\rm
eff}(C)$. For coarsening systems $\Te(C)$ only takes two values:
$T$ and infinite for the stationary and aging regimes respectively. 
\footnote{We note however that in some special cases coarsening
systems can display more complex behaviour of $\Te(C)$, see
Sections~\ref{ferro},\ref{coars} and the discussion in Section 13.1 in \cite{Cugliandolo02}.}.  For structural glasses $\Te(C)$ also takes two values: $T$ and $\Te(s)>T$ for the
stationary and aging regimes respectively. These are often referred as
two-timescales systems. Finally for spin-glass systems $\Te(C)$ takes
a continuous spectrum of values extending from a lower bound $T^*>T$
up to infinity. These are known as many-timescales systems. All these
three limit cases correspond to a well known static low-temperature
description in the framework of spin-glass theory \cite{MezParVir87}
in terms of replica-symmetry breaking (RSB): coarsening systems are
those where replica symmetry (RS) is unbroken, structural glass
systems correspond to one-step of RSB while spin-glass systems
corresponds to full RSB.  As a particular example of one-step systems
there are some models (such as entropy barrier models, see section
\ref{esm:nodisorder}) that display glassy behavior only at zero
temperature. The stationary regime is then absent in these models and
their non-equilibrium dynamics is characterized by a single effective
temperature, the bath temperature being zero. Along this review, we
will often refer to them as one-timescale models. The three possible
scenarios are depicted in figure \ref{fig:IRF:1} (see section
\ref{IRF:FD} for a more detailed exposition). The experimental
challenge of these ideas remains one of the most awaited results.

\subsection{The unbiased component ensemble and the master free energy
equation}
\label{micro:EXTME}
One of the key ideas behind the existence of a quasi
fluctuation-dissipation theorem (QFDT) in aging systems is the emergence
of a non-equilibrium ensemble in the asymptotic long-time regime of the
relaxation process. A related non-equilibrium ensemble could also emerge
in driven stationary systems.   Although the nature of this ensemble
is yet to be understood we can anticipate some of its main
properties. Some of these ideas have been already presented \cite{CriRit02}. Here we present more elaborated work.

 In a description of glassy phenomena
where the system is kept in contact with a thermal bath at temperature
$T$ the energy is not constant and equipartition does not necessarily
hold.  Therefore this ensemble is neither microcanonical or canonical
but of a more complicated nature.

The possible existence of a non-equilibrium ensemble traces back to Palmer \cite{Palmer82}
who introduced the {\em unbiased component ensemble} to characterize the
equilibrium sampling of phase space components. Let us consider a given
partition of phase space into components (see Section \ref{gen:EXTME}) and
let us define the free energy $F(\RR)$ of a given component $\RR$ by,
\be
F(\RR)=-T\log\Bigl(\sum_{\CC\in\RR}\exp(-\beta \HH(\CC))\Bigr)
\label{micro:EXTME:2}
\ee
It is possible to extend
this idea to the non-equilibrium regime by assuming an equal probability
hypothesis: different components with identical free energy $F(\RR)$ do
have the same probability, 
\be
\PP(\RR,t)={\bf P}(F(\RR),t)   .
\label{micro:EXTME:3}
\ee
In what follows, we will use the letter $\FF$ (as opposed to $F(\RR)$)
to denote component free energies defined in \eq{micro:EXTME:2} after
dropping off the explicit argument $\RR$.  The existence of the unbiased
ensemble tantamount to the appearance of a new measure based on free
energy rather than on energy.
Contrarily to Palmer \cite{Palmer82} (who assumes an equilibrium
probability distribution for ${\bf P}(\FF,t)$, see \eq{eq9d} below)
the probability distribution ${\bf P}(\FF,t)$ is unknown, time dependent and must be found
as a solution of a master equation (ME) as follows.
To derive the free-energy ME we define the probability density,
\be
{\bf P}(\FF,t)=\sum_{\RR}\PP(\RR,t)\delta(\FF-F(\RR))=\PP(\FF,t)\Omega(\FF,T)
\label{eq1d}
\ee
where we have used \eq{micro:EXTME:3} and the definition,
\be
\Omega(\FF,T)=\sum_{\RR}\delta(\FF-F(\RR))       .
\label{eq5d}
\ee
and we have introduced explicitly the temperature dependence in $\Omega$
to stress the temperature dependence of the free energy \eq{micro:EXTME:2}. 
Although consistency requires to add the $T$ dependence also to $\FF$
here we drop this dependence in order to lighten the notation.
Equation \eq{eq1d} describes
the probability for the system to be in a component of free energy $\FF$
at time $t$. We have indicated it in bold to distinguish it from the
probability $\PP(\RR,t)$.  Summing (\ref{eq3b}) over components
having identical free energy $\FF$ we get,
\be
\frac{\partial {\bf P}(\FF,t)}{\partial t}=\sum_{\FF'}{\bf P}(\FF',t)Z_t(\FF|\FF')-
\sum_{\FF'}{\bf P(\FF,t)}Z_t(\FF'|\FF)
\label{eq2d}
\ee
with the conditioned probabilities $Z_t$ defined by,
\begin{eqnarray}
\fl
Z_t(\FF|\FF') = \frac{1}{{\bf P}(\FF',t)}\, 
 \sum_{\RR,\RR'}\, \WW(\RR|\RR';t)\delta(\FF(\RR)-\FF)
                                  \delta(\FF'(\RR')-\FF')P(\RR',t)
\label{eq3d}
\end{eqnarray}
where the $\WW$ have been already defined in \eq{eq6b}.  Note that both
$\WW$ and $Z_t$ are time-dependent rates.  Again, as for the transition
probabilities (\ref{eq4b}), the new rates $Z_t(\FF|\FF')$ do not satisfy
detailed balance but satisfy the other requirements (non-negativeness,
ergodicity and causality). Expression (\ref{eq3d}) is exact but
intractable. As we are postulating the existence of the unbiased
component ensemble \eq{micro:EXTME:3}, consistency in the component
master equation (\ref{eq3b}) implies that the rate $\WW(\RR,\RR';t)$ is
a time-dependent function of the initial and final components, only
through the value of their free-energies $\FF(\RR'),\FF(\RR)$,
\be
\WW(\RR,\RR';t)=\WW_t(\FF(\RR),\FF(\RR'))\qquad .
\label{nova}
\ee
The transition rates (\ref{eq3d}) can be further simplified,
\be
Z_t(\FF|\FF')=\WW_t(\FF|\FF')\Omega(\FF,T)
\label{eq4d}
\ee
The quantity $\Omega(\FF,T)$ is exponentially large with the volume of
the system and defines what we will denote as the configurational
entropy or complexity $S_c(\FF,T)$ \footnote[1]{The term configurational
entropy has often been used with different meanings, leading to
confusion. Originally, as used by Adam and Gibbs for their thermodynamic
theory, it denotes the part of the total entropy including only the
configurational degrees of freedom. More recently, in the context of
spin-glass theory, this concept has been coined to denote that part of
the configurational entropy that counts the number of metastable states
rather than configurations. It is with this last meaning that we
understand it here. For a throughout discussion of this concept see
section \ref{conf:entr}},
\be
\Omega(\FF,T)=\exp(S_c(\FF,T))
\label{eq6d}
\ee
Therefore, all the information on the master equation (\ref{eq2d}) goes
into the density of components $S_c(F,T)$ and the rates
$\WW_t(\FF|\FF')$. These contain all the information about the
properties of the unbiased ensemble. 

The description of glassy dynamics in terms of a master free-energy
equation such as \eq{eq2d} has been wandering around for many years in
the literature of the field.  Several equations have appeared scattered
in the literature during the last decades, but generally written in
terms of the energy instead of the free energy, see for instance
\cite{Brawer84}. These equations describe what are usually known as trap
models (see Section \ref{trap}). Examples master equations proposed by
Dyre \cite{Dyre87} and Bouchaud \cite{Bouchaud92}. Other attempts
include granular media \cite{ConNic01}.

\subsubsection{Complexity and the effective temperature}
\label{interlude}
Before closing the present discussion let us note that, in equilibrium,
both the transition rates $Z_t(\FF|\FF')$ and $\WW_t(\FF|\FF')$ are
time-independent and satisfy detailed balance,
\bea
\frac{Z^{\rm eq}(\FF'|\FF)}{Z^{\rm eq}(\FF|\FF')}=\exp(-\beta( \Phi(\FF',T)- \Phi(\FF,T)))
\label{eq7d}\\
\frac{\WW^{\rm eq}(\FF'|\FF)}{\WW^{\rm eq}(\FF|\FF')}=\exp(-\beta(\FF'-\FF))
\label{eq8d}
\eea
with 
\be
\Phi(\FF,T)=\FF-TS_c(\FF,T)
\label{eq8d2}
\ee
where $\Phi(\FF,T)$ is a new potential where the free energy
appears balanced by the complexity. The probability (\ref{eq1d})
assumes the simple form,
\be
{\bf P}^{\rm eq}(\FF)=\frac{\exp(-\beta\Phi(\FF,T))}{{\cal Z}(\beta)}
\label{eq9d}
\ee
where ${\cal Z}(\beta)$ is given by \eq{basic:micro:15}. We will describe in
Section \ref{conf:entr:num} how this relation provides us with a tool to obtain the
configurational entropy as function of both the free energy and the
temperature.

The hints at the existence of this new potential were found by
Kirkpatrick, Thirumalai and Wolyness
\cite{KirWolb87b,KirThi87a,KirThi87b} who identified the marginal
transition $T_{\rm A}$ (``A'' standing for activated) as the
temperature below which different metastable states concur in such a
number to compensate the lower equilibrium free energy of the
paramagnetic or liquid state. The subindex in $T_{\rm A}$ stands for
the fact that below that temperature activation is dominant and
relaxation occurs in the form of activated jumps from one metastable
state to another. $T_{\rm A}$ is identified with the mode-coupling
transition temperature $T_c$ of mode-coupling theories (see section
\ref{mct}) and corresponds to a spinodal instability
\cite{KirWolb87b,KirThi87a,KirThi87b}. The mathematical argument
behind the compensation of the free-energy of metastable states by the
complexity is as follows. Let us decompose the canonical partition
function of the system as a sum over a set of non-overlapping
components $\RR$ (as explained in section \ref{gen:EXTME}),
\bea
{\cal Z}(T)=\sum_{\CC}\exp(-\beta\HH(\CC))=\sum_{\RR}\exp(-\beta
\FF(\RR))\nonumber\\
=\sum_{\FF}\Omega(\FF,T)\exp(-\beta \FF)=\sum_{\FF}\exp(-\beta\Phi(\FF,T))
\label{eq3f}
\eea
where we have used the definition \eq{micro:EXTME:2} for the free energy
of components $\FF(\RR)$ and \eqq{eq6d}{eq8d2}. Due to the
extensive character of the variables $\FF,S_c(\FF,T)$ and $\Phi(\FF,T)$, the dominant
contribution to the sum in (\ref{eq3f}) is evaluated through the saddle
point method. At each temperature $T$ there is a free energy $F^*(T)$
such that its contribution to the exponent (\ref{eq3f}) is dominant,
i.e. ${\cal Z}(T)\sim
\exp(-\beta\Phi^*(T))$ where we have defined
$\Phi^*(T)=\Phi(F^*(T),T)$. The behavior of this solution depends on the
shape of the function $S_c(\FF,T)$. In general this function is a
monotonically increasing function of the free energy $\FF$. Because the
exponential is a positive definite function we have $\Phi(F,T)\ge F_{\rm para}(T)$
where ${\cal Z}(T)=\exp(-\beta F_{\rm para}(T))$ and $F_{\rm para}(T)$
denotes the paramagnetic free energy.

Above $T_A$ there is no solution $F^*(T)$ which can
compensate the equilibrium paramagnetic free-energy and $\Phi(F,T)>
F_{\rm para}(T)$.  Below $T_A$ a solution appears $F^*(T)$
such that $\Phi^*(T)=F_{\rm para}(T)$ and gives the dominant
contribution to (\ref{eq3f}) so $F^*(T)$ satisfies the saddle-point
relation,
\be
\frac{1}{T}=\left.
    \frac{\partial S_c(\FF,T)}{\partial \FF}\right|_{\FF=F^*(T)}
\label{eq4f}
\ee
The identity $\Phi^*(T)=F_{\rm para}(T)$ implies $F_{\rm
para}(T)=F^*(T)-TS_c(F^*(T),T)$. This means that for $T\le \Tc$ there
is a band of components with free energy $F^*(T)\ge F_{\rm para}(T)$
(therefore with free energy above the equilibrium one) whose difference
with $F_{\rm para}(T)$ is compensated by the complexity
$S^*(T)=S_c(F^*(T),T)$. This solution exists as long as
$S^*(T)>0$. Because $S_c(F,T)$ is a monotonically increasing function of
$F$ and both $F^*(T),S^*(T)$ decrease with $T$ there is a temperature
$\TK$ at which $S^*(\TK)=0$. Below this temperature, the complexity
vanishes and the solution $F^*(T)$ ceases to change with temperature (so
equation (\ref{eq4f}) does not hold anymore) but sticks to its minimum
value $F^*(\TK)$. This is the entropy crisis scenario where $\TK$
corresponds to the Kauzmann temperature \cite{Kauzmann48}.\\

In mean-field models it has been shown \cite{FraVir00} that the
complexity $S_c(F,T)$ defines a free-energy dependent effective temperature through the
relation,
\be
\frac{1}{\Te(\FF)}=\frac{\partial S_c(\FF,T)}{\partial \FF}
\label{sc_tempeff}
\ee
From this relation, it emerges that the configurational entropy plays
the role of the thermodynamic potential associated to the effective
temperature. Similarly, the entropy is the potential conjugated to bath
temperature in the microcanonical ensemble. However, \eq{sc_tempeff} has
been only derived in mean-field models close to the asymptotic
free-energy threshold where lower free-energy states are inaccessible
\cite{FraVir00,BirKur01}. In models that do not have a marginal
free-energy threshold above the equilibrium value, it is possible to
show that free-energies are uncorrelated random variables exponentially
distributed \cite{MezParVir85,MezParVir87}. The extension of these
results beyond mean-field where all free energy states are accessible
through activated processes is at the roots of the existence of the
unbiased component ensemble. Some attempts have been proposed in
\cite{Nieuwenhuizen98b,CriRit01,CriRit01a}.  In particular it would be
very interesting to understand the general form of the transition
probabilities $Z_t$ as these lead to very specific
predictions amenable of numerical checks. The exponential character of
the free-energy distribution of the lower free-energy states below the
threshold, and the fact that this distribution is time dependent (as
shown by the fact that $Z_t$ is itself time dependent) appear as the two
crucial ingredients to understand the emergence of effective
temperatures in glassy systems. These features are present in many
models of glasses such as, mean-field spin-glass models, trap models
(see Section \ref{trap}) or entropy barrier models (see Section
\ref{esm:nodisorder:osc}).

\subsection{The integrated response function (IRF) and fluctuation-dissipation 
 (FD) plots}
\label{IRF:FD}

When a system is in a non-equilibrium state its response to a external
perturbation cannot be described, in general, by FDT relations such as
\eq{FDT} since this has been derived assuming that the system is in a stationary
state. However, the glassy state is a particular non-equilibrium state
characterized by extremely slow relaxation processes.  Hence, while the
system is not in a real equilibrium state, it may be thought as being in
a sort of quasi-equilibrium regime over time-scales much longer than the
microscopic time-scales but still smaller than the typical relaxation
time-scales of the slow processes. In this quasi-equilibrium regime the
evolution of the system is quasi-stationary since non-stationary effects
are seen only for times of the order of the time-scales of the slow
processes. In this situation we may think that relations similar to
\eq{FDT} can still be valid, even if TTI cannot be assumed anymore. Thus
a possible generalization of FDT to the glassy regime requires to
introduce the following nondimensional quantity:
\begin{equation}
  X_{A,B}(t,s) = \frac{T\,R_{A,B}(t,s)}
                      {\frac{\partial}{\partial s}\, C_{A,B}(t,s)},
		   \qquad   t>s
\label{eq2eb}
\end{equation}
where $X_{A,B}(t,s)$ is called the Fluctuation-Dissipation Ratio
(FDR). In equilibrium $X_{A,B}(t,s)=1$ whatever times $t,s$ and
observables $A,B$ are used [Cfr. (\ref{FDT})].  Thus $X$ is a measure
of the violation of the true equilibrium in the quasi-equilibrium
glassy state. The validity of \eq{eq2eb}, i.e. the proportionality between
the response and the time derivative of correlation function,
can only be checked a
posteriori since it is based on the quasi-equilibrium hypothesis that
up to now it has been proved only in mean-field models.  Eliminating
one time in favour of the correlation function, the time dependence of
$X_{A,B}(t,s)$ can be recast in the form
\begin{equation}
  X_{A,B}(t,s) \equiv X_{A,B}[C_{A,B}(t,s),s]. 
\label{eq2ec}
\end{equation}
The FDR was first studied in spin glasses where 
analytical results have shown that glassy systems in
general satisfy the weak ergodicity breaking
scenario \cite{Bouchaud92}, discussed in Section
\ref{sgm}.
For the present purpose it is enough to note that 
calculations in mean-field spin-glass models 
\cite{CugKur93,CugKur94} have shown that 
in the limit $s\to\infty$ the FDR is a non-trivial function which depends 
on the relation between the times $t$ and $s$ only through the 
correlation function $C_{A,B}(t,s)$. 
The following specific form of FDT violations has been proposed
to be generically valid in the non-equilibrium regime of glassy systems,
\begin{equation}
  \lim_{s\to\infty}X_{A,B}(t,s)=X_{A,B}[C_{A,B}(t,s)]
  \label{eq2ed}
\end{equation}
Using \eq{eq2ed} and \eq{eq2eb} we obtain the differential form of the
Quasi-FDT (QFDT) relation \cite{Horner84},
\begin{equation}
  X_{A,B}(C) = \left[\frac{T\,R_{A,B}(t,s)}
                          {\frac{\partial}{\partial s}\,C_{A,B}(t,s)}
               \right]_{C_{A,B}(t,s)=C}
  \qquad t>s.
\label{eq4e}
\end{equation}
which describes the response of the system in the (quasi)-equilibrium state
to a impulsive perturbation at time $s<t$.
Experiments and numerical simulations usually measure integrated response
functions\footnote[1]{The integrated response functions are also 
  called time dependent or non-equilibrium susceptibilities.}
(IRF), i.e., the response at time $t$ to
a perturbation switched on or off at time $s<t$.
According to the definition (\ref{eq11a})
the variation of the observable $\la A(t)\ra_{\epsilon}$ 
to linear order in the perturbation intensity $\epsilon$ is given by,
\begin{equation}
  \la A(t)\ra_{\epsilon_s} =  \la A(t)\ra_{0}
                           +\epsilon\,R_{A,B}(t,s)
			   +{\cal O}(\epsilon^2), 
\qquad t > s.
\label{eq1e}
\end{equation}
Assuming that the perturbation acts for all times $t>s$ and that
its intensity is small enough for the accumulated
response to be linear in $\epsilon$ we get
\begin{eqnarray}
  \la A(t)\ra_{\epsilon_s} &= \la A(t) \ra_{0}
                +\sum_{t'=s}^{t}\,\epsilon_{t'}\,R_{A,B}(t,t')    
\nonumber\\
              &= \la A(t) \ra_{0}
                +\int_{s}^{t}dt'\,\epsilon(t')\,R_{A,B}(t,t')    
  \label{eq2e}
\end{eqnarray}
where $\epsilon_{t'}$ (or $\epsilon(t')$) is the perturbation
at time $t'$. In the particular case of
$\epsilon(t') = \epsilon\,\theta(t'-s)$, i.e., of a constant perturbation,
one gets
\begin{equation}
  \chi_{A,B}(t,s) = \lim_{\epsilon\to 0}
                        \frac{\la A(t)\ra_{\epsilon_s} - \la A(t)\ra_{0}}
                             {\epsilon}
                       =\int_{s}^{t}dt'\,R_{A,B}(t,t').
\label{eq3e}
\end{equation}
which is also called zero-field cooled (ZFC) susceptibility.
The name zero-field cooled follows from the 
experimental protocol used in spin glass measurements
to distinguish it from the thermoremanent magnetization (TRM). 
To measure the TRM susceptibility a constant 
external perturbation is applied to the system at time $t=0$
and removed at time $s>0$ and the subsequent decay
of $\la A(t)\ra$ is recorded. The TRM susceptibility is
given by
\begin{equation}
  \chi_{A,B}^{\rm TRM}(t,s) = \lim_{\epsilon\to 0}
         \frac{\la A(t)\ra_{\epsilon_s} - \la A(t)\ra_{0}}
              {\epsilon} 
           = \int_{0}^{s}dt'\,R_{A,B}(t,t')
\label{eq3e2}
\end{equation}
In the large $t,s$ limit the
ZFC (\ref{eq3e}) and TRM (\ref{eq3e2}) susceptibilities 
are equivalent since 
they are related by,
\begin{equation}
  \chi_{A,B}(t,s)+\chi_{A,B}^{\rm TRM}(t,s) = \chi_{A,B}(t,0)
\label{eq3e3}
\end{equation}
and $\chi_{A,B}(t\to\infty,0) = \chi_{A,B}^{\rm eq}$.  

Inserting the QFDT relation \eq{eq4e} in (\ref{eq3e}) we obtain
the formula
\begin{equation}
  \chi_{A,B}(t,s) = \frac{1}{T}\int_{C_{A,B}(t,s)}^{C_{A,B}(t,t)}dC'\,
                          X_{A,B}(C')
\label{eq5e}
\end{equation}
which relates $\chi_{A,B}(t,s)$ to
the FDR $X_{A,B}$ and provides a simple way to 
calculate $X_{A,B}$ from 
measurements of $C_{A,B}(t,s)$ and $\chi_{A,B}(t,s)$ in the time sector $t>s$. 
Suppose indeed we fix the lowest time $s$ and plot $\chi_{A,B}(t,s)$ 
as function of $C_{A,B}(t,s)$ for different values of $t$, then the
value of the FDR can be obtained from the slope
of the resulting curve.
In many of the examples considered in this review  
the equal times correlation function is time-independent, 
for instance $C_{A,B}(t,t)=1$. In this case the slope 
can be simply obtained by derivation of $\chi_{A,B}(t,s)$ with
respect to $C_{A,B}(t,s)$ for fixed $s$
which from (\ref{eq5e}) yields,
\begin{equation}
  X_{A,B}(C) = -\beta\left.
             \frac{\partial \chi_{A,B}(t,s)} 
                  {\partial C_{A,B}(t,s)}
                     \right|_{C_{A,B}(t,t)={\rm const},\ s\ {\rm fixed}}
\label{eq6e}
\end{equation}
Typical FD plots are shown in Figure \ref{fig:IRF:1} for the three
possible scenarios (see section \ref{aging}).  In the general case in
which $C_{A,B}(t,t)$ is time-dependent one needs to be more careful in
computing the FDR. Sollich and coworkers have proposed
\cite{FieSol02,SolFieMay02} to construct FD plots $\chi$ versus $C$ with
$t$ kept constant and varying the lowest time $s$. From \eq{eq2eb} we
have,
\begin{equation}
  X_{A,B}(C) = -\beta\left.
                 \frac{\partial\chi_{A,B}(t,s)} 
                      {\partial C_{A,B}(t,s)}
                    \right|_{ t\ {\rm fixed}}
\label{eq7e}
\end{equation}
If $C(t,t)$
changes with time it is convenient to normalize correlations and the IRF
by the equal times correlation $C_{A,B}(t,t)$:
\begin{equation}
  \label{eq7e3}
  \widetilde{C}_{A,B}(t,s)    = \frac{C_{A,B}(t,s)}{C_{A,B}(t,t)},
  \qquad
  \widetilde{\chi}_{A,B}(t,s) = \frac{\chi_{A,B}(t,s)}{C_{A,B}(t,t)}.
\end{equation}
With these definitions,
\begin{equation}
  X_{A,B}(C) = -\beta\left.
                 \frac{\partial\widetilde{\chi}_{A,B}(t,s)} 
                      {\partial\widetilde{C}_{A,B}(t,s)}
                      \right|_{t\ {\rm fixed}}
\label{eq7e2}
\end{equation}

The importance of normalizing the raw FD plots \eq{eq7e} is well
appreciated in trap models discussed in section \ref{trap} or in
kinetically constrained models discussed in section \ref{kcm}. In this last
case, for example, 
raw FD plots can lead to awkward representations 
as the one shown in Figure \ref{fig:kcm:1}.
\begin{figure}
  \centering 
  \includegraphics[scale=0.35]{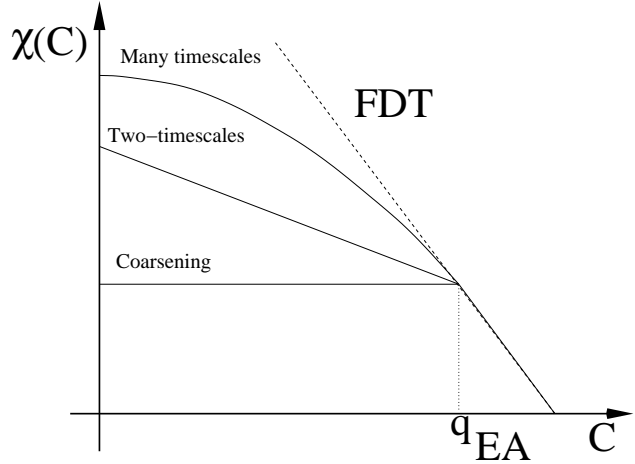} 
  \caption{FD plots for the
  three possible scenarios. From bottom to top: 1) RS models or
  coarsening systems with two-timescales and two-temperatures (one
  identical to the bath, the other infinite), 2) one-step RSB models with
  two-timescales and two-temperatures (one identical to the bath, the
  other finite but higher than that of the bath), 3) full RSB models
  with many timescales and temperatures. For $C>q_{\rm EA}$ (the
  Edwards-Anderson parameter), i.e. the
  stationary regime,  all models
  satisfy FDT (the dashed line).}
  \label{fig:IRF:1}
\end{figure}

\subsection{The concept of neutral observables}
\label{neutral:obs}

If the FDR \eqq{eq2eb}{eq4e} has the physical interpretation
of a temperature (as has been suggested, see the discussion in
Section \ref{fdr:efftemp}) then one would expect the FDR to be independent
on the observables $A,B$ used to construct correlations and
responses. In fact, this is true in equilibrium where $X_{A,B}=1$
whatever $A,B$. However, although $X_{A,B}\ne 1$ observable independence
is not at all required in the glassy regime. In this section we present
a brief digression on which conditions the observables $A,B$ must satisfy
for the FDR to be observable independent. This issue is yet unresolved, so
the present discussion is quite speculative.

Albeit restricted, for simplicity we will consider here the case of a
glassy system with only two-timescales where $A=B$ (so we will denote
$X_{A,A}$ simply by $X_A$) in the time sector where $(t-s)/s\sim {\cal
O}(1)$ or $X_A\neq 1$. In equilibrium one could argue that the equality
$X_A=1$ is related to the fact that the entropy $S(E,A)$, as defined in
the microcanonical ensemble Section \ref{basic:micro}, is maximum
\eq{basic:micro:12} for $A$ equal to its equilibrium value. This
property, is observable independent as well as it is the identity
$X_A=1$. A similar argument, but extended to the glassy regime, would
require to define the configurational entropy $S_c(\FF,{\cal A})$
(i.e. the equivalent generalization \eq{basic:micro:1} of \eq{eq6d})
where $\FF,{\cal A}$ denote the component averaged values of the
free-energy \eq{micro:EXTME:2} and the corresponding restricted Gibbs
average for the observable $A$. We could then say that $A$ is a neutral
observable if its dynamically averaged value at all times $<A(t)>$
coincides with the stationary maximum of the function $S_c(\FF,{\cal
A})$ as ${\cal A}$ is varied. Of course, if this were not true the value
of $X_A$ would then depend on the value of $A$ in the same way that the
value of the temperature $1/T=\beta$ in the microcanonical ensemble, and
for a given value of the energy $E$, would depend on the value of the
observable $A$ if $\mu$ in \eq{eq6d} were not zero.

For instance, the magnetization in mean-field spin-glass models is known
to be a neutral observable and nearly all computations of the FDR have
used this observable (see Sections \ref{sgm}). In fact, in the framework of
the TAP approach it can be shown that the configurational entropy
evaluated as a function of the free-energy and magnetization of the TAP states is
maximum at zero magnetization. Indeed, the fact that the magnetization
is not a good order parameter in these models (it vanishes in both the paramagnetic and
the spin-glass phase) is related to its neutral character. Not by
chance the majority of numerical studies in glassy systems use the
magnetization as the central observable to investigate and measure FDT
violations.

A non-neutral observable $A$ would correspond to a situation where,
starting at time zero from a non-equilibrium initial state where the
value of $A$ is taken to coincide with its equilibrium value (this is
not a contradiction, since $A$ could coincide with its equilibrium value
but not the value of any other observable) the subsequent evolution of
$A$ deviates from its initial equilibrium value. In this case $A$ is not
neutral because its time evolution is correlated to that of the other
observables. On the contrary, if $A$ were a neutral observable, then it
would stick to the value $A$ forever (as happens for the magnetization
in mean-field spin-glass models). Fig \ref{neutral:obs:fig1} illustrates
this behavior. The definition of neutrality can be easily adapted in
trap models by assuming that observable values of traps are uncorrelated
with their energies~\cite{FieSol02} (see also Section \ref{trap}).

\begin{figure}
  \centering
  \includegraphics[scale=0.35]{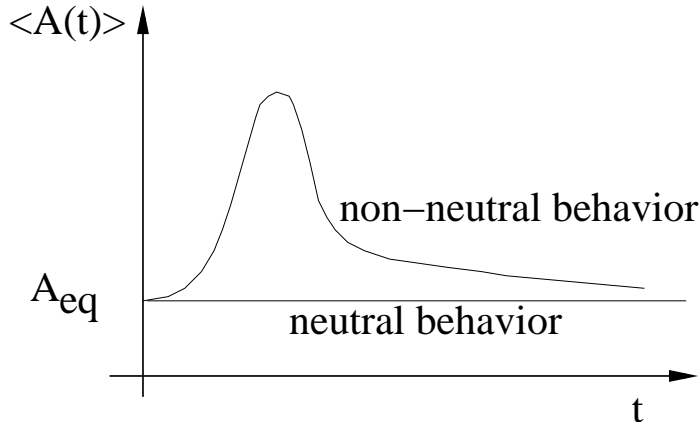}
  \caption{Illustration of the neutrality property of observables. The
    value of a neutral observable remains fixed to its initial (equilibrium)
    value despite of the fact that the initial state is non-equilibrium. 
  }
  \label{neutral:obs:fig1}
\end{figure}

\subsection{Numerical approach to component dynamics: 
            Stillinger-Weber decomposition} 
\label{is} 

The increase of computational power and the recent developments in the 
theory of disordered systems has pushed forward an approach to the
glass transition based on the analysis of a reduced dynamics in the component
space. The underlying ideas date back to more than 30 years ago to
a seminal, but talkative, paper of Goldstein \cite{Gol69}. The glass 
transition is of purely dynamical origin and hence must reflect the 
properties of the dynamical evolution of the system. 
Goldstein suggested that the dynamics of a supercooled liquid can be
understood in terms of a diffusive process between different {\it basins}
of the potential energy surface. At low temperature the dynamics slows down 
since the system gets trapped for a long time in a basin.
This approach, which focus on the topological properties of the energy 
surface is rather appealing since it naturally leads to a convenient framework 
for understanding the complex phenomenology of glassy systems. 

The implementation of these (qualitative) ideas came some
years later by Stillinger and Weber\cite{StiWeb82,StiWeb84,Stillinger95b} (SW) 
who formalized the concept of basin in configuration space
identifying it with a component in the component space,
and proposed a procedure for identifying them: the set of all configurations
connected to the same local energy minimum by a 
steepest-descent path on the energy surface uniquely defines the basins 
of the minimum. Stillinger and Weber (SW) called the local minimum
{\it inherent structure} (IS) to stress its intrinsic nature. 
Since the identification of ISs is unique, the mapping from configurations 
to local minima gives a {\it unique well defined} decomposition of the phase 
space into a {\it disjoint} set of basins. 
The SW decomposition defines a mapping of the phase space to the 
component space in which each basin,
usually labeled by the energy $E_{\rm IS}$ of the local minimum, 
is a component.

This decomposition does not cover
completely the configuration space since it leaves out the boundaries between 
different basins. 
However, under the
assumption that those configurations do non contribute to the thermodynamic
of the system, e.g., the boundaries between basins are sub-extensive, 
it does cover {\it almost all} the phase space and the partition function
can be written as a  sum of contributions from different 
components:
\begin{equation}
{\cal Z}(T) \simeq \sum_{E_{\rm IS}}\, {\cal Z}_{\rm IS}(E_{\rm IS},T)
\label{eq:5.2-1}
\end{equation}
Let $\Omega(E)$ denote the number of IS with energy $E_{\rm IS} = E$, 
then collecting all components with the same value of $E_{\rm IS}$
\begin{eqnarray}
{\cal Z}(T) &\simeq \sum_{E}\, \Omega(E) 
	\sum_{E_{\rm IS} = E} {\cal Z}_{\rm IS}(T) /  \Omega(E) 
\nonumber\\
             &= \sum_{E} \exp\left[ S_c (E) - \beta\,F_b(T,E)\right].
\label{eq:5.2-2}
\end{eqnarray}
The term 
\begin{equation}
S_c(E) = \log \Omega(E),
\label{eq:5.2-3}
\end{equation}
which accounts from the entropic contribution arising from the number 
of different basins with the same IS energy, is called the SW 
{\it configurational entropy} of {\it complexity}. This quantity is strongly 
related to the partitioning, so we add the 
adjective SW to distinguish it to other definitions of configurational
entropy taken from mean-field concepts.

The second term $F_b(T,E)$ is defined as
\begin{equation}
F_b(T,E) = -T\, \log\left[ 
	\frac{1}{\Omega(E)}\, \sum_{E_{\rm IS} = E} {\cal Z}_{\rm IS}(T)
                             \right].
\label{eq:5.2-4}
\end{equation}
In general this quantity differs from the average free energy of components
with $E_{\rm IS} = E$, however if all these components have similar 
statistical properties, then $F_b(T,E)$ is the free energy of the system when 
constrained to any one of the components with $E_{\rm IS} = E$.
In the thermodynamic limit the system populates components with energy
$E_{\rm IS} = E_{\rm IS}(T)$ fixed by the condition
\begin{equation}
-\beta F(E)=S_c (E) - \beta\,F_b(T,E) = \mbox{\rm maximum over $E$}
\label{eq:5.2-5}
\end{equation}
and the free energy of the system can be calculated using
\begin{equation}
F(T) = F_b[T,E_{\rm IS}(T)] - T\,S_c[E_{\rm IS}(T)]
\label{eq:5.2-6}
\end{equation}
The condition (\ref{eq:5.2-5}) is equivalent to that of $F(T)$ being
minimal, i.e., 
\begin{equation}
\label{eq:5.2-7}
\frac{\partial F}{\partial E} = \frac{\partial F_b(T,E)}{\partial E} 
                               -T\, \frac{\partial S_c(E)}{\partial E} 
                              = 0.
\end{equation}
Note that the minimum condition follows from the balance between
the contribution from the change with the energy of the shape of the basins 
($\partial F_b(T,E) / \partial E$) and 
its corresponding number ($\partial S_c(E) / \partial E$).
Often the free energy is written as
\begin{equation}
\label{eq:5.2-8}
F_b(T,E_{\rm IS}) = E_{\rm IS} + F_v(T,E_{\rm IS})
\end{equation}
The first term in \eq{eq:5.2-8} takes into account the average energy of IS visited
in equilibrium at temperature $T$,
as can be seen from (\ref{eq:5.2-6}):
$U(T) = \partial (\beta F_b) / \partial \beta =$
$E_{\rm IS}(T) + \partial (\beta F_v) / \partial \beta$.
It can be shown \cite{Heu97,BucHeu99} that if the density of states
$\Omega(E)$ is Gaussian and the basins have approximately the same
shape then $E_{\rm IS} \propto 1/T$.
%
%
The second term in \eq{eq:5.2-8} describes the volume of the corresponding 
components and is called the ``vibrational'' contribution.

To understand the success and limitation 
of the IS approach we have to 
analyze the idea behind the SW approach. 
It is clear that even if the phase space can be always partitioned, not
all possible partitions will lead to a physically 
relevant dynamics in the component space.
This is a well known problem in the theory of dynamical systems, where
the component dynamics is called {\it symbolic dynamics}, 
see e.g. Ref. \cite{BecSch93}.
To prove that the SW  is a physically good partition for a 
given system is a problem of the same
hardness as proving ergodicity.
One then adopts a constructive
point of view, along the same lines of equilibrium statistical theory:
based on some reasonable hypothesis one first assumes that 
the SW partition is a good partition and then check if this reproduces
the desired features of the dynamics. 

The physical motivation behind the SW proposal follows from the
observation that the potential energy surface of a 
super-cooled liquid contains a large number of local minima
and that the time evolution can be separated into two 
different processes: thermal relaxation into basins
({\it intra-basin} motion)  and 
thermally activated potential energy barrier crossing between different
basins ({\it inter-basin} motion). 
This scenario has been recently 
confirmed from numerical analysis 
\cite{Heu97,BucHeu99,SasDebStiSchDyrGlo99,ScrSasDyrGlo00,AngLeoRuoScaSci00}. 
The time-scales separation of the two 
processes strongly depends on temperature. 
When the temperature is lowered down to the order of the critical
mode-coupling theory (MCT) temperature $\Tc$ 
the typical barrier height is of the order of the
thermal energy $k_{B} \Tc$, and 
the slow inter-basin motion dominates the 
relaxation dynamics.
If the temperature is further reduced the
relaxation time eventually becomes of the same order of the 
physical observation time and the system falls to a non-equilibrium state 
since there is not enough time to cross barriers and equilibrate. 
With this picture in mind it is natural to view the IS partitioning as the
natural elements to describe the slow glassy dynamics. 
This approach is rather appealing since naturally leads to 
universality: all glassy systems with similar IS dynamics must have 
similar glassy behaviour. 
Recent IS analysis performed on disordered spin systems displaying
a transition of fragile glass type do support this conclusion 
\cite{CriRit00,CriRit00b,CriRit00c,CriRit00d,CriRit02}.
It should be noted \cite{BirMon00} that the definition of IS for spin 
systems is more subtle than for systems with continuous variables. 
Indeed usually for spin systems IS are defined as one-spin flip stable 
states, however these may be not stable for two-spin or higher number of 
spin flips. One possibility of making IS well defined also for spin systems 
is to define them directly from the $T=0$ limit of the dynamics, i.e., 
as states which are stable under the $T=0$ dynamics \cite{CriRitRocSel00}. 
This is the definition used in this review when discussing IS for 
spin systems.

\section{Thermodynamic description of the aging state}
\label{thermo}

We saw in section \ref{micro:EXTME} how the self-generated dynamical
measure allows a description of the aging dynamics in terms of a
probabilistic master equation with transition rates characterized by
an extensive quantity that was defined as a configurational entropy or
complexity (\ref{eq6d}). This quantity has received considerable
attention in studies of spin-glasses since the seminal paper 
of Thouless, Anderson and Palmer (TAP) \cite{ThoAndPal77} on the SK model 
where a way to compute the configurational entropy was proposed 
\cite{BraMoo80}.  Later studies in the context of
structural glasses \cite{KirWolb87b,KirThi87a,KirThi87b} have shown
its importance as the mechanism for an entropy crisis of the supercooled
liquid as proposed by Kauzmann many years ago \cite{Kauzmann48}.

\subsection{Methods to compute the complexity}
\label{conf:entr}

In this section we present a schematic overview of some of the
analytical and numerical methods that have appeared in the literature to
compute the configurational entropy. In the absence of a full solution
of the dynamics in many systems, and under the assumption that there is
a connection between the effective temperature and the configurational
entropy (see the discussion in Section \ref{interlude}), the calculation
of the later, by using equilibrium methods taken from statistical
physics, appears as an alternative way of quantifying FDT violations. In
mean-field theories metastable states give a natural partition of the
phase space since their life-time diverges in the thermodynamic
limit. For systems with short-range interactions, however, metastable
states can be defined unambiguously only referring to some reference
time-scale. Therefore the identification of metastable states for real
systems can be a very hard task. In section \ref{is} we have presented a
partition scheme, proposed by Stillinger and Weber, which in principle
can be applied to any system. The scheme essentially uses a
zero-temperature dynamics and thus it is free from the ambiguities due
to the finite metastable life-time. The results
described in next sections must be seen as instructive attempts to
evaluate a quantity (the complexity) that governs the slow dynamics of
relaxational glassy systems. The extension of these equilibrium concepts
to other non-equilibrium systems beyond aging systems (e.g. driven
systems) remains an open problem.

\subsubsection{Analytical methods.}
\label{conf:entr:ana}

Bray and Moore \cite{BraMoo80} calculated $S_c(\FF,T)$ for the
Sherrington-Kirkpatrick model within the TAP approach.  The TAP
equations give the local magnetization $m_i$ in a system confined to a
metastable state, which for mean-field models have infinite
life-time. As a consequence the number of metastable states (i.e.,
components) can be readily obtained just counting the number ${\cal
N}_s(\FF,T)$ of solutions of the TAP equations at temperature $T$ with a
free energy $\FF$:
\begin{eqnarray}
   {\cal N}_s(\FF,T) = \int_{-1}^1\prod_{i=1}^N\,dm_i\,
      &\bigl|\det H(\lbrace m_i\rbrace)\bigr|\,
\nonumber\\
      &\times
      \delta\bigl(\FF-\FF(\lbrace m_i\rbrace, T)\bigr)\,
      \delta\bigl(g_i(\lbrace m_i\rbrace)\bigr)\,
\label{eq1f}
\end{eqnarray}
where $\FF(\lbrace m_i\rbrace,T)$ if the TAP free energy at temperature $T$ 
as function of the local magnetizations $m_i$,
$g_i(\lbrace m_i\rbrace)=\partial \FF(\{m_i\},T)/\partial m_i = 0$
are the TAP equations and $H_{ij}=\partial g_i(\{m_i\}) /\partial m_j$
the Hessian. This type of calculation has been
done for other  mean-field models, such as 
$p$-spin \cite{CrisSom95} and Random Orthogonal Model (ROM)
 \cite{ParPot95,ParPot95b}),
finding in all cases that ${\cal N}_s(\FF,T)$ increases exponentially
fast with the system size system $N$. 
This remains true if the number of free-energy minima 
${\cal N}_m(\FF,T)$, instead of the number of
stationary points ${\cal N}_s(\FF,T)$,  is considered \cite{CavGiaPar98}. 
Although these type of calculations can be done only in exactly solvable 
mean-field models the 
exponential growth with the system size of the number of free energy 
local minima or stationary points is generally applicable to any system
(mean-field or not) displaying glassy behaviour. The
knowledge of the number of minima allows to define the complexity 
\eq{eq5d} as,
\begin{equation}
  S_c(\FF,T)=\log{\cal N}_m(\FF,T)
\label{eq2f}
\end{equation}
and hence the thermodynamic potential $\Phi(\FF,T)=\FF-TS_c(\FF,T)$ 
as described in section \ref{micro:EXTME}.

A general framework to evaluate the complexity has been devised by
Monasson \cite{Monasson95}. The starting point in his procedure is to
consider $m$ interacting copies or replicas of the original system,
with an attractive interaction term of the form
$\eps\sum_{a,b=1}^{m}Q(\CC_a,\CC_b)$ where $Q(\CC_a,\CC_b)$ is a
suitable overlap function which takes its maximum value only if
$\CC_a=\CC_b$.  The free energy of the replicated system is then,
\begin{equation}
e^{-\beta\,F(T,m)} =\sum_{\CC_1,\ldots, \CC_m}
                  \exp\Bigl[- \beta\sum_{a=1}^m\HH(\CC_a)
                            + \beta\eps\sum_{a,b=1}^mQ(\CC_a,\CC_b)
                      \Bigr]
\label{eq5f}
\ee
If the thermodynamic limit is taken before the limit $\eps\to 0^+$ then
the configurations $\CC_a$, $\CC_b$ tend to lie as close as possible
since maximization of the coupling term minimize the global
free energy $F^{(m)}(T)$ and hence, given a phase space partition,
the replicas tend to ``condensate''  into the same component.
Thus collecting all components with the same free energy
the partition function ${\cal Z}(T,m)$ can be decomposed as
\begin{eqnarray}
  {\cal Z}(T,m) &= \sum_{\FF}\Omega(\FF,T)\exp(-m\beta \FF)
\nonumber\\
                &=  \sum_{\FF}\exp[-\beta\Phi(\FF,T,m)]
\label{eq6f}
\end{eqnarray}
where $\Phi(\FF,T,m)=m\FF-TS_c(\FF,T)$ is basically the potential
$\Phi(\FF,T)$ discussed in section \ref{micro:EXTME} with the term $\FF$
multiplied by $m$ (the order of limits, first volume $\to\infty$ and
then $\eps\to 0^+$, enforces the $m$ replicas to occupy the same
component $\RR$).  In the limit $m\to 1$ we recover the potential
$\Phi(\FF,T)$: $\Phi(\FF,T,m=1)=\Phi(\FF,T)$. The knowledge of
$\Phi(\FF,T,m)$ allows to compute the configurational entropy.  In the
thermodynamic limit the sum in (\ref{eq6f}) is dominated by the free
energy $F^*(T,m)$ that satisfies the relation,
\begin{equation}
  \frac{m}{T} = \left.\frac{\partial S_c(\FF,T)}{\partial \FF}
                \right|_{\FF=F^*(T,m)}
\label{eq7f}
\end{equation}
Inserting the solution $F^*(T,m)$ into $\Phi(\FF,T,m)$ we obtain the 
free-energy potential $\Phi^*(T,m)=\Phi(F^*(T,m),T,m)$. 
If $m$ (originally an integer value) is continued to real values then it
can be show that the following relations are satsified,
\begin{eqnarray}
 \frac{\partial}{\partial m}\,\Phi^*(T,m) &=& F^*(T,m),
 \nonumber\\
 \frac{\partial}{\partial m}\, \left[\frac{\Phi^*(T,m)}{m}\right] &=&
                         \frac{T}{m^2}\,S_c(F^*,T)~~~~.
\label{eq8f}
\end{eqnarray}
Varying $m$ allows to compute the configurational entropy $S_c(\FF,T)$
as function of the two variables $\FF$ and $T$.
The potential $\Phi(\FF,T,m)$ can be explicitly evaluated with the
sole knowledge of the microscopic Hamiltonian of the system and using
the replica method. Although this procedure was initially applied only to
mean-field disordered systems \cite{Monasson95,Mezard98}, more recently 
it has been extended to more realistic interacting potentials
such as Lennard-Jones liquids \cite{MezPar99,MezPar99b} and binary
mixtures  \cite{ColParVer00,ColParVer00b}.

This method of computing the configurational entropy can be easily
implemented in the framework of the standard replica method for
mean-field disordered systems. This has been worked in some detail
in \cite{ParRitSan95,Ritort96}.
The starting point is free energy at
one-step level of replica symmetry breaking $F(q_0,q_1,m)$, where 
$q_0$, $q_1$ and $m$ are the parameters that describe the Parisi
matrix \cite{MezParVir87} in the one-step replica symmetry breaking scheme. 
By expanding the free energy around $m=1$ one
gets, $F(q_0,q_1,m)= F_{RS}(q_0)+F^{(1)}(q_0,q_1)(m-1)+{\cal O}((m-1)^2)$
where $q_0$ stands for the overlap among replicas belonging to different
subboxes and $F_{RS}(q_0)$ is the free energy in the replica symmetric
approximation, i.e., in the limit $m\to 1$.
Extremization of $F_{RS}(q_0)$ yields $q_0(\beta)$ which 
inserted into $F(q_0,q_1,m)$ allows to find $F_{RS}(\beta),F^{(1)}(\beta,q_1)$.
The knowledge of these functions fully determines 
the configurational entropy of the system
for temperature $\Trsb<T<\Tc$ (where $\Tc$ corresponds to the MCT, see
Section \ref{mct}). 
Indeed the dynamical transition $\Tc$ is found solving the equations 
$(\partial/ \partial q_1) F^{(1)}(\beta,q_1) = 
 (\partial^2 / \partial q_1^2) F^{(1)}(\beta,q_1)=0$ 
while the static transition
$\Trsb$, where the configurational entropy vanishes, follows from the
solution of 
$F^{(1)}(\beta,q_1) = (\partial / \partial q_1) F^{(1)}(\beta,q_1)=0$. 
Finally the complexity in the region $\Trsb<T<\Tc$ is 
given by the value of
$F^{(1)}(\beta,q_1)$ evaluated for $q_1(\beta)$
solution of the equation 
$(\partial / \partial q_1)F^{(1)}(\beta,q_1)=0$. 
This approach gives a detailed
description of the metastable properties in the range
$\Trsb<T<\Tc$. Below $\Trsb$ more sophisticated methods 
are needed to describe the metastable behavior.

The potential method has been proposed by Franz and Parisi
\cite{FraPar95} in the framework of the replica approach.  The starting
point in this procedure is to write down the partition function of a
generic system at temperature $T$ whose configurations $\CC$ are
constrained to have an overlap $Q(\CC,\CC_0)$ with a reference
configuration $\CC_0$. The free energy of the constrained system is then
averaged, using the replica method, over the reference configuration
$\CC_0$ thermalized at a temperature $T'$ in general different from
$T$. This yields the potential $V(Q,T,T')$.  For $T=T'$, and in a given
range of temperatures, the potential $V$, as a function of $Q$, has two
local minima. The difference of the potential at these two values yields
the configurational entropy at that temperature.  The method has been
applied to evaluate the configurational entropy in the hypernetted chain
approximation usually employed for liquids 
\cite{CarFraPar98,CarFraPar99}.

\subsubsection{Numerical methods}
\label{conf:entr:num}

Among numerical approaches, Speedy \cite{Speedy93} has proposed a
method that consists in estimating what he defines as the {\it
statistical entropy} (basically identical to the intrastate entropy in
the inherent-structures approach discussed in section \ref{is}) and
comparing it to the thermal or total entropy obtained from integration
of the specific heat. The difference between the thermal entropy and
the statistical entropy is the complexity. To compute the statistical
entropy the method considers different reference configurations
representative of an amorphous glass state and introduces a coupling
term between a reference configuration and the system that forces it
to stay within a given distance of that reference configuration. By
progressively slowly changing the intensity of the coupling the energy
of the system can be evaluated for each value of the coupling.  The
entropy associated to a particular reference state is then estimated
by integrating the energy as function of the intensity of the
coupling.  Speedy has applied this approach
\cite{Speedy93,Speedy01b,Speedy01} to hard sphere systems where the
center of the hard spheres are tethered to a spherical region with a
variable diameter that regulates the intensity of the coupling.

Probably up to now the most powerful method to compute numerically 
the configuration  entropy numerically is the one based on the IS formalism.
Moreover, due to its relatively simple implementation,
the IS formalism has become an important tool in the 
numerical analysis of models. For this reason we shall give a more detailed
presentation.
The calculation of IS, summarized in Figure \ref{fig:sw-dec}, follows 
directly from the definition. 
First the system is equilibrated at a given temperature $T$, 
then starting from an equilibrium configuration the system is
instantaneously quenched down to $T=0$ by decreasing the
energy along the steepest descent path. The procedure is repeated 
several times starting from uncorrelated equilibrium configurations.
In this way the IS are identified and their probability distribution
can be computed.
\begin{figure}
  \centering
  \includegraphics{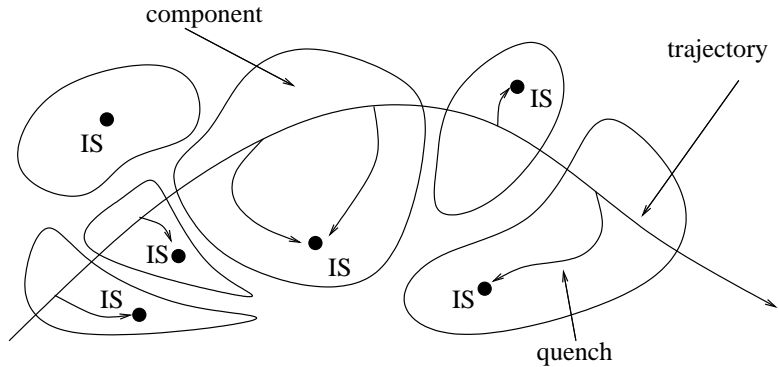}
  \caption{Stillinger and Weber decomposition}
  \label{fig:sw-dec}
\end{figure}
In equilibrium at temperature $T$ the system explores the IS 
of energy $E_{\rm IS} = E$ 
with probability [see (\ref{eq:5.2-2}), (\ref{eq:5.2-8}] 
\begin{equation}
\label{eq:5.2-10}
 {\cal P}(E,T) =  \exp \left[S_c(E) 
                                   -\beta E
                                   -\beta F_v(T,E)
                                   +\beta\, F(T)
                               \right]
\end{equation}
where $F(T)$ is the equilibrium free energy.
Then the 
SW configurational entropy can be computed just inverting this relation:
\begin{equation}
\label{eq:5.2-11}
  S_c(E) = \ln {\cal P}(E,T) + \beta E + \beta\, F_v(T,E) 
               -\beta\, F(T)
\end{equation}
and using the computed IS probability distribution.
If the energy dependence of $F_v(T,E)$ can be neglected, then curves 
for different temperatures can be superimposed and
the resulting curve is, except for 
an unknown constant, the SW complexity $S_c(E)$. 
The unknown constant can be fixed either comparing the numerical results 
with known theoretical predictions, or using the method described below.
This method works rather well for some disordered spin systems
as the Random Orthogonal Model (ROM), where the data collapse is rather
good for a quite large energy interval \cite{CriRit00}. 
The vibrational contribution $F_v$
follows from the motion inside the component. Its independence
from the energy of IS means that all components are equivalent, i.e., have
similar shapes. In general this is not the case and its contribution
must be included. For systems with continuous variables
$F_v$ can be calculated at low $T$ in the harmonic approximation 
by expanding the energy about the IS configuration 
\cite{SciKobTar99,MosETAL02}.

An alternative numerical 
method consists in computing directly the configurational 
entropy as function of temperature. This method is free from unknown 
constants but does not resolve the configurational entropy as function
of IS energy.
Considering (\ref{eq:5.2-6}) and (\ref{eq:5.2-8}) we have
\begin{eqnarray}
\label{eq:5.2-13}
F(T) &=& E(T) - T\,S(T) \nonumber\\
     &=& E_{\rm IS}(T) + E_v(T)  - T\,S_c(T) - T\,S_v(T).
\end{eqnarray}
where $E_{\rm IS}(T)$ is the average energy of IS seen at equilibrium at
temperature $T$.
The total entropy is then the sum of two contributions
\begin{equation}
\label{eq:5.2-14}
 S(T)  = S_c(T) + S_v(T)
\end{equation}
The first term accounts for the multiplicity of components of energy
$E_{\rm IS}(T)$  while the second for their ``volume''.
The SW configurational entropy can then be computed as difference between the
total and the vibrational entropy.

The total entropy $S$ can be evaluated via thermodynamic integration
of the total energy at temperature $T$ from a known reference point:
\begin{equation}
\label{eq:5.2-15}
\Delta S = S(T) - S(T^*) 
         = \int_{T^*}^{T}\, \frac{d E}{T}.
\end{equation}
To compute the vibrational 
contribution is more difficult, however at low temperature 
the system mainly explores the bottom of the components, near the IS.
If the system is described by continuous variables
then the vibrational contribution can be computed in the harmonic
approximation by expanding about the IS. This leads to
\begin{equation}
\label{eq:5.2-16}
S_v(T) \simeq S_{\rm harm}(T) 
       = {\cal N} - \sum_{i=1}^{\cal N}\, \log\left[
	 \frac{\hbar \omega_i(T)}{k_{B}\,T}
	\right]
\end{equation}
where $\omega_i$ is the (average) frequency of the $i$-th normal mode and
${\cal N}$ the number of normal modes.
It is possible to refine this approximation by adding terms which take 
into account the basins anharmonicities, however usually these are 
negligible when compared with (\ref{eq:5.2-16}) \cite{MosETAL02}.
For systems with discrete variables, as for example disordered 
Ising-spin systems,
the vibrational contribution can be estimated from the $T\to 0$ expansion
of the Thouless Anderson Palmer entropy, which leads to
\begin{equation}
\label{eq:5.2-17}
 S_v(T) \simeq \sum_{i=1}^{N}\, 2\beta |h_i|\, \exp(-2\beta |h_i|)
\end{equation}
where $N$ is the number of spins, and $h_i$ is the local field acting on
$i$-th spin evaluated at the IS configuration \cite{CriRit02u}.
	
These methods have been successfully applied to several model systems with both 
continuous variables such as 
Lennard-Jones glasses \cite{SciKobTar99,Sastry01,MosETAL02} or
discrete variables like the ROM \cite{CriRit00} or the SK model
\cite{CriRit00,ColMarParRie00}.

Recently in \cite{ColCriMarRitRoc01} it has been introduced
a numerical method to compute directly $S_c(\FF,T)$
within the IS decomposition scheme
based on the probabilistic 
definition of the component free energy. The dynamical evolution 
of the system in equilibrium defines a probability measure $p_{\RR}$ over the
components.
In the case of an ergodic dynamics,
and assuming that the observation time 
$\tau_{\rm obs}$ is larger than the equilibration time, 
the statistical weight of the component $\RR$ is:
\begin{equation}
  \label{pval}
  {p}_{\RR}(T)= \frac{\tau_{\RR}}{\tau_{\rm obs}}
              = \exp{\bigl[-\beta F(\RR)+\beta F(T) \bigr]}
\end{equation}
where $\tau_{\RR}$ denote the time spent by 
system in the component $\RR$ during the total observation time
$\tau_{\rm obs}$, $F(T)$ the equilibrium free energy and
$F(\RR)$ the component free energy [See (\ref{micro:EXTME:2})].
The probability to find at temperature $T$ a component with free
energy equal to $\FF$ is
\begin{eqnarray} 
{\cal P}(\FF,T) &= 
                 \sum_{\RR}\,p_{\RR}(T)\,\delta\left(\FF-F(\RR)\right)
\nonumber\\
               &= \exp\bigl[S_c(\FF,T)-\beta \FF+\beta F(T)\bigr]
\label{eq3}
\end{eqnarray}
so that
\begin{equation}
  S_c(\FF,T) = \ln {\cal P}(\FF,T)+ \beta\bigl[\FF-F(T)\bigr]\ .
\label{eq4}
\end{equation}
If the number of different components is not too large 
$F(\RR)$ can be estimated directly 
using (\ref{pval}) and the frequency with which a given component
$\RR$ appears in a (long) simulation at temperature $T$:
\begin{equation}
 F(\RR) = - T \ln\left(\frac{\tau_{\RR}}{\tau_{\rm obs}}\right) +
                   F(T).
\label{eq5}
\end{equation}
The equilibrium free energy $F(T)$ can be computed 
by performing simulations at different temperatures and integrating the 
energy $E(T)$ of the system from $T=\infty$ down to $T$:
\begin{equation}
  \beta F(T)= \int_{0}^{\beta} d\beta'\, E_{\rm eq}(\beta') - 
  S(\beta=0).
\end{equation}
where $S(\beta=0)$ is the infinite temperature entropy of the
system. From the value of $F(\RR)$ it is now easy to construct the
histogram ${\cal P}(\FF,T)$ and using (\ref{eq4}) compute $S_c(\FF,T)$.
Because the system is equilibrated, in this approach components with
identical free energy are sampled with the same probability. This
differs from the previous method where components with the same energy
$E_{\rm IS}$ are assumed to be equally probable which is clearly an
approximation. The two methods coincide only if components have similar
volumes so that the component entropy is the same.

This method has been successfully applied in \cite{ColCriMarRitRoc01} to
the study of the ROM and the SK model, two cases with completely
different critical behaviour.  In both cases the computed potential
$\Phi$ allows for a very precise calculation of critical temperatures
using relatively small systems giving confidence on
$S_c(\FF,T)$. Moreover the form of $\Phi$ clearly discriminates between
the two different types of transitions.

\subsection{The concept of the effective temperature} 
\label{fdr:efftemp}

There is the long standing idea
that the non-equilibrium regime in aging or driven systems can be
characterized by the FDR $X_{A,B}(C)$ \eq{eq4e}
that has the meaning of a temperature in the thermodynamic sense. 
This
suggestion stems from the observation that  \eq{eq4e} can be recast in
the following form,
\begin{equation}
  \frac{1}{\Te^{(A,B)}(C)}=
          \frac{X_{A,B}(C)}{T}=
           \left.\frac{R_{A,B}(t,s)}
                      {\frac{\partial}{\partial s}\,C_{A,B}(t,s)}
           \right|_{C_{A,B}(t,s)=C}, 
\qquad     t>s
\label{fdr:1}
\end{equation}
defining an effective temperature through the relation
\be
\Te^{(A,B)}(C)=\frac{T}{X_{A,B}(C)}
\label{fdr:2}
\ee
As defined in (\ref{fdr:1}), (\ref{fdr:2}) the effective temperature is nothing
else than a suitable parameter which tells that the QFDT becomes the
usual FDT by replacing the bath temperature with the effective 
temperature. 
In many cases $X_{A,B}(C)<1$ so the effective temperature 
$\Te^{(A,B)}$ is larger than the bath temperature. 

The idea that some concepts of thermodynamic systems can be applied also
to non-equilibrium systems has been wandering around in the literature
for a long time (in the context of turbulence see \cite{HohShr89} or in
the context of structural glasses \cite{Tool46,Hodge87,Hodge94}). The
statement that the effective temperature \eq{fdr:2} has indeed a
thermodynamic meaning faces some conceptual problems and difficulties
not found in equilibrium theory.  $\Te^{(A,B)}$ should satisfy the
following properties,

\begin{itemize}

\item{Observable independence.}  $\Te^{(A,B)}(C)$ must be independent of
the observables $A,B$ used to construct correlations and responses. If
this is not always possible, as the present numerical evidence suggests,
at least one would still like to know beforehand which set of ``good''
observables endow \eq{fdr:2} with a physical meaning. These observables
have received the name of neutral observables and have been discussed in
section \ref{neutral:obs}. From many perspectives, this condition
appears quite strong. It could be relaxed by only requiring independency
of $\Te$ upon the measured observable $A$ for a given perturbation $B$
(rather than on both $A,B$).

\item{Zero$^{\rm th}$ law.} 
  If the slow degrees of freedom of a system
  described by effective temperature $\Te(C)$
  (we have dropped the $A,B$ dependence)
  are put in contact with a thermal bath
  at temperature $T$, the net heat flow between the system and the bath should
  vanish only if $\Te(C)=T$, where $C$ determines the
  relevant timescale (as described in section \ref{aging}) at which the
  thermal bath, acting as a thermometer, responds. The same conclusion must hold between two glassy
  systems described by two effective temperatures 
  $\Te^{(1)}(C)$, $\Te^{(2)}(C)$. After putting them in contact the net heat
  current between them, at the relevant timescale defined by the
  correlation $C$, vanishes only if 
  $\Te^{(1)}(C)=\Te^{(2)}(C)$. 
  This definition, apparently reasonable, encounters some
  difficulties that we will describe below. In particular, systems with
  identical effective temperatures $\Te^{(1)}(C)=\Te^{(2)}(C')$
  but at different timescales ($C\ne C'$) cannot be in mutual equilibrium.

\item{Existence of a non-equilibrium measure.} The zero$^{\rm th}$ law
carries associated a maximum principle. In standard thermodynamics the
zero$^{\rm th}$ law establishes that after putting in contact two
systems at different temperatures the global system reaches a stationary
state with a unique temperature. This stationary state is the one that
maximizes the global entropy of the compound system compatible with a
given total energy. Moreover fluctuations around this maximum entropy
state are ruled by the temperature. By the same token, the aging state
of relaxational systems and the stationary state of driven systems must
exhibit some fluctuations or deviations around the aging (or driven)
state that are described by the effective temperature $T^{\rm eff}(C)$. 
The full characterization of these fluctuations is presently
unknown.

\end{itemize}

In Section \ref{micro:EXTME} we tried to fortify the idea that a
thermodynamic description is indeed possible and that the effective
temperature shares some properties of thermodynamic temperatures.
Cugliandolo, Kurchan and Peliti
\cite{CugKurPel97} have emphasized these aspects showing that the
effective temperature can be defined only regarding the timescale under
consideration. They considered a small thermometer that can be mimicked
by a single harmonic oscillator of frequency $\omega$ that is put in
contact with the original system. For definitiveness let us consider
that $x$ denotes the oscillator coordinate and $O(y)$ an observable of
the system (described by the variable $y$) to which the oscillator is
coupled by an interaction term, $-\eps x O(y)$ where $\eps$ is the
intensity of the coupling. If $\eps$ is small enough, then the interaction
between the oscillator and the system can be treated within the
linear-response theory and the energy of the oscillator evaluated in the
stationary state.  The equipartition theorem relates this energy to the
temperature measured by the oscillator acting as a thermometer. In aging
systems, the effective temperature is given by the FDT in the frequency
domain (also called Nyquist formula),
\be
\Te^{(O)}(\omega,t_w)=\frac{\pi}{2} \frac{\omega S_O(\omega,t_w)}{\chi_O''(\omega,t_w)}
\label{fdr:3}
\ee
where $S_O(\omega,t_w)$ is the power spectrum or correlation $\la
O(t)O(t_w)\ra$ expressed in Fourier space (see \eq{exp1a}) and
$\chi_O''(\omega,t_w)$ the corresponding out-of-phase susceptibility.  A
similar expression is valid for the stationary non-equilibrium state of
driven systems, however because TTI holds the $t_w$ dependence in
\eq{fdr:3} drops off.  The connection between \eq{fdr:2} and \eq{fdr:3}
appears when translating the meaning of $\omega$ and $t_w$ into the many
timescales scenario. According to that $\omega$ corresponds to
$1/(t-t_w)$ and therefore we can define $C^*\equiv
C(t,t_w)=C(t_w+1/\omega,t_w)$. This means that a thermometer put in
contact with the system at time $t_w$ and responding at a given
frequency $\omega$ measures the temperature
$\Te(C^*)=T/X(C^*)$. Equivalently, to measure the temperature
$\Te(C^*)=T/X(C^*)$ in an aging system a thermometer responding to a
time scale $t^*=1/\omega$ with $C^*\equiv C(t_w+1/\omega,t_w)$ should be
used. In aging systems with two-timescales (such as structural glasses)
characterized by the full aging $t/t_w$, in order to measure the
effective temperature associated to the slow process, the frequency of
the thermometer must be $\omega \sim 1/t_w$. The thermometer must
respond in a timescale of the order of the waiting time!!. In this
scenario, effective temperatures can be extremely difficult to measure
and this raises the question about their true meaning as the system
drifts away from that state in the time required for a single
measurement. To cope with this problem it has been proposed
\cite{CugKur99b} that an ensemble of small thermometers is needed for
the measurement.  However, this does not solve the problem of how to
measure, using this procedure, the effective temperature of a vitrified
piece of glass quenched one thousand years ago. These difficulties are
inherent to aging systems but not necessarily in driven systems that
reach a stationary TTI state. For these reasons, experimental
measurements of FDT violations and effective temperatures could be more
suitable in driven rather than aging systems (concerning experiments see
discussion in section \ref{exp}).

Zero$^{\rm th}$ law aspects of the effective temperature have been
considered in \cite{CugKurPel97,CugKur99a} within the spherical $p$-spin
model.  It has been shown that in the large $t_w\to \infty$ limit and
low frequency limit $\omega\to 0$ there are only two possible Ansatz
solutions that close the dynamical equations in the aging state: either
the two systems remain uncoupled with different effective temperatures
or they thermalize and reach a common temperature. These results have
been endorsed by a systematic study of these {\em coupled} solutions in
the framework of the oscillator (OSC) model \cite{GarRit01b}. The OSC
model is characterized by a single timescale corresponding to the slow
process at zero temperature (see section \ref{esm:nodisorder:osc}). It
has been shown that, in the presence of an interaction term in the
Hamiltonian that describes a compound system formed by two ensembles of
the OSC model, the dynamics behaves in two different ways: either the
effective temperatures equalize or they are different as if the systems
were uncoupled. Dynamics remains always uncoupled (independently of the
coupling intensity) if dynamics is sequential on both systems, i.e., the
updating is done sequentially over the two ensembles $1\to 2\to 1\to 2\
....$. In this case the effective temperatures of the two systems
differ, each one corresponding to that of the non-interacting OSC
ensemble.  If dynamics is parallel, i.e., updating is done
simultaneously over the two OSC ensembles $1+2\to 1+2\to 1+2\ ....$ and
so on, then the effective temperatures of both models coincide even for
a zero value of the coupling $\eps$ in the Hamiltonian. This result
shows that, in general, two glassy systems interacting through a
coupling term in the Hamiltonian do not necessarily reach the same
effective temperature on timescales of the order of the waiting time.

In all these studies the same question remains always unanswered: why
fast and slow degrees of freedom decouple into different effective
temperatures (in two-timescales systems, one is the bath temperature,
the other the (higher) effective temperature). A necessary condition
is that the relaxation rate of the energy (or entropy production)
decays to zero slow enough \cite{CugDeaKur97}.  To better understand
this question, in \cite{GarRit01} the thermal current between the
oscillator model and a thermometer was analyzed.  There it was shown
that the measured temperature $T_{\rm m}$, which makes the net current
flow between system and the thermometer vanish, coincides with the
effective temperature if $\omega t\sim 1$. However, in the limit $\omega
t\ll 1$ the measured temperature is much smaller than the effective
temperature, while in the other extreme $\omega t\gg 1$ the thermometer
measures tolerably well the effective temperature. Discrepancies between
the measured and the effective temperature have been also reported in
the SK model in the presence of asymmetric interactions as an example of
a driven system \cite{ExaPel00}.  Another important aspect is that the
zero$^{\rm th}$ law is hardly effective as transport coefficients (such
as the thermal conductivity) are exceedingly small, in agreement with
the uncoupling of degrees of freedom occurring in glassy systems with
many timescales.
\begin{figure}
\centering
\includegraphics[scale=0.4]{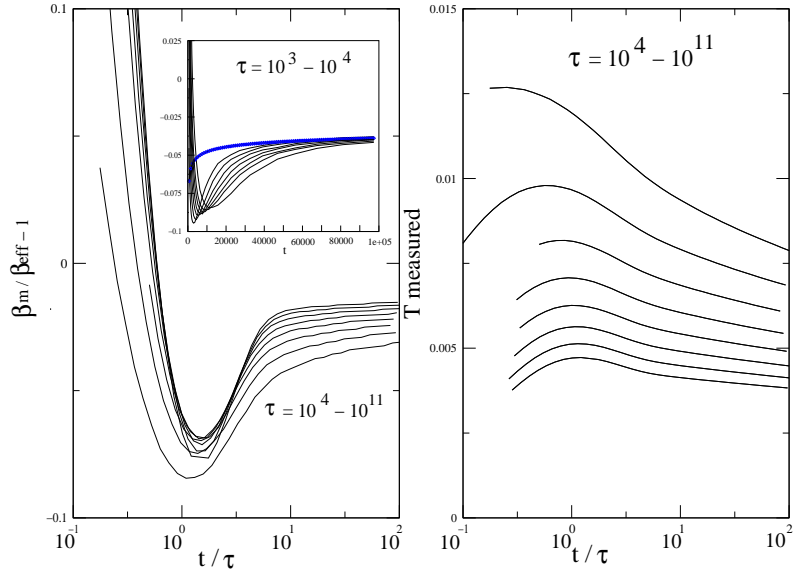}
\caption{Left panel: Relative difference between the measured temperature
  by a thermometer responding in a timescale $\tau=1/\omega$ and the
  glassy OSC model aged at $t$, plotted against $t/\tau$ for different 
  values of
  $\tau=10^4,10^5,10^6,10^7,10^8,10^9,10^10,10^11$ (from bottom to
  top). The measured temperature coincides with $\Te$ for
  $t/\tau\sim {\cal O}(1)$. In the inset, the same relative difference but
  plotted against $t$
  for $\tau=10^3,10^4$. Right panel: Measured temperature plotted as a 
  function of $t/\tau$.
  From \protect\cite{GarRit01}.
}
\label{fdr:fig1}
\end{figure}
From another point of view
Nieuwenhuizen \cite{Nieuwenhuizen98a,Nieuwenhuizen98b,Nieuwenhuizen98c} has formulated a
theory to describe the aging regime of glassy systems with
two-timescales assuming, right from the beginning, that the effective
temperature is indeed associated to a thermodynamic potential. In this
formulation, the configurational entropy is an extensive thermodynamic
potential conjugated to the effective temperature. The first law of
thermodynamics, that expresses energy conservation reads 
$dE=dW+TdS_{\rm eq}+\Te dS_c$ where $S=S_{\rm eq}+S_c$ 
is the total entropy that
receives contributions from the equilibrium (or intrastate) entropy and
the configurational entropy. The configurational entropy and the
effective temperature manifest in the reported experimental failure of
the second Ehrenfest relation (while the first is automatically
satisfied by construction) leading to values of the Prigogine-Defay
ratio larger than 1 \cite{Nieuwenhuizen97a}.

\subsection{The Edwards' measure for granular materials}
\label{edwards}

Granular materials are systems made of a large number of individual
grains such as sands or powders. At first sight granular systems look
quite different from thermodynamical systems since for example they
interact mainly through frictional forces and hence the energy is not
conserved. Moreover power is supplied to them by tapping, shearing or
shaking, all mechanisms quite different from thermal contact with a
thermal bath.  However, despite these differences, granular materials
share with thermal systems the property that their properties are
reproducible given the same set of extensive operations, i.e.,
operations acting upon the system as a whole and not on individual
grains.  For example if some sand is poured uniformly and at low density
into a container one expects to have a sand of a certain reproducible
density.  Based on these facts it is then reasonable to hypothesize that
granular systems can be described at a macroscopic level by a small
number of parameter analogous, e.g., to temperature or pressure, using
some ideas of statistical mechanics
\cite{Edwards89,EdwOak89,MehEdw89,Edwards90b,MehEdw90}.

The most important variable describing the state of a granular system
is its density, or equivalently its volume $V$. The
thermal energy of a granular system at room temperature is indeed negligible. The volume $V$ is the
actual volume occupied by the system, for example measured by the position
of a piston in the container, and hence it depends on the configurations
(position and orientation) of the grains. In principle other variables 
should be necessary to describe the state of the system, but if we assume that 
the grains are rigid then all other microscopic details can be neglected.
Thus the only valid configurations of grains are 
stable arrangements where grains can remain at rest under 
the influence of confining forces, and with no overlap among them.
We are then led to a statistical description of all possible 
stable configurations of grains in the real space, i.e., 
to a {\it configurational} statistical
mechanical theory for the random packing of grains.
Since in a stable configuration grains cannot move 
these configurations are also called {\it blocked states}.

The next step to develop a statistical mechanical description 
for granular systems consists in the introduction of a 
{\it volume function} $W$ \cite{Edwards89,EdwOak89} 
which plays the role of energy in statistical mechanics. The function
$W$ specifies the volume of the system in term of the 
configuration of grains: $V=W(\bi{q})$, where $\bi{q}$ denotes
all the grain positions and orientations in a blocked state.
Under the assumption that for a given volume $V$ all configurations 
for which $V=W(\bi{q})$ are equally probable a statistical description
can be developed through a process completely analogous to that of
conventional statistical mechanics. It is then possible to introduce a 
microcanonical ensemble with distribution function 
\begin{equation}
 {\rm e}^{-S}\,\delta(V-W(\bi{q}))
\label{eq:edw1}
\end{equation}
where $S=S(V)$ is the entropy defined as usual in terms of the total number
of blocked states
\begin{equation}
 \Omega = \int d\bi{q}\, \delta(V-W(\bi{q})), \qquad
 S = \ln\Omega
\label{eq:edw2}
\end{equation}
Similar to its thermodynamic counterpart, the entropy $S$ is an extensive
quantity as can be seen, for example, in simple toy models 
\cite{EdwOak89,MonPou97}.
The measure (\ref{eq:edw1})-(\ref{eq:edw2}) is usually called
{\it Edwards' measure} for granular systems.

To define a canonical ensemble it is necessary to define a parameter 
analogous to temperature which characterizes the state of the system.
This parameter is the {\it compactivity} $X$ defined as
\begin{equation}
 \frac{1}{X} = \frac{\partial S}{\partial V}
\label{eq:edw3}
\end{equation}
and thus the partition function is
\begin{equation}
\label{eq:edw4}
{\cal Z} = {\rm e}^{-Y(X)/X} = \int\, d\bi{q}\, {\rm e}^{-W(\bi{q})/X}
\end{equation}
where potential $Y(X)$ is called the {\it effective volume} and plays 
the role of a free-energy:
\begin{equation}
 Y(X) = V(X) - X\,S
\label{eq:edw5}
\end{equation}
The analogy can be pushed forward and many other relations similar to
that of conventional statistical mechanics can be derived. We shall not
pursuit this here, however, before concluding we shall spend some more
words on the compactivity $X$. The compactivity $X$ measures the packing
of grains, indeed from its definition it is clear that it may be
interpreted as being characteristic of the number of possible ways of
arranging the grains of a system by changing the volume in an amount
$\Delta V$, the change in entropy being equal to $\Delta
S$. Consequently the two limits of $X$ are $0$ and $\infty$
corresponding to the most compact the least compact arrangements,
respectively.  The compactivity $X$ also describes the balance between
the tendency of the system to increase or decrease its volume and the
tendency to increase or reduce its entropy.

At first sight blocked states in granular systems resemble IS discussed
for glasses. Indeed a connection between the two can be drawn
\cite{BerFraSel01,BerBarKur01,ConFieNic01,ConNic01} introducing a
``tapping''dynamics for glasses, i.e., a dynamics in which each tap
consists in raising the temperature and, after a short time, quenching
it to zero.  Similar to what has been done for other glassy systems one
can try to describe the dynamics of the slow degrees of freedom through
an effective temperature defined from the FDR
\cite{CugKurPel97,Kurchan00}.  At the mean field level this temperature
turns out to coincide with the Edwards compactivity, which is related to
the derivative of the entropy of {\it blocked configurations} of a given
density. The Edwards ensemble immediately leads to the definition of an
entropy $S_{\rm Edw}(\rho)$ as the logarithm of the number of blocked
configurations of given $\rho$.  The soundness of the Edwards approach
encounters difficulties reminiscent of those present in the IS approach.
At the present the correspondence
between the Edwards' construction and the long-time slow dynamics for
non mean-field models can only be checked ``a posteriori'' and it is
presently unknown how to derive it from first principles.

\section{QFDT from exactly solvable models} 
\label{esm}

In the structural glass problem the spatial randomness is
self-generated rather than put in by hand as in random spin glass
models.  This suggest that there should be a connection also with
frustrated but regular models.  In the last years several spin models,
both with and without randomness, displaying structural glass
transition like properties have been found. Interestingly, some of
them can be solved in closed form, offering an important tool for
understanding the glass transition.

In this section we shall summarize the main results on violation of FDT 
obtained from the exact solution of some spin models.

\subsection{The Mode-Coupling Theory}
\label{mct}

A model Hamiltonian or an effective Lagrangian capable of describing
relaxation processes in supercooled liquids and structural glasses is
difficult to obtain.  Early studies based on both dynamical
mode-coupling theories or equilibrium density-functional theories
suggested that there may be a close connection with mean-field
spin-glass models \cite{KirWolb87a}.  They thus provide a set of
microscopical models where glassy dynamics can be studied analytically.
The basic simplification occurring in mean-field models is that after
averaging over the disorder and making the number of spins very large
($N\to\infty$) one is left with a closed set of equations for the
two-times correlation and response functions.  Above the a critical
temperature $\Tc$ those equations admit a TTI
solution satisfying the FDT. In this regime they are basically
equivalent to the schematic mode-coupling equations introduced by
Leutheusser, G\"otze and others \cite{Leu84,BenGotSoj84,Got91} as a
model for the ideal glass transition. Below $\Tc$ the ergodicity is
broken and the FDT is violated.  This is signaled by the appearance of a
finite Edwards-Anderson order parameter $q_{\rm EA}$ at $\Tc$.

The fundamental quantities in the dynamical mode-coupling theory (MCT) are the 
local particle density correlation functions
$\langle\delta \rho(\bi{x},t)\,\delta \rho(\bi{x})\rangle$
where $\delta \rho(\bi{x}) = \rho(\bi{x}) - \rho_0$, 
with
\begin{equation}
\label{eq:6.0}
 \rho(\bi{x}) = \sum_{i=1}^{N}\, \delta(\bi{x} - \bi{x}_i)
\end{equation}
the local particle density, and $\rho_0$ the 
uniform fluid density
\begin{equation}
 \langle\rho(\bi{x})\rangle = \rho_0 \qquad\mbox{(homogeneous state)}
\end{equation}
where the angular brackets denote an ensemble average.
In the glassy phase the system is trapped into metastable states
with nonuniform (average) local density field
$\langle\rho(\bi{x})\rangle\not= \rho_0$ and the density-fluctuations 
correlation functions do not decay to zero for $t\to\infty$:
\begin{equation}
  \label{eq:6.1}
  \lim_{t\to\infty} \langle\delta \rho(\bi{x},t)\,\delta \rho(\bi{x})\rangle
  \not= 0.
\end{equation}
The complete mode-coupling theories lead 
to the time-evolution equations 
for the normalized correlation functions
\begin{equation}
\label{eq:6.3}
\phi_q(t) = \frac{\langle\delta\rho(\bi{q},t)^*\,\delta\rho(\bi{q},0)\rangle}
                 {N\,S_q}
\end{equation}
where $S_q= \langle|\delta\rho(\bi{q})|^2\rangle / N$ is the static
structure factor, and $\rho(\bi{q})$ are the Fourier components of the density field $\rho(\bi{x})$
\begin{equation}
\rho(\bi{q})  = \int\, \exp(-i\bi{q}\cdot\bi{x})\,\rho(\bi{x})\, d\bi{x}
  = \sum_{i=1}^{N} \, \exp(-i\bi{q}\cdot\bi{x}_i).
\label{eq:6.4}
\end{equation}
The basic idea of MCT is to derive the equations of motion for the slow 
relaxing modes integrating out all fast modes. This leads to a set of
self-consistent equations involving only slow modes variables in which all
informations from fast modes are buried into density-fluctuations 
memory kernels of the form 
\begin{equation}
 M_{q}(t) = i\nu_q + \Omega_q^2\,m_q(t)
\end{equation}
where $\nu_q$ is a (white-noise) frictional term arising from fast modes,
$\Omega_q>0$ gives the frequency or time-scale of microscopic motion and
$m_q(t)$ accounts for slow modes couplings arising from the integration
of the fast modes. The general form of the MCT equations is \cite{Got91}
\begin{equation}
\fl
\partial_t^2 \phi_q(t) + \nu_q\,\partial_t\phi_q(t) + \Omega_q^2\,\phi_q(t)
+ \Omega_q^2\int_{0}^{t}\, ds\, m_q(t-s)\, \partial_s\phi_q(s) = 0
\label{eq:6.2}
\end{equation}
which must be solved with initial conditions:
\begin{equation}
 \phi_q(t=0) = 1, \qquad \partial_t\phi_q(t=0) = 0.
\end{equation}
The fundamental mechanism for the glass transition in the MCT is the feedback 
between slow density fluctuations expressed through $m_q(t)$.
The solution of these equations is a formidable task since the kernel
$m_q(t)$ involves higher-order correlations between density-fluctuation
modes.  Therefore when these theories are implemented approximations are
generally made.  The simplest approximation consists of replacing the
average of products with products of averages to obtain a set of closed
equations. This is some sort of mean-field approximation.  Indeed
within this scheme the memory-kernel $m_q(t)$ can be expressed as a
functional of the $\phi_q$
\begin{equation}
\label{eq:6.5}
 m_q(t) = {\cal F}_{q}[{\bf V},\{\phi_q\}].
\end{equation}
with some vertex functions $\bf{V}$. Despite this rather strong approximation,
similar to a mean-field approach, the theory contains the basic mechanism
of the glass transition. We note that due to this approximation, the MCT is
not capable of describing activated process, in the same way they cannot be
discussed within mean-field theories. Therefore the appearance of 
activated-process dominated regimes  is signaled in this theory by the
divergence of some quantities. Activated process could in principle
be included as perturbative terms, however consistent theories which 
account for them are  not yet available.

The main properties of the MCT can best be seen using a simplified version 
of the theory called {\it schematic mode-coupling theory} in which only one 
relaxation function is considered \cite{Leu84,BenGotSoj84,Got91} :
\begin{equation}
\label{eq:6.6}
\partial_t^2 \phi(t) + \nu\,\partial_t\phi(t) + \Omega^2\,\phi(t)
+ \Omega^2\int_{0}^{t}\, ds\, m(t-s)\, \partial_s\phi(s) = 0.
\end{equation}
The simplest model describing an idealized structural glass transition is the 
one
specified by the two coupling constants $(v_1,v_2)$:
\begin{equation}
\label{eq:6.7}
  m(t) = v_1\,\phi(t) + v_2\,\phi(t)^2.
\end{equation}
This theory predicts a transition from an ergodic liquid phase, where 
$\phi(t\to\infty) \to 0$, to a glass phase,
where the ergodicity is broken and $\phi(t\to\infty) \to f > 0$,
as the parameter $(v_1,v_2)$ are varied. Depending on the 
values of $(v_1,v_2)$ the nature of the transition
can be either continuous (type A)
with $f$ growing continuously from zero
or discontinuous (type B) with $f$ jumping from zero to a finite value 
as the transition line is crossed.

\subsection{Disordered spin-glass models}
\label{sgm}

Mean-field spin-glass models can be classified into two broad classes
depending on the value of the Edwards-Anderson parameter ($q_{\rm EA}$)
at the transition (for the structural glass transition this can be
identified with the long-time limit of the density correlation
functions). The first class, called {\it discontinuous} models, includes
models for which a finite $q_{\rm EA}$ appears discontinuously at
$\Tc$.  The prototype model in this class, which we discuss here,
is the spherical $p$-spin model. Other models included in this family
are, Potts-glasses with more than four components
\cite{GroKanSom85,KirWolb87b}, quadrupolar glass models
\cite{GroKanSom85,GolShe85} and $p$-spin interaction spin-glass models
\cite{Der81,GroMez84,Gar85,KirThi87a,KirThi87b}. The second class
includes models, such as the Sherrington-Kirkpatric (SK) model
\cite{SheKir75,KirShe78}, for which $q_{\rm EA}$ starts continuously
from $0$ at $\Tc$. Those models are termed {\it continuous}
models.

\subsubsection{$p$-spin spherical model.}
\label{pspin}

Among the mean-field spin-glass model with a discontinuous spin glass 
transition 
an important role has been played by the spherical $p$-spin spin-glass model.
Spin-glass models with multispin interactions were first considered in the 
eighties
for both Ising \cite{Der81,GroMez84,Gar85}
as well as soft spins \cite{KirThi87a,KirThi87b}. 
However, while the static properties could be 
computed for arbitrary values of $p$, dynamical properties were limited to 
values of $p$ close to $2$ \cite{KirThi87a,KirThi87b}.   
An important step forward came with the introduction of
the spherical $p$-spin spin-glass model \cite{CriSom92,CriHorSom93}, 
since its statics and dynamics can be solved in closed form for any value 
of $p$.

The spherical $p$-spin interaction spin-glass model is defined by the
Hamiltonian
\begin{equation}
\label{eq:6.8}
\HH= -\sum_{1\leq i_1<\cdots<i_p\leq N}\, J_{i_1\ldots i_p}\,
              \sigma_{i_1}\cdots\sigma_{i_p}
        -h\sum_{i=1}^N\sigma_i
\end{equation}
where $h$ is an external field, which in the following we shall take equal to 
zero for simplicity. It describes a system of $N$ continuous spins
$\sigma_i$ interacting via randomly quenched infinite range $p$-spin
interactions $J_{i_1\ldots i_p}$ which are taken to be
Gaussian with zero mean and variance
\begin{equation}
\label{eq:6.9}
\overline{(J_{i_1\ldots i_p})^2}= {J^2 p!\over 2N^{p-1}}\;,\qquad
      1\leq i_1<\cdots< i_p\leq N\;.
\end{equation}
The overbar stands for the average over the couplings.
The scaling with $N$ is chosen such that there is a well defined
thermodynamic limit $N\to\infty$. 
The spins can vary continuously from $-\infty$ to $+\infty$, but are subject
to the global spherical constraint 
\begin{equation}
\label{eq:6.10}
\sum_{i=1}^N\sigma_i^2= N\;
\end{equation}
which must be satisfied at any time.
The relaxational dynamics for $\sigma_i(t)$ is given by the set of
Langevin equations \cite{HohHal77}
\begin{equation}
\label{eq:6.11}
  \Gamma_0^{-1}\partial_t\sigma_{i}(t)=
                - r(t)\,\sigma_i(t) -{\delta\beta \HH \over\delta\sigma_i(t)}
                        + \eta_{i}(t)
\end{equation}

The kinetic coefficient $\Gamma_0$ sets the time scale of the microscopic
dynamics, and will be henceforth set to $1$ without loss of generality, while
$\beta = 1 / T$. 
The last term in (\ref{eq:6.11}) 
$\eta_i(t)$ is a Gaussian random field with zero mean and
variance
\begin{equation}
\label{eq:6.12}
\langle \eta_i(t)\,\eta_j(t') \rangle= 
    2\Gamma_0^{-1}\,\delta_{ij}\,\delta(t-t')
\end{equation}
representing the effects of thermal noise. The average over thermal noise 
is denoted as usual by angular brackets 
$\langle\cdots\rangle$.
 The first term in (\ref{eq:6.11}) enforces the spherical constraint
at any time, and must be fixed self-consistently. 
In the mean-field limit $N\to\infty$ the sample-averaged dynamics is entirely 
described by the evolution of the two-times correlation and linear 
response functions
\begin{equation}
\label{eq:6.14}
C(t,s)= \overline{\langle\sigma_i(t)\,\sigma_i(s)\rangle}
\end{equation}
\begin{equation}
\label{eq:6.15}
 R(t,s)= \left.\frac{\delta\,\overline{\langle\sigma_i(t)\rangle}}
               {\delta h_i(s)}\right|_{h_i=0}
\end{equation}
which vanishes for $t<s$ due to causality.  The dynamical equations for
$C$ and $R$ are obtained from \eq{eq:6.11} \cite{CriHorSom93} 
through standard
functional methods \cite{SomZip82,KirThi87a,KirThi87b}.
At high temperatures the system is ergodic, thus for initial time
$t_{\rm i}\to-\infty$ both TTI and FDT hold.
Using the FDT relation the equation for $C(t-s)$
reads
\cite{CriHorSom93}
\begin{equation}
\label{eq:6.22}
\partial_\tau C(\tau) + C(\tau)+ 
  \int_0^\tau d\tau'\, m(\tau-\tau') \, \partial_{\tau'} C(\tau') = 0
\end{equation}
where  $\tau=t-s$ and
\begin{equation}
\label{eq:6.23}
m(\tau) = \mu\, C(\tau)^{p-1}
\end{equation}
which has the same structure of the schematic MCT equation 
(\ref{eq:6.6}). The correlation 
always decays to zero for large $t$. However slightly 
above $\Tc$ is develops a plateau at $C\sim q_{\rm EA}$ before eventually 
decaying to zero, see Figure \ref{fig:ct-p3}. 
The length $\tau_p(T)$ of the plateau increases as a power of 
$T-\Tc$ and diverges at $\Tc$. Near the
plateau one finds that:
\begin{eqnarray}
 C(\tau) &\sim q_{\rm EA} + c_a\,\tau^{-a}, \qquad 
      &C \gtrsim q_{\rm EA}
\label{eq:6.16a}\\
 C(\tau) &\sim q_{\rm EA} - c_b\,\tau^{b}, \qquad 
      &C \lesssim  q_{\rm EA}
\label{eq:6.16b}
\end{eqnarray}
where the exponent $a$ and $b$ are related by:
\begin{equation}
\label{eq:6.19}
 \frac{\Gamma^2(1-a)}{\Gamma(1-2a)} = 
 \frac{\Gamma^2(1+b)}{\Gamma(1+2b)} =  
  \frac{(p-2)(1-q_{\rm EA})}{2q_{\rm EA}}
\end{equation}
The plateau length scale $\tau_p(T)$ sets the equilibration
time-scale. Therefore, as the temperature is lowered down to $\Tc$ the
system undergoes a transition since the length of the plateau diverges
and the correlation fails to decay to zero.  In the low-temperature
spin-glass phase ($T<\Tc$) the system cannot equilibrate
and the ergodicity is broken so the state of the system may depend on its
initial state.  In this scenario it is clear that even a mean-field
theory can be highly non-trivial.
\begin{figure}
\centering
\includegraphics{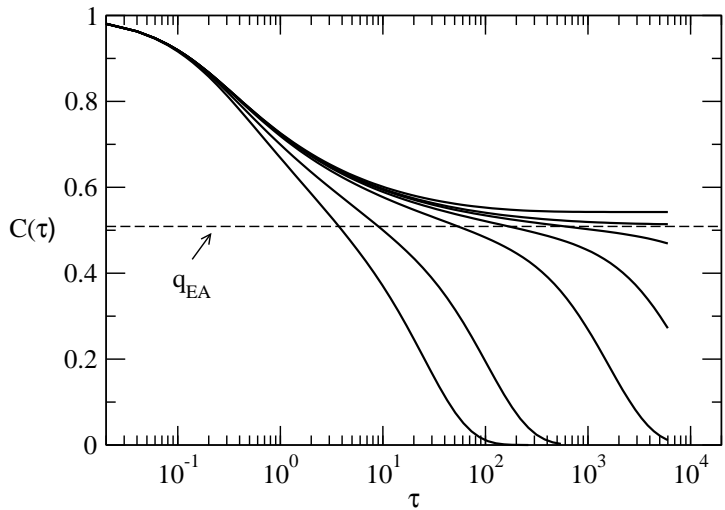}
\caption{$C(\tau)$ as function of time for the spherical $p$-spin model
  with $p=3$. Left to right: $T=0.7$, $0.65$, $0.62$, $0.615$,
  $0.613$, $0.612$ and $0.61$.
The horizontal line is $q_{\rm EA}(T=\Tc)\sim 0.509034$
  and $\Tc=0.612372$
}
\label{fig:ct-p3}
\end{figure}
To discuss the dynamics in the low-temperature 
phase it is convenient to have an
understanding of the low-temperature structure of the phase-space.  A
standard method to deal with such a problem is the so called 
{\it replica trick}, see e.g. \cite{MezParVir87}.  
The breaking of ergodicity 
results in a breaking of the permutation symmetry between
replicas \cite{Parisi83}. The general form of the breaking is, however,
not known. For $p$-spin models it has been found that the solution is 
given by the so called {\it one-step replica symmetry breaking} (1RSB) form
\cite{GroMez84,Gar85,KirThi87b,CriSom92}. Physically this means that the
phase space is broken into equivalent ergodic components separated by
infinite (for $N\to\infty$) barriers and the equilibrium solution 
is described by only three parameters: the overlap between two different 
ergodic components, the overlap inside an ergodic component 
(the Edwards-Anderson order parameter) and the probability that two 
different replicas will be found in the same ergodic component.  
For the $p$-spin spherical model
this solution is valid everywhere in the low-temperature phase
\cite{CriSom92}. Ising-spin $p$-spin models present an even lower temperature
phase with a more complex structure described by an {\it infinite step
replica symmetry breaking} ($\infty$-RSB) \cite{Gar85}. A similar
scenario is also found in Potts glasses with more than four components
\cite{GroKanSom85}.

Besides the replica approach, which gives essentially information on the
lowest-lying states which mostly contribute to the equilibrium measure,
a good understanding of the landscape topology is given by the
TAP approach. Using these methods we have now a
rather good knowledge of the landscape
\cite{KurParVir93,CrisSom95,FraPar95,FraPar97,CavGiaPar97,BarFraPar97,CavGiaPar97b}.
Roughly speaking the picture that emerges for the $p$-spin spherical
model is that equilibrium states, whose energy and free energy
difference is $O(1)$, are separated by $O(N)$ barriers. As the
temperature is changed the solutions neither merge nor bifurcate, and
their free energy smoothly changes.  The TAP equations for the $p$-spin
spherical model have non-trivial solutions in the free energy range
$(F_{\rm 1rsb}, F_{\rm thr})$, where the threshold free energy $F_{\rm thr}$ 
is larger that the equilibrium free energy $F_{\rm 1rsb}$ by an $O(N)$ 
quantity,
the difference being the complexity, see section \ref{conf:entr}, which is 
maximal for $F=F_{\rm thr}$ and vanishes when $F=F_{\rm 1rsb}$.
Below
the threshold the equilibrium states are local minima of the TAP free
energy separated by $O(N)$ barriers, while above the
threshold there are no minima.

To have a meaningful investigation of the non-ergodic phase some
regularization scheme of the dynamics on a very long time-scale is
required.  One possibility is to refer to a large finite system.  
The finiteness of $N$ guarantees ergodicity by
allowing the penetration of barriers whose height would diverge for
$N\to\infty$ limit, and the system can equilibrate. 
This approach was first proposed by Sompolinsky for
the Sherrington-Kirkpatrick model \cite{Sompolinsky81}. 
In the Sompolinsky's scheme TTI holds and FDT is satisfied up to some 
time-scale $t_0$, which diverges as $N\to\infty$, 
related to the (free)energy barrier crossing
but it is violated for time-scales larger than $t_0$
where it is replaced by a modified form called
{\it quasi fluctuation-dissipation theorem} (QFDT).
See section \ref{sk} for a more detailed discussion of the Sompolinsky's 
scheme.
In order to consider the motion on the two different 
time regime ($t\ll t_0$ and $t\gg t_0$) one writes \cite{Horner84,CriHorSom93} 
\begin{equation}
\label{eq:6.25}
C(\tau) = C_1(\tau) + C_0(\xi), \quad
R(\tau) = R_1(\tau) + \frac{1}{t_{0}}\, R_0(\xi)
\end{equation}
where $\xi = \tau/t_0$.
The functions $C_1(\tau)$ and $R_1(\tau)$ 
describe the motion in a single ergodic 
component and vary on time-scales $\ll t_0$, while the functions
$C_0(\xi)$ and $R_0(\xi)$ describe the motion among different minima and 
hence vary on time-scales $\gg t_0$.
 Continuity imposes 
\begin{eqnarray}
\label{eq:6.26}
 C_1(\tau=0) &= 1 - q_{\rm EA}, \quad &C_1(\tau\to\infty) = 0
\\
\label{eq:6.27}
 C_0(\xi=0) &= q_{\rm EA}, \quad     &C_0(\xi\to\infty) = 0
\end{eqnarray}
thus $C_0$ describes the slow decay of correlation function 
from $q_{\rm EA}$ to zero. The initial conditions for $R_0,R_1$ are
obtained from the FDT, the QFDT and the continuity condition for $C$.

On time-scales $\tau\ll t_0$ FDT holds, so $C_1(\tau)$ obeys a dynamical
equations similar to (\ref{eq:6.22}). The equations for $C_0(\xi)$ and
$R_0(\xi)$ are more complex, however it can be proved \cite{CriHorSom93}
that they are related by the QFDT relation
\begin{equation}
\label{eq:6.30}
 R_0(\xi) = -\beta x\,\theta(\xi)\,\partial_{\xi}C_0(\xi)
\end{equation}
with $0\leq x\leq 1$ given by:
\begin{equation}
\label{eq:6.31}
 x = \frac{(p-2)(1-q_{\rm EA})}{q_{\rm EA}}
\end{equation}
Within the replica formalism the parameter $x$ corresponds to the location of
the discontinuity in the order parameter $q(x)$. However at difference with the
static calculation, where $x$ is fixed by the requirement of stationarity
of the replica free energy $F$ with respect to $x$, the dynamical calculation
requires $\partial_x F$ be maximal (marginal condition)
\cite{CriHorSom93}. This condition is equivalent to the condition of 
a maximal configurational entropy
\cite{KirWolb87b,CrisSom95,Monasson95}, so that the dynamics is dominated by 
the states with the largest degeneracy (threshold states).

The Sompolinsky approach has several similarities with the static approach,
and indeed in the static limit it correctly reproduces the static
results obtained within the Parisi scheme.
However it suffers of some problems which are difficult
to amend since it would require the detailed knowledge of the
dynamics for a large but finite system. 
One among the most serious inconsistencies of the
Sompolinsky dynamics is that FDT violations, as given in
\eq{eq:6.30}, satisfy TTI. This is untenable in the aging regime of
purely relaxational systems (although not necessarily in driven systems)
where TTI is clearly violated. It must be noted though, that the
Sompolinsky approach was never proposed to explain aging, since it is
eliminated from the theory at the beginning.

For this reason there have been various attempts to
amend the solution. One possibility, proposed by Horner
\cite{Horner84,Horner86,Ioffe88,CriHorSom93}, consists in cooling the system, 
at finite cooling rate the system from $T>\Tc$ to $T<\Tc$. This introduces a
regularization time scale, the inverse of cooling rate, which is eventually 
sent to infinity at the end. Another approach \cite{Horner84}, consist
in making the disorder time-dependent hence restoring ergodicity 
on time-scales larger than the disorder typical time-scale. 

The above methods assume in one way or another equilibrium, thus cannot
describe non-equilibrium properties typical of glasses such as aging,
see section \ref{aging}.  To tackle them a different scheme has been
proposed by Cugliandolo and Kurchan \cite{CugKur93}.  The main
difference lies in that the thermodynamic limit is taken {\it before}
any large time limit, including the initial limit $t_{\rm i}\to
-\infty$. At difference with the Sompolinsky's approach this is a
non-equilibrium scheme since the system evolves from a non-equilibrated
initial configuration.
As the system evolves in time the dynamical
 free-energy density decreases, and the systems explores an ever
 decreasing portion of phase space. 
The {\it weak ergodicity breaking} \cite{Bouchaud92} describes this 
non-equilibrium regime before equilibrium is reached.
In this scenario the important parameter is the {\it waiting time} $t_w$,
i.e.,  the time elapsed since the quench into the
 low-temperature phase. For longer waiting times the system can explore
 deeper minima becoming less susceptible to external perturbations an
 hence ages. The weak ergodicity breaking scenario can be summarized in 
the following assumptions.

\begin{enumerate}

\item 
After any time $t_w$ the system continues to drift away and asymptotically 
reaches the 
maximum  allowable distance. Thus the correlation functions satisfy
\begin{equation}
\label{eq:6.32}
 \partial_\tau\,C(\tau+t_w, t_w) \leq 0, \quad
 \partial_{s}\,C(t,s) \geq 0, \quad (t>s)
\end{equation}
and in the absence of external magnetic fields
\begin{equation}
\label{6.33}
  \lim_{\tau\to\infty} C(\tau+t_w,t_w) = 0 \quad \mbox{for any fixed $t_w$}
\end{equation}
\item
The response to a constant small magnetic field applied from $s=0$ to
$s=t_w$ -- i.e., the TRM $M^{\rm TRM}(t,t_w)$-- decays to 
zero after long enough times
\begin{equation}
\label{eq:6.34}
 \lim_{t\to\infty}\int_{0}^{t_w} dt'\, R(t,t') = 
 \lim_{t\to\infty}M^{\rm TRM}(t,t_w)  = 0 
\end{equation}
for any fixed $t_w$.
\item
The evolution of the two-times correlation function presents two distinct 
regimes. After a long time $s$, but $\tau = t-s$ small,
the correlation decay from $1$ at equal time to a plateau value $q_{\rm EA}$
defined as:
\begin{equation} 
\label{eq:6.24}
 q_{\rm EA} = \lim_{\tau\to\infty}\lim_{s\to\infty} C(\tau+s,s).
\end{equation}
This {\it fast} decay corresponds to a fast relaxation toward a local
minimum.  In this time-sector the system behaves as if it were in a local
equilibrium, and both TTI and FDT hold.  The value of $q_{\rm EA}$
measures the size of the local minima or, equivalently, the width of the
channel through which the system evolves.  This fast relaxation is
followed by a {\it slow} decay of $C$ below $q_{\rm EA}$ and the
exploration of different minima. Since the depth of minima increases with
time, $C$ decays from the plateau in a manner that depends on both $s$
and $\tau$.  
To show the two processes (i.e. the contribution from fast and
slow motion) the response and correlation functions are split in two
different terms, in a fashion similar to that used in the Sompolinsky scheme:
\begin{equation}
\label{eq:6.35}
\fl  \qquad C(t,s) = C_{\rm st}(t-s) + C_{\rm ag}(t,s), \quad
  R(t,s) = R_{\rm st}(t-s) + R_{\rm ag}(t,s)
\end{equation}
with
\begin{eqnarray}
\label{eq:6.36}
 &C_{\rm st}(t-s=0) = 1 - q_{\rm EA}, \quad 
 &C_{\rm st}(t-s\to\infty) = 0
\\
\label{eq:6.37}
 &C_{\rm ag}(t,t) = q_{\rm EA}, \quad 
 &\lim_{t\to\infty} C_{\rm ag}(t,s) = 0.
\end{eqnarray}
\end{enumerate}
The assumption of local equilibrium implies that FDT is satisfied by the
fast motion:
\begin{equation}
\label{eq:6.38}
 R_{\rm st} (t-s) = \beta\, \theta(t-s)\,\partial_{s}\, C_{\rm st}(t-s).
\end{equation}
On long time-scales, however, FDT is violated and replaced by 
[cfr. eq (\ref{eq:6.30})]
\begin{equation}
\label{eq:6.39}
 R_{\rm ag}(t,s) = \beta\, X[C_{\rm ag}(t,s)]\,\theta(t-s)\,
                     \partial_{s}\,C_{\rm ag}(t,s)
\end{equation}
with the {\it Ansatz} that $X$ depends on times only through $C_{\rm ag}$.
The two forms of FDT can conveniently condensed into one extending the
definition of $X$ as $X(z) = 1$ for $q_{\rm EA}\leq z\leq 1$. Then we can write
\begin{equation}
\label{eq:6.40}
 R(t,s) = \beta\, X[C(t,s)]\,\theta(t-s)\,
                     \partial_{s}\, C(t,s)
\end{equation}
where $C$ and $R$ are the full correlation and response functions.
For the FDT part the MCT-like equations (\ref{eq:6.22}) are recovered.
To derive the equations for the non-FDT part the time derivatives 
are neglected since $C_{\rm ag}$ and $R_{\rm ag}$ are slow varying 
functions:
\begin{equation}
\label{eq:6.43}
 \partial_t\, C_{\rm ag}(t,s) \sim \partial_{s}\, C_{\rm ag}(t,s) \sim 0 
 \quad \mbox{for large $t,s$}.
\end{equation}
As a consequence the solutions are time-reparametrization
invariant, i.e., the solutions are invariant under the transformation
\begin{equation}
\label{eq:6.44}
\fl \qquad   C_{\rm ag}(t,s) \Rightarrow C_{\rm ag}\bigl(h(t),h(s)\bigr), \quad
  R_{\rm ag}(t,s) \Rightarrow \left[\frac{dh(s)}{ds}\right]\, 
                  R_{\rm ag}\bigl(h(t),h(s)\bigr)
\end{equation}
where $h(t)$ is an arbitrary (well-behaved) function.  The full
dynamical solution obviously does not have such an invariance. 
This ambiguity stems from the fact that equations 
are only the first order equations of an asymptotic
expansion. If higher order terms are included the ambiguity is removed, 
however we shall not discuss this problem.
Motivated by the fact
that the relevant time scale in slow relaxation motion is 
given by the waiting time, one looks for a solution of the form 
\cite{CugKur93}:
\begin{equation}
\label{eq:6.45}
C_{\rm ag}(t,s) \to C_{\rm ag}\Bigl(h(t) / h(s) \Bigr).
\end{equation}
The selection of the correct function $h(t)$ is still an open problem
that requires the matching of the short and long-time regimes. 
Numerical  solution of the dynamical equations \cite{CugKur93}
suggest a power law $h(t)\sim t^{\lambda}$.

An analysis of the correlation 
function near the plateau $q_{\rm EA}$ \cite{CugDou96}, similar to that done in
the high-temperature phase \cite{CriHorSom93}, reveals the following scenario.
As found in the high-temperature phase, the decay of the correlation function 
to the plateau $q_{\rm EA}$ is a power law with a temperature
dependent exponent. The subsequent departure from the plateau is still 
a power law with another temperature dependent exponent, 
but at difference with the high-temperature phase it is also
$t_w$ dependent:
\begin{eqnarray}
 C(\tau + t_w,t_w) &\sim q_{\rm EA} + c_a\,\tau^{-a}, \qquad 
      &C \gtrsim q_{\rm EA}
\label{eq:6.48a}\\
 C(\tau + t_w,t_w) &\sim q_{\rm EA} - 
               c_b\,\left(\frac{\tau}{{\cal T}_w}\right)^{b}, 
         \qquad 
      &C \lesssim q_{\rm EA}
\label{eq:6.48b}
\end{eqnarray}
with ${\cal T}_w = [ d\ln h(t_w)/ dt_w]^{-1}$ an effective waiting time.
The exponent $a$ and $b$ are related:
\begin{equation}
\label{eq:6.49}
 \frac{\Gamma^2(1-a)}{\Gamma(1-2a)} = 
 X\, \frac{\Gamma^2(1+b)}{\Gamma(1+2b)} =  
  \frac{(p-2)(1-q_{\rm EA})}{2q_{\rm EA}}
\end{equation}
where $q_{\rm EA}$ is determined by the marginal condition 
\begin{equation}
\label{eq:6.46}
 \mu (p-1)\,q_{\rm EA}^{p-2}= \frac{1}{(1-q_{\rm EA})^2}
\end{equation}
\begin{equation}
\label{eq:6.47}
 X(C) = \frac{(p-2)(1-q_{\rm EA})}{q_{\rm EA}} \quad \mbox{if $C<q_{\rm EA}$}
\end{equation} 
and $X(C) = 1$ otherwise.

Here we did not consider the case of non-zero external field, this was
studied in \cite{CriSom92,CriHorSom93}. Another extension of the 
$p$-spin spherical model is the case of multiple phases 
treated in \cite{Nieuwenhuizen95b,HerSheNie99,CiuCri00,CriHerNieShe03}.
It is interesting to note that depending on the degree of non-linearity of the
interaction three different scenarios for the transition can be observed
\cite{CiuCri00}. Finally, it is worth to note that the 
$p=2$ version of the model has the property of being solvable even for intermediate 
time-scales \cite{CiuDep88,CugDea95, CugDea95b,CamParRan98,ZipKuhHor00}
or finite sizes \cite{CiudePasMon90}.

\subsubsection{The Sherrington-Kirkpatrick model.}
\label{sk}
The Sherrington-Kirkpatrick (SK) model \cite{SheKir75,KirShe78} 
belongs to the class of continuous 
spin-glass models characterized by a low-temperature spin-glass (SG) phase of
$\infty$-RSB type with a continuous order parameter function
$q(x)$. The transition to the SG phase is continuous with a $q(x)$ which grows
continuously from zero as $T$ is lowered below $\Tc$. Other models in this 
class are, e.g.,  the case of a particle in a long-range correlated random
potential \cite{KinHor93} or spherical models with mixture of $p=2$ and 
$p>3$ interactions \cite{Nieuwenhuizen95b,CiuCri00}.
For $T>\Tc$ and large times the correlation function decays to zero 
in the absence of external fields. However, at difference
with the discontinuous SG models, it does not exhibit a plateau at
$q_{\rm EA}$ for $T$ slight above $\Tc$.
The SK model without external fields is defined by the
Hamiltonian
\begin{equation}
\label{eq:6.50}
\HH= -\sum_{i<j}\, J_{ij}\,
              \sigma_{i}\sigma_{j}
\end{equation}
where the interaction $J_{ij}$ are 
independent quenched Gaussian variables of zero mean and variance
\begin{equation}
\label{eq:6.51}
\overline{(J_{ij})^2}= {J^2\over N}.
\end{equation}
As done previously $J$ can be set equal to $1$ by rescaling the temperature.
The spin variables can be either Ising spins ($\sigma=\pm 1$) or soft spins 
in which case and extra term may be added to the Hamiltonian to
control fluctuations. The first choice is used in static calculations,
while the second is preferred in a dynamical approach.

The $\infty$-RSB structure of the SG phase reflects a complete different
topology of the phase space. Again below $\Tc$ the phase space is broken into
a large number (exponentially in $N$) states, but now 
the overlap between states can take any value from $0$ to $q_{\rm EA}$.
The equilibrium states are organized in an ultrametric fashion with 
non-extensive barriers -- $O(N^{\alpha})$ with $\alpha\sim 1/3$ -- 
between them.
The TAP analysis shows that the TAP solutions tends to split as the 
temperature is lowered is a fashion similar to a second-order transition.
Moreover the spectrum of the Hessian matrix of the solution extends down to
zero, leaving the possibility of finite free-energy barriers.  
All these facts lead to a dynamical scenario quite 
different from that discussed above in section \ref{pspin}. 

The relaxational dynamics of the soft-spin version is given by 
Langevin equations similar to (\ref{eq:6.11})-(\ref{eq:6.12}) and
the self-consistent 
dynamical equations for the two-times correlation and response functions
were first derived by Sompolinsky and Zippelius \cite{SomZip81,SomZip82}. 
These are more involved than those of the spherical case because spin
variables cannot be integrated away and hence will not be reported here.
However, near the critical point the dynamical equations can be written
in the MCT form (\ref{eq:6.22}), (\ref{eq:6.5}) with suitable 
$v_1$ and $v_2$ \cite{SomFis85}. The two parameters $v_1$ and $v_2$ are
not independent and their particular form leaves only 
the possibility for the type A transition  with $q_{\rm EA}$
growing continuously from zero at $\Tc$.

Above the critical temperature  $\Tc$ no ergodicity breaking occurs, 
and the solution are TTI and satisfy FDT.
Below $\Tc$ the ergodicity
is broken and some scheme must be adopted for the long-time dynamics.
In the Sompolinsky approach \cite{Sompolinsky81}, as discussed in the
preceding section, one assumes that the initial time is sent to
$-\infty$ keeping the system size large but finite so that the
system can equilibrate, and only then $N$ is sent to $\infty$.  
Two-time quantities as correlations and response are then trivially 
TTI, but FDT may not be necessarily satisfied for the infinite
size system due to the emergence of infinitely high (free)energy barriers
for $N\to\infty$ where freezing of some degrees of freedom 
confine the system to a portion of the available phase space. 
In the finite system, on the contrary, 
all barriers can be surmounted in a finite (but large) time, so the 
degrees of freedom can be frozen only for times smaller than the typical
time-scale for barrier crossing.
The large number and complex structure of states in the phase space 
led Sompolinsky
to postulate a set of very long time-scales,
eventually diverging for $N\to\infty$, for
(free)energy barrier crossing. 
The times are organized hierarchically, i.e., denoting them with  
$t_x$, where $x$ is an index varying for convenience in
$[0,1]$, then 
\begin{equation}
\label{eq:6.53}
  \lim_{N\to\infty} t_x = \infty, \quad \mbox{but}\quad
  \frac{t_{x'}}{t_x}\to\infty \quad\mbox{if $x  > x'$}.
\end{equation}
to account for the ultrametric organization of states.
When the system is observed at time $t_x \ll t \ll t_{x'}$ 
all degrees of freedom with relaxation times $t_{x}$ shorter then $t$ 
will have relaxed completely, while those with longer relaxation times will 
remain essentially  frozen and cannot contribute to the response 
at time $t$ to an external
perturbation at time zero. The FDT must then be modified to account for
the missing contribution of the frozen degrees of freedom. This leads to an
{\it anomalous response term} in the response function 
which measures the degree of FDT violation.
Since barriers with time-scale larger than $t=t_x$ cannot be crossed the
correlation function cannot decay to zero but relaxes to  
\begin{equation}
 \label{eq:6.54}
 q(t_x) = \overline{\langle \sigma_i(t_x)\,\sigma_i(0)\rangle}
\end{equation}
In the thermodynamic limit these partially relaxed states
will become stable states of typical size
$q(x) = q(t_x)$ since barriers with time-scale larger than $t_x$ 
cannot be surmounted anymore, 
while those with time-scale smaller than $t_x$ have been already 
crossed several times.
The overlap $q(x)$ is a non-decreasing function of 
$x$ and corresponds in the thermodynamic limit to the Parisi's order 
parameter function of the static calculation.
The correlation time-persistent part $q(t_x)$ can be
written as a sum of contributions from unrelaxed degrees of freedom
\cite{Sompolinsky81} [cfr. (\ref{eq:6.25}) for one time-scale]:
\begin{equation}
\label{eq:somp3}
 q(t_x) = \sum_{x'<x} q'_{x'}
\end{equation}
Since the system is equilibrated for time-scales smaller that $t_x$ 
FDT must be satisfied on that time-scales. The presence of a 
time-persistent part in the correlation function and the requirement 
of FDT for time-scales smaller that $t_x$ leads to 
an extra term [cfr. (\ref{eq:6.25}) for one time-scale], 
called {\it anomalous response term} and denoted by $\Delta'_x$,
in the response function \cite{Sompolinsky81,Sommers83}.

Like Parisi's, Sompolinsky's derivation of the self-consistent equations
for the overlap and the anomalous response term is heuristic but allows
for a dynamical theory which presents many similarities with the static
solution.  At difference with the usual Parisi's solution, however, the
Sompolinsky's solution is expressed in terms of two order parameter
functions: the overlap $q(x)$ and the integrated response function
$\Delta(x)$, sum of the anomalous response function terms.  This extra
freedom, called ``gauge invariance'', reflects the
time-reparametrization invariance of the Sompolinsky's solution: any
reparametrization $t_x\to h(t_x)$ where $h(t)$ is an arbitrary well
behaved function preserving the relations (\ref{eq:6.53}) will lead to
an acceptable solution.  We have already encountered this properties
when discussing non-equilibrium solutions of the spherical $p$-spin
model in section \ref{pspin}. This fact should not be surprising since
it just reflects our missing of knowledge on how the barrier-crossing
time-scales diverge in the thermodynamic limit. Viceversa we may also
say that this invariance is intrinsic in any mean-field solution since
the details on how barriers diverge are irrelevant, the only important
point is that they diverge. We note indeed that the gauge-invariance of
the Sompolinsky's solution has its static counterpart in the invariance
of the replica solution under replica permutations. The Parisi's
solution corresponds to the ``special'' gauge \cite{Sompolinsky81}
\begin{equation}
\label{eq:somp4}
 \Delta'(x) = -\beta\,x\,q'(x)
\end{equation}
where the prime means derivation, 
which relates the anomalous response term to the derivative of the
overlap (or time-persistent correlation) at time-scale $t_x$
[cfr. (\ref{eq:6.39})]. Equation (\ref{eq:somp4}) known with the name
of {\it Parisi's gauge} is actually an FDT relation.

As already noted in section \ref{pspin}  the Sompolinsky's solution, 
and its variants, are equilibrium solutions and cannot account for 
aging phenomena found in spin-glass experiments. To deal with them
Cugliandolo and Kurchan have proposed a 
non-equilibrium scheme in which the thermodynamic 
limit $N\to\infty$ 
is taken before any large-time limit, including the initial
time limit $t_{\rm i} \to-\infty$, to force the system
to a non-equilibrium state \cite{CugKur93}.  
This procedure drives the system 
to a non-equilibrium regime named
weak ergodicity breaking
in which TTI is lost and the system displays aging, 
see section \ref{pspin} for more details.
The TTI is recovered only for small time separations (the so called
FDT regime) where the dynamics is described by 
Sompolinsky-like equations and the correlation function 
approaches the plateau $C= q_{\rm EA}$ with the power law  
form (\ref{eq:6.48a}) with a temperature dependent 
exponent $a$. 

The departure from the plateau, i.e., the aging or non-FDT regime, 
is more complex since the presence of many different 
time-scales related to different (free)energy barriers
must be taken into account, and 
representations like (\ref{eq:6.45}) or (\ref{eq:6.48b}) cannot be
adequate.
Adapting the Sompolinsky's picture of many time-scales to the weak ergodicity 
breaking regime the non-equilibrium 
relaxation from time $s$ to time $t$ is due to the crossing of
(free)energy barriers with time-scales between $s$ and $t$. 
Then, using the time-reparametrization invariance of mean-field solutions,
(\ref{eq:6.45}) is replaced by 
the asymptotic form valid for large times \cite{CugKur94}
[cfr. (\ref{eq:somp3})]:
\begin{equation}
 \label{eq:6.55}
 C_{\rm ag}(t,s)  \sim \sum_i\, C_{\rm ag}^{(i)}\Bigl(
                           h_i(t) / h_i(s)
                                          \Bigr)
\end{equation}
where each contributing term $C_{\rm ag}^{(i)}$ will vary in each
separate time sector defined by two successive barrier crossing and, as
in the case of one time-scale, the functions $h_i(t)$ could be power law
with a time sector dependent exponent $h_i(t) \sim t^{\lambda_i}$.
Using the fact that correlations decrease as times become more and more
separated, it is possible to show that for large times the following
relation must hold
\begin{equation}
\label{eq:6.56a}
 C(t_3,t_1) = f[C(t_3,t_2), C(t_2,t_1)], \quad 
 t_1\leq t_2\leq t_3
\end{equation}
where $f$ is an associative function which defines the geometry of the 
triangles described by the trajectory in the phase space \cite{CugKur94}.
Next one defines the fixed point $q_i$ of $f$ as $f(q_i,q_i) = q_i$.
The intermediate value of the correlation between two successive fixed points 
defines a time-sector. 
We note that triangles
whose side belongs to different time-scales, e.g., $C(t_3,t_2) < q_i$ but
$C(t_2,t_1) > q_i$, are isosceles with 
$C(t_3,t_1) = \min[C(t_3,t_2), C(t_2,t_1)]$. This defines
an ultrametric geometry analogous to what is found in equilibrium calculations:
\begin{equation}
\label{eq:6.57a}
 C(t_3,t_1) = \min [C(t_3,t_2), C(t_2,t_1)], 
\end{equation}
if at least one $C(t_3,t_2)$, $C(t_2,t_1)$ is less then $q_{\rm EA}$.

The set of fixed points $q_i$ can be either discrete or continuous.  In
the latter case the correlation \eq{eq:6.55} is the limit case of a
continuous sum of infinitely many scaling functions $C_{\rm ag}^i$.  As
for the Sompolinsky's approach, the term (\ref{eq:6.55}) in the
correlation function implies an analogous term in the response function
[see (\ref{eq:6.35})] which can be related to the correlation function
through (\ref{eq:6.39}) with an $X(C)$ no more constant for $C<q_{\rm
EA}$ but which coincides with the function $x(q)$ of the static
treatment \cite{BalCugKurPar95}.

We note that for the spherical $p$-spin model $x(q)$ evaluated from
statics is different from that evaluated from dynamics
\cite{CriSom92,CriHorSom93,CugKur93}.  It can be shown that a sufficient
condition for the equality between the static $x(q)$ and $X(q)$ is that
the system is stochastically stable
\cite{FraMezParPel98,FraMezParPel99a}, i.e., the overlap probability
distribution $P_{\epsilon}(q)$ of the system in the presence of a small
perturbation must smoothly converge towards the probability distribution
of the unperturbed system when $\epsilon\to 0$.  Moreover the limit
$\epsilon\to 0$ must also commute with the limit of large times in the
dynamics. If this is the case, $x(q)$ and $X(q)$ are then equal. This
result holds for mean-field spin glasses with continuous RSB such as the
SK model, but not for the spherical $p$-spin model where the dynamics is
dominated by long lived metastable states.

\subsection{Random manifolds and diffusive models}
\label{other:nes}

The basic ingredient of a glass behavior is the appearance of a 
multitude of long lived states, that prevents exploration of the whole
phase space. 
This situation is not restricted to glasses but may be present in several,
apparently unrelated, far from equilibrium problems. Typical examples have 
been discussed in  the previous sections, here we shall comment a bit on 
random manifolds and diffusion models.

A typical situation where glassy behavior shows up is when studying
the dynamics of an elastic manifold, with or without internal structure,
in a random quenched environment. This problem appears, for example, 
in flux lattices in high-$\Tc$ superconductors \cite{BlaFeiGesLarVin94},
interfaces in random fields \cite{NatRuj89}, charge density waves, and 
surface growth on disordered substrates \cite{CulSha95,Nieuwenhuizen97b}.
The competition between elastic stress and disorder produces a state
with many characteristics of a glass: slow dynamics, nonlinear 
macroscopic response, aging and so on \cite{CugKurDou96,CugDou96}.
In the mean-field limit a manifold in a random media is described by 
a field theory with a large number of components. In this case it is 
possible to derive a closed set of dynamical equations of the type
discussed for spin-glass models. The model of a manifold of 
dimension $d$ embedded into a random medium of dimension $N$ is
described by the Hamiltonian
\begin{equation}
\label{eq:6.56}
 {\cal H}[\phi(\bi{x})] = \int\, d^dx \left(
              \frac{c}{2}|\nabla\bphi(\bi{x})|^2
              + V[\bphi(\bi{x}),\bi{x}] 
              + \frac{\mu}{2}\bphi(\bi{x})^2
	      \right)
\end{equation}
where the $N$ component field $\bphi = (\phi_1,\phi_2,\ldots,\phi_N)$ 
gives the displacement of the manifold. The mass term $\mu$ constrains the
fluctuations of the manifold to a restricted volume of the embedding space.
The potential term $V$ is a Gaussian variable of zero mean and
correlations
\begin{equation}
\label{eq:6.57}
 \overline{ V[\bphi,\bi{x}]\,V[\bphi',\bi{x}'] } =
 - N\,\delta(\bi{x}-\bi{x}')\, {\cal V}\left[
                                \frac{(\bphi - \bphi')^2}{N}
                                         \right]
\end{equation}
A common choice for ${\cal V}$ is 
\begin{equation}
\label{eq:6.58}
 {\cal V}(z) = \frac{ (\theta + z)^{1-\gamma} }
                    { 2(1-\gamma) }
\end{equation}
where $\theta$ is the free-energy fluctuation exponent.
The models are divided into ``long-range'' models if
$\gamma < 1$ and ``short range'' models for $\gamma > 1$
since in the first case correlations grow with distance, while in 
the second case decay.

The study of the static (equilibrium) properties 
of the $d=0$ limit \cite{Engel93}, i.e. the case of a particle moving in a 
random potential, reveals that the short-range case is solved by 
a 1RSB Ansatz, while in the opposite case of long range the 
full RSB scheme is needed.

The manifold dynamics is given by the usual Langevin equations:
\begin{equation}
\label{eq:6.59}
 \frac{\partial}{\partial t}\,\bphi(\bi{x},t) = 
 -\frac{\delta {\cal H}[\phi(\bi{x})]}{\delta \bphi(\bi{x},t)} +
  \bfeta(\bi{x},t)
\end{equation} 
where $\bfeta$ is a Gaussian random process of zero mean and 
variance
\begin{equation}
\label{eq:6.60}
 \langle \eta_{\mu}(\bi{x},t)\, \eta_{\nu}(\bi{x}',t') \rangle =
    2\,T\,\delta_{\mu\nu}\,\delta(\bi{x}-\bi{x}')\,\delta(t-t').
\end{equation}
To study the long-time dynamics one introduces two-time
quantities, which for the simple case of $d=0$ are the
usual correlation and response functions
\begin{equation}
\label{eq:6.61}
 C(t,s) = \frac{1}{N}\,\overline{\langle \bphi(t)\cdot\bphi(s)\rangle},  
 \qquad
 R(t,s) = \frac{1}{N}\, \frac{\overline{ \delta\langle\bphi(t)\rangle }   }
                              {\delta\bi{h}(s)}
\end{equation}
where $\bi{h}(s)$ is a small perturbation applied at time $s< t$.
In addition one also considers the mean-squared
displacement correlation function
\begin{eqnarray}
\label{eq:6.62}
 B(t,s) &=& \frac{1}{N}\,
           \overline{ \langle \left[\bphi(t) - \bphi(s) \right]^2 \rangle } 
\nonumber\\
         &=& C(t,t) + C(s,s) - 2\,C(t,s)
\end{eqnarray}

The analysis of the long-time dynamics for $N\to\infty$ reveals that in
both cases (long and short range models), two regimes can be defined in
the relaxation from an initial random configuration: (i) FDT regime for
large waiting time $t_w$ and not too large time difference; (ii) non-FDT
regime for large $t_w$ and time differences.  Under the assumption of a
weak ergodicity breaking scenario, the FDT in the non-FDT regime can be
replaced by the generalized form (\ref{eq:6.40}) with a function $X$
which is different for short and long-range models \cite{CugDou96}. For
short range models, $X(C)$ is solved with the two-timescales Ansatz,
i.e., it is 1 in the stationary sector while it is a constant smaller
than 1 in the aging sector. For long-range models many time-scales are
needed, and $X(C)$ is a non-trivial non-constant function, as found for
the SK model.  This scenario has been extended to the $d>0$ case in
reference \cite{CugKurDou96}.

Related studies have analyzed FDT violations in polymer models. Yoshino
 \cite{Yoshino98} considered the directed polymer model in random media
 (i.e. the random manifold \eq{eq:6.56} with $c=0$ and $N=d=1$) and
 through numerical simulations confirmed the two-timescales character of
 aging dynamics, see figure \ref{manifolds:1}.  Pittard and Shakhnovich
 \cite{PitSha01} have considered a heteropolymer model with
 uncorrelated monomer-monomer interactions.  By analyzing the
 mode-coupling equations they found a two-timescales solution that
 violates FDT as reported for the random manifold model in the
 short-range case.
\begin{figure}
 \centering
 \includegraphics[scale=0.55]{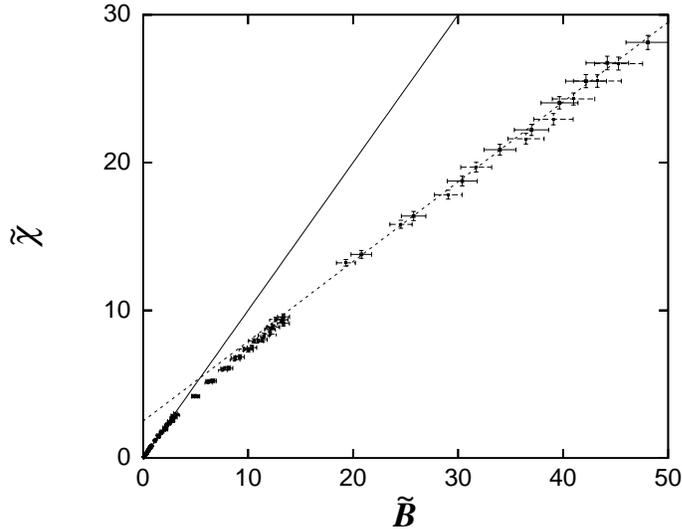}
\caption{FD plot in the model of directed polymer in random media
revealing the 1RSB or two-timescales character of the dynamics. From
\cite{Yoshino98}.}
\label{manifolds:1}
\end{figure}

The dynamics of the directed polymer reduces to pure diffusion in the
absence of disorder. This is the well known random walk which
in the continuum limit is represented by a stochastic variable $x(t)$
satisfying the (stochastic) differential equation
\begin{equation}
\label{eq:6.63}
 \frac{d}{dt}\, x(t) = \eta(t)
\end{equation}
where $\eta$ is a Gaussian random process of zero mean and correlation
\begin{equation}
\label{eq:6.64}
 \langle \eta(t)\,\eta(s) \rangle = 2\,T\,\delta(t-s)
\end{equation}
and $T$ is the bath temperature. A simple calculation
gives for the two-time correlation and response functions
\begin{equation}
\label{eq:6.65}
  C(t,s) =  \langle x(t)\, x(s) \rangle 
          = 2\,T\,\min(t,s)
\end{equation}
\begin{equation}
 \label{eq:6.66}
  R(t,s) =  \frac{\delta\langle x(t)\rangle }
                  {\delta h(s)} 
          = \theta(t-s)
\end{equation}
where $h(t)$ is an external field added to the l.h.s of
eq. (\ref{eq:6.63}).  As a consequence we see that for any $t$ and $s$
\begin{equation}
\label{eq:6.67}
 X(t,s) \equiv T\,R(t,s)\, \big/\, [\partial C(t,s) / \partial s]
         = 1 / 2
\end{equation}
a constant value different from $1$ so that the FDT is violated 
for all times. Despite violation of FDT, the model is simpler than those
discussed above, indeed  the correlation and the IRF show a rather simple
form as a function of the waiting time:
\begin{equation}
\label{eq:6.68}
\fl \qquad C(t,s) = 2\,T\,s,\qquad
 \chi(t,s) = \int_{0}^{s}\, dt' R(t,t') = s.
\end{equation}
They both depend on $s$ but not on $t$ (as required by causality).  This
is a rather extreme example, however other less trivial cases, such as
spinodal decomposition, scalar fields at criticality and so on also do
exhibit non trivial non-equilibrium behavior \cite{CugKurPar94}.  More
complicated non-linear diffusion effective models have been shown
\cite{CorNicPicCon99,CorNicPicCon01} to display FDT violations
compatible with a one-timescale aging scenario with a single valued FDR $X(t_w)<1$
that monotonically converges to $1$ like in entropic models (Section
\ref{esm:nodisorder}).

\subsection{Trap models}

\label{trap}
A successful family of models to describe the glass transition are
phenomenological trap models. The dynamics in the aging regime can be
understood in terms of jumps among different phase space components,
each jump corresponding to a new rearrangement of a cooperative
spatially localized region. The dynamics of the system can then be
viewed as an intermittent motion where some regions remain inactive for
a long time (and no net heat current is present between the system and
its surroundings) until an activated jump occurs and thermal heat is
released from the system to the surroundings, and from there, to the
thermal bath. Phenomenological trap models, contrarily to mode-coupling
theories, are based on the activated nature of glassy dynamics.
Although trap models have appeared from time to time in the
literature (see \cite{Dyre87} and references therein) the most
recent and successful study is due to Bouchaud \cite{Bouchaud92} who has
considered its relevance to describe aging phenomena in glassy systems.

The trap model corresponds to a set of unstructured energy (or free
energy) traps that live in a ``free energy space'' without any explicit
reference to real-space configurations. 
It corresponds, in many aspects, to the
coarse-grained description developed in section \ref{micro:EXTME} where
activated processes are represented as transitions between different 
phase space
components $\RR$ that here could be visualized as traps. The number of
traps, like the number of components $\RR$ in the coarse grained
description of the phase space, is exponentially large with the volume
of the system. 
The model is defined by a set of traps of different
depths $E$ (with $E>0$) with a density $\rho(E)$ and a distribution of
escape times given by the Arrhenius expression $\tau(E)=\tau_0\exp(E/T)$
where $\tau_0$ is a microscopic time and $T$ is the temperature of the
bath. Note that, in this last expression, the top-level for all barriers is
fixed at zero-height. 
The dynamics of
the trap model is then described by the ME \eq{eq2d} discussed
in section \ref{micro:EXTME} in terms of the probability function $P(E,t)$
that specifies the probability that the system stays in a trap of energy
$E$ at time $t$,
\be
\frac{\partial P(E,t)}{\partial t}=\sum_{E'}P(E',t)Z(E|E')-\sum_{E'}P(E,t)Z(E'|E)
\label{trap:1}
\ee
The rates $Z(E|E')$ are assumed to be given by \eq{eq4d},
\be
Z(E|E')=\WW(E|E')\,\rho(E)=\frac{\rho(E)}{\tau(E')};
 \qquad \int_0^{\infty}\rho(E)dE=1
\label{trap:2}
\ee
where we have identified $\Omega(\FF',T)\equiv \rho(E')$ and where the
bare rate $\WW(E|E')=1/\tau(E')$ has the dimensions of a frequency. Note
that this bare rate only depends on the energy of the departure trap but
not on the energy of the arrival trap. Other versions of the trap model
include a dependence on the energy on the arrival trap, see for instance \cite{BarMez95}.
Inserting \eq{trap:2} into
\eq{trap:1} we obtain,
\be
\frac{\partial P(E,t)}{\partial t}=\omega(t)\rho(E)-\frac{P(E,t)}{\tau(E)}
\label{trap:3}
\ee
where 
\be
\omega(t)=\int_0^{\infty}dE'\,\frac{P(E',t)}{\tau(E')}.
\label{trap:4}
\ee
The rates \eq{trap:2} satisfy detailed balance if $P^{\rm eq}(E)\propto
\rho(E)\tau(E)$, where $\lim_{t\to\infty}P(E,t)=P^{\rm eq}(E)$. The
Bouchaud trap model (BTM) \cite{Bouchaud92} is described by the distribution
$\rho(E)=(1/\Tg)\exp(-E/\Tg)$. 
The static formulation of such a model
corresponds to the random-energy model (REM) of
Derrida \cite{Derrida80,Derrida81}. 
Other trap models have considered a Gaussian
distribution of energies \cite{Dyre87,MonBou96}. However, the main
interest of the model proposed by Bouchaud is the existence of a
critical temperature $\Tg$ where the distribution 
$P^{\rm eq}(E)$ ceases
to be normalizable. In general, for any distribution $\rho(E)$ the
temperature $T_0$ that marks the onset of the 
non-normalizability of
$P^{\rm eq}(E)$ is given by \cite{BouComMon95}, 
\be
\frac{1}{T_0}= - \lim_{E\to\infty} \frac{\log(\rho(E))}{E}
\label{trap:5}
\ee
the BTM corresponding to the case $T_0=\Tg$. Let us note that, in the
trap model, energies are not extensive but finite. Comparing the BTM
with the REM, where energies are extensive, we observe that the finite
$T$ dynamics in the Bouchaud model occurs in a range of finite energies
around $E_0$, the value at which the energy freezes in the REM below
$\Tg$. The same exponential distribution of states, over a finite
free-energy interval, is found in the SK model
\cite{MezParVir85,MezParVir87}.

Dynamics in the trap model has been exhaustively investigated in many
works. In particular, it offers a rather good explanation of magnetic
relaxation phenomena observed in spin-glasses
\cite{Bouchaud92,BouDea95,BerBou02}, viscosity anomalies in glasses
\cite{BouComMon95,MonBou96,Odagaki95} and, more recently, it has been
used as a test model to check whether FDT violations are well described
by the Ansatz (\ref{eq2eb}) and whether FD plots are meaningful
\cite{FieSol02,SolFieMay02}.  Correlation and response functions can be
defined in the BTM by assigning magnetizations to the different traps as
is done to analyze the statics of the REM. The effect of
the magnetic field on the traps is to modify the escape time by the
relation, $\tau_h(E)=\tau_0\exp((E+mh)/T)=\tau(E)\exp(mh/T)$.  The
resulting FD plots have been analyzed by Sollich and coworkers
\cite{FieSol02,SolFieMay02}.  As remarked in reference
\cite{FieSol02,SolFieMay02}, equal time correlations can be  unbounded so
proper FD plots are constructed from the raw plots \eq{eq7e} by
normalizing correlations and IRF by the equal times correlation at the
later time $C_{A,B}(t,t)$ as described in \eq{eq7e2}. The ME
\eq{trap:3}, modified to include the effect of the field, is described
by a probability distribution $P(E,m,t)=P(E,t)\,\sigma(m|E)$ where
$\sigma(m|E)$ is the probability that a trap of depth $E$ has
magnetization $m$. This probability is assumed to be a Gaussian
parametrized by its mean $\overline{m}(E)$ and variance $\Delta^2(E)$.
As there is no specific meaning attached to the observable $m$ one can
think of the two functions $\overline{m}(E),\Delta^2(E)$ as describing
different class of observables.  Therefore, observable dependence in the
BTM refers to dependence of the FD plots on the choice of these
functions. The following cases have been considered
\protect\cite{FieSol02,SolFieMay02}:
$\overline{m}(E)=\exp(nE/2T),\Delta^2(E)=0$ or
$\overline{m}(E)=0,\Delta^2(E)=\exp(nE/T)$ with $n>T-1$ in both cases.
Figure \ref{fig:trap:1} shows some typical FD plots.

\begin{figure}
  \centering
  \includegraphics[scale=0.6]{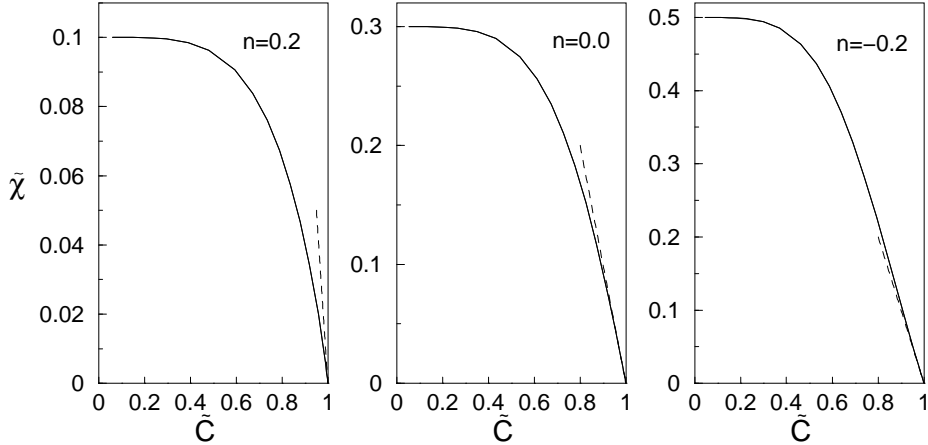}
  \caption{Normalized FD plots \eq{eq7e2}  for the BTM for a Gaussian
    distribution of trap magnetizations with zero mean and variance
    $\Delta^(E)=\exp(nE/T)$
    for different values of $n$ and $T=0.3$ ($\tau_0=\Tg=1$). For each $n$
    times shown are $10^6,10^7$ which are indistinguishable since the
    limiting FD plot has been reached. Note, however, that a
    temperature factor has been absorbed in $\tilde{\chi}$ in such a way
    that the slope is -1 at equilibrium (dashed line). 
    From \protect\cite{FieSol02,SolFieMay02}.
  }
  \label{fig:trap:1}
\end{figure}
Three are the main results of this study: 1) The FD plots strongly
depend on the average and the variance of the Gaussian distribution,
therefore the FDR and the effective temperature are observable
dependent; 2) Most importantly, for a given choice of observables the
effective temperature \eq{fdr:2} smoothly changes along time within a
given time sector. In fact, in the trap model the scaling $t/t_w$ is
fulfilled in the glassy phase $T<\Tg$ but two straight lines (typical of
two-timescales glassy systems) are not observed in the FD plots; 3) For
all observables $X_{\infty}=0$, see \eq{ferro:1}, supporting the
conjecture that this quantity is indeed universal and may have some
physical meaning (see
section \ref{ferro}).

Among these results 2) seems particularly interesting.  Why FD plots do
not display the characteristic two-step form of relaxational systems
with two-timescales?  Still for observables with zero mean and finite
(but $E$ independent) variance where $C(t,t)=\Delta^2$, i.e. for
observables that can be considered neutral (see discussion in section
\ref{neutral:obs}), the one-step shape of the limiting FD plot is
absent.  The origin of this discrepancy is presently unknown and the
finding of a trap model that shows a two straight-lines FD plot remains
an interesting open problem~\footnote{After completion of this work, it
has been shown that the influence of the dependence of the perturbed
rates in a field (upon the observable value taken at the arrival trap)
are crucial to get well defined limiting FDRs and effective
temperatures~\cite{Ritort03}}. Let us finish these considerations by
noting that FD plots, such as the one depicted in Figure \ref{fig:trap:1},
are more characteristic of systems with full RSB where $X(C)$ is a
non-trivial function. Quite
interestingly, it has been shown \cite{BerBou02} that the BTM at the
critical temperature $T=\Tg$ has correlations that do not fulfill the
simple $t/t_w$ scaling but a more complicated dependence with many time
sectors and ultrametricity. However, the resulting FD plot at
criticality shows only very small deviations from the equilibrium
behavior $X=1$ being very similar to the FD plot observed in the Ising
chain \cite{LipZan00} (see section \ref{ferro}).

Just before ending this section, let us remind that these results have
been endorsed by considering the corresponding driven version of the BTM
introduced in \cite{SolLeqHebCat97,Sollich98} in the context of
rheology. In this case, as explained in section \ref{sheared}, the
equivalent of the waiting time is the inverse of the shearing rate. In
the stationary state TTI is satisfied but FDT violations persist. For the
driven model \cite{FieSol02b}, as well as for the purely relaxational
model, the same relationship between correlations and responses holds
and the two models the non-driven and the driven one) show similar
behavior.

\subsection{Models with entropy barriers} 
\label{esm:nodisorder}

Many of the results described in the previous subsections deal with
disordered models with a complex thermodynamics. However, many aspects
of the violations of the equilibrium FDT as well as the existence of
an QFDT can be investigated in the framework of simpler models that
are exactly solvable but still retain the key ingredients for the
emergence of these new properties. In turn, this can help to
identify the basic ingredient that any sensible general theory must incorporate. 

The scope if this section is mainly illustrative as it intends to
present some of the basic ideas of previous section \ref{micro:EXTME}
applied to very simple examples. We will focus our discussion on the
oscillator (OSC) and the Backgammon (BG) models.  A comprehensive
account of other results about these models can be found in a recent
review \cite{RitSol02}.  Both models have a simple energy landscape and
dynamics is determined by the presence of entropy barriers. The
intuitive meaning of this term is the following. In general, relaxation
in glassy models proceeds by activated jumps over energy barriers that
allow the system to escape from a given trap after reaching a barrier of
height $B$, the typical time needed in this process being given by the
Arrhenius law $\tau\sim \exp(B/T)$.  Activated dynamics is strongly
temperature dependent, and for $T=0$ the dynamics is completely
arrested, the system remains trapped forever and correlations do not
decay to zero anymore (ergodicity is broken). When the dynamics is
dominated by entropy barriers the relaxational mechanism is
different. The system escapes from a trap through a process which
involves a timescale $\tau$ which does not directly depend on the
temperature but, for instance, on the typical energy $E$ of the trap
itself, $\tau(E)$ which usually is a decreasing function of the energy
$E$ (see Figure \ref{esm:nodisorder:fig1})\footnote{Now we assume the
standard convention for energies being negative rather than positive as
in the trap model.}. At $T=0$ relaxation is not arrested but proceeds
slower as the energy decreases. This type of dynamics is sometimes
called marginal dynamics \cite{KurLal95} as the system wanders around
saddle point configurations, hence is never arrested.  Of course, the
temperature dependence of the relaxation time in entropic models can
appears as effectively activated: at finite (but low) temperatures the
maximum value of $\tau(E)$ corresponds to the lowest energy reached,
i.e. $\tau(E_{\rm eq}(T))$, $E_{\rm eq}(T)$ denoting the equilibrium
energy. The temperature dependence of the relaxation time in equilibrium
$\tau(E_{\rm eq}(T))$ is activated in most entropy barrier models. A
phenomenological description of these entropy models has been introduced
by Barrat and Mezard \cite{BarMez95} who have generalized the BTM to the
case where the distribution of trapping times is itself a time-dependent
function. The oscillator and Backgammon models described in this section
are microscopic versions of this entropic trap model.
\begin{figure}
  \centering
  \includegraphics[scale=0.4,angle=-90]{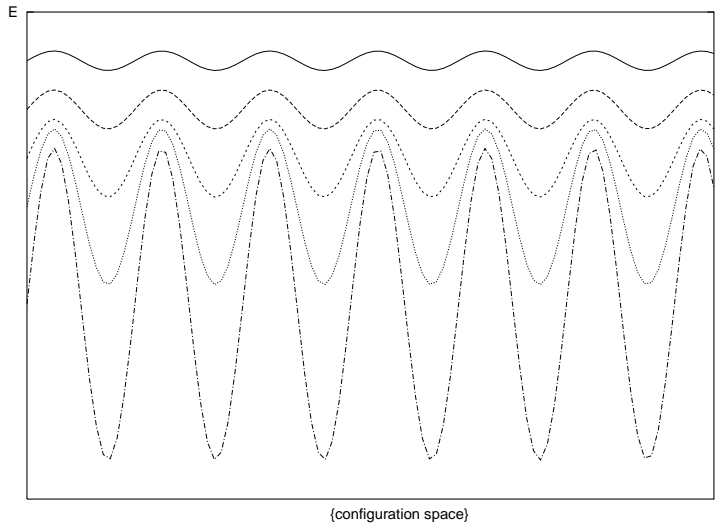}
  \caption{A typical energy landscape in a glassy model determined by
    the presence of entropy barriers. The {\em effective} barrier
    $\log(\tau(E))$
    increases as $E$ decreases.
    From \protect\cite{LeuRit02}.
  }
  \label{esm:nodisorder:fig1}
\end{figure}
According to the scenario presented in section \ref{micro:EXTME}, the role of
entropy appears to be important as relaxation in many glassy systems
is accompanied by the emergence of a non-equilibrium microcanonical ensemble 
which determines fluctuations and responses in the aging state,
leading to the existence QFDT and an effective temperature.  This fact suggests
that a deep comprehension of the glassy dynamics in exactly solvable
entropy barrier models can be a first step toward grasping the
leading aspects behind the behavior of more realistic systems, where
both entropy and energy barriers simultaneously intervene.

\subsubsection{Oscillator models} 
\label{esm:nodisorder:osc}

We begin our discussion by describing the OSC model in its simplest
version. A review of some results can be found
in \cite{CriRit00d,Garriga02,RitSol02}. Originally, oscillator
models where introduced indirectly in the analysis of the Monte Carlo
dynamics of the spherical Sherrington-Kirkpatrick model, which can be
mapped to a set of disordered harmonic
oscillators \cite{BonPadParRit96,BonPadParRit96b}. The OSC model is
obtained by simplifying the previous one to an ensemble of identical harmonic
oscillators \cite{BonPadRit98}. The OSC model is defined by the
following energy function 
\be
E =\frac{K}{2}\sum_{i} x_i^2
\label{eqESM:OSC:1}
\ee
where the $x_i$ are real-valued displacement variables for the $N$
oscillators and $K>0$ is the Hooke constant. The equilibrium properties
are trivial due to the absence of interactions, but the Monte Carlo
dynamics couples the oscillators in a nontrivial way. Moves are
proposed according to the following rule,
\be
x_i\to x'_i=x_i+\frac{r_i}{\sqrt{N}}
\label{eqESM:OSC:2}
\ee
where the $r_i$ are Gaussian random variables with zero mean and
variance $\Delta^2$. Moves are accepted according to the usual
Metropolis rule. Each move is a parallel update of the whole set of
oscillators. Both the energy function \eq{eqESM:OSC:1} and the dynamics
\eq{eqESM:OSC:2} are invariant under rotations in the
$N$-dimensional space of the $x_i$. This symmetry makes the dynamics
exactly solvable, and many non-equilibrium quantities \eg, 
effective temperatures and FDT violations can be tackled
analytically.

The OSC model is a classical model where the equilibrium entropy is
given by $S(T)=\frac{1}{2}\log(T)$, thereby diverging when $T\to 0$ as
expected for a continuous model (the ground state is a zero-measure
state corresponding to the absolute global minimum of \eq{eqESM:OSC:1},
\ie\ the configuration $x_i=0,~ \forall i$). At $T=0$ only those moves
that decrease the energy are accepted, therefore as the system
approaches the global minimum the frequency of accepted
moves \eq{eqESM:OSC:2} dramatically decreases. However, that frequency
never vanishes so dynamics is never arrested. Dynamics slows down
because phase space directions where energy decreases are exceedingly
difficult to find.
The full solution of the OSC model has been presented
in \cite{BonPadRit98}.

The main physical quantity containing detailed information about the
dynamical evolution is the probability density of energy changes
$P(\Delta E)$. This quantity expresses the probability density that a
proposed move \eq{eqESM:OSC:2} changes the total energy of the
ensemble by the amount $\Delta E$.  $P(\Delta E)$ was originally
derived in \cite{BonPadRit98} using standard integration tools.  Here
we present two other alternative derivations which help to understand the
mechanisms behind slow relaxation.

The first method relies on the Gaussian form of the distribution while
the second one uses a microcanonical argument to count the number of
accessible configurations from a reference configuration with a given
energy $E$. The first derivation is rather simple as the distribution
for $\Delta E$ can be easily obtained.  Indeed using \eq{eqESM:OSC:1}
and \eq{eqESM:OSC:2}
\be
\Delta E=\frac{K}{\sqrt{N}}\sum_{i}\,x_ir_i+\frac{K}{2N}\sum_ir_i^2     .
\label{eqESM:OSC:2b}
\ee
and the Gaussian property of $r_i$, it follows that $\Delta E$ has a 
Gaussian distribution.
The mean and variance of the distribution are $M_{\Delta
E}=\overline{\Delta E}=K\Delta^2/2$, $\sigma_{\Delta
E}=\overline{(\Delta E)^2}-(\overline{\Delta E})^2=2EK\Delta^2/N$,
yielding \cite{BonPadRit98},
\be P(\Delta E)=\bigl(4\pi eK\Delta^2\bigr)^{-\frac{1}{2}}
      \exp\left[-\frac{(\Delta E-\frac{K\Delta^2}{2})^2}{4eK\Delta^2}
           \right]
\label{eqESM:OSC:3}
\ee
where $e=E/N$ is the energy per oscillator.

The second method to derive \eq{eqESM:OSC:3} is based on a microcanonical computation.
In Figure \ref{figESM:OSC:2} we depict a schematic two-dimensional
representation of the motion of a representative configuration in phase
space. The reference configuration $\lbrace x_i^0\rbrace$ at a given
time has an energy $E$ and lies on the spherical hypersurface of radius
$R=\sqrt{2E/K}$ (depicted as the point $P$ in the figure) with center at
the origin $\lbrace x_i=0\rbrace$ (point $O$ in the figure). The smaller
dashed circle represents the region of points accessible from $\lbrace
x_i^0\rbrace$ according to the dynamics \eq{eqESM:OSC:2}. All accessible
points $\lbrace x_i \rbrace$ satisfy $\sum_i(x_i-x_i^0)^2=\Delta^2$
i.e. lie at a distance $\Delta$ from $\lbrace x_i^0\rbrace$ which is the
radius of the smaller dashed circle. The accessible configurations in a
single move lie in a spherical hypersurface of dimension $N-2$
corresponding to the intersection of the hypersurface of energy $E'$
and the smaller spherical hypersurface
of radius $\Delta$. We call this region the intersecting region $I$ as
shown in Figure \ref{figESM:OSC:2}. The final configurations contained
in $I$ lie at a distance $R'=\sqrt{2E'/K}$ to the origin $O$. The change
in energy associated to this transition is $\Delta E=E'-E$. The
probability of this jump is therefore proportional to the surface of the
intersection region, $P(\Delta E)\propto C^{N-2}$ where $C$ is the
radius of the intersecting region.  The computation of $C$ is quite
straightforward as can be deduced from the triangle including the points
P,O as vertices and whose three sides are $R,R',\Delta$. In terms of
$R,R'$ and $\Delta$, the distance $C$ is given by the relation,
\begin{equation}
  C^2=\Delta^2-\frac{K}{8E}\left( \frac{2\Delta E}{K}+\Delta^2
                           \right)^2\qquad .
\label{eqESM:OSC:6a}
\end{equation}
\begin{figure}
  \centering \includegraphics[scale=0.3]{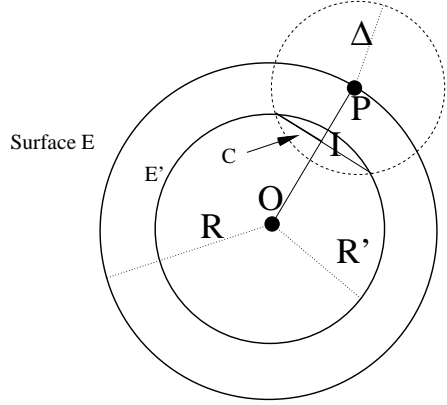} 
  \caption{Geometrical
  construction to compute $P(\Delta E)$. The thick lines denote the
  departing and final energy hypersurfaces centered at $O$. The dashed
  circle indicates the hypersurface accessible from point $P$. The intersection
  region between the accessible hypersphere centered at $P$ and the final
  hypersurface of energy $E'$ defines a hypersurface $I$ of radius $C$
  (the radius is represented by a thick line). See
  the text for explanation.}  
  \label{figESM:OSC:2}
\end{figure}
The surface  $\Omega(E,\Delta E)$ corresponding to the
region $I$ of radius $C$, relative to the energy $E$ of the reference
configuration $\bi{x}^0$ is,
\begin{equation} 
  \Omega(E,\Delta E)\propto C^{N-2} = 
              \left[\Delta^2-\frac{K}{8E}\left(
 		\frac{2\Delta E}{K}+\Delta^2\right)^2
	      \right]
	      ^{\frac{N-2}{2}}\qquad .
\label{eqESM:OSC:7a}
\end{equation}
Using the fact that $E$ is extensive with $N$ this expression can be
rewritten as,
\begin{equation}
  \Omega(E,\Delta E)\propto \exp\left[ 
             -\frac{(\Delta E-\frac{K\Delta^2}{2})^2}{4(E/N)K\Delta^2} 
	     \right]
\label{eqESM:OSC:8}
\end{equation}
which is proportional to the probability distribution
\eq{eqESM:OSC:3}.

The distribution $P(\Delta E)$ is depicted in figure
\ref{figESM:OSC:1} for different values of the energy.  As $E$
decreases the number of moves with $\Delta E<0$ shrinks as the total
area under the curve with $\Delta E<0$ decreases.
\begin{figure}
  \centering
  \includegraphics[scale=0.3]{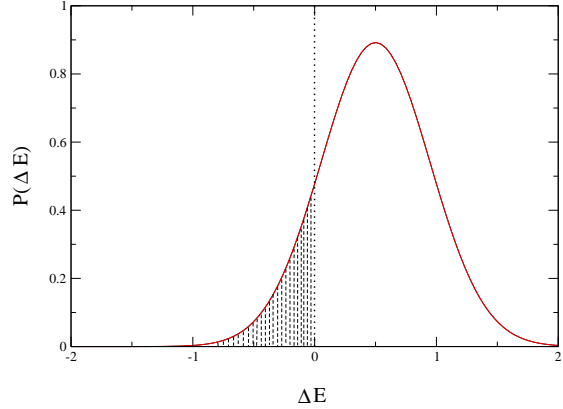}
  \caption{Probability density of the energy of the proposed moves for
    different values of the energy $E$ as defined in Figure
    \protect\ref{figESM:OSC:2}
  }
    \label{figESM:OSC:1}
\end{figure}
From \eq{eqESM:OSC:3} the dynamical equations immediately follow by
defining a Monte Carlo step as a collection of $M$ elementary moves
(each elementary move corresponds to a global change of all oscillator
coordinates as described in \eq{eqESM:OSC:2}). Because the average
change of energy $\Delta E$ is finite in an elementary move and the
total energy (\ref{eqESM:OSC:1}) is extensive, the number of moves $M$ in a
Monte Carlo step must be proportional to $N$. For simplicity we will
take $M=N$. In the limit $N\to\infty$,  
time becomes a continuous variable. This allows to
write a closed equation for the energy $E$ and acceptance rate $A$, i.e., 
the average number of accepted moves in a Monte Carlo step,
\be
\frac{\partial E}{\partial t}=\int_{-\infty}^{\infty}xP(x)W(x)\,dx
\label{eqESM:OSC:4}
\ee
\be
\frac{\partial A}{\partial t}=\int_{-\infty}^{\infty} P(x)W(x)\,dx
\label{eqESM:OSC:5}
\ee
where $P(x)$ is the probability distribution (\ref{eqESM:OSC:3}) and
$W(x)$ is the transition rate which ensures that detailed balance is
obeyed. For instance, according to the Metropolis rule
$W(x)=\min[1,\exp(-\beta x)]$ or in heat-bath 
$W(x)=1/[1+\exp(\beta x)]$. 
At zero temperature the transition rate rules
$W(x)$ converge to $W(x)=\theta(x)$. In this limit  (\ref{eqESM:OSC:4}),
(\ref{eqESM:OSC:5}) become
\be
\frac{\partial E}{\partial t}=\int_{-\infty}^{0}xP(x)dx
\label{eqESM:OSC:6}
\ee
\be
A(t)=\int_{-\infty}^{0} P(x)dx
\label{eqESM:OSC:7}
\ee
As shown in Figure \ref{figESM:OSC:1} the acceptance is given by the
shaded area enclosed in the negative tail of the distribution.  At zero
temperature, according to \eq{eqESM:OSC:3}, both the mean and the width
of the Gaussian decrease as well as the shaded area in
Figure \ref{figESM:OSC:1}, implying a systematic decrease of the
acceptance rate.  

We do not want to dwell here on all results one can learn by solving the
dynamical equations \eqq{eqESM:OSC:6}{eqESM:OSC:7} (see
\cite{BonPadRit98,Nieuwenhuizen98c}). Interestingly, in the OSC model
the FDR \eq{fdr:1} for the magnetization $M=\sum_{i=1}^N x_i$ can be
exactly computed at any temperature \cite{BonPadRit98},
\be
\Te(s) = 2E(s)+\frac{1}{f(s)}\frac{\partial E(s)}{\partial s}
\label{oscx}
\ee
and the QFDT is satisfied. At $T=0$ in the large time limit one gets
$\Te(s)=2E(s)$ plus subleading corrections. This result shows that FD
plots are straight-lines starting at $C(s,s)=(2E(s)/K),\chi(s,s)=0$ and
finishing at $C=0,\chi=1/K$. 
The relation
between the effective temperature and the dynamical energy is exactly
the same as the equilibrium relation given by the equipartition
theorem. The aging system is in a quasistationary state where relations
between dynamical quantities are formally the same as in
equilibrium. This allows to define a time-dependent configurational
entropy $S_c(E)$ through the relation \eq{sc_tempeff}, 
\be
\frac{1}{\Te(E)}=\frac{\partial S_c(E)}{\partial E}\quad ;\quad
S_c(t)=S_c(E(t))=\frac{1}{2}\ln(E(t))\quad .
\label{oscsc}
\ee

Until now we have discussed some of the analytical results obtained by
solving the dynamical equations of the OSC model.  However, an
interesting question is the following: can we determine the value of
the effective temperature from the sole evolution of the energy $E$
without having to analyze correlations and responses in the framework
of the QFDT as described in \eq{oscx}?  
Ideally we would like to apply the ideas presented in
section \ref{micro:EXTME} to identify the
value of $\Te$ solely from the off-equilibrium transition rates
$Z(F|F')$. In that description dynamics proceeds by activated jumps
over different states, whose dynamics is described by the free-energy ME
\eq{eq2d}. What are the states in the reduced description of the OSC
model? As the energy landscape is a single parabolic well it appears
that a reduced description is not possible. The clue to this question
is easy to understand if one realizes that at zero temperature the
acceptance of the dynamics goes to zero with time, therefore each time
a proposed elementary move is accepted we can effectively consider
that the system has jumped from one state to another, the typical time
for this jump steadily growing with the time elapsed since the system
was quenched. In this view, each state corresponds to a configuration
and the reduced dynamics simplifies. Moreover, the free energy of the
state is simply the energy of the corresponding configuration. In the
asymptotic regime $(dE/dt)/E<<1$ where finite size effects are not
important, i.e. $\Delta\ll \sqrt{2E/K}$, the probability distribution
describing the energy change after the first jump is given by,
\be
P(\Delta E)\sim \exp\Bigl(\frac{\Delta E}{4e}\Bigr)\theta(-\Delta E)
\label{eqESM:OSC:6b}
\ee 
where we used \eq{eqESM:OSC:3} and expanded it around $\Delta E=0$ up to
the linear order.  Using relation \eq{oscsc} we can recast
\eq{eqESM:OSC:6b} in the following form,
\be
P(\Delta E)=\frac{1}{2\Te(E)}\exp\left(\frac{\Delta E}{2\Te(E)}\right)\theta(-\Delta E)
\label{eqESM:OSC:5a}
\ee
showing that the statistics of energy jumps is an exponential with a
width that directly depends on the effective temperature. This result
has two consequences: 1) It shows that the OSC model is a microscopic
version of the trap model proposed by Barrat and Mezard
\cite{BarMez95}; 2) The effective temperature could be computed from
the statistics of the first free-energy jumps among components (here
corresponding to configurations).

Before finishing, let us note that a number of variants of the
oscillator model have been considered, all sharing the feature that
oscillators do not interact. For example, Nieuwenhuizen and
Leuzzi \cite{Nieuwenhuizen98c,LeuNie01,LeuNie01b,LeuNie02} have
considered a model with spin variables in addition to oscillators. The
new variables are discrete and used to mimic fast relaxational
processes not contained in the original formulation; aging, slow
relaxation in these models, can be still described in terms of entropy
barriers.

\subsubsection{The Backgammon and urn models} 
\label{backgammon}

Another instructive model where relaxation is determined by entropic
barriers is the Backgammon (BG) model \cite{Ritort95}. The model
belong to a large class of models under the name of urn models where
$N$ particles or balls are distributed among $M$ urns or boxes. The BG
model is defined by the energy function,
\be
E=-\sum_{r=1}^M\delta_{n_r,0}=-N_{\rm empty}
\label{eqESM:BG:1}
\ee
where $n_r$ stand for the occupancies for each box and $N_{\rm empty}$
stands for the number of empty boxes. The model has different versions 
\cite{GodBouMez95}
according to whether particles are distinguishable (Maxwell-Boltzmann
statistics) or not (Bose-Einstein statistics). The easiest way to
compute the thermodynamic properties of the BG model is by expressing
the partition function in terms of the occupancies $n_r$ rather than in
terms of the set of boxes occupied by the particles.  In the
Maxwell-Boltzmann case, thermodynamics needs to be corrected by
dividing the partition function by the usual $N!$ 
term to avoid Gibbs' paradox. In the model with Bose-Einstein statistics 
this is
not necessary. The dynamical rules of the BG model directly depend on
the type of statistics considered. In the Maxwell-Boltzmann case, a
departure box $d$ is selected with a probability proportional to the
occupancy $n_d$ of that box and a new arrival box $n_a$ is selected
with uniform probability. The proposed move is given by $n_d\to
n_d-1,n_a\to n_a+1$ and accepted according to the standard Metropolis
rule. Instead, in the Bose-Einstein case, the proposed move and the
transition rate are the same as for the Maxwell-Boltzmann case but
both the departure and the arrival box $d$ and $a$ are selected with a
uniform probability $1/M$.

The resulting dynamics of the model is strongly dependent on the type of
statistics considered, the interesting one corresponding to the
Maxwell-Boltzmann case which displays a strong separation of time-scales.
In what follows, although otherwise stated, we will concentrate in the
analysis of the model with $M=N$. 

The thermodynamics of the model is exactly solvable and, the entropy
being given by $S(e)=\log(1+e)$ where $e=E/N$ denotes the entropy per
box (or per particle). To analyze the dynamics from the perspective of
the scenario described in section \ref{micro:EXTME} we follow a similar
reasoning as we did for the OSC model.  Let us consider a system
quenched down to $T=0$ and the aging regime reached in the asymptotic
long-time regime where $(dE/dt)/E<<1$.  In that limit, the system has a
number of empty boxes $N_{\rm empty}=-E$ and further decrease of that
number by one unit $\Delta E=-1$ requires a time that exceedingly grows
as $E$ decreases toward its minimum value $E=-N+1$ (all particles
condensed into a single box) \cite{AroBhaPra99,Lipowski97}. Therefore,
as relaxation slows down, for a long time the system wanders through the
constant energy surface by exchanging particles among occupied
boxes. When a new box is emptied the energy decreases by one unit.  As
we did in the OSC model, also in the $T=0$ dynamics of the BG model each
component $\RR$ corresponds to a single configuration with free energy
equal to the energy of that configuration.

The transition rate for going from a configuration of energy $E$ to a
configuration of energy $E-1$ is sole function of the number of
available configurations in the initial and final state,
\be
\frac{Z_t(E-1|E)}{Z_t(E|E)}=\frac{\Omega(E-1)}{\Omega(E)}
\label{eqESM:BG:2}
\ee
where the rate frequency $1/t(E)$ associated to $Z_t$ (for its
definition, see Section \ref{micro:EXTME}) has canceled out from the
numerator and denominator in the left hand side of (\ref{eqESM:BG:2}).
As $E$ is extensive with $N$, and using (\ref{eq6d}) we obtain,
\be
\frac{Z_t(E-1|E)}{Z_t(E|E)} =\exp\Bigl(-\frac{\partial S(E)}{\partial E}\Bigr)=\exp\Bigl(-\frac{1}{\Te(t)}\Bigr) \qquad . 
\label{eqESM:BG:3}
\ee
Equation \eq{eqESM:BG:3} provides a way to estimate the effective temperature
by looking at the number of moves required for a move to decrease the
energy by one unit. Inverting (\ref{eqESM:BG:3}) yields
\be
\Te(t)=\left(\ln\left[\frac{Z_t(E|E)}{Z_t(E-1|E)}\right]\right)^{-1}
     =\frac{1}{N^*(t)}\qquad . 
\label{eqESM:BG:4}
\ee
Therefore the ratio $Z_t(E|E)/Z_t(E-1|E)$ is simply given by the
inverse of the number of {\em accepted} moves $N^*$ required to
decrease the energy by one unit. We stress that $N^*$ is not the
number of MC steps but rather the number of accepted elementary moves
with $\Delta E=0$ required until a move with $\Delta E=-1$ is found.
This number $N^*$ is independent of the size of the system. Note also
that the rate of rejected moves (which gives the acceptance) does not
enter into the expression of the effective temperature but rather into
the value of the typical timescale $t(E)$ associated to the jump. This
is an essential ingredient required in any ``quasi-equilibrium'' or
microcanonical description of the relaxation. The value of the
effective temperature at a given time only depends on the number of
accessible states with lower free energy (energy in the present model)
and not on the time necessary to escape from that state.  This
expression is very amenable to numerical calculations.

Note that the energy levels in the BG model are discrete, therefore the
relation \eq{eqESM:OSC:5a} cannot be applied. However, one can consider a
non-degenerate disordered BG model where boxes are assigned different
energies \cite{LeuRit02}. In this case the distribution $P(\Delta
E)$ has been derived and shown to have a exponential time-dependent behavior 
characteristic of trap models. However, the computation of the
effective temperature in that model, and the verification of
\eq{eqESM:OSC:5a} has not yet been done.

The Backgammon model as well as some of its variants has been solved by
the technique of the generating function
\cite{FraRit95,GodBouMez95,FraRit96,GodLuc96,FraRit97,PraBreSan97c,GodLuc99,GodLuc00b}.
A discussion of some of the main results and analytical techniques has
been presented in \cite{RitSol02}. The origin of the effective
temperature as derived from the QFDT remains yet not completely
understood. No neutral observable has been yet identified in the BG
model. Although all studied observables show that the QFDT \eq{eq4e} is
verified in a one-timescale scenario, there seem to be different
effective temperatures depending upon the observable considered
\cite{FraRit97,GodLuc99,GarPagRit03}. On the other hand, none of the
different effective temperatures associated to these observables appears
to be linked to the {\em effective} temperature obtained within the
adiabatic approximation \cite{FraRit95}. An interest variant of urn
models are zeta urn models introduced in the context of quantum gravity
that show a finite-temperature Bose-Einstein condensation transition
\cite{GodLuc01,GodLuc02}, see section \ref{ferro}.

\subsection{Ferromagnetic models at criticality} 
\label{ferro}

A new emerging line of research related to the mainstream of
researchs on FDT violations in glassy systems, is the study of FDT
violations in critical points. In critical points the relaxation time
diverges and one can investigate, for instance through field theoretical
methods, the time-dependence of correlations, responses and the
resulting FDR \eq{eq2eb} in the asymptotic regime $s\to\infty$. These
type of investigations are not different of those undertaken in the
study of coarsening behavior in ordered phases.  The main difference
between the slow dynamics in a critical point and coarsening is that, in
the former case, critical slowing down is at the origin of the slow
dynamics. Interfaces have no stiffness tension and their motion is
subdiffusive (or diffusive) and only consequence of the curvature of the
interface. Growing domains are not islands of up or down spins but
regions of spatially and temporally correlated spins of zero net
magnetization.  At $\Tc$ dynamics is described by the
renormalization-group dynamical equations of the corresponding
finite-temperature fixed point.  However, in coarsening systems below
$\Tc$ activated processes are important and interfaces have
non-zero tension as they separate domains of up and down spins.
Competition between the curvature of the surface and its tension leads
to different growth laws, described by the zero-temperature fixed point.
Because the origin of slow dynamics at $\Tc$ is different than in
standard glassy systems, the main Ansatz \eq{eq2eb} may not hold. An
indeed, many studies of the FDR at criticality reveal that $X(t,s)$ is
not a single function of the correlation $C(t,s)$.

One among the first studies of ferromagnetic models at criticality is
the Ising chain solved by Glauber in 1963 \cite{Glauber63}.  Strictly
speaking the dynamics of this model is that of a coarsening model as the
system orders at the critical point $T=0$ where the magnetization discontinuously jumps
from 0 (for $T>0$) to 1 (at $T=0$), i.e. the critical exponent $\beta$
vanishes. We will see below that
the ferromagnetic Ising chain has some peculiar properties. Starting
from a random initial configuration the coarsening dynamics at $T=0$ has
long been studied in \cite{Bray89} and revisited in
\cite{PraBreSan97}. In the asymptotic long-time limit, the aging part of
the two-times correlation functions scales like $C_{\rm ag}(t,s)\sim
F(L(t)/L(s))$ \eq{inserted:coars:1} with $L(t)\sim t^{1/2}$
corresponding to a diffusive process of the interfaces (see section
\ref{coars}).  More recently, and nearly at the same time, the FDR has
been analytically computed \cite{GodLuc00,LipZan00} for a random
staggered perturbation finding $X(C)=1/[2-\sin^2(\pi C/2)]$ in agreement
with the Ansatz \eq{eq2eb}. Or, in terms of times, $X(t,s)=(1/2)(1+s/t)$
showing that $X\to 1/2$ if $t\to \infty$ or $C\to 0$. This last result
coincides with that found in the random walk or in the Gaussian model
\cite{CugKurPar94}, all models characterized by a diffusive
dynamics. This has led to the proposal \cite{GodLuc00b,GodLuc02b} that,
in systems at criticality, the limiting value $X(t,s)$ for $t\to \infty$
is a universal quantity,
\be
X_{\infty}=\lim_{s\to\infty}\lim_{t\to\infty}X(t,s).
\label{ferro:1}
\ee
This conclusion has been substantiated by the exact solution of the
ferromagnetic spherical model in general $d$ dimensions
\cite{GodLuc00b,GodLuc02b}. The authors have noticed how $X_{\infty}$ is
only a function of amplitude ratios describing the leading scaling
behavior of correlations and responses. This conclusion is endorsed by
the following result: $X_{\infty}=1/2$ for $d>d_{\rm ucd}=4$ and
$X_{\infty}=1-2/d$ for $2<d<d_{\rm ucd}=4$ where $d_{\rm ucd}$ stands
for the upper critical dimension (below $d=2$ there is no finite $T$
transition in the model).  Interestingly, for $d=2$ the FDTR $X(C)$ has
a non-trivial form \cite{CorLipZan02}, similar to what is found for the
ferromagnetic Ising chain.  
Numerical results in the Ising model show that $X_{\infty}\simeq 0.26$
in 2d~\footnote{After completion of this review we learned from recent
results finding a slightly different value $X_{\infty}\simeq 0.34$ in
2d~\cite{MayBerGarSol03}. More work is necessary to accurately estimate
that number.}  and preliminary results in 3d give $X_{\infty}\simeq
0.4$.  In figure \ref{fig:ferro:1} we show the the IRF corresponding to
the TRM susceptibility \eq{eq3e2} as function of $C$ in the 2d Ising
model at the critical point.
\begin{figure}
  \centering \includegraphics[scale=0.9]{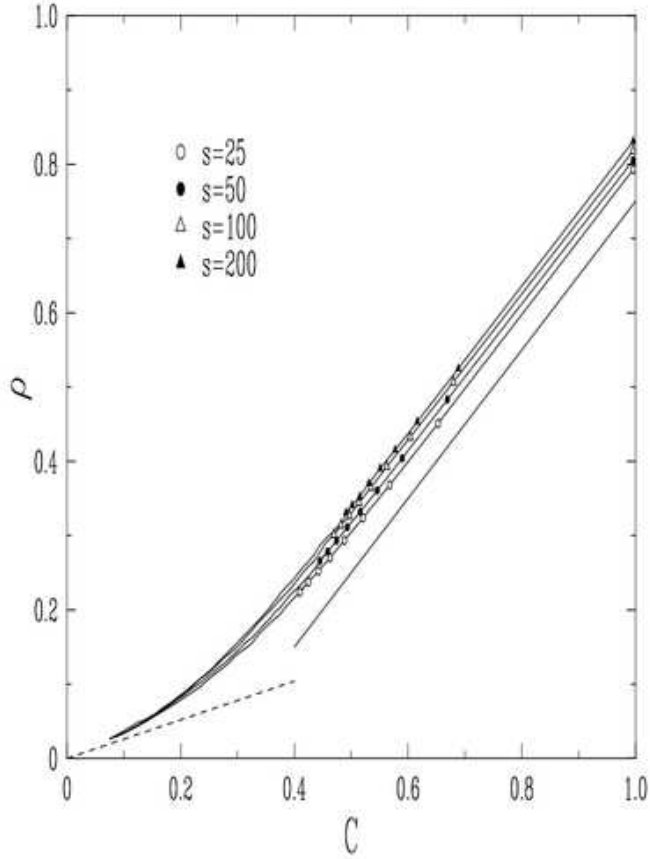}
  \caption{IRF \eq{eq3e2} corresponding to the thermoremanent
  magnetisation in the 2d Ising model at criticality. The full and
  broken line correspond to the quasi stationary regime $X=1$ and
  $X_{\infty}=0.26$ respectively.  From \protect\cite{GodLuc00b}.  }
  \label{fig:ferro:1}
\end{figure}
The general scenario about ferromagnetic models at criticality is as
follows \cite{GodLuc02b}. Let us consider a system that is quenched from
a random configuration to a temperature below $\Tc$. If $\tau=t-s\ll s$
the system is in equilibrium, TTI holds and FDT is not
violated ($X(t,s)=1$). 
However, if $t/s\sim {\cal O}(1)$ the system
ages, and both TTI and FDT are violated. A solution for the correlation
$C_{\rm ag}(t,s)$  and response $R_{\rm ag}(t,s)$ 
that matches the intermediate regime between the stationary $X=1$ and the 
aging regimes is given by,
\be
C_{\rm ag}(t,s)=m^2_{\rm eq}\,\widehat{C}\left(\frac{t}{s}\right),
\qquad R_{\rm ag}(t,s)=s^{-a-1}\,\widehat{R}\left(\frac{t}{s}\right)
\label{ferro:2}
\ee
where $a>0$ is a coarsening exponent. From \eq{ferro:2} $X(t,s)\sim
s^{-a}/m^2_{\rm eq}$, therefore $X\to 0$ for coarsening systems where
$m_{\rm eq}$ is finite. The
same expression is valid for the critical point but replacing $m_{\rm
eq}$ by its time-dependence at $\Tc$. Using $m_{\rm eq}\sim
(\Tc-T)^{\beta},\xi\sim (\Tc-T)^{-\nu}, t\sim \xi^z$ where
$\beta$, $\nu$, $z_c$ are the, correlation length and dynamical
exponents respectively. Substituting these relations into 
\eq{ferro:2} and using $a_c={2\beta}/{z_c\nu}$ gives at $\Tc$,
\be
C_{\rm ag}(t,s)=s^{-a_c}\,\widehat{C}\left( \frac{t}{s}\right),
\qquad R_{\rm ag}(t,s)=s^{-a_c-1}\,\widehat{R}\left( \frac{t}{s}\right)
\label{ferro:3}
\ee
leading to $X(t,s)=\hat{X}(t/s)$.  This result has two consequences: 1)
for $t\to\infty$, $X(t,s)\to X_{\infty}$ and 2) only for $a_c=0$,
according to the left expression in \eq{ferro:3}, $X(t,s)$ can be
expressed as sole function of $C_{\rm ag}$. This is the case of the
aforementioned Ising chain where the Ansatz \eq{eq2eb} is valid because
$\beta=0$.  It has been suggested \cite{HenPleGodLuc01a} that $R_{\rm
ag}(t,s)$ covariantly scales under conformal transformations of time
leading to specific predictions for the scaling function $\widehat{R}$.
An important implication of the result $a_c=0$ is that the aging part of
the IRF \eq{eq3e} does not vanish in the asymptotic long-time limit. As
noted in \cite{CorLipZan01} the non-vanishing of the aging part of IRF
in the large $s$ limit is related to the failure of the scheme that
links static and dynamical properties
\cite{FraMezParPel98,FraMezParPel99a} in the Ising chain. In fact, as
discussed in section \ref{sk} the stochastic stability property links
the equilibrium $P(q)$ with the behavior of IRF,
\begin{equation}
  P(q)=-T\left.\frac{d^2\chi(C)}{dC^2}\right|_{C=q}
      = \left. \frac{dX(C)}{dC}\right|_{C=q}
\label{ferro:4}
\end{equation}
where we have used \eq{eq5e}. In the Ising chain
$P(q)=\delta(q-1)$. Equation \eq{ferro:4} is not fulfilled in the Ising
chain as can be easily checked inserting the result
$X(C)=1/(2-\sin^2(\pi C/2))$ (discussed in the paragraph preceding
\eq{ferro:1}).  As soon as $\Tc$ is finite ($d>1$),
$X(C)=\theta(C-m_{\rm eq}^2)$ for $T<\Tc$ and \eq{ferro:4} is again
satisfied.  Despite its simplicity, the Ising chain appears to be an
interesting solvable example that allows to check many results.  For
instance, observable independence has been also recently addressed
\cite{SolFieMay02} showing that $X_{\infty}=0$ for a large class of
observables.  Recent progress in the study of the FDR at criticality has
been achieved by Calabrese and Gambassi \cite{CalGam02a,CalGam02b} who
have considered the FDR in momentum space $X_{\vec{q}}(t,s)$. The study
of $O(N)$ models using field-theoretical techniques yields estimates for
the value of $X_{\infty}$ \eq{ferro:1} in an $\eps=4-d$
expansion. Two-loops computations\cite{CalGam02b} give, for the Ising
case $N=1$, the following values:
$X_{\infty}(3d)=0.429(6),X_{\infty}(2d)=0.30(5)$ compatible with the
results obtained from numerical
simulations~\cite{GodLuc00b,MayBerGarSol03}.

Other studies of models at criticality include the XY model with a
Kosterlitz-Thouless transition \cite{BerHolSel01}. This model has a
low-temperature phase where correlation functions decay algebraically,
therefore correlations are critical below $\Tc$. A failure of the
stochastic stability property, similar to that reported in the Ising
chain, appears in the 2d XY model at low temperatures where the density
of vortices increases.

Finally, let us comment on results for zeta-urn models that have
confirmed the validity of the relation \eq{ferro:1}
\cite{GodLuc01,GodLuc02,RitSol02}. Zeta urn models show a quite rich phase
diagram described in terms of the density $\rho$ of balls (i.e. the number of
balls divided by number of boxes) and the temperature. There is a
critical line $\rho_c(\beta)$ that separates a fluid regime
($\rho<\rho_c(\beta)$) from a condensed regime ($\rho>\rho_c(\beta)$) with
glassy dynamics. Along
the so called regular part of the critical line ($\beta>3$)
$X_{\infty}=\frac{\beta+1}{\beta+2}$ is temperature dependent. This
number lies in the interval $[4/5,1]$ quite far from the values
$X_{\infty}<1/2$ found in ferromagnets. In the condensed phase
$X_{\infty}\to T^{1/2}$ at low temperatures. It vanishes at $T=0$ as in
coarsening systems, however the condensation dynamics in urn models is
totally inhomogeneous in contrast to the homogeneous character of
coarsening in ferromagnets.

\section{QFDT: the numerical evidence}
\label{qfdt:numerics}

Computer simulations have the great advantage, over the real experiment,
of direct access to the microscopic level, even if only relatively small
timescales and system sizes can be studied. Although this
can be a serious limitations, the true fact is
that the numerical study of aging phenomena and FDT violations has been
successfully done during the last years for several systems.

\subsection{Structural glasses}
\label{structural:glasses}

Using ideas developed in the field of
spin glasses, many conjectures have been formulated concerning
the structure of the phase space of glassy systems. However, obtaining
direct information, either from experiments or from numerical 
simulations, is a difficult challenge.
Relaxation times in a glass are so long as to preclude equilibration 
within experimental times. Numerical or experimental exploration of the
phase space in these systems is therefore necessarily incomplete.
The increase in computational power and the recent developments in the
theory of disordered systems has pushed forward an approach which should
not suffer from these limitations. The idea, which was actively
developed in the study of spin glasses, is that relevant information on
the phase space structure should be hard-encoded into the
non-equilibrium dynamics of glassy systems.  

According to the conjecture of the similarity between structural glasses
and some spin-glass models, $X(C)$ for structural glasses is a two
valued function with $X(C)=1$ at short times, and $X(C) = m <1$ in the
long-time aging regime. This scenario has been largely studied using
numerical simulations.

In a numerical investigation of aging effects not only the 
waiting time must be 
changed over several order of magnitude, but for a given waiting time the
subsequent dynamics must be studied over a long time.  
For these reasons aging phenomena have been studied for models that are
simple enough to be simulated over long times, but at the same time 
still catch the essential features of real glasses.  Moreover,
to maintain the systems in a non-equilibrium state for very long times, 
crystallization must be strongly inhibited. This is obtained either with a 
particular choice of interaction potential parameters or by 
adding a (small) extra term in the potential.
In the following we discuss results obtained for several models.

\subsubsection{Mixtures of soft particles of different sizes}
\label{soft}
\cite{BerHanHiwPas87,BarRouHan90,HanYip95,Parisi97a,Parisi97b,Parisi97c}.

This system consists of $N$ particles 
half of which are of type $A$ and half of type $B$ interacting via the
Hamiltonian 
\begin{equation}
\label{eq:7.1-1}
 {\cal H} = \sum_{i<j} \,\left( \frac{r_i + r_j}
                                     {|\bi{x}_i - \bi{x}_j|}
                         \right)^{12}
\end{equation}
where the radius $r_i$ depends on the type of particle. It is known that
the choice of two different types of radius such that $r_B / r_A = 1.2$
prevents crystallization and the system can be brought into a glassy
state.

Due to the simple scaling of the potential, the thermodynamic quantities
depend only on $\Gamma = \rho\beta^4$, where $\rho$ is the density which
can be taken equal to one. This model presents a glass transition at
about $\Gamma_c = 1.45$ \cite{BarRouHan90}.  In figure \ref{Fig:7.1-1}
we report the response of the particle to a force of strength $\epsilon$
\begin{equation}
\label{eq:7.1-2}
 \chi(t_w+\tau,t_w) \simeq \frac{1}{N\epsilon} \sum_{i=1}^{N}
                    \left\langle 
		           \bi{f}_i\cdot\bi{x}_i(t_w+\tau)
                    \right\rangle
\end{equation}
where $\bi{f}_i$ is a random Gaussian vector of squared length equal to 
the space dimension $d$, 
versus the self-diffusion function
\begin{equation}
\label{eq:7.1-3}
 \Delta(t_w+\tau,t_w) = \frac{1}{N} \sum_{i=1}^{N}
                   \left\langle |\bi{x}_i(t_w+\tau) - \bi{x}_i(t_w)|^2
                   \right\rangle
\end{equation}
\begin{figure}
  \centering
  \includegraphics[scale=0.7]{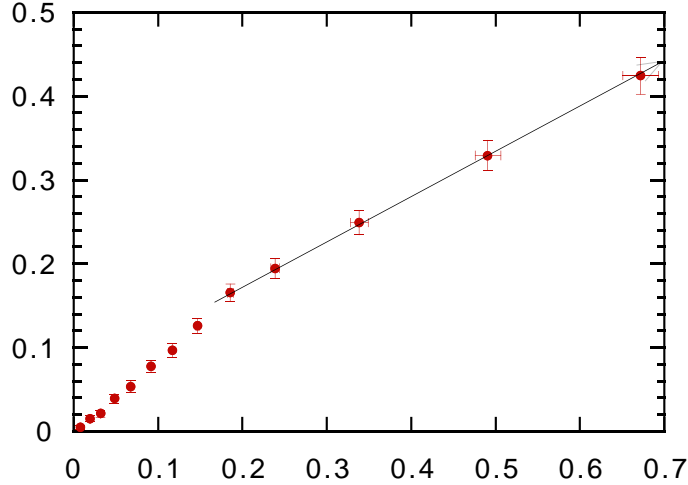}
  \caption{$\chi$ versus $\beta\Delta$ at $\Gamma=1.6$ for
    $t_{w}=t_{w}=8192.$ and $t_{w}=2048$. From \cite{Parisi97b}.}
  \label{Fig:7.1-1}
\end{figure}
The average is over different initial states at $t_w$ and realization of 
$\bi{f}$. Two linear regions with different slopes, one with 
$X(C) =1$ and one with $X(C)=m<1$  are clearly 
seen, in agreement with the two-timescales scenario. The dependence of $m$ 
with $T$ is well fitted by the spin-glass model prediction
$m(T) = T/\Tc$, see figure \ref{Fig:7.1-2}. Similar results have been
obtained in monoatomic Lennard-Jones
glasses \cite{LeoAngParRuo00,AngLeoParRuo01}, see section \eq{monoatomic}.

\begin{figure}
  \centering
  \includegraphics[scale=.7]{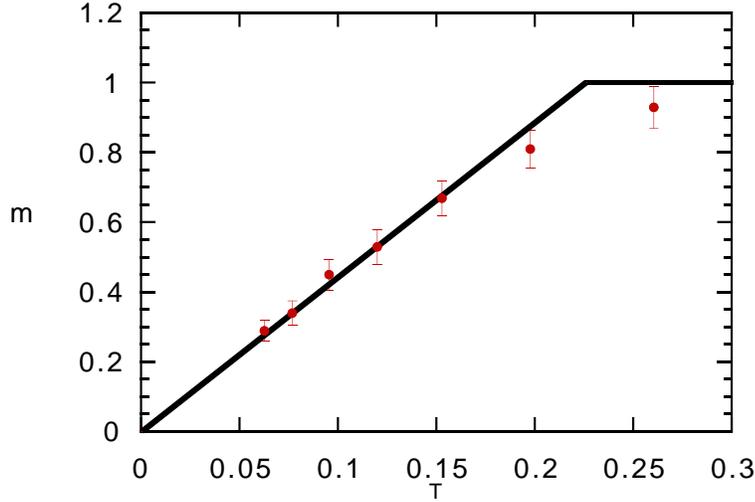}
\caption{The quantity $m\equiv {\partial \chi\over \partial \beta \Delta}$ as $t_{
    w}=2048$ as 
  function of the temperature. The straight line is 
  the prediction of the approximation $m(T)=T/\Tc$. 
  From \cite{Parisi97b}.}
\label{Fig:7.1-2}
\end{figure}

\subsubsection{Lennard-Jones binary-mixtures}
\label{lennard}
\cite{KobAnd94,KobAnd95a,KobAnd95b,GleKobBin98} 

The system consist of a mixture of particles of type $A$ and $B$
of equal mass $m$
interacting via a $12-6$ Lennard-Jones 
potential of the form
\begin{equation}
\label{eq.7.9}
 V_{\alpha\beta}(r) = 4\epsilon_{\alpha\beta}\,\left[
              \left(\frac{\sigma_{\alpha\beta}}{r}\right)^{12} -
              \left(\frac{\sigma_{\alpha\beta}}{r}\right)^{6}\,
                                             \right]
\end{equation}
where $\epsilon_{\alpha\beta}$ and $\sigma_{\alpha\beta}$ depend on the 
particle pair type and are chosen to prevent 
crystallization. For a $80:20$ mixture, and using  $\epsilon_{AA}$ and 
$\sigma_{AA}$ as units of energy and length and 
$(m\sigma_{AA}^2/48\epsilon_{AA})^{1/2}$ as the unit of time, these are
$\epsilon_{AA} = 1$, 
$\sigma_{AA}=1$, $\epsilon_{AB} = 1.5$, $\sigma_{AB}=0.8$,
$\epsilon_{BB}=0.5$ and $\sigma_{BB}=0.88$. The atomic dynamics of this model 
is well described by the mode-coupling theory with a critical temperature
of $\Tc = 0.435$ in reduced units.

Typical FD plots numerically obtained by Barrat and Kob
\cite{BarKob99a,BarKob99b,KobBar99a} are shown in figure \ref{Fig:7.1-3}. The
correlation is given by the incoherent scattering function for a wave
vector $\bi{k}$
\begin{equation}
\label{eq:7.1-4}
C_k(t_w+\tau,t_w) = \frac{1}{N}\sum_{j}\,
                   e^{i\bi{k}\cdot[\bi{r}_j(t_w+\tau) - \bi{r}_j(t_w)]}
\end{equation}
while the response is measured by adding to the potential a term
of the form
\begin{equation}
 \label{eq:7.1-5}
  \delta V = V_0\,\sum_j \epsilon_j\,\cos(\bi{k}\cdot\bi{r}_j)
\end{equation}
where $\epsilon_j=\pm 1$ with equal probability and $V_0 < T$.  Again a
two-timescales scenario is clearly seen. Moreover the effective
temperature in the slow regime where $m<1$ is in reasonable agreement
with the glass transition temperature $\Tg$ of the system.%
\footnote[1]{The glass transition $\Tg$ is defined as the temperature
below which the system fails to equilibrate on the experimental time-scale.
For structural glasses $\Tg$ is defined as the temperature at which the 
viscosity is equal to $10^{13}$ Poise or, equivalently, a relaxation time
of $100\,s$.}

Thus the FDT is broken as the system fails to equilibrate, as
expected from spin-glass models.
\begin{figure}
 \centering
 \includegraphics[scale=0.5]{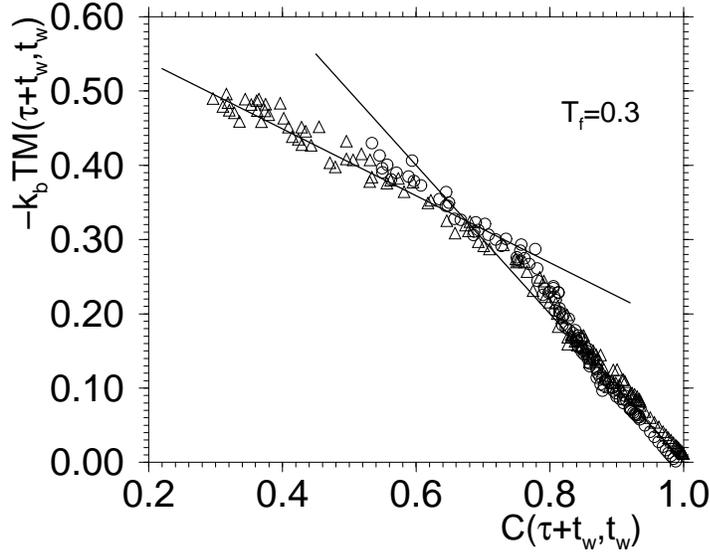}
\caption{Parametric plot of the integrated response function
  $M(t_w+\tau,t_w)$ and the correlation function $C(t_w+\tau,t_w)$ for
  $k=7.25$. Final quench temperature $T_{\rm f}=0.3$,  $t_w=1000$. 
  Circles: $t_w=10000$. The straight
  lines have slopes $-1.0$ and $-0..45$. 
  From \cite{BarKob99b}.}
\label{Fig:7.1-3}
\end{figure}

\subsubsection{Monoatomic Lennard-Jones systems}
\label{monoatomic}
\cite{AngParRuoVil98a,AngParRuoVil99a,LeoAngParRuo00,AngLeoParRuo01}.

The system consists of equal particles interacting via the
potential $V = V_{\rm LJ} + \delta V$, where 
$V_{\rm LJ}$ is the usual $12-6$ Lennard-Jones potential 
(expressed in reduced units), and $\delta V$ is a
many-body term that inhibits crystallization:
\begin{equation}
\label{eq:7.1-6} 
 \delta V = \frac{\alpha}{2}\, \sum_{\bi{q}}\,
          \theta(S(\bi{q}) - S_0)\, (S(\bi{q}) - S_0)^2
\end{equation}
where $S(\bi{q})$ is the static structure function, $\alpha = 0.8$  and
$S_0=1$. The sum is made over all $\bi{q}$ with 
$q_{\rm max} - \Delta < |\bi{q}| < q_{\rm max} + \Delta$, where
$q_{\rm max} = 7.12\,\rho^{1/3}$ and $\Delta = 0.34$, $\rho$ being the
particle density.

In figure \ref{Fig:7.1-4} we report the parametric plot of the mean
square displacement $\Delta$ and IRF in a crunch experiment. It is
important to note that the temperature below which $m<1$ does coincide
with the glass transition temperature of the system at the density
reached after the crunch. This shows that the breaking of FDT does
not depend on the initial state nor on the path followed in the
$(T,\rho)$ plane, but only on the final (non-equilibrium) state to which
the system is brought.

\begin{figure}
  \centering
  \includegraphics[scale=0.4]{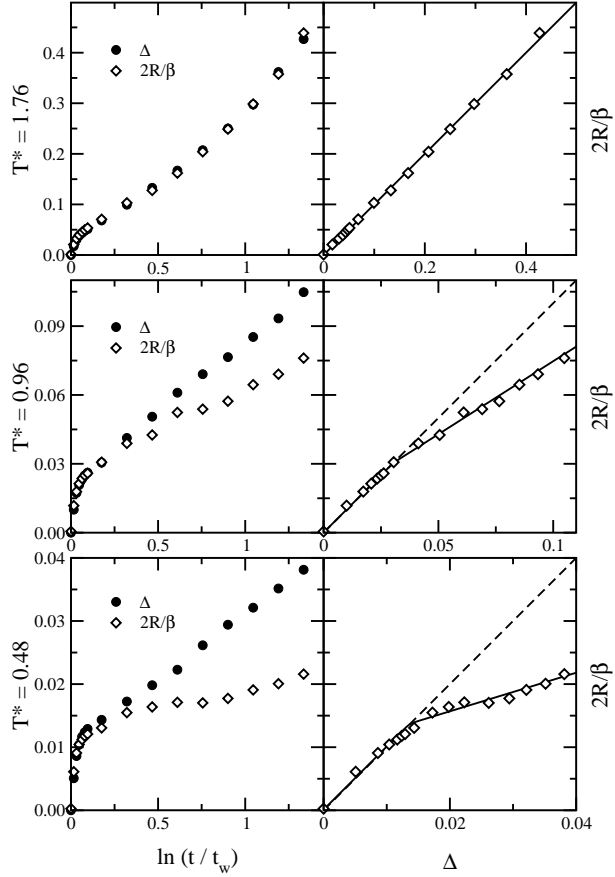}
\caption{Mean square displacement $\Delta$ and integrated response function
  $2R/\beta$ in reduced units at $\rho^*_2$=$1.24$ for three temperatures
  $T^*$=$0.48, 0.96, 1.76$.
  The left side shows the $\log$ time dependence of the two quantities.
  In the right side the response function vs $\Delta$. Dashed lines indicate
  equilibrium FDT, while full lines fit the
  off-equilibrium aging region.
  From \cite{LeoAngParRuo00}.}
\label{Fig:7.1-4}
\end{figure}
The FD plots for glass-forming liquids discussed here 
reveal the typical two-timescales (or 1RSB) scenario found in some spin-glass models.
This supports the original Goldstein's idea \cite{Gol69} that the 
phase space of supercooled liquids is divided by high barriers into different
valleys each with its own statistical properties. 
This picture has been recently confirmed by a 
a direct analysis of the motion of a glass forming liquids in terms of 
IS \cite{SasDebStiSchDyrGlo99}.

In the right panel of figure \ref{Fig:7.1-5} \cite{SciTar01} we show the
  temporal behavior of the average energy of minima for a binary mixture
  Lennard-Jones system visited in the non-equilibrium motion following a
  quench to a low temperature.
\begin{figure}
  \centering
  \includegraphics[scale=0.6]{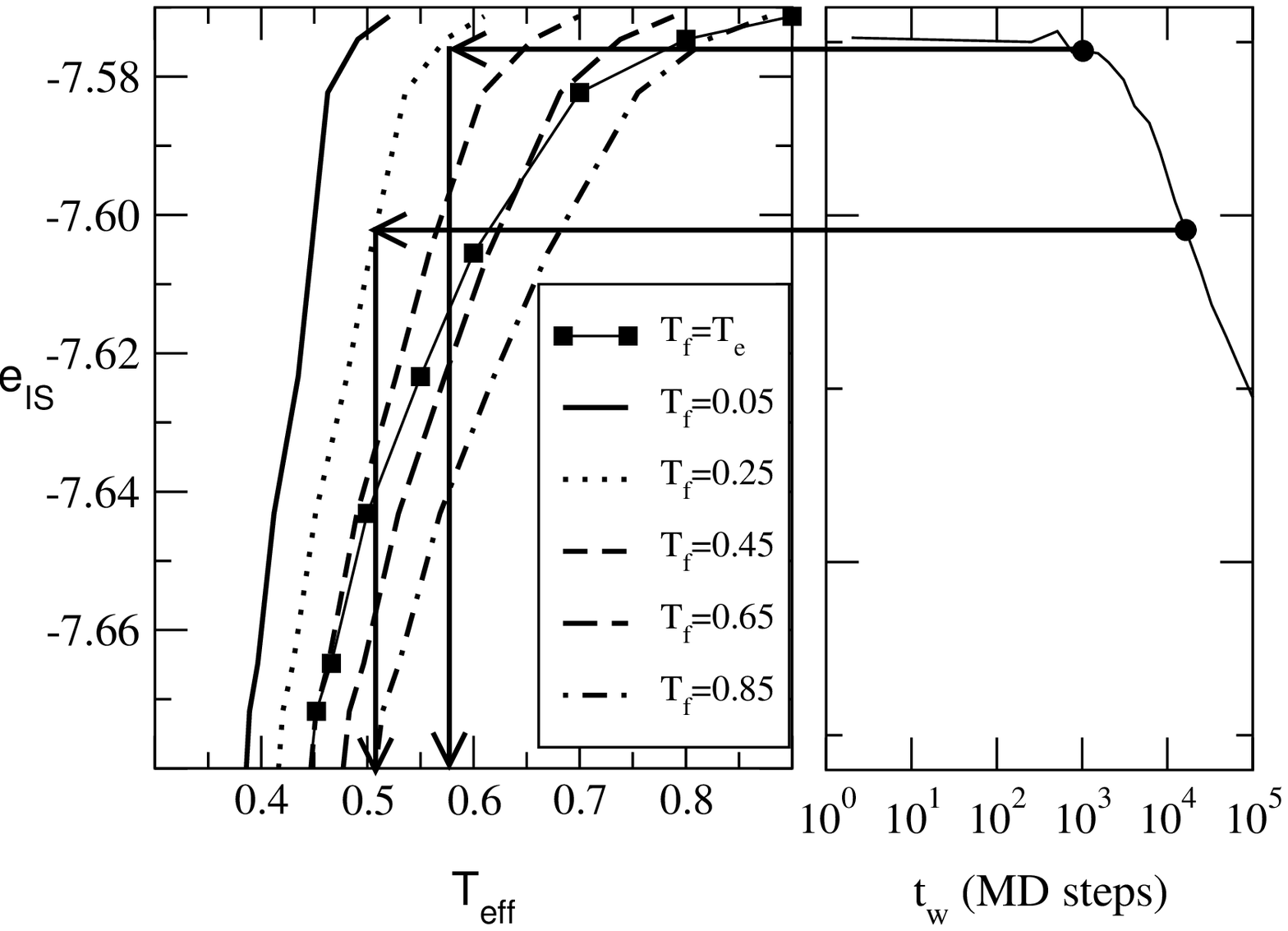}
\caption{Left: Solutions of  (\ref{eq:7.1-7}) for several  values of 
         of the final quench temperature $T_{\rm f}$ 
         for the BMLJ system. 
         Right: $e_{\rm IS}$ as a function of time, following the 
         temperature quench.
         The arrows show graphically the procedure which connects 
         the $e_{\rm IS}(t)$ value to $\Te$ value, once $T_{\rm f}$ 
         is known. 
         [Data courtesy of F. Sciortino and P. Tartaglia,
          see also Ref. \cite{SciTar01}].
	 From \cite{CriRit02}. 
}
\label{Fig:7.1-5}
\end{figure}
As a consequence when the system is quenched from an high-temperature
state to temperature $T$, 
the fast intra component degrees of freedom will
quickly equilibrate with the thermal bath temperature $T$.
Applying the condition of minimum free energy to the system, 
constrained to stay in components of depth $E_{\rm IS}$, allows to
define an effective temperature
\begin{equation}
\label{eq:7.1-7}
\Te(E_{\rm IS},T) = \frac {(\partial / \partial E_{\rm IS})\, 
                                    F_v(T,E_{\rm IS}) } 
                                   { (\partial / \partial E_{\rm IS})\, 
                                    S_c(E_{\rm IS}) }
\end{equation}
which reflects the non-equilibrium net heat flow from the system to the 
thermal bath \cite{CugKurPel97}. 
This expression coincides with that proposed in \cite{FraVir00}
in the context of $p$-spin models, once the components are identified 
with the Thouless-Andreson-Palmer states \cite{ThoAndPal77,CrisSom95},
see also the discussion after \eq{sc_tempeff}.
Inserting into (\ref{eq:7.1-7}) the value of $E_{\rm IS}$ as function of
time one finally gets the function $\Te(t)$.
The definition of $\Te$ is shown graphically in 
figure \ref{Fig:7.1-5} \cite{SciTar01}.

The two-timescales scenario is rather well confirmed by numerical
results, as shown in the FD plots of figure \ref{Fig:7.1-6}
\cite{SciTar01}, where the FD plot for the binary mixture Lennard-Jones
system is reported. The full lines are the prediction from
(\ref{eq:7.1-7}), the agreement is rather good.
\begin{figure}
  \centering
  \includegraphics{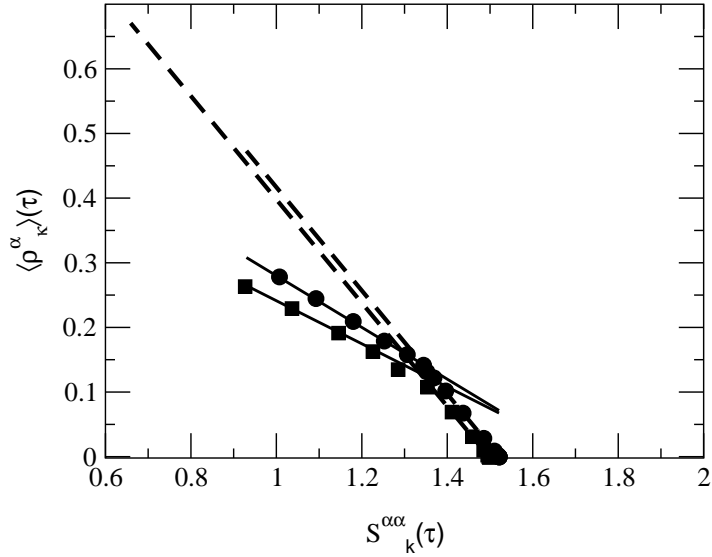}
\caption{Response $\langle  \rho^{\alpha}_{\bf k}(\tau)\rangle$ 
         versus the dynamical structure factor 
         $S_{\bf k}^{\alpha\alpha}(t) \equiv \langle \rho^{\alpha}_{\bf k}(t)
         \rho^{\alpha*}_{\bf k}(0) \rangle$, where 
         $\rho^{\alpha}_{\bf k}$ is the Fourier transform component of 
         the density of $\alpha=A,B$ particles at wave-vector ${\bf k}$,
         for the binary mixture Lennard-Jones 
         particles system for a quench from initial temperature $
         T_{\rm i} = 0.8$ to final temperature  $T_{\rm f}=0.25$ and two 
         waiting times $t_w=1024$ (square) and $t_w=16384$ (circle).
         Dashed lines have slope $T_{\rm f}^{-1}$ while thick lines have slope
         $\Te^{-1}$. 
         [Data courtesy of F. Sciortino and P. Tartaglia, 
          see also Ref. \cite{SciTar01}]. From \cite{CriRit02}}
\label{Fig:7.1-6}
\end{figure}

\subsubsection{Finite-size mean-field glasses}
\label{rom}
\cite{CriRit00,CriRit00b,CriRit00c,CriRit02}.

We have seen in the previous sections that the essential features of MCT
for glass-forming systems are also common to some fully connected
spin-glass models, called mean-field $p$-spin glasses with $p>2$.  In
the thermodynamic limit, the high-temperature paramagnetic phase is
described by the schematic mode MCT for super-cooled liquids.  At the
critical temperature $\Tc$ an ergodic to non-ergodic transition takes
place.  In mean-field models the relaxation time diverges at $\Tc$
as barriers separating different ergodic
components become proportional to the system size, thereby
diverging in the thermodynamic limit.
In real systems, or glass models just described, the barriers are of
finite height and the transition to a glassy state appears at the glass
transition temperature $\Tg<\Tc$, where the typical activation time over
barriers is of the same order as the observation time.

Despite these differences mean-field models, having the clear advantage
of being analytically tractable, are a very useful tool to study the
phase space structure of glassy systems, especially between the
dynamical temperature $\Tc$ and the static temperature $\Trsb$ 
(Kauzmann temperature $\TK$ in glass language).  
The main drawback is that, since
activated processes are not captured by mean-field models, the picture
that emerges is not complete.  To go beyond mean-field it is necessary
to include activated processes, a very difficult task since it implies
the knowledge of the excitations involved in the dynamics.  A simple
approach is to include {\it finite-size effects} in the dynamics of an
infinite mean-field system just extending the analysis to {\em
finite-size} mean-field.  This approach has been suggested by
Nieuwenhuizen \cite{Nieuwenhuizen98b} and is somehow reminiscent of the
dynamical approach of Sompolinsky \cite{Sompolinsky81},
see also section \ref{sk}.  We stress that
the assumption that finite-size mean-field models capture the physics of
glasses beyond MCT is not trivial.  In fact, activated process in
finite-size mean-field models could be different from those of
supercooled liquids, making the behavior different. This, for example,
seems to be the case of the Potts Glass model, where recent studies on a
finite-size version indicate some differences with the fragile-glass
scenario \cite{BraKobBin01,BraKobBin02}.

A spin-glass model in the $p$-spin universality class, that displays a
fragile-glass behavior, is the Random Orthogonal Model (ROM)
\cite{MarParRit94,ParPot95}. The model is defined by the
Hamiltonian\footnote[1]{The factor 2 in \eq{eq:7.1-9} is set only for
convenience to match the values of all relevant temperatures with those
reported in the original paper~\cite{MarParRit94}. The Hamiltonian
studied in \cite{ParPot95} differs in a factor $4$ from the present
definition. To compare the results discussed here with those in
\cite{ParPot95} temperatures and energies must be properly scaled by a
factor 4.}
\begin{equation}
\label{eq:7.1-9}
 {\cal H} = -2\sum_{ij}\, J_{ij}\, \sigma_i\, \sigma_j
\end{equation}
where $\sigma_i$ are $N$ Ising spin variables ($\sigma= \pm 1$) and
$J_{ij}$ is a random $N\times N$ symmetric orthogonal matrix with zero
diagonal elements. We note that at difference with previously discussed
spin-models, the condition of orthogonality leads to a strong
correlation among the matrix-elements. In the limit $N\to\infty$ this
model has the same thermodynamic properties as the $p$-spin model. The
dynamical transition is at $\Tc = 0.536$ with threshold energy per spin
$e_{\rm th} = -1.87$. A static transition occurs at $\Trsb = 0.256$
and the critical energy per spin is $e_{\rm 1rsb} = -1.936$ where the
complexity vanishes \cite{MarParRit94,ParPot95}.  The analysis in the
mean-field limit gives a rather clear ``geometrical'' interpretation of
the two transitions. The phase space is composed by an exponentially
large (in $N$) number of components, separated by infinitely large (for
$N\to\infty$) barriers.  Each component is labeled by the energy density
$e$ of its minimum and the largest allowable value of $e$ is $e_{\rm
th}$ \cite{CavGiaPar98,BarFra98}.  Components with $e$ equal to $e_{\rm
th}$ have the largest (exponentially with $N$) statistical weight and
become dominant, in thermodynamic sense, at the dynamical transition
$T=\Tc$.  Since components with $e$ smaller than $e_{1rsb}$ have
negligible statistical weight \cite{CrisSom95,ParPot95}, the static
transition is ruled by components with $e=e_{1rsb}$, i.e., the lowest
accessible ones \cite{KirWolb87b,CrisSom95}.

For finite $N$ the scenario is different since not only basins with 
$e < e_{\rm th}$  acquire statistical weight, but basins with 
$e>e_{\rm th}$ with few negative directions \cite{CavGiaPar98} 
may become stable, 
simply because for finite $N$ there are not enough degrees of freedom to 
hit them. 
The ROM  for finite $N$ has been largely studied during the last years
\cite{CriRit00,CriRit00b,CriRit00c,CriRit02,RaoCriRit02a} and 
its behavior has been compared with that of supercooled liquids
finding a remarkable agreement.

In the left panel of figure \ref{Fig:7.1-7}  it is shown 
the average energy-minima of basins (IS) as function of temperature for
the ROM with $N=300$ obtained from a Monte Carlo simulation.
It can be shown \cite{Heu97,BucHeu99} that if the density of states
$\Omega(E)$ is Gaussian and if the basins have approximately the same
shape or are, to a good degree, harmonic, then the IS energy density
$e_{\rm IS}\propto 1/T$.
The data in the figure can be well fitted by
\begin{equation}
\label{eq:7.1-10} 
e_{\rm IS} =  e_{\infty} + e_1\,T^{-1} + O(T^{-2})
\end{equation}
indicating that for a relatively large energy range 
the basins are roughly of the same shape.  This means that 
the contribution $f_{v}(T,e_{\rm IS})$ to the free energy density 
of the system is of the form
$f_{v}(T,e_{\rm IS}) = e_{\rm IS} + \delta f_{v}(T)$, with the second term
independent of $e_{\rm IS}$, i.e., of the component \cite{CriRit00}.  
This in turn implies that the effective temperature for the ROM is
completely determined once the complexity (density) $s_c(e)$ is known 
\cite{CriRit00b}. 
Indeed from \eqq{sc_tempeff}{eq:7.1-7} we have
\begin{equation}
\label{eq:7.1-11}
\frac{1}{\Te(e_{\rm IS})} = \frac
                        {\partial s_c(e_{\rm IS})} 
                        {\partial e_{\rm IS}}
\end{equation}
Furthermore, in an aging experiment $\Te$ depends on time only through
$e_{\rm IS}(t)$.
For each time $t$ the effective temperature $\Te$ can be obtained 
graphically as shown in figure \ref{Fig:7.1-7}.
The left panel in that figure  shows 
the average $e_{\rm IS}$ energy as a function of time in a typical
aging experiment. We note the two-regimes decay also observed in 
supercooled liquids \cite{KobSciTar00}. 
The two regimes are associated with different relaxation processes.
In the first part the system has enough energy and
relaxation is mainly due to {\it path search} out of basins through
saddles of energy lower than $T$, where $T$ is the temperature after 
the quench.
This part depends only on the temperature of the equilibrium state
from which the system has been quenched. 
This process stops when all barrier heights
become of $O(T)$ and relaxation slows down since it must
proceed via activated inter component processes.
\begin{figure}
  \centering
  \includegraphics[scale=0.8]{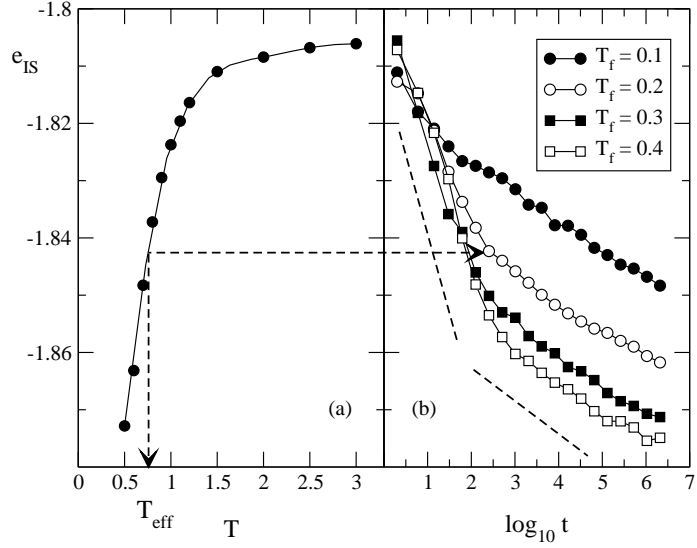}
  \caption{Panel (a): Equilibrium average $e_{\rm IS}$ a function of 
         temperature. The arrows indicate the construction of the
         effective temperature $\Te(e_{\rm IS})$. Panel (b): 
         Average inherent structure energy for the ROM as function of time for
         initial equilibrium temperatures $T_{\rm i}= 3.0$ and final
         quench temperatures $T_{\rm f}= 0.1$, $0.2$, $0.3$ and $0.4$. 
	 The average is over
         $300$ initial configurations. The system size is $N=300$. 
         The lines denotes the two regimes.
                  [see also Ref. \cite{CriRit00b}]
	 From \cite{CriRit02}.}
\label{Fig:7.1-7}
\end{figure}
In figure \ref{Fig:7.1-8} it is shown the response versus correlation
plot for the ROM. Correlations and responses were computed by projecting
over the IS, the corresponding FDT also holds in equilibrium as
discussed in section \ref{gen:EXTME}. The figure clearly show the
two-timescales scenario with $X=1$ at short times and 
$X=T/\Te<1$ at later times, with $\Te$ in very good 
agreement with
the value predicted by (\ref{eq:7.1-7}). Also notorious is the fact that
the effective temperature shifts with time as expected.
\begin{figure}
  \centering
  \includegraphics[scale=0.8]{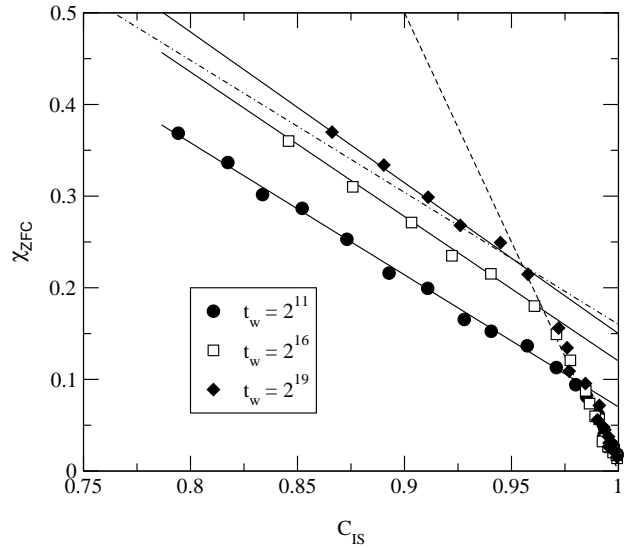}
  \caption{Integrated response function as a function of IS correlation
         function, i.e, the correlation between different IS configurations,
         for the ROM.  The dash line has slope $T_{\rm f}^{-1} = 5.0$,
	 where $T_{\rm f}$ is the final quench temperature, 
         while the full lines are the prediction
         (\protect\ref{eq:7.1-11}):
         $\Te(2^{11})\simeq 0.694$, 
         $\Te(2^{16})\simeq 0.634$ and
         $\Te(2^{19})\simeq 0.608$.
         The dot-dashed line is $\Te$ for
         $t_{\rm w}=2^{11}$ drawn for comparison.
         [see also Ref. \cite{CriRit00b}].
	 From \cite{CriRit02}.}
\label{Fig:7.1-8}
\end{figure}

\subsection{Spin glasses and other random systems} 
\label{disor}

As we have explained in section \ref{esm}, spin glasses represent the
most important motif of many results regarding FDT violations.
Particularly, numerical simulations have been the most widespread tools
to investigate many aspects of the equilibrium behavior of spin glasses
that cannot be tackled by analytic means (for a review see
\cite{MarParRui97}).  It is usually said that the advantage that
numerical simulations offer in the study of non-equilibrium properties,
as compared to equilibrium ones, relies on the fact that systems do
not need to be equilibrated.  However this observation is naive and deceitful
as many dynamical aspects cannot be observed in the range of accessible
time-scales. Indeed, it is widely believed that many dynamical results
in spin glasses are suspect because the asymptotic dynamical regime,
defined as that regime where the dynamic correlation length $\xi$ is
many lattice spacings, is usually not reached. Establishing whether the
range of simulated time-scales reaches the asymptotic long-time regime
is at the heart of a present controversy in the field. Indeed, not by
chance, this controversy is quite reminiscent of another parallel
ongoing discussion concerning the magnitude of finite-size effects in
the equilibrium properties.  Stochastic stability arguments linking
non-equilibrium with static properties
\cite{FraMezParPel98,FraMezParPel99a} confirm that any strong
finite-size corrections to the equilibrium properties should manifest
as strong finite time-scales corrections in dynamical
experiments. Unfortunately, a precise theory that quantifies (even in an
approximate way) these corrections is presently unknown.  We will not
deal here with the difficult issue of ascertaining in which cases
simulations do reach the asymptotic time regime, but present the
evidence on FDT violations for the accessible simulated time-scales.

\subsubsection{Spin glasses}
\label{disor:sg}

We begin our tour by reporting the first numerical
evidence of FDT violations in 3d EA spin glasses \cite{AndMatSve92}
and their representation in the form of FD plots \cite{FraRig95}. In
\cite{AndMatSve92} it was shown that deviations from the equilibrium
FDT appear at timescales comparable or larger than the age of the
system. In figure \ref{fig6DSSG} we show TRM measurements by
Franz and Rieger \cite{FraRig95} on the 3d EA model. In those
measurements the systems starts from a random initial configuration
and a magnetic field $h$ is applied for time $t_w$. The field is cut off
at $t_w$ and the subsequent decay of the 
$M_{\rm TRM}(t,t_w)$ recorded. In the linear response regime this
experiment is equivalent \cite{CugKurRit94} to a ZFC setup where the
field is initially zero and switched on at $t_w$, the resulting
 $M_{\rm ZFC}(t,t_w)$ being given by
$M_{\rm TRM}(t,t_w)+M_{\rm ZFC}(t,t_w)=M_{FC}$ for $t$ large enough (see
\eq{eq3e3}) where $M_{FC}$ is the
equilibrium magnetization. 
Most of the numerical simulations use
the ZFC procedure.
\begin{figure}
  \centering
  \includegraphics[scale=0.7]{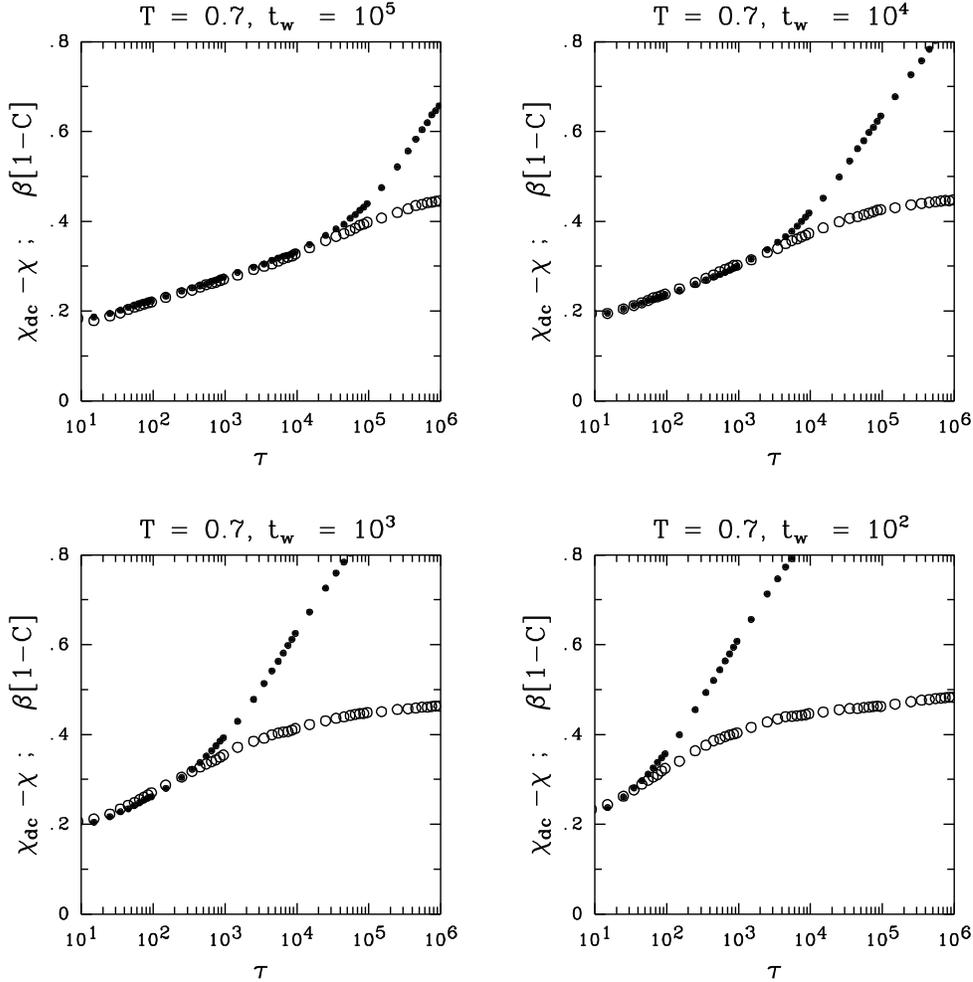}
  \caption{Zero-field cooled IRF
    $\chi_{\rm ZFC}(t,t_w)=M_{\rm ZFC}(t,t_w)/h=\chi_{dc}-\chi_{FC}(t,t_w)$ 
    (empty circles) in the 3d EA model plotted as function of the time $t\equiv
    \tau$ for different values of $t_w$ at $T=0.7$. For times $t\equiv
    \tau> t_w$ deviations from the FD relation
    $\chi_{\rm ZFC}(t,t_w)=\beta(1-C(t,t_w))$ (filled circles) are noticeable.
    From \protect\cite{FraRig95}.
  }
  \label{fig6DSSG}
\end{figure}
The most extensive simulations and the most clarifying FD
plots, as described in section \ref{IRF:FD}, have been done in $d=3$ and
$d=4$ 
\cite{MarParRicRui98a,ParRicRui98,ParRicRui98b,ParRicRui99,MarParRicRui00}. In
those papers, the authors consider the IRF associated to the global
 and the spin-spin autocorrelation function as described in
section \ref{IRF:FD}. The system is quenched at low temperatures for a time
$t_w$ and a small magnetic field is subsequently applied and the
 $M(t)$ measured. 
Typical FD plots in the 3d 
Edwards-Anderson (EA) model
using this construction are shown in figure \ref{fig4DSSG}.  There we show
$S(C)$ at two different magnetic field intensities as well as two
different waiting times. FD plots reveal that a constant slope $X(C)$
for $C<q_{EA}$ is a good approximation to the data (although more
accurate data in 4d hint at the existence of a curvature in
$S(C)$ \cite{ParRicRui98b,ParRicRui98}). This constancy of the slope
$X(C)$ (i.e. the linearity of $S(C)$ in the region where FDT is
violated) can be also interpreted as complementary evidence of the
accuracy of the $t/t_w$ scaling in the correlation
function \cite{PicRicRit01}.  
Stochastic stability
arguments \cite{FraMezParPel98,FraMezParPel99a} state that the dynamical
$X(C)$ is related to the static function $x(q)$ by the relation
$P(q)=x'(q)$ or,
\begin{equation}
\label{eq:7.5.3-1}
 x(q) = \int_{0}^{q}\, dq'\, P(q')
\end{equation}
where $P(q)$ is the probability distribution of overlaps between 
replicas of the same system. This identity offers a way to obtain $S(C)$ from equilibrium data. The
derivative of the relation (\ref{eq6e}) respect to $C$ yields
$P(C)=-{d^2S(C)}/{dC^2}$. Inverting this identity and inserting
an estimate of $P(C)$ as obtained from equilibrium simulations allows an
alternative way to compute $S(C)$. The applicability of this method is
shown in figure \ref{fig4DSSG}.
\begin{figure}
  \centering
  \includegraphics[scale=0.7]{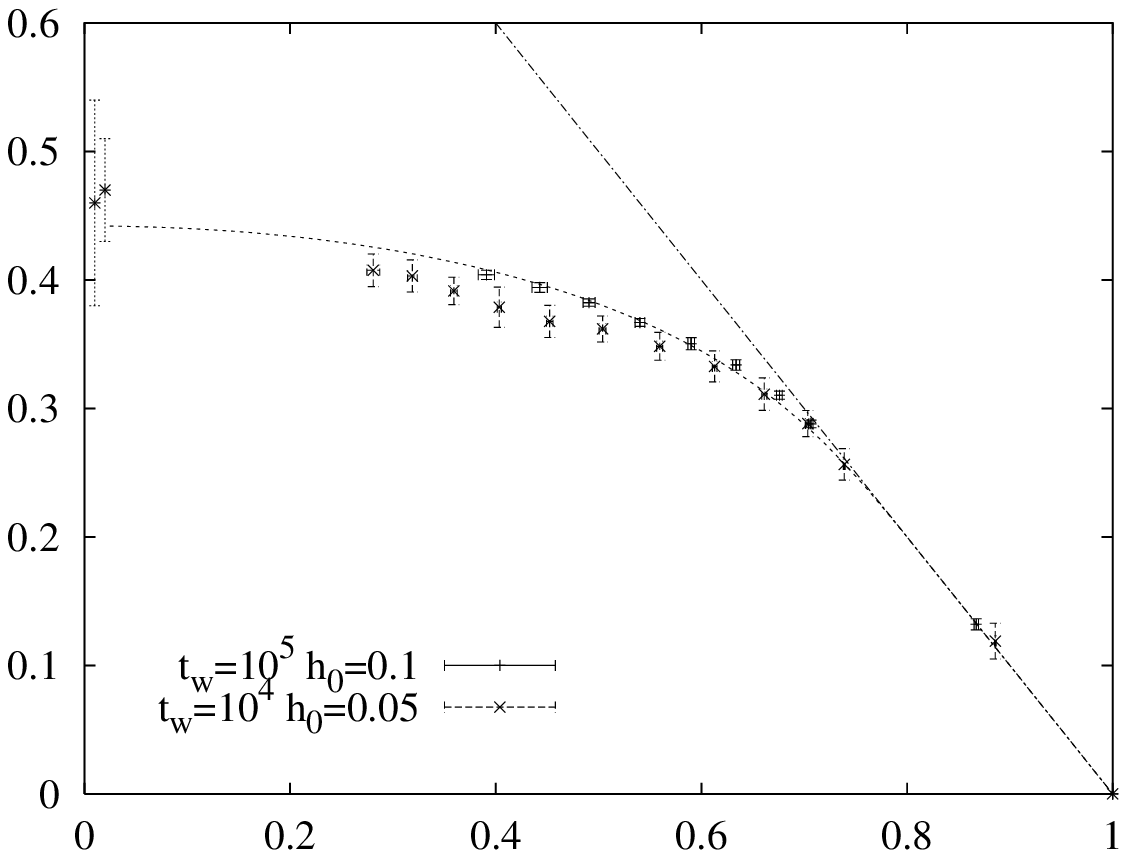}
  \caption{$S(C)$ versus $C$ in the 3d EA model at $T=0.7<\Tc\simeq 1$ for
    $L=64$. The continuous line is the prediction obtained from equilibrium
    data for $L=16$ (averaged over 900 samples) as explained in the text.
    The straight line is the FDT
    prediction. From \protect\cite{MarParRicRui98a}.
  }
  \label{fig4DSSG}
\end{figure}
A quantitative evidence of the mean-field character of the FDT violations
has been also reported by checking the accuracy of the Parisi-Toulouse
approximation in spin glasses (this approximation states that the
order-parameter function $q(x)$ is a function of the argument $x/T$, see
\cite{ParTou80,VanTouPar81} for an exposition). Within this
approximation (which works pretty well in MF spin glasses) it can be
shown \cite{MarParRicRui98a,MarParRicRui00} that $\chi(t,t_w)\equiv
\chi(C)=\beta S(C)$ is independent of temperature in the region
$C<q_{EA}$ where FDT is violated. In the SK model it can be proved that
this function $\chi(C)\simeq \sqrt{(1-C)}$. In general, for any
short-range system one can assume the following behavior: a)
$\chi(C)=\beta(1-C)$ for $C>q_{EA}$ when FDT holds and b)
$\chi(C)=A(1-C)^B$ for $C<q_{EA}$. Multiplying $\chi(C)$ by $T^{1-\phi}$
with $\phi=1/(1-B)$ one finds that the resulting quantity is a sole
function of the argument $(1-C)T^{-\phi}$:
$T^{1-\phi}\chi(C)=\hat{\chi}((1-C)T^{-\phi})$ thereby showing that data
for different waiting times and temperatures should collapse on a single
master curve. Bot in $d=3,4$ a best collapse is obtained taking $B=0.41$
\cite{MarParRicRui98a,MarParRicRui00} which is quite close to the
mean-field result $B=1/2$, see figure \ref{fig5DSSG}.
\begin{figure}
  \centering
  \includegraphics[scale=0.7]{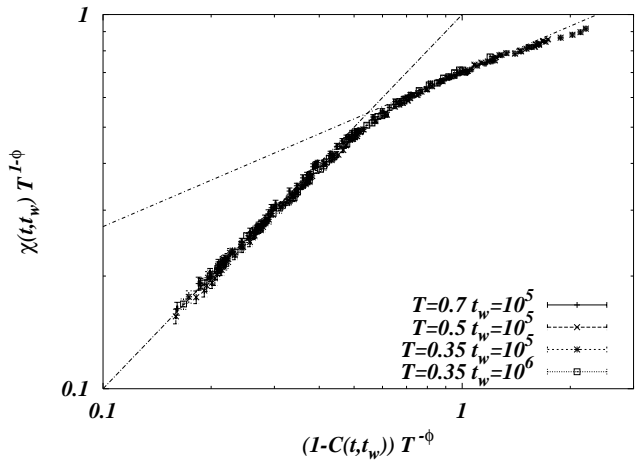}
  \caption{Scaling plot as described in the text for a cubic lattice,
    $d=3$ with $L=64$ showing the existence of two dynamical regimes
    described by two different scalings for the scaling function
    $\hat{\chi}(x)$ where $x=(1-C)T^{-\phi}$.  In the FDT regime
    $\hat{\chi}(x)\sim x$ while for $x>1$, $\hat{\chi}(x)\sim x^B$ with
    $B\simeq 0.41$. From \protect\cite{MarParRicRui00}.
  }
  \label{fig5DSSG}
\end{figure}

Apart form the EA model many other results have been obtained studying
short-range versions of the disordered p-spin model. Two different
models have been considered. 
On the one hand there is the so called {\em disordered plaquette model} 
\cite{AlvFraRit96a} where spins occupy the
vertexes of a finite-dimensional lattice and the interaction occurs
between $p$ spins belonging to a given plaquette. The Hamiltonian of
this model reads,
\be
{\cal H}=-\sum_{\square}J_{\square}\prod_{i\in\square}\s_i
\label{eqDSSG1}
\ee
where '$\square$' denotes a plaquette (not necessarily a square plaquette)
that connects different spins.
As usual, $J_{\square}$ are quenched
variables of zero mean and finite variance and $\s_i=\pm 1$ denote
Ising variables. The simplest case, the one considered in
\cite{AlvFraRit96a}, is a disordered version of the $p=4$ model and
consists of a regular lattice of side $L$ and dimension $D$ where spins
occupy the vertexes and plaquettes correspond to the different faces of
the lattice. Each face contains four spins and each spin belongs to
$4{D\choose 2}$ plaquettes.
A schematic picture of the lattice in $d=2$ is shown in figure \ref{fig1DSSG}.
\begin{figure}
  \centering
  \includegraphics[scale=0.4]{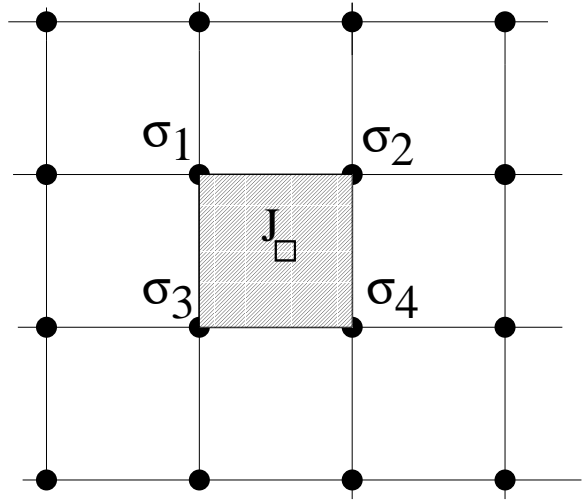}
  \caption{Schematic figure of the disordered plaquette model in $d=2$.
  }
  \label{fig1DSSG}
\end{figure}
The study of the static and dynamics properties of this model revealed
that, although there was no compelling evidence in favour of a finite
$T$ spin-glass transition, the relaxation time shows superactivation
effects and stretching of correlation functions characteristic of
fragile glasses. The relaxation time can be fitted both to a VTF law
with $T_0=0.65$ or to an exponential inverse temperature squared law
with $T_0=0$. The study of the equilibrium properties confirmed both
possible scenarios ($\Tc\sim T_0=0.65$ or $T_c\simeq 0$) but show that,
whatever scenario holds, the relaxation time $\tau$ and the equilibrium
correlation length $\xi$ are linked by the relation,
$\tau=A\exp(B\xi/T)$ supporting a scenario of cooperative dynamics.
Also, the trap-like character of the dynamics was confirmed by studying
the overlap among to identical replicas at $t_w$ but evolving with
different noises, $Q(t_w,t_w+t)$. This quantity should coincide with
$C(t_w,t_w+2t)$ if jumps among configurations are uncorrelated or
entropically driven. Numerical results are compatible with this
prediction. Accordingly, the FD plot (figure \ref{fig2DSSG}) measured at
$T=0.7$ (just above the suspected finite $\Tc$) showed strong deviations
from the equilibrium line $\chi=\beta(1-C)$ with a FDR $x\sim 0.4$, and
in agreement with the one-step pattern characteristic of structural
glasses (see section \ref{structural:glasses}).
\begin{figure}
  \centering
  \includegraphics[scale=0.9,angle=-90]{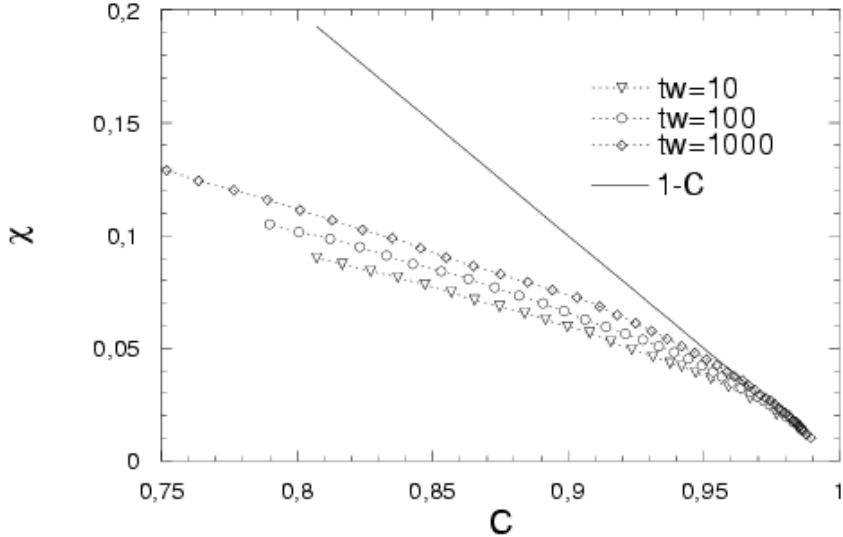}
  \caption{FD plot for the disordered plaquette model (with $p=4$) in a
    cubic lattice of lattice size $L=20$ at $T=0.7$ and three 
    different waiting time values. From \protect\cite{AlvFraRit96a}.
  }
  \label{fig2DSSG}
\end{figure}

On the other hand there is another short-range version of the $p$-spin
model \cite{MarNaiParPicRit98,FraPar99,CamParRan98,CamColPar98a,ParPicRit99}
where $M$ spins ($s_i^1,..,s_i^M$) occupy the different sites $i$ of a
cubic lattice. For each two adjacent sites $i$ and $j$ one considers all
possible groupings of different $p$ spins that can be formed by taking $k$
spins 
from site $i$ and $p-k$ spins
from site $j$. In an obvious abuse of notation we can write,
\be
{\cal H}=-\sum_{(i,j)}\sum_{g\in (i,j)}J_g\prod_{k\in g}\s_k
\label{eqDSSG2}
\ee
%
where the sum runs over all possible nearest neighbours $(i,j)$ and all
different groups of $p$ spins as described above.  Again, the $J's$ are
quenched variables with zero mean and finite variance. Two cases have
been considered: $M=2,p=3$ \cite{MarNaiParPicRit98,ParPicRit99} and
$M=3,4, p=4$ \cite{FraPar99,CamColPar98a} (for $M=2,p=4$ the model
reduces to the standard EA model \cite{FraPar99}). In the first case the
model is not time-reversal invariant while it is in the second case.  FD
violations have been measured in this last case
\cite{CamColPar98a}. Finite-size scaling studies of the model shows some
evidence of a second order phase transition at $\Tc=2.6$ characterized
by a divergent spin-glass susceptibility. This result is confirmed by a
study of the FDT violations in this model that show the existence of a
non-trivial $X(C)$ characteristic of a full RSB scenario. As for the EA
model, the $X(C)$ appears to be linked to the static $P(q)$ via the
relation $P(q)=X'(q)$, see figure \ref{fig3DSSG}.
\begin{figure}
  \centering
  \includegraphics[scale=0.4,angle=-90]{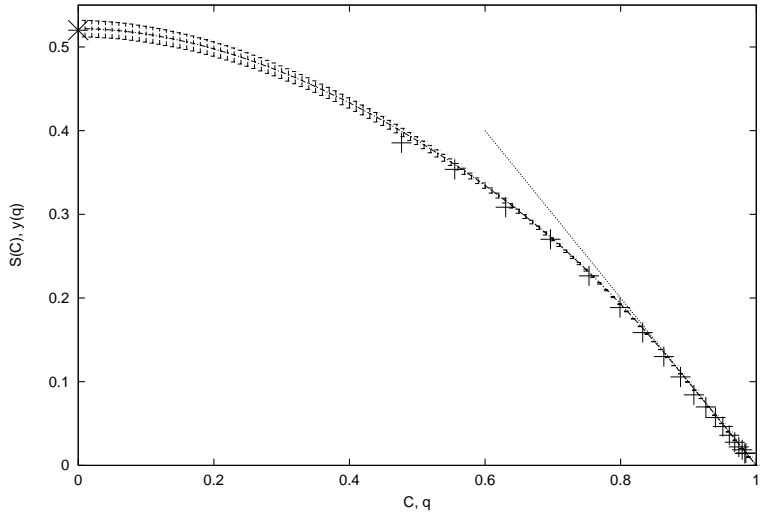}
  \caption{FD plot for the short-range version of the $p$-spin model (with
    $p=4,M=3$) in a cubic lattice of lattice size $L=16$ at $T=2.0=0.77 \Tc$
    and $t_w=2^17$. The continuous line with error bars is the shape of
    $X(C)$ derived by numerical integration of the relation $P(q)=x'(q)$ and
    measuring the equilibrium $P(q)$ for $L=5$, the dotted line is the
    equilibrium line $S(C)=1-C$ and the isolated cross at $C=0$ correspond
    to the FC magnetization. From \protect\cite{CamColPar98a}.
  }
  \label{fig3DSSG}
\end{figure}
The main message conveyed by most of these results is that FDT
violations are qualitatively and quantitatively well described in the
framework of MF theories of spin glasses. However, the
implications of these similarities must not be taken too far. In
particular, the already old but recurrent issue about the validity of
the many state picture in finite-dimensional spin glasses cannot be
answered from such a point of view. As discussed previously, the precise
link between statics and dynamics proposed by stochastic stability
arguments confirms that the dynamic $X(C)$ is related to the static
$x(q)$ after introducing a coherence length $l(t_w)$ (related to the
spin-glass correlation length $\xi(t_w)$ obtained from the two-point
replica correlation function) which depends on the waiting time $t_w$
according to the relation $x(q,l(t_w))=X(C(t,t_w))$. A numerical study
of FDT violations in $d=2$ \cite{BarBer01a} shows that such an assumption
is indeed true and $l(t_w)\propto \xi(t_w)$. More important, the
resulting FD plots are extremely similar to those found in $d=3,4$ and
the scaling Ansatz for $S(C)$, as derived from the assumption that the
low $T$ phase has many states, also works
pretty well. However, as in $d=2$ there is no finite $T$ transition
these results show that statements in favour of the validity (or not) of the
mean-field picture in the accessible range of timescales (in
off-equilibrium experiments) or sizes (in equilibrium measurements) are
inconclusive. The stochastic stability
property is well satisfied also in the XY model at not too low temperatures 
\cite{BerHolSel01} where critical fluctuations dominate, the overall resulting
behavior being quite similar to that of the 3d EA model.
From a completely different perspective, the overall
presence of mean-field aspects in the analysis of off-equilibrium data
suggests that the many state picture is effectively valid and that only
for experimentally inaccessible sizes or timescales (therefore
irrelevant from a practical point of view) the true scenario (whatever
mean-field or droplet) is recovered.

We finish this subsection by commenting some recent results
\cite{CasChamCugKen01} aiming to identify and quantify the low-energy
fluctuations that locally result in deviations from the average QFDT as
measured in the bulk.  Local deviations from the bulk QFDT curve
$\chi(C)$ are the equivalent of fluctuations from the average
magnetization in a Heisenberg magnet, where transverse fluctuations
correspond to low-energy spin-wave excitations and the longitudinal
fluctuations that modify the length of the magnetization vector being
the massive ones. Numerical results in the $3d$ EA model show that local
correlations $C_l(t,t_w)$ and IRFs $\chi_l(t,t_w)$ measured over local
boxes spread over the whole lattice generate a two-dimensional surface
$\rho(C_l,\chi_l)$ with a prominent maximum centered around the bulk
curve. The contour lines of this density map gently deform along the
QFDT bulk curve $\chi(C)$ and deviations far away from that curve appear
to be penalized. This study offers the possibility to understand the
connections between the mean-field character of the FDT violations and
the existence of deviations due to short scale cooperative processes in
an eventual (but yet unclear) heterogeneous scenario.

\subsubsection{Other random systems}
\label{disor:other}

Apart from spin glasses other lattice models with quenched randomness have been
considered in the literature aiming to elucidate whether
off-equilibrium studies can tell something about the character of the
low-temperature phase. Many of the conclusions of these
numerical studies need to be taken cautiously as no conclusive
evidence in support of a given scenario or in refusal of other ones is
ever reached. 

Let us start the discussion with the ferromagnetic diluted and
random-field Ising models (RFIMs). The 3d version of both models has
been investigated in \cite{ParRicRui98}. Simulations in the low $T$
phase and in the Griffiths phase (i.e. the region of temperatures
between the critical temperature of the pure system and the transition
of the random system) show that FD plots in ferromagnetic diluted and
RFIMs are very similar one each other but quite different to those
measured in spin-glass systems (see the preceding section
\ref{disor:sg}). The former ones are characteristic of a ferromagnetic
phase with $X=0$ while the latter are described by a non-trivial
function $X(C)$. These studies exclude the possibility of a spin-glass
and Griffiths phase in both models described by a mean-field like RSB
solutions.

Other finite-dimensional model with interesting behavior is the frustrated 
Ising lattice gas (FILG) \cite{NicCon97} defined by the Hamiltonian,
\be
\beta{\cal H}=-J\sum_{\langle i,j\rangle}(\eps_{ij}\s_i\s_j-1)n_in_j-\mu\sum_{i=1}^Vn_i
\label{eqDSSG3}
\ee
where the $\s_i=\pm 1$ are Ising spins and $n_i=0,1$ are occupancies
which may take the value 1 or 0 depending whether site $i$ is occupied
by a spin or not.  The sum is over near-neighbours on a $d$-dimensional
lattice.  The $\eps_{i}$ are quenched random variables that may take the
values $\pm 1$ and $\mu$ stands for a chemical potential. The average
particle density $\rho=\frac{1}{V}\sum_in_i$ is a monotonically
increasing function of $\mu$. In the limit $J\to\infty$ the model
converges to the site frustrated percolation problem \cite{NicCon97} (a
variant of the standard percolation problem where clusters are made out
of sites connected by non frustrated links).  This model has been
simulated in 3d \cite{StaAre99,AreRicSta00} where different regimes have
been singled out.  The percolation transition occurs at a given value
of the chemical potential $\mu_p$ and manifests in the onset of two
different relaxational regimes (a fast exponential relaxation followed
by a slow stretched decay). A second transition is observed at a higher
value of $\mu$, $\mu_d>\mu_p$ where the relaxation time grows extremely
fast and dynamics arrests. However, it is unclear whether the relaxation
time diverges at $\mu_d$. Less clear is, in the case of the existence
of a dynamical singularity at $\mu_d$, whether this is associated to a
thermodynamic singularity.  For the FILG \eq{eqDSSG3} different
correlation functions can be constructed depending on whether spin
variables $\s$ or occupancies $n$ are considered. A study of FDT
violations in this model leads to the following conclusions
\cite{AreRicSta00}: 1) In the glassy regime $\mu>\mu_d$ dynamics is
one-step like, i.e. a two-timescale scenario with two temperatures
describes pretty well the relaxational behavior; 2) The effective
temperature $\Te$, as derived from the slope of the FD plots, is pretty
independent on the observable, whether this corresponds to spin
variables ($C(t,t_w)=\la \s(t)\s(t_w)\ra$) or mixed spin-occupancy
variables ($C(t,t_w)=\la \s(t)n(t)\s(t_w)n(t_w)\ra$) and 3) $\Te$ is
apparently independent on the waiting time. However, this last result
has to be taken with caution as the range of waiting times considered in
\cite{AreRicSta00} may not be large enough~\footnote{In fact, the range
of waiting times explored does not even cover one order of magnitude.} to
display such a small effect (compare for instance with the results
described in section \ref{structural:glasses}, figure
\ref{Fig:7.1-8}). The observable independence of $\Te$ is shown in
figure \ref{fig7DSSG}.  It should be noted that the above scenario is
reminiscent of 1RSB behavior and occurs in finite $D$ rather than in the
mean-field limit $D\to\infty$. Actually the mean-field version of the
model does not have a 1RSB low temperature phase \cite{CriLeu02}.
\begin{figure}
  \centering
  \includegraphics[scale=0.75]{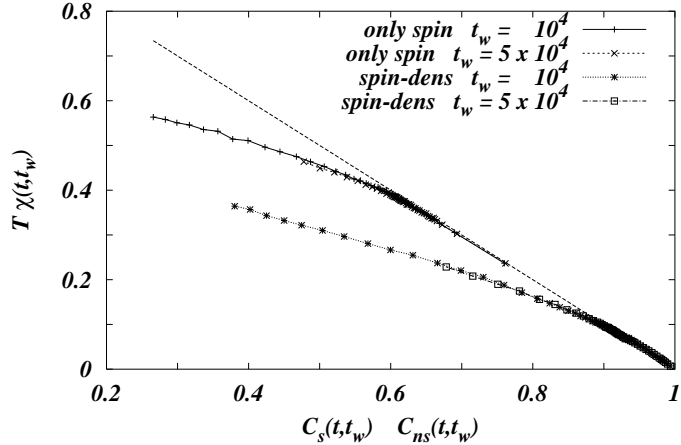}
  \caption{FD plot for the 3d FILG \eq{eqDSSG3} with lattice size $L=30$
    at $T=1,J=10,\mu\simeq 5.5$. Spin-density and spin autocorrelations (see
    the text) yield FD plots compatible with the same effective temperature.
    From \protect\cite{AreRicSta00}.
  }
  \label{fig7DSSG}
\end{figure}
\subsection{Coarsening systems}

\label{coars}

Although coarsening has been briefly sketched in section \ref{ferro}
here we present a more detailed account of results.
Coarsening systems are the paradigm of systems which do not reach
equilibrium.  In such systems TTI does not hold, and all time-dependent
correlation functions for large times are of the form $C(t,s) =
C(L(t)/L(s))$ where $L(t)\propto t^{1/z}$ is the typical size of the
coarsening regions \cite{Bray94}. The dynamic exponent $z$ is
characteristic of the universality class of the system and its value depends
on whether dynamics conserves or not the value of the order parameter.
This functional form is similar to that found for the long-time
correlation functions of glasses in the aging regime, and indeed a
certain type of coarsening has been advocated as responsible for slow
relaxation in glasses \cite{FisHus88,Fisher97,KivTar98,CriRit01a}.  The
difference between the two systems only becomes manifest when one also
considers the response functions associated to the correlation
functions.  Glasses, such as spin glasses or molecular glasses, are
characterized by long term memory which results in a non-zero
FDR $X$. On the other hand, for systems
stochastically stable \cite{FraMezParPel98,FraMezParPel99a} the FDR
$X(C)$ coincides with the static Parisi's function \eq{eq:7.5.3-1}. For
a ferromagnetic system $P(q)$ is trivial
\begin{equation}
 \label{eq:7.5.3-2}
  P(q) = \delta(q - m^2)
\end{equation}
where $m = m(T)$ is the magnetization a temperature $T$, and hence $X$
is $1$ if $1 > C > m^2$ and $0$ if $ C < m^2$.  The argument can be
easily extended to the case of few separate phases. Therefore we expect
that in systems in which two (or few) phases separate, $X$ should vanish
for long times, signaling the presence of weak long term memory.

The simplest model displaying domain-growth is a ferromagnetic
Ising model on a square or cubic lattice of linear size $L$
with a single-spin-flip Glauber dynamics.
When the system is quenched at time $t=0$ from a random configuration 
($T=\infty$)  to a finite temperature $T$ below the critical temperature
$\Tc$
domains of
``up'' and ``down'' spins start to form and grow.
This is well reflected by the  behavior of the two-times spin-spin
correlation function
\begin{equation}
\label{eq:7.5.3-3}
 C(t,s) = \frac{1}{N}\,\sum_{i=1}^{N} \langle \s_i(t)\, \s_i(s)\rangle
\end{equation}
which for times $t-s\ll s$ (assuming $s<t$) is TTI and
rapidly decays from $1$ to $m^2$, $m$ being the average magnetization at 
temperature $T$. Later, for more separate times $t-s\gg s$ the TTI
is lost, the aging part of the correlation 
scaling like 
\be
C_{\rm ag}(t,s)=F\left[\frac{L(t)}{L(s)}\right]
\label{inserted:coars:1}
\ee
where $L(t)$ is the typical size of the domains at time $t$.
The calculation of the linear response proceeds as usual, i.e., at a certain
waiting time $t_w$ a small magnetic field $h_i$ is applied and the 
induced magnetization is computed. For disordered systems, such as spin 
glasses, the applied field can be either uniform or random. The advantage of 
an uniform field is that averaging over different realizations of the 
field is avoided. However, for systems without disorder, such as 
ferromagnetic systems, an uniform field would favor one of the phases
making it growing faster. In this case a random field must be used
and the correct quantity to measure is the staggered magnetization
\cite{Barrat98}:
\begin{equation}
 \label{eq:7.3.5-4}
 M(t,s) = \frac{1}{N}\, \sum_{i=1}^{N}\, 
         \overline{\langle \s_i(t)\,h_i\rangle}
\end{equation}
where $h_i$ is the local (quenched) random field, and the over-bar denote
overage over the field realizations. 

In figure \ref{Fig:7.5.3-1} we show the
curves $\chi(t,t_w) = M(t,t_w)/h$ versus $C(t,t_w)$ 
obtained with a bimodal field
distribution $h_i = \pm h$ \cite{Barrat98}.
The FDT region and the flattering of the curve are well evident. 
Two aspect of these curves are worth to be noted. First of all 
the value of the plateau reached by the magnetization decreases as
$t_w$ increases. Moreover, for fixed $t_w$ the magnetization first 
grows in the non-aging part as $h(1-C)/T$, then saturates and eventually
goes down again. Indeed, the comparison 
\cite{BerBarKur99} with the equilibrium response function
shows that the equilibrium value of the response lies rather below the plateau.
The study of a soft spin version with Langevin dynamics
\cite{Barrat98,BerBarKur99} leads to similar results.
\begin{figure}
  \centering
  \includegraphics[scale=0.25,angle=-90]{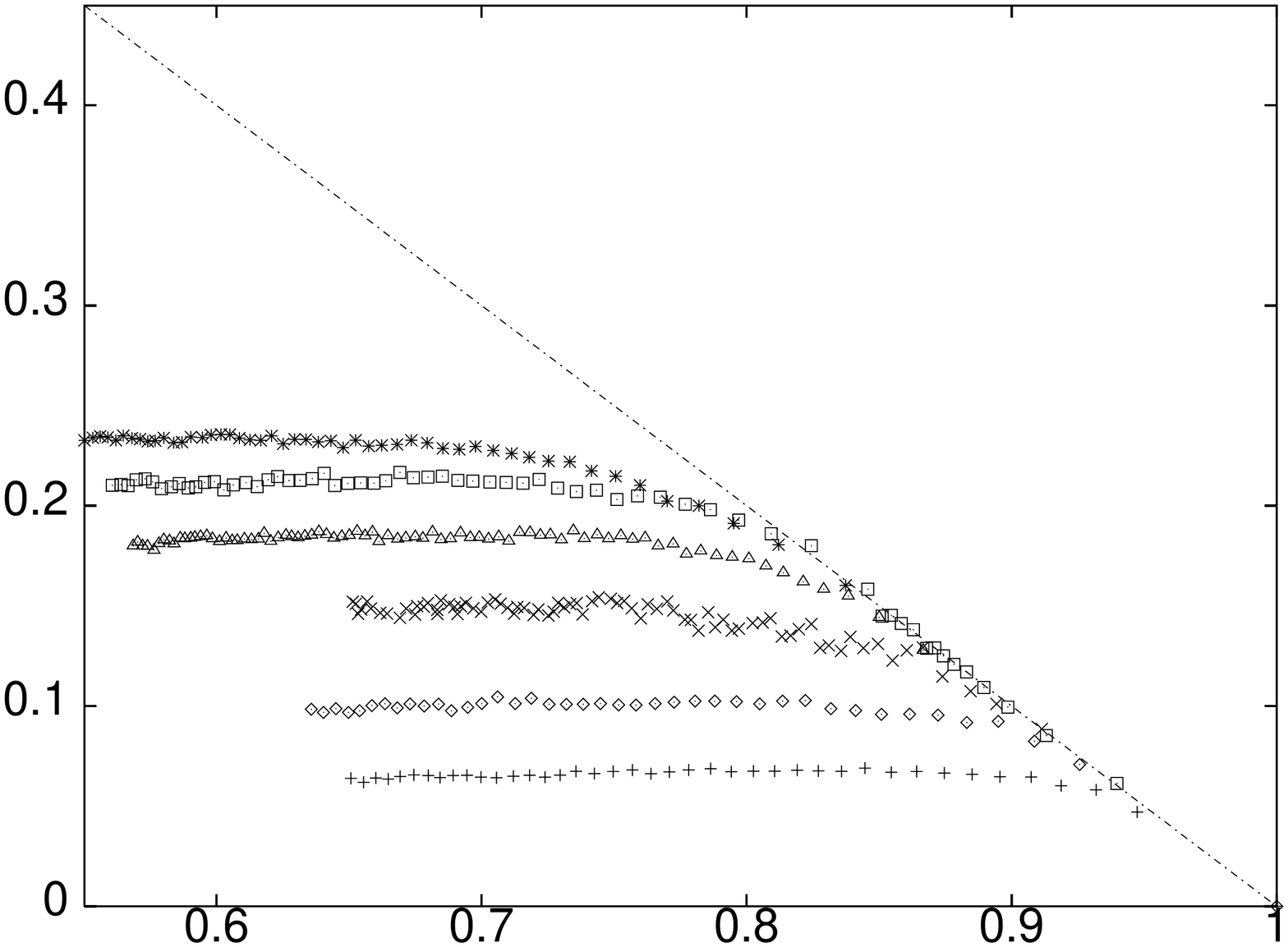}\includegraphics[scale=0.25,angle=-90]{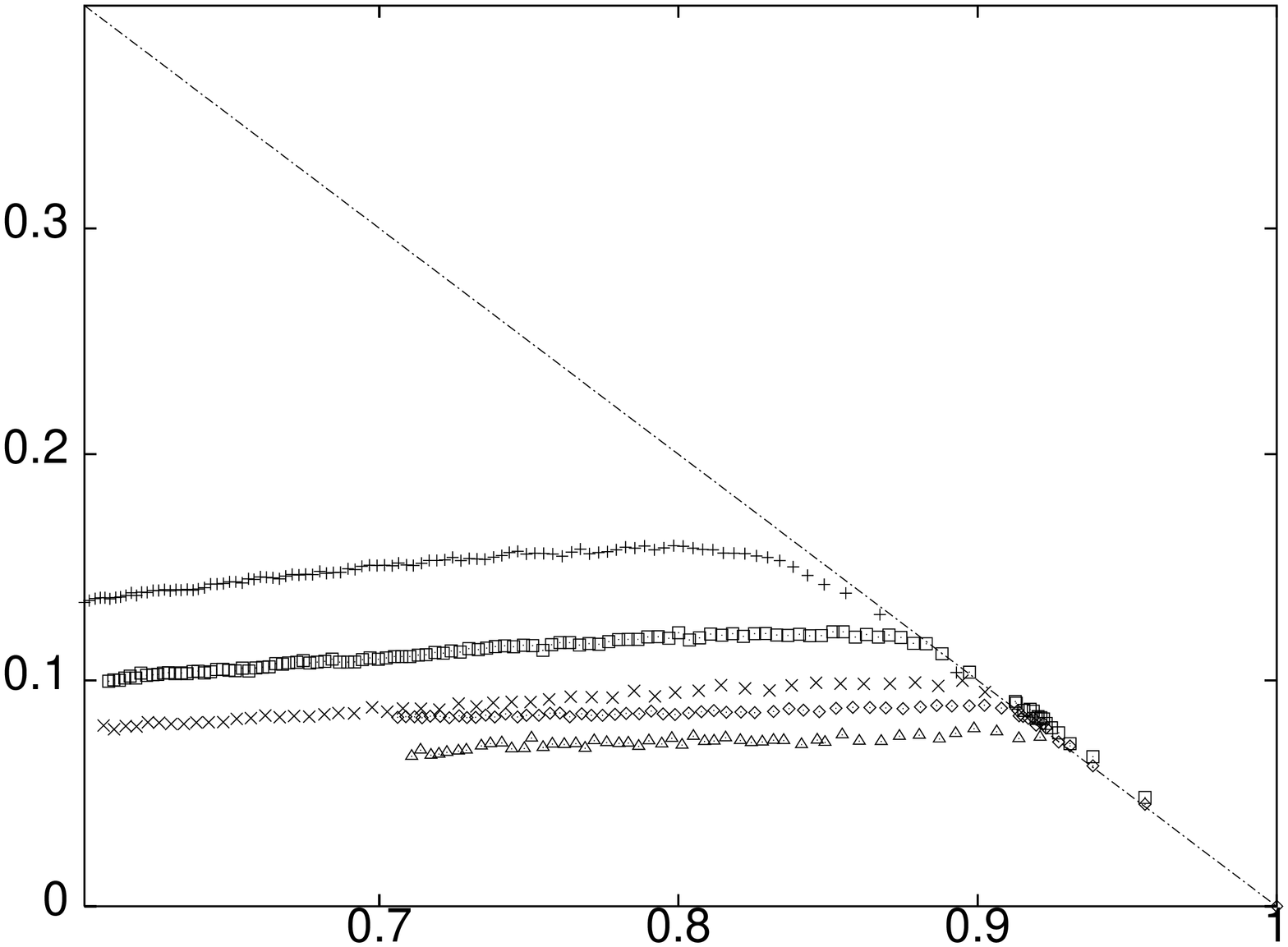}
  \caption{$T M(t,t_w)/h$ versus $C(t,t_w)$ domain growth. Left Panel:
  2d case with $\Tc=2.27$, at temperatures (from top to bottom) $T=1.7$
  and $t_w=200,\ 400,\ 800,\ 2000$, $T=1.3$ and $t_w=800$, $T=1$ and
  $t_w=800$. Right panel: 3d case with $T=2.5$ ($\Tc \approx 3.5$),
  $t_w=100,\ 300,\ 600,\ 1000,\ 1500$. The straight line is $M=1-C$: we
  see that FDT holds at short times $t$, and the violation of FDT with
  $X=0$ at longer time separation.  From \cite{Barrat98}}
\label{Fig:7.5.3-1}
\end{figure}

There are two contribution to the staggered magnetization: one from the
domain walls, the other from the domain bulks.  The difference between
the plateau and the equilibrium value of $M(t,t_w)$ can be attributed to
the domain wall response.  After a time $t_w$ the domains have reached a
certain typical size, and the domain walls have a certain total
length. The effect of the random field is to try to flip some
spins. Clearly the flipping is easier at the domain walls where the
spins are less constrained by their neighbors.  As the time proceeds the
domains grow increasing the bulks at the expense of the total domain
wall length. Therefore the contribution from the interfaces decreases
with time. On the other hand, the contribution of the bulk is almost
independent of $t_w$ since the effect of a random field on ``up'' and
``down'' domains is the same on average.  Therefore, after the initial
quasi-equilibrium growth, the total staggered magnetization decreases as
$t_w$ increases and, for a fixed value of $t_w$ it decreases as $t$
increases.  Analytical results \cite{BerBarKur99} show that the aging
part of the IRF is of the form
\begin{equation}
\label{eq:7.5.3-4}
 M_{\rm ag}(t,s) \sim A(s)\, F\left[\frac{L(t)}{L(s)}\right]\qquad 
C_{\rm ag}(t,s) \sim F\left[\frac{L(t)}{L(s)}\right]
\end{equation}
where
\begin{eqnarray}
\label{eq:7.5.3-5}
 A(t) &\sim& \frac{1}{L(t)}\qquad d > 2
\nonumber \\
      &\sim& \frac{\ln(L(t))}{L(t)}\qquad d =2~~~.
\end{eqnarray}
Because $X(s)\sim |\frac{\partial A(L(s))}{\partial L(s)}|$ the results
\eq{eq:7.5.3-5} explain the slower decrease of the IRF for the
two-dimensional case observed in figure \ref{Fig:7.5.3-1}. The numerical
test of the scaling law (\ref{eq:7.5.3-5}) in 2d is shown in figure
\ref{Fig:7.5.3-4}.

This scenario leads to the conclusion that coarsening systems do not
display a non-trivial $X(C)$. Recent results \cite{CorLipZan01},
however, indicate the possibility of non-trivial $X(C)$ also in these
systems.  The motion of the domain wall in the presence of an external
random field follows from two competing processes: the tendency to
reduce the interface curvature due to surface tension and the pinning of
the domain wall in favorable positions introduced by the external
field. This introduces a dependence on the space dimension since the
curvature process, which dominates at large enough dimensions, weakens
as the dimension decreases. When dimension reaches the lower critical
dimension the curvature process disappears and the response of the
system becomes non-trivial \cite{CorLipZan01,CorLipZan02} as it is
indeed seen for the ferromagnetic Ising chain ($d=1$)
\cite{GodLuc00,LipZan00} \footnote{See also the discussion in
section~\ref{ferro}). Similar effects are also observed in kinetically
constrained models described in Sec.~\ref{kcm}}.

The domain walls may give a large contribution to the response also at
the early FDT part, but almost exclusively from their deformation on
relatively short lengths. These fluctuations can be considered
thermalized and hence do not spoil the $1/T$ behavior but, on the
contrary, contribute to make longer the $1/T$ slope of the initial part
in the FD plot.
\begin{figure}
  \centering \includegraphics[scale=0.7]{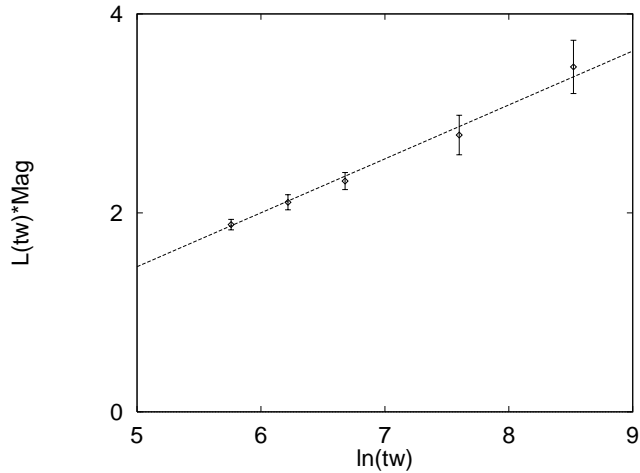}
  \caption{Test of the scaling (\protect\ref{eq:7.5.3-4}) -
  (\ref{eq:7.5.3-5}) which predicts a linear dependence of
  $L(s)M_{ag}(t,s)$ w.r.t. $\ln (s)$ with $s=t_w$. The dashed line fits
  this dependence very well.  From \cite{BerBarKur99}.}
\label{Fig:7.5.3-4}
\end{figure}
A generalization of the Ising model to include some frustration via
long-range anti-ferromagnetic interactions has been studied in
\cite{StaCan99}.  Models of this type have been proposed to study, among
others, the avoided phase transition in supercooled liquids
\cite{KivKivZhaNus95} and charge density waves in doped antiferromagnets
\cite{PryKivHon98}~\footnote{These models are of interest for the
information storage in ultra-thin ferromagnetic films
\cite{SamAlbMen96}}. They are described by the Hamiltonian
\begin{equation}
\label{eq:7.5.3-5a}
 {\cal H} = -\delta\sum_{\langle i,j\rangle} \s_i\, \s_j 
            + \sum_{(i,j)} \frac{\s_i\,\s_j}{r_{ij}^3}
\end{equation}
where $\s_i$ are Ising spins, the first sum runs over all pairs of
nearest-neighbor sites of the lattice, the second over all distinct
pairs, and $r_{ij}$ is the distance between sites $i$ and $j$.  The
parameter $\delta$ represents the local/non-local exchange ratio.  It
is known \cite{MacWhiRobBel95} that for a two-dimensional square lattice
the ground state of the model is anti-ferromagnetic for $\delta <
0.85$. For $\delta>0.85$ the anti-ferromagnetic state becomes unstable
with respect to formations of striped domains.  The study of
the non-equilibrium dynamics reveals a crossover from a logarithmic
decay for $\delta<\delta_c \sim 2.7$ to an algebraic decay for
$\delta>\delta_c$ \cite{StaCan99}.  However despite this richer
scenario, the FD plot leads in both cases to a vanishing $X$, and hence
to a coarsening scenario. Similar results are found in other
coarsening-like systems such as the Migdal-Kadanoff spin glass
\cite{RicRit00} although in that model the definition of the dynamics
appears rather tricky as most of the spins occupy the deepest
hierarchical layer in the model.

We conclude this section on coarsening discussing a hard-sphere lattice
gas model in the spherical approximation originally introduced by
Lebowitz and Percus (LP) \cite{LebPer66}. The interest in this model is
twofold. On the one hand, its dynamical behavior can be solved
analytically, hence allowing a detailed investigation of the
non-equilibrium behavior. On the other hand, it is known that
hard-sphere models have a fragile-glass behavior \cite{Speedy01}.

Lattice-gas models are defined on a lattice of finite dimensions. On each site
there can be a density $\rho(\bi{x})$ of particles, which in the limit of
hard-spheres can take only the values $0$ ({\it empty}) or $1$
({\it occupied}).  In the LP model this restriction is relaxed 
and $\rho$ takes any continuous value allowed by a spherical constraint:
\begin{equation}
\label{eq:7.5.3-6}
 \sigma_1 \equiv \sum_{\bi{x}} \rho(\bi{x})^2 - 
                 \sum_{\bi{x}} \rho(\bi{x}) 
          = 0
\end{equation}
There is the additional restriction that the density-density correlation
function between nearest neighbors vanishes. This is added to mimic some
kind of extended hard-core:
\begin{equation}
\label{eq:7.5.3-7}
 \sigma_2 \equiv \sum_{\bi{x}} \sum_{\bi{q}} \rho(\bi{x})\,\rho(\bi{x}+\bi{q})
          = 0
\end{equation}
where $\bi{q}$ are the vectors that join a lattice site $\bi{x}$ 
to its nearest neighbors.
In \cite{PadRit97} the following Langevin dynamics for an open system 
has been studied
\begin{equation}
\label{eq:7.5.3-8}
 \frac{\partial}{\partial t}\,\rho(\bi{x}',t) = \mu
        -\frac{\partial}{\partial\rho(\bi{x}')} {\cal H}[\rho(\bi{x}'),t] 
	+ \eta(\bi{x}',t)
\end{equation}
where $\mu$ is the chemical potential, $\eta$ the thermal noise, and
${\cal H}$ the Hamiltonian
\begin{equation}
\label{eq:7.5.3-9}
 {\cal H} = \lambda_0(t)\sum_{\bi{x}} \left[
                    \rho(\bi{x})^2 - \rho(\bi{x})
                                      \right]
            + \lambda_1(t) 
            \sum_{\bi{x}} \sum_{\bi{q}} \rho(\bi{x})\,\rho(\bi{x}+\bi{q})
\end{equation}
Strictly speaking in this model there is no energy, but only entropy.
The role of $\lambda_0(t)$ and $\lambda_1(t)$ in the Hamiltonian is to
make the dynamics to fulfill the constraints (\ref{eq:7.5.3-6}) and
(\ref{eq:7.5.3-7}) at all times. The non-equilibrium dynamics of this
model shares a large number of features with that of the spherical SK
model \cite{KosThoJon76,CiuDep88,CugDea95} where the dynamics is driven
by the macroscopic condensation of the system onto the disordered ground
state.  A positive chemical potential would increase the local density,
thus starting from an empty state the local density relaxes toward the
equilibrium value. The relaxation can be divided into two regimes, the
first one where the system is filled in a spatially uncorrelated way.
The typical time for this process is order $t^*\sim O(1)$.  It is only
later that slow relaxation starts when the system is spatially
correlated and needs to reorganize large regions in order to increase
its density.  The model has no built-in disorder, and the slowing down
is a purely entropic, direct consequence of the decrease of the number
of available configurations imposed by the the short-range
constraints. When the temperature is below the critical temperature
$\Tc$ the two motions have well separated timescales and the two-times
correlation function shows the usual two-steps form with a first part
TTI and the second part scaling as $t/t_w$, see figure \ref{aging:fig1}
(that plot corresponds to the 3d model at $T=0.1$. Different waiting
times from top to bottom are $t_w=10000$,1000,300,100,30,10,3,1) The
study of the FDR reveals that there is no anomaly in the response
function \cite{PadRit97} and $X$ vanishes for values of $C$ below the
plateau value of the correlation.  The glassy scenario of this model
corresponds to that of phase-ordering kinetics with non-conserved order
parameter. Similar results have been reported in spherical models with
long-range ferromagnetic interactions \cite{CanStaTam01}.

\subsection{Non-relaxational driven systems}
\label{nonrelax}

We have been underlining throughout this review that aging systems are
characterized by a non-equilibrium behavior with lack of TTI and by
the presence of FDT violations. The relevant parameter which {\em
controls} the aging non-equilibrium state is the waiting time or time
elapsed since the system was quenched. However, there
is another way to generate a non-equilibrium state that can be
characterized by a given timescale in the same fashion as the waiting
time characterizes the aging state. For instance, adding a
time-dependent perturbation of frequency $\omega$ to a
time-independent Hamiltonian. In the regime where $\omega t_w\ll 1$
the perturbation oscillates slow enough to probe only slow processes
occurring at time-scales $\tau\gg t_w$. While in the opposite regime
$\omega t_w\gg 1$ the oscillatory perturbation probes fast relaxation
processes occurring at timescales $\tau\ll t_w$. The line $\omega t_w\sim 1$ marks the onset
of glassy behavior, the shape of this line depending also on the intensity of the
perturbation. 

Driven systems have advantages when compared to aging systems. One of
the most important differences is that driven systems, in the stationary
state, are TTI but FDT is still violated. Again the concept of an
effective temperature can be introduced as a measure of these
violations. However, as the non-equilibrium stationary state of driven
systems can be described by the intensity of the driving force, they are
more experimentally accessible than aging systems, where the waiting
time appears as an external parameter difficult to control. For this
reason, it has been advocated that experimental measurements of FDT
violations and the effective temperature should be done in driven
systems rather than aging systems.

There are many ways to put a system into a driven stationary state and
these have been investigated in the literature for different types of
models. Driven systems can be classified into two main groups:

\begin{itemize}

\item{Sheared systems.} In this case one
considers systems where, in addition to conservative forces, other
non-conservative forces (i.e. that cannot be derived from a potential
function) act upon the system. In these systems the non-conservative
forces maybe time-dependent or not. In both cases, the non-conservative forces
do net work along a given closed dynamical path. This implies that
energy power is continually supplied to the system by the driving
force. The parameter which describes the stationary state is the
intensity of the shearing or driving force which we will identify by the symbol
$\eps$ or $\dot{\gamma}$. These systems include models that violate the action-reaction
principle (such as models with asymmetric couplings) and sheared fluids. 
These systems are described in section \ref{sheared}.

\item{Tapped systems.} In this class, systems are driven to a
non-equilibrium state by a time-dependent force which, however, derives
from a time-dependent potential. This means that the driving force, if
constant in time, does not exert work upon the system whatever its
intensity $\eps$. This class of systems includes spin-glass models in a
oscillating magnetic field and tapped granular systems where the
relevant parameter is the frequency of the driving force. These systems are
described in section \ref{tapped}. 

\end{itemize}

\subsubsection{Sheared systems}
\label{sheared}
Studies of models described by non-conservative forces go back to the
study of neural network models described by synaptic interactions that
are non-symmetric \cite{Parisi86}. This has inspired future
investigations of disordered models where couplings among spins include
an important degree of asymmetry. In \cite{CriSom87,CriSom88} it was considered
the relaxational dynamics of the $p=2$ spherical spin-glass with
pairwise interactions $J_{ij}$ given by $J_{ij}=J_{ij}^S+\eps J_{ij}^A$
where $J_{ij}^S=J_{ji}^S$ denotes the symmetric (therefore conservative)
part and $J_{ij}^A=- J_{ij}^A$ stands for the antisymmetric (therefore
non-conservative) part. It was shown
that any finite amount of asymmetry is enough to destroy the spin-glass
phase. The relaxational time of the system was found to diverge as
$\eps^{-6}$ for $\eps\to 0$. Importantly, this result suggests the
following general scenario: whatever the intensity of the
non-conservative force, the stationary state has a finite relaxation
time, therefore the stationary state, although of non-equilibrium
nature, must be TTI (and therefore,
correlations and response functions do not display aging). Subsequent
investigations have confirmed this result showing that this is a generic
feature of driven systems.

Among this family of asymmetric spin-glass models, one
which has been intensively investigated in the past years is the $p$-spin
spherical spin-glass Hamiltonian \eqq{eq:6.8}{eq:6.9} with asymmetry in the
interactions. The model is defined according to the following
Langevin dynamics \cite{CugKurDouPel97,BerBarKur00a}
\be
\frac{\partial \s_i(t)}{\partial t}=-r(t)\s_i(t)+F_i(\lbrace \s \rbrace )
  +\eta_i.
\label{sheared:1}
\ee
This dynamics is the same as  \eq{eq:6.11} but now the force $F_i(\lbrace
\s\rbrace)$ is replaced by,
\begin{eqnarray}
F_i(\lbrace \s\rbrace) &=-\frac{\delta{\cal H}}{\delta\s_i}
           +F_i^{\rm drive}(\lbrace \s\rbrace)
\nonumber\\
 &=-\frac{\delta{\cal H}}{\delta\s_i}
     +\eps\sum_{j_1<..<j_{k-1}}K_i^{j_1,j_2,..,j_{k-1}}\s_{j_1}..\s_{j_{k-1}}
\label{sheared:2}
\end{eqnarray}
where ${\cal H}$ is the Hamiltonian given in \eq{eq:6.8} and the driving
force $F_i^{\rm drive}(\lbrace \s\rbrace)$ describes a $k$-spin
interaction term where the couplings $K_i^{j_1,j_2,..,j_{k-1}}$ are
uncorrelated among all permutations of the $k$ different indexes
$(i,j_1,j_2,..,j_{k-1})$,
\be
\overline{K_i^{j_1,j_2,..,j_{k-1}}\,
          K_{j_r}^{j_1,j_2,..,j_{r-1},i,j_{r+1},..,j_{k-1}}}=0
\label{sheared:3}
\ee
These models show the following behavior. In the regime above the
mode-coupling temperature $\Tc$, the relaxation time is finite for the
unsheared model $\eps=0$. Therefore both TTI and FDT hold in the
stationary state. A small driving force $\eps>0$ puts the system in a
new stationary state where TTI holds but FDT is violated. The numerical
analysis of the mean-field equations \cite{BerBarKur00a} reveals that
both FD plots and the value of the FDR are very similar to those found
in aging systems. Figure \ref{fig:sheared:1} shows these quantities
above $\Tc$ for the model with parameters $k=p=3$. Below $\Tc$ a new
phenomenon, called ``shear thinning'' occurs. At $\eps=0$ the relaxation
time diverges (as is common in mean-field (MF) models where activated
processes are neglected, see discussion in section \ref{mct}). However,
as $\eps>0$ the relaxation time becomes finite and decreases with $\eps$
(shear thinning). Again, numerical analysis of the mean-field equations
reveals that for finite $\eps$ the resulting FD plots are the same as
for an aging system with a waiting time $t_w$ given by $t_w\sim
\eps^{-\alpha(T)}$ where $\alpha(T)$ is a temperature dependent exponent
that takes the value $2$ at $\Tc$ and slowly increases as $T$ decreases.
%
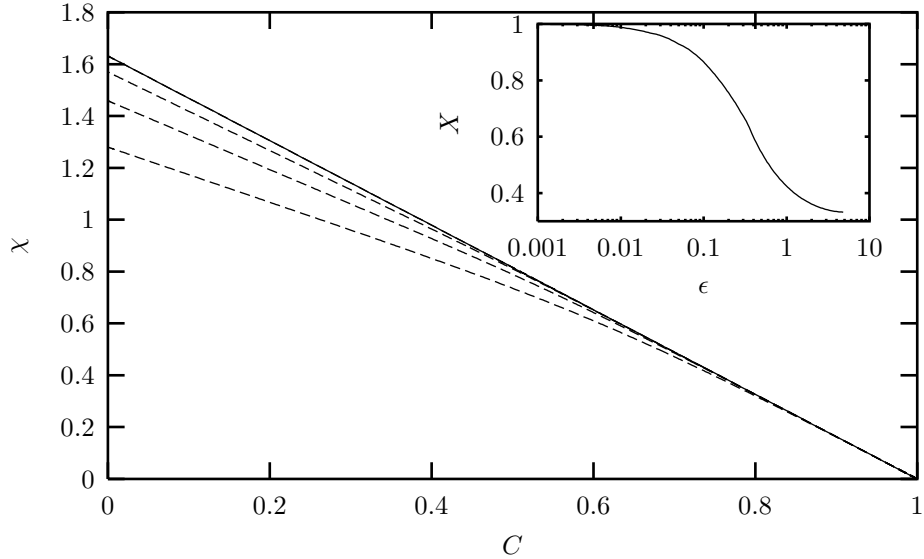
\begin{figure}[t]
\begin{center}
\vspace*{0.5cm}
\begingroup%
  \makeatletter%
  \newcommand{\GNUPLOTspecial}{%
    \@sanitize\catcode`\%=14\relax\special}%
  \setlength{\unitlength}{0.1bp}%
\begin{picture}(3600,2160)(0,0)%
\special{psfile=figsheared1 llx=0 lly=0 urx=720 ury=504 rwi=7200}
\put(2645,1022){\makebox(0,0){$\epsilon$}}%
\put(1720,1644){%
\special{ps: gsave currentpoint currentpoint translate
270 rotate neg exch neg exch translate}%
\makebox(0,0)[b]{\shortstack{$X$}}%
\special{ps: currentpoint grestore moveto}%
}%
\put(3270,1172){\makebox(0,0){10}}%
\put(2958,1172){\makebox(0,0){1}}%
\put(2645,1172){\makebox(0,0){0.1}}%
\put(2333,1172){\makebox(0,0){0.01}}%
\put(2020,1172){\makebox(0,0){0.001}}%
\put(1970,2016){\makebox(0,0)[r]{1}}%
\put(1970,1803){\makebox(0,0)[r]{0.8}}%
\put(1970,1591){\makebox(0,0)[r]{0.6}}%
\put(1970,1378){\makebox(0,0)[r]{0.4}}%
\put(1925,50){\makebox(0,0){$C$}}%
\put(100,1180){%
\special{ps: gsave currentpoint currentpoint translate
270 rotate neg exch neg exch translate}%
\makebox(0,0)[b]{\shortstack{$\chi$}}%
\special{ps: currentpoint grestore moveto}%
}%
\put(3450,200){\makebox(0,0){1}}%
\put(2840,200){\makebox(0,0){0.8}}%
\put(2230,200){\makebox(0,0){0.6}}%
\put(1620,200){\makebox(0,0){0.4}}%
\put(1010,200){\makebox(0,0){0.2}}%
\put(400,200){\makebox(0,0){0}}%
\put(350,2060){\makebox(0,0)[r]{1.8}}%
\put(350,1864){\makebox(0,0)[r]{1.6}}%
\put(350,1669){\makebox(0,0)[r]{1.4}}%
\put(350,1473){\makebox(0,0)[r]{1.2}}%
\put(350,1278){\makebox(0,0)[r]{1}}%
\put(350,1082){\makebox(0,0)[r]{0.8}}%
\put(350,887){\makebox(0,0)[r]{0.6}}%
\put(350,691){\makebox(0,0)[r]{0.4}}%
\put(350,496){\makebox(0,0)[r]{0.2}}%
\put(350,300){\makebox(0,0)[r]{0}}%
\end{picture}%
\endgroup
 
  \caption{FD plot for the model \eqqq{sheared:1}{sheared:2}{sheared:3}
    with $k=p=3$ and $T=0.613>\Tc=0.612$. The full line is the $\eps=0$
    equilibrium curve. Dashed lines correspond (from bottom to top) 
    $\eps=0.333,0.143,0.05,0$. The inset is the FDR as function
    of the intensity of the perturbation. From \protect\cite{BerBarKur00a}.
  }
  \label{fig:sheared:1}
\end{center}
\end{figure}
This scenario, as derived from the study of MF spin-glass models, has
been confirmed in numerical studies of binary mixtures of Lennard-Jones
(LJ) sheared fluids (see section \ref{lennard}). In a series of papers,
Barrat and Berthier \cite{BerBar01a,BerBar01b,BarBer01} have shown that
driven short-range systems display the same features as their equivalent
MF disordered models.  These similarities have been confirmed by other
studies that measure the temperature dependence of the shearing rate at which the
average potential energy deviates from its equilibrium
value \cite{AngRuoSciTarZam02}.  
The similarity between LJ models and
MF spin glasses is striking concerning rheological properties. In a
sheared fluid the velocity field $\bi{v}=\dot{\gamma}y\bi{e}_x$
induces a stress $\s_{xy}$ that is well described by the
phenomenological law: $\s_{xy}=\s_0+a\dot{\gamma}^n$. This law is
commonly found in rheological systems \cite{Larson99}. The viscosity is
defined by $\eta=\s/\dot{\gamma}$: for Newtonian fluids $n=1$ and
$\s_0=0$, the viscosity is therefore independent of the stress. However,
this is known to be inaccurate as many complex fluids show transport
coefficients (such as the viscosity) that depend on the shear rate (for
a discussion of these effects in the framework of non-equilibrium
thermodynamic theories see \cite{BedRub02,JouCasLeb99}). 
In figure \ref{fig:sheared:2} we show the flow curves for the viscosity as
function of the shear rate for both the LJ fluid and the MF spin-glass,
the viscosity $\eta$ and the shear $\dot{\gamma}$ in the LJ model
corresponding to the terminal relaxation time $\tau_{\alpha}$ and
$\eps/\tau_{\alpha}$ in the MF spin glass respectively. The similarity
is noteworthy. There are two regimes depending whether $T>\Tc$ or
$T<\Tc$. In the first regime the fluid is Newtonian at low shearing
rates so the viscosity is shear independent. In this regime, standard
non-equilibrium thermodynamics \cite{GroMaz} is applicable. However, for
$T<\Tc$ the fluid is non Newtonian and the viscosity diverges at zero
shear, $\eta\sim \dot{\gamma}^{-\alpha(T)}$ with $\alpha(T)$ between 2/3
(at $\Tc$) and 1 (for $T\to 0$) in agreement with the aforementioned
results found in the MF spin glass.
\begin{figure}
  \centering
  \includegraphics[scale=0.7]{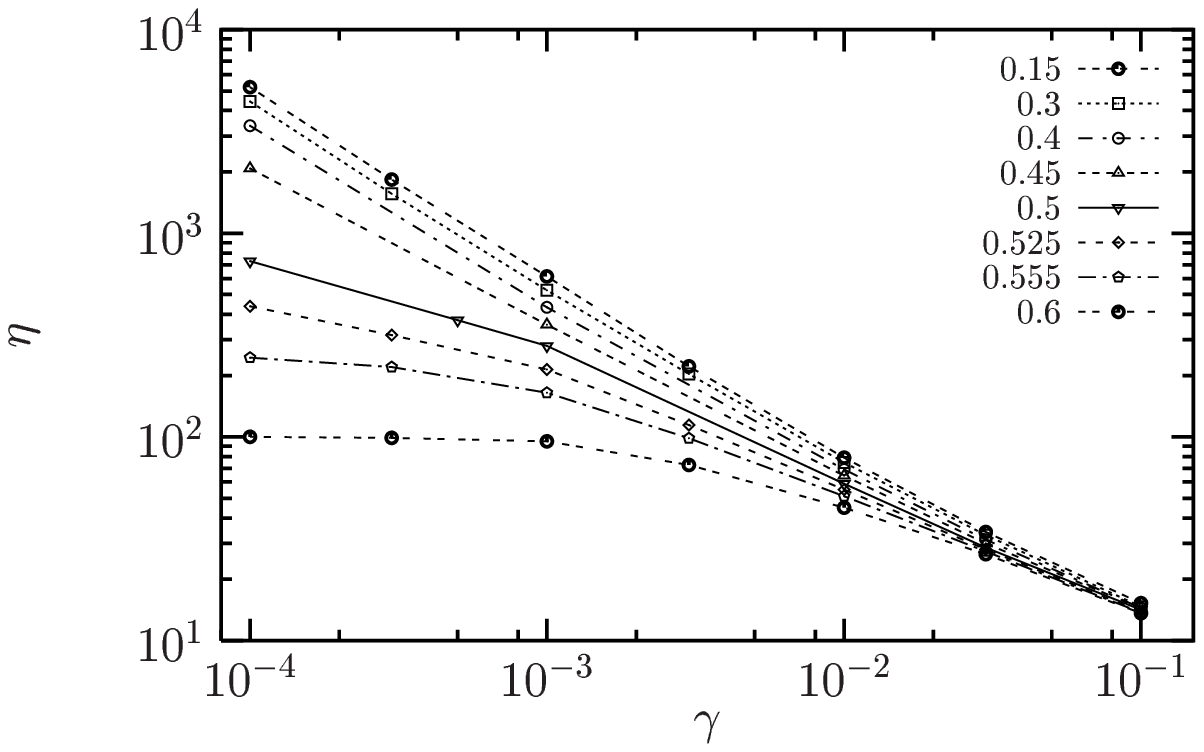}
  \includegraphics[scale=0.7]{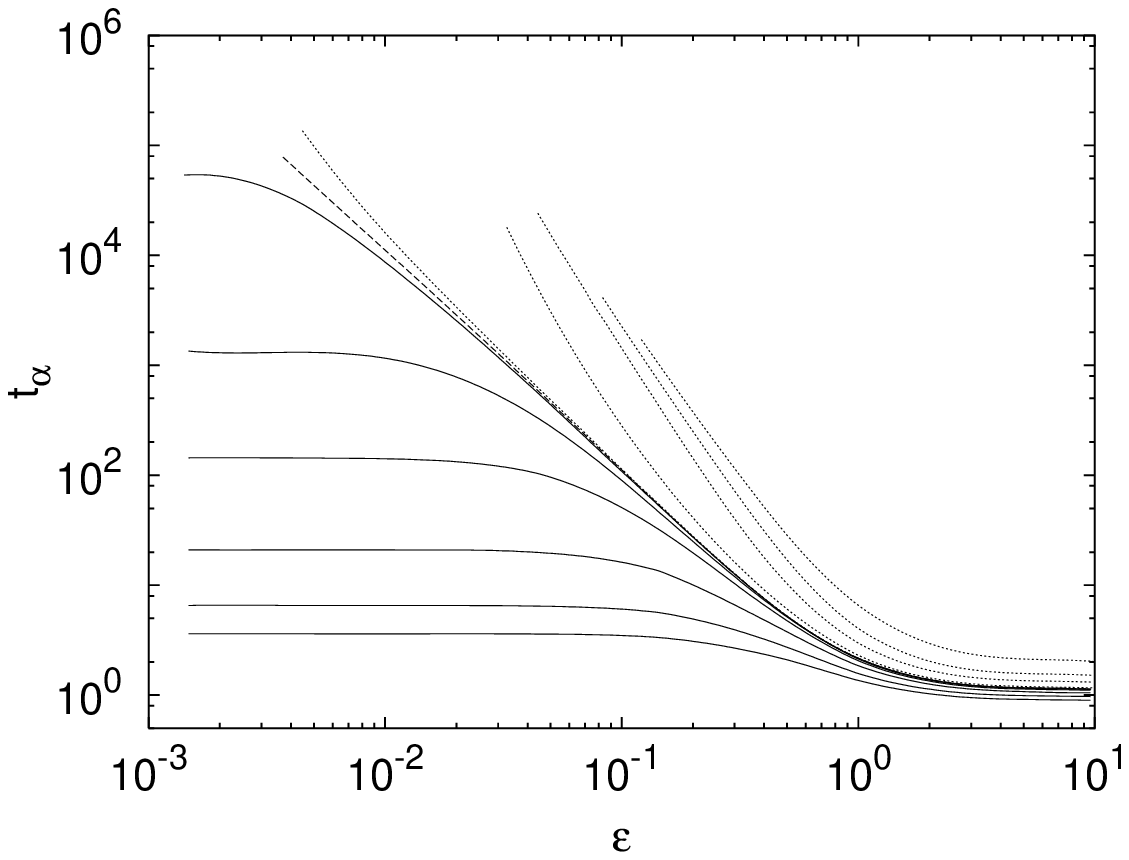}
  \caption{Viscosity as a function of the shear rate for the LJ (top
    panel) and MF
    spin-glass models (bottom panel). 
    The viscosity $\eta$ and shear rate $\dot{\gamma}$ in
    the LJ fluid correspond to the terminal time $\tau_{\alpha}$ and
    $\eps/\tau_{\alpha}$ respectively in the MF model. $\Tc\simeq 0.435$ in
    the LJ model and $\Tc\simeq 0.612$ in the MF model. Temperatures are
    (from bottom to top): (bottom panel)
    0.9,0.8.0.7,0.64,0.62,0.613,$\Tc$,0.6115,0.58,0.45,0.3,0.01 and (upper
    panel) as indicated in the box. From \protect\cite{BerBarKur00a} (left
    panel) and \protect\cite{BerBar01b}.
  }
  \label{fig:sheared:2}
\end{figure}
The central question addressed in \cite{BerBar01a,BerBar01b,BarBer01}
was the dependence of the resulting FD plots on the type of observable
used as a perturbation. They considered 5 different classes
of observables: 1) The incoherent part of scattering functions
(corresponding to single particle density fluctuations), 2) the coherent
part or the correlations, 3) The ``chemical'' observables associated to
correlations of a single species of particles in the binary mixture, 4)
The mean-square displacement of particles associated to a constant small
force transverse to the flow and 5) the stress in the transverse
direction after compression of the box. 1), 2) and 3) were measured at
different wavevectors. In all cases, the effective temperature was found
to be the same within numerical accuracy. A compendium of their results
is shown in figure \ref{fig:sheared:3}. 
\begin{figure}
  \centering
  \includegraphics[scale=0.8]{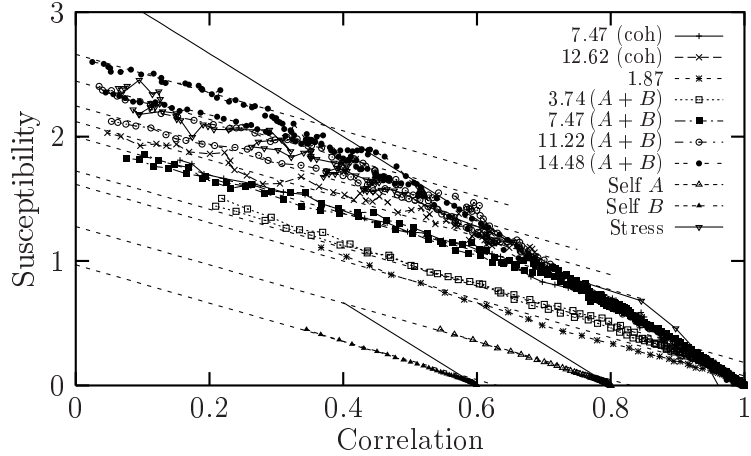}
  \caption{14 FD parametric plots in the LJ model showing the independence
    of the effective temperature on the observable. The numbers in the key
    of the figure indicate different values of the wavevectors. 1.87 refers
    to class 1 in the text. {\em coh} stands for coherent or global
    scattering functions (class 2). $A$ and $B$ refer to the two different
    chemical species, (A+B) standing for incoherent scattering observables
    independently measured for $A$ and $B$ species (class 3). {\em self}
    indicates the use of the mean-squared displacement (class 4). {\em
      stress} refers t class 5. The value of the effective temperature is
    compatible with $\Te=0.65>T=0.3$ for all cases. 
    From \protect\cite{BerBar01b}.
  }
  \label{fig:sheared:3}
\end{figure}
There have been other studies on driven systems that have investigated
the shear thinning effect, i.e. whether a driving force of rate
$\dot{\gamma}$ stops aging up to a timescale proportional to
$1/\dot{\gamma}$ by restoring TTI. Corberi et al. \cite{CorbGonLipZan02}
have investigated coarsening models in the presence of a driving
force. In particular, they have considered a $N$-component ferromagnetic
model with non-conserved order parameter in the large $N$ limit. The
authors find that, above the ferro-paramagnetic transition temperature
$\Tc$, the inverse of the driving rate $1/\dot{\gamma}$ sets the time
scale after which aging stops and the system becomes TTI. However, below
$\Tc$, contrarily to what is observed in MF spin glasses, aging never
stops even in the presence of shearing. The origin of this difference is
unclear.  It could be related to the fact that in coarsening systems the
mechanism responsible of aging is different than in MF spin glasses or
LJ glasses. This difference can be traced back to the
absence of complexity in the coarsening model as compared to the other
cases. Moreover, this behavior has to be contrasted with what is found
in tapped systems such as MF spin glasses in an AC field
\cite{BerCugIgu01} where aging survives below $\Tc$ in a certain range
of magnetic field values (see further discussion on this model in
section \ref{tapped}).  Other, studies have considered models such
as vortex glasses with random pinning centers in two dimensions
\cite{KolExaCugDomGro02} where the external driving force is uniform
over all vortices. Also in this case FDT violations appear to be
described by an effective temperature related to the slow motion of
vortices. Trapping dynamics due to pinning defects has also been
considered in the study of a driven classical particle submitted to a
force \cite{Horner96,Thalmann98}

A common feature of all studies reported here, that has not been
emphasized enough in the existing literature, is that measured
correlations must be transverse to the direction of the shear flow to
yield a meaningful effective temperature. FDT violations for longitudinal
observables are apparently not described by an effective
temperature. This suggests that neutral observables describing an
effective temperature are restricted to the transverse direction, a fact
that lacks yet a clear explanation.

\subsubsection{Tapped systems} 
\label{tapped}

As already said, another class of driven
systems correspond to those where the external force derives
from a time-dependent potential that pumps energy into the system. 
This includes also driven granular media that have recently received
considerable attention.

In \cite{BerCugIgu01}, the authors have studied the MF $p$-spin glass model
in an oscillating AC field.  This problem is interesting as it shows
a behavior different to that observed in sheared MF spin glasses or LJ
glasses (see section \ref{sheared}). The model considered was
again \eqq{eq:6.8}{eq:6.9} but in the presence of an uniform AC field of
frequency $\omega$ and intensity $h$. The phase diagram of the model is
described by three parameters $T,h,\omega$. Below $\Tc$, at fixed
$\omega$, the glassy phase survives below a critical field
$h^*_{\omega}(T)$ meaning that, contrarily to sheared systems, a
small driving force does not destroy the glassy phase and aging never
stops, hence TTI is not restored. This justifies the use of AC fields in
experiments to explore the aging regime within the linear response
region. A striking result in these studies is the presence of reentrant
behavior at constant field $h$, as $\omega$ is varied, indicating that
$\lim_{\omega\to 0}h^*_{\omega}(T,\omega)<h^*_{\omega}(T,\omega=0)$, a
result that needs yet to be clarified. All over the glassy phase of the
driven model FDT is violated with the characteristic FD plots of the
corresponding relaxational model.

Granular systems may present very slow processes, analogous to what is
seen in glasses. Similar to
what has been done for other glassy systems one can try to describe
the dynamics of the slow degrees of freedom through an effective
temperature defined from the FDR \cite{CugKurPel97}.
The first example we consider is the kinetically constrained 
Kob-Andersen model \cite{KobAnd93} that, even if very schematic, 
reproduces rather well several aspects of glasses \cite{KurPelSel97} and
of granular compaction \cite{SelAre00}. Although next section \ref{kcm}
is devoted to kinetically constrained models we prefer to describe 
the Kob-Andersen model here since it is a good model for granular media. 
This model consist of $N$ particles in a cubic lattice, with periodic boundary
conditions. There can be at most one particle per site. 
Apart from the hard-core repulsion there are no other static interactions 
among particles.
At each time step a particle can move
to a neighboring empty site of a three-dimensional lattice only if it has
less than four neighbors in the initial and final position. 
In its simpler version there is no gravity, but the system is subject to a 
constant pressure on its surface, obtained by adding or destroying particles
on the topmost layer with a chemical potential $\mu$.
The dynamic rule guarantees that at equilibrium all configurations
of a given density are equally probable. 
Indeed when the density are of the order of the 
the jamming density $\rho_g\simeq 0.88$ 
the particle diffusion becomes extremely slow, due to the kinetic constraints,
slowing down the whole compaction process. 
As done for glasses, one can try to describe the slow non-equilibrium motion
through an effective dynamical temperature $\Te$ 
which can be defined via a generalization of the 
Einstein-Stokes relation between the diffusion coefficient and the viscosity
to a non-equilibrium (aging) situation:
\begin{equation}
\label{eq:7.3-3}
 \frac{\partial}{\partial \bi{f}} 
    \langle\bi{r}(t) - \bi{r}(t_w)\rangle= -\frac{X(B)}{2T} 
\frac{\partial}{\partial t_w} \langle(\bi{r}(t) - \bi{r}(t_w))^2\rangle
\end{equation}
where $\bi{r}$ is the particle position, $\bi{f}$ 
a (small) perturbing field, $t>t_w$ two widely separated times,
and $B(t,t_w)$ the mean square displacement:
\begin{eqnarray}
\label{eq:7.3-4}
   B(t,t_w) &=& \langle(\bi{r}(t) - \bi{r}(t_w))^2\rangle\
\nonumber\\
                &=& \frac{1}{3N} \sum_{a=1}^{3} \sum_{i=1}^{N}
                   \langle[r^{a}_{i}(t) - r^{a}_{i}(t_w)]^2\rangle
\end{eqnarray}
The linear response function can be computed numerically by applying a
small random perturbation at time $t_w$ of the form \cite{Sellitto98}:
\begin{equation}
\label{eq:7.3-5}
 \delta {\cal H}_{\epsilon} = \epsilon\sum_{a=1}^{3} \sum_{i=1}^{N}
            f^{a}_{i} r^{a}_{i}
\end{equation}
where $f^{a}_{i}$ are independent quenched random variables which 
take the value $\pm 1$ with equal probability.
With this choice the IRF is defined as:
\begin{equation}
\label{eq:7.3-6}
 \chi(t,t_w) = \frac{1}{3N} \sum_{a=1}^{3} \sum_{i=1}^{N}
    (\langle f^{a}_{i} \Delta r^{a}_{i}(t)\rangle
\end{equation}
where $\Delta\bi{r}(t+t_w)$ is the difference between the displacement
of the same particle in two identical copies of the system one evolving
in presence of the external perturbation and one without.
From (\ref{eq:7.3-3}) $\chi$ is related to $B$ through;
\begin{eqnarray}
 \label{eq:7.3-7}
 \chi(t,t_w) &=& \frac{\epsilon}{2T}\,\int_{0}^{B(t,t_w)}
                X(B)\,dB
\nonumber\\
            &=& \frac{\epsilon}{2\Te}\,B(t,t_w),
           \qquad \mbox{if $X(B) = m = const$}
\end{eqnarray}
with $\Te = T/m$.
The numerical simulations show a two regime scenario \cite{Sellitto98} 
similar to what is seen in glasses, i.e., for $t-t_w$ smaller than $t_w$
$m = 1$ and FDT holds, while for $t-t_w \gg t_w$ FDT is violated with
$m<1$, figure \ref{Fig:7.3-1}.
\begin{figure}
  \centering
  \includegraphics[scale=0.7]{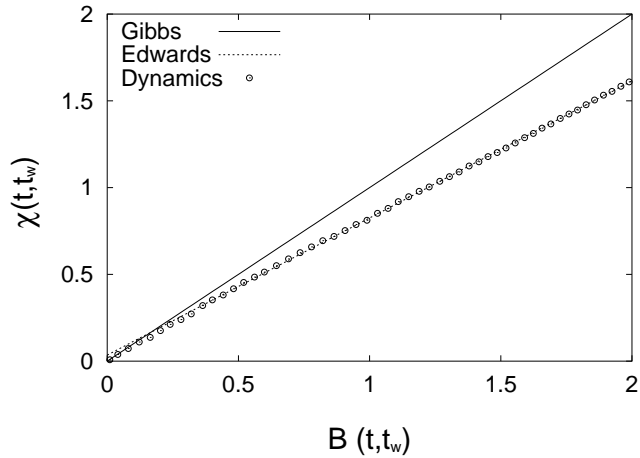}
  \caption{Einstein relation in the Kob-Andersen model: plot of  
    the mobility   $\chi(t,t_w)$ vs. the mean-square displacement  
    $B(t,t_w)$ (data shown as circles). 
    The slope of the full straight line corresponds to the 
    equilibrium temperature
    ($T=1$), and the slope of the dashed one to Edwards' prescription obtained 
    from figure \ref{Fig:7.3-2} at $\rho(t_w)=0.848$.
    From \cite{BarKurLorSel01}.}
\label{Fig:7.3-1}
\end{figure}
As discussed in section \ref{edwards} the value of the observables
attained dynamically in a granular system could be computed from the
usual equilibrium microcanonical distribution at the corresponding
density $\rho$ restricted to the subset of {\it blocked} configurations
(Edwards ensemble), i.e., only to those configurations in which every
grain is unable to move.  Much alike to the IS analysis in glassy
systems (section \ref{is}), the Edwards ensemble leads to
the definition a temperature, called the Edwards' temperature, as
\begin{equation}
 \label{eq:7.3-10}
  \frac{1}{T_{\rm Edw}(\rho)} = -\frac{1}{\mu}\frac{\partial s_{\rm Edw}(\rho)}
                                        {\partial \rho}
\end{equation}
where $s_{\rm Edw}(\rho)$ is the Edwards' entropy density obtained
from the logarithm of the number of blocked configurations of given $\rho$,
see (\ref{eq:edw2}). The chemical potential fixes the dimension
\cite{BarKurLorSel00,BarKurLorSel01}.

The Edwards' entropy for this
model has been computed in \cite{BarKurLorSel00,BarKurLorSel01} through
the use of an auxiliary model in which each particle has energy equal to
one if the dynamic rules allow it to move, and zero otherwise. The
auxiliary energy $E_{\rm aux}$ is hence equal to the number of mobile
particles. The configurations of the auxiliary model are sampled with
a Monte Carlo procedure with non-local moves at the auxiliary
temperature $1/\beta_{\rm aux}$. These non-local moves have nothing to
do with the true dynamics of the original model, and the auxiliary
model is not glassy. This allows to obtain equilibrium properties such as
the auxiliary energy density $e_{\rm aux}$ from which the entropy
density can be obtained via thermodynamic integration:
\begin{eqnarray}
\label{eq:7.3-8}
 s_{\rm aux}(\beta_{\rm aux},\rho) = && s_{\rm equil}(\rho) 
            +\beta_{\rm aux}\,e_{\rm aux}( \beta_{\rm aux},\rho)
\nonumber\\
            &&-\int_{0}^{\beta_{\rm aux}} e_{\rm aux}(\beta_{\rm aux}',\rho)
                          \,d\beta_{\rm aux}'
\end{eqnarray}
since $s_{\rm aux}(0,\rho) = s_{\rm equil}(\rho)$.
The blocked configurations at a given density can be computed
performing a simulated annealing down to $\beta_{\rm aux}\to\infty$ 
of the auxiliary model at fixed particle density.
The Edwards' entropy density is then obtained as
\begin{eqnarray}
 \label{eq:7.3-9}
 s_{\rm Edw}(\rho) &=& \lim_{\beta_{\rm aux}\to\infty} 
                     s_{\rm aux}(\beta_{\rm aux},\rho)
\nonumber\\
                   &=& = s_{\rm equil}(\rho)
                         -\int_{0}^{\infty} e_{\rm aux}(\beta_{\rm aux},\rho)
                          \,d\beta_{\rm aux}.
\end{eqnarray}
since $e(\beta_{\rm aux},\rho)$ vanishes for $\beta_{\rm aux}\to\infty$.
The Edwards' entropy is shown in figure \ref{Fig:7.3-2} as function 
of density. 
\begin{figure}
  \centering
  \includegraphics[scale=0.8]{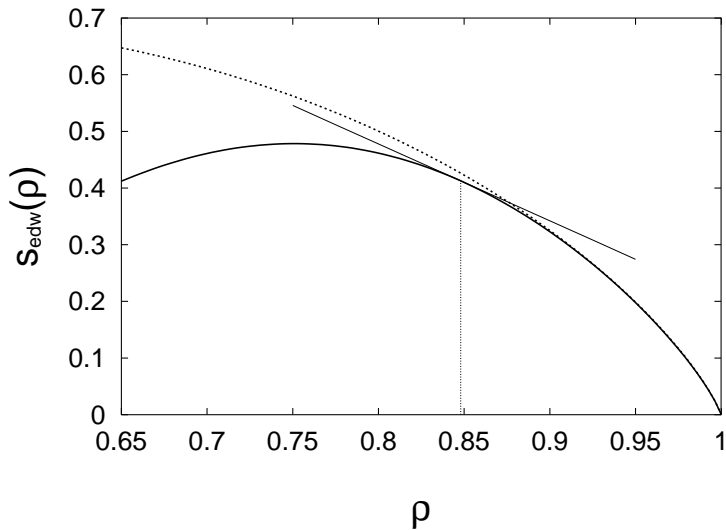}
  \caption{Edwards entropy per particle of the Kob-Andersen model vs. density
    (full curve).
    For comparison we also show the equilibrium entropy (dashed curve).
    At high enough density  the curves are indistinguishable,
    and join exactly only at $\rho=1$.
    The slope of the tangent to $s_{\rm Edw}(\rho)$ for a generic 
    $\rho$ allows to
    extract $T_{\rm Edw}(\rho)$ from the relation
    $T_{\rm Edw}(\rho) \frac{ds_{\rm Edw}}{d\rho}= \mu$.
    From \cite{BarKurLorSel01}.}
\label{Fig:7.3-2}
\end{figure}
We are now in a position to compare the long-time non-equilibrium results with 
those obtained from the Edwards' measure. In figure \ref{Fig:7.3-1} it 
is shown the mobility $\chi(t,t_w)$ versus the mean square displacement
$B(t,t_w)$. The agreement between $\Te$ and $T_{\rm Edw}$ is 
clearly satisfactory.

Similar results have been found
\cite{BarKurLorSel00,BarKurLorSel01,BarColLor02} also for models with
geometrical, rather than kinetical, constrains, the so called ``Tetris''
models \cite{CagLorHerNic97,Nicodemi99}. These models are defined on a
two-dimensional lattice with particles of randomly chosen shapes an
sizes. The only constraint is that particles cannot overlap: for two
nearest neighbor particles the sum of the arms oriented along the bond
connecting them has to be smaller that the bond length.

Tapped granular matter \cite{EdwGri98,EdwGri99} has been considered in
several models such as one-dimensional model ferromagnet \cite{LefDea01}
showing that the Edwards measure provides a very good description of the
stationary state. A similar agreement is found in simulations of sheared
granular matter \cite{MaaKur02}.  Real compaction occurs in presence of
gravity \cite{LevAreSel01}.  The gravity introduces an extra term in the
Hamiltonian
\begin{equation}
\label{eq:7.3-12}
 {\cal H}_g = mg\sum_{i=1}^{N} h_i
\end{equation}
where $g$ is the gravity constant, $m$ the particle mass and $h_i$ the
eight of $i$-th particle.  The simplest way of including gravity in
the above models is assuming that particles can move up and down (if
they respect the geometrical or kinetical constraints) with different
probability, $p = \min[1,\exp(-mg\Delta h/T)]$ where $\Delta h = -1,
0, 1$ is the elementary vertical displacement.  A closed boundary is
situated at the bottom of the system.  The control parameter $x =
\exp(-mg/T)$ represents the ``vibration''.  The presence of gravity
introduces a preferred direction in the diffusive motion (downward)
which in turn may produce inhomogeneities in the vertical density
profile making the situation more complicated, 
moreover horizontal (transversal) and vertical
(longitudinal) quantities must be treated separately.

Inhomogeneities are only along the vertical direction, so transversal
observables when measured well inside the bulk are not too sensitive to
the detailed form of the density profile. Indeed if homogeneity of the
bulk is imposed, the dynamical temperature obtained from the FDT ratio
for the horizontal displacement-mobility coincides with the Edwards'
temperature \cite{BarColLor02}.

The analysis of the FDT ratio for the longitudinal motion
 \cite{Sellitto01,Sellitto02} also shows a two-slope scenario, however
 the comparison with the Edwards' measure is more complex due
to the presence of inhomogeneities.
 In \cite{Sellitto02}, based on the observation that the density
 profile far from the top and bottom layers is rather flat, the
 effective dynamical temperature for longitudinal observables and the
 Edwards' temperature have been compared assuming an homogeneous
 density.  The comparison, however, reveal strong deviations when
 inhomogeneities are stronger indicating that inhomogeneities of the
 density profile must be included. This could be done by using the
 recent introduced restricted Edwards measure
 \cite{BerFraSel01,Lefevre02}, a route not yet explored.

We stress that the Edwards' measure is constructed from a white sampling of blocked
configurations, hence it reproduces the physical quantities at large times
only if this condition is satisfied by the long-time dynamics. In other words
the Edwards' measure can be inappropriate even though the system 
presents a slow dynamics. This is for example the case of 
the three-dimensional Ising model in a weak random magnetic field
\cite{BarKurLorSel00,BarKurLorSel01}. Here the long-time configurations
at low temperatures
are made of domains of ``up'' and ``down'' spins of similar volumes, so that 
the global magnetization is zero. This is quite different from either the
equilibrium or blocked configurations since both of them are magnetized.
Other studies have extended this method to the study of kinetically
constrained models (KCMs) (see section \ref{kcm}) such as the
Kob-Andersen model \cite{FraMulPar02}.

\subsection{Kinetically constrained models} 
\label{kcm}

A category of statistical models that has received considerably
attention during the last years are kinetically constrained models
(KCMs). In a nutshell, KCMs are models with trivial thermodynamics but
complicated dynamics arising from a, put by hand, set of forbidden
transitions in configurational space. Allowed transitions are selected
according to a given transition rule that constrains the dynamics of the
system. There are different ways to justify KCMs as valuable models for
glassy dynamics.  One can think of KCMs as effective models
in which the slowest degrees of freedom are idealized as quenched
variables that manifest as dynamical constraints.  Therefore, the
statistical variables of KCMs can be seen as the fast dynamical
variables that are slaved, through the dynamical constraints, to the
motion of the slowest ones.  These constraints are of local nature as 
also is the interaction between the original degrees of freedom.

KCMs display most (if not all) of the features characteristic of glassy
systems including slow relaxation, activated behavior, cooperativity and
non-equilibrium phenomena such as aging and FDT violations. Recently, a
review on KCMs has been written that covers all these aspects. For this
reason, here in this review we will not dwell much on discussing FDT
violations on these models but content ourselves on underlining some of
the most important results.  We refer the reader to
\cite{RitSol02} for a comprehensive and exhaustive survey.

The most representative families of KCMs are spin-facilitated models
\cite{FreAnd84,FreAnd85}, lattice gases
\cite{KobAnd93,KobAnd93b,KurPelSel97,PelSel98,Sellitto98}, topological
cellular models \cite{DavShe00,DavSheGarBuh01} and plaquette models
\cite{LipJoh00,LipJoh00b,BuhGar01b}. We already discussed in section
\ref{tapped} the Kob-Andersen model as an example of kinetically
constrained lattice gas for granular matter.
There are two aspects of
KCM's that makes them specially attractive from a theoretical point of
view. On the one hand, the thermodynamics of KCMs is straightforward,
there is no underlying thermodynamic singularity, and even more
astonishing, often the model is purely non-interacting from the point
of view of its equilibrium properties. On the other hand, the
non-interacting character of the energy function, entails that the
slow dynamics is described by representing the original variables in
terms of a new set of effective variables. Relaxation can be
visualized as a dynamical process where defects diffuse (in either a
free or cooperative way) and annihilate each other, leading to a more
tractable problem from the analytical point of view.

We begin our tour presenting the simplest among these type models. Our
intention is to illustrate with an example what type of models KCMs
are. Maybe the simplest KCM is the spin facilitated model (SFM)
introduced by Fredrickson and Andersen (FA)
model \cite{FreAnd84,FreAnd85} consisting of free ``spins'' in a
field. The model is defined by,
\be
E=h\sum_{i=1}^Nn_i
\label{kcm:1}
\ee
where $n_i=0,1$ corresponding to two possible orientations of the spin
($n=1$ up, $n=0$ down)
\footnote[1]{The variables $n$ cannot be considered as occupancies and
\eq{kcm:1} does not define a lattice gas model. In lattice gas models
the total number of particles (i.e. the energy $E$ in \eq{kcm:1}) is
conserved while in the SFM it is not. See \cite{RitSol02} for a
throughout discussion. However, we will continue denoting the spins by
$n$.}  that occupy a D-dimensional lattice. This model would be trivial
if it were not by the dynamical rules that describe how spins can
flip. In the standard SFM model transitions $n_i\to 1-n_i$ are allowed
if at least $f$ among the possible nearest neighbours are up or $n=1$.
This model defines the $f,d$-SFM that shows different behaviors
according to whether $f=1$ (diffusive) of $f>2$
(cooperative). Variations of the SFM include KCMs with directed
constraints such as the 1d East model \cite{JaeEis91} or the 2d
North-East model \cite{ReiMauJae92}, defined again by \eq{kcm:1}, but
where a spin $n_i$ can flip only if the spin on its left is up (East
model) and if both spins, that are nearest neighbours in two fixed
orthogonal directions (North-East model), point up. In 1d, it is also
possible to define a model that interpolates between the $1,1$-SFM and
the East model \cite{BuhGar01}. 

In what follows we concentrate our attention in FDT violations in
diffusive KCMs where some understanding has been recently gained.  A
generic aspect of KCM's is that dynamics is determined by the motion and
annihilation of isolated defects. Slow dynamics strictly occurs at
temperatures close to $T=0$ (and timescales larger than an initial fast
transient). The relevant relaxing variable is the number of isolated
defects $c(t)$ (e.g. in the $1,1$-SFM this corresponds to the number of
up spins). This has interesting consequences
\cite{BuhGar01b,CorLipZan01,CorCanLipZan02} in the behavior of the
IRF. The IRF in the presence of an external random staggered field
$\chi(t,t_w)$ is then proportional to the product of the number of
isolated defects $c(t)$ and the individual local response typical of
one defect $\chi_{\rm eff}$. This local response is TTI as the
defect is isolated leading to,
\be
\chi(t,t_w)\simeq \chi_{\rm ag}(t,t_w)=c(t)\chi_{\rm eff}\Bigl( \frac{t-t_w}{\tau(T)} \Bigr)
\label{kcm:2}
\ee
where $\tau(T)$ is the relaxation time of the defect. This relation has
two important consequences: 1) The function $\chi(t,t_w)/c(t)$ is TTI, a
result which is characteristic of coarsening systems (see below) but
not of other (cooperative) glassy models where the IRF is a genuine
aging function depending, for instance, on the ratio $h(t)/h(t_w)$ and 2) $\chi(t,t_w)$
has a maximum as function of $t-t_w$ when the defect concentration (that
decreases with time) compensates the growth of the monotonically
increasing effective response $\chi_{\rm eff}$. 

Non-equilibrium measurements for the
$1,1$-SFM \cite{CriRitRocSel00,CriRitRocSel02} and related models such
as the 2d triangle model \cite{NewMoo99,GarNew00} or topological
cellular models \cite{DavShe00,DavSheGarBuh01} show an IRF \eq{kcm:2}
displaying a maximum as a function of $t-t_w$ and leading to awkward FD
plots.  One example is shown in figure \ref{fig:kcm:1}.
\begin{figure}
  \centering
  \includegraphics[scale=0.8]{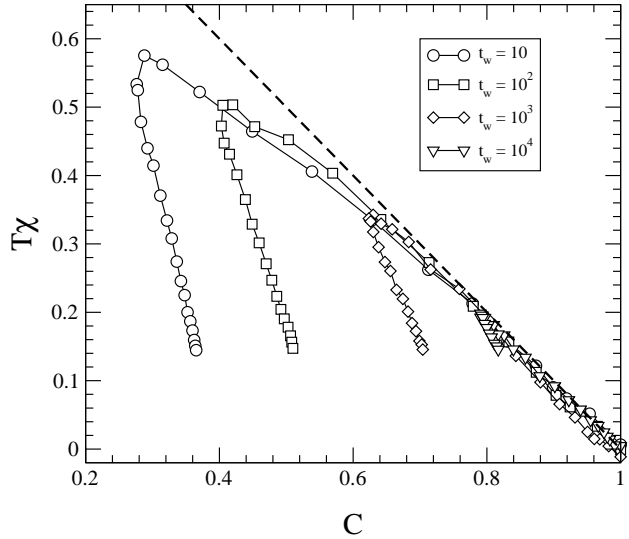}
  \caption{FD plots for the $1,1$-SFM \eq{kcm:1} with $N=10^5$, 
    quench temperature $T=0.3$
    and different waiting times (see the box). The straight line is the FDT
    relation. From \protect\cite{CriRitRocSel00}.
  }
  \label{fig:kcm:1}
\end{figure}
Buhot and Garrahan \cite{BuhGar01b} have explained how to recover well
defined FD plots by using \eq{kcm:2} and plotting $\chi(t,t_w)$ as
function of the difference $C_c(t,t)-C_c(t,t_w)$ where $C_c$ is the
standard connected correlation function $C_c(t,s)=\la (n(t)-c(t))(n(s)-c(s))\ra$
with $c(t)=\la n(t)\ra$ for the $1,1$-SFM. In this case, the resulting
FD curve corresponds to a straight line (corresponding to the
equilibrium result) casting doubts on the
usefulness of FD plots for non-cooperative models.

The origin of the maximum in the IRF has also been considered in the 1d
RFIM with infinite-ferromagnetic coupling $J=\infty$
\cite{CorCanLipZan02}. Actually, in this limit the model turns out to be
a KCM as transitions are only allowed on the spins sitting on the
interfaces of the ferromagnetic domains, i.e. a spin can flip only if
its left and right neighbours point in different directions (meaning
that the ferromagnetic contribution to the local field acting on that
spin vanishes). If quenched at low enough temperature this model
displays a coarsening behavior with two well separated regimes. If the
average distance between interfaces (or average domain length) $L(t)$ is
smaller than a length scale $L_g\sim (T/h_0)^2$ (where $h_0$ is the mean
square deviation of the intensity of the random field) then dynamics is
diffusive. However, if $L(t)>L_g$ dynamics becomes activated of the
Sinai type. In the Sinai regime the IRF is well described by the
relation \eq{kcm:2}, $\chi(t,t_w)=\chi_{\rm ag}(t,t_w)=c(t)\chi_{\rm
eff}(t,t_w)$ leading to a maximum of the IRF at intermediate times where
$L(t)\sim L_g$. This relation sets a crossover timescale where pure
diffusion takes over to Sinai diffusion. Comparing this result with
\eq{aging1} we note that there is no stationary contribution to the IRF
because for $J=\infty$ thermal fluctuations within domains are
suppressed. Again this leads to FD plots similar to figure
\ref{fig:kcm:1}. For the 1d RFIM in the asymptotic long-time limit
$L(t)\gg L_g$ the $\chi(t,t_w)$ decays to zero, a property required to
establish a link between static and dynamic properties.  As remarked in
section \ref{ferro}, however, this property does not hold for the 1d
Ising model.

A description of the IRF for the 2d plaquette model along the same lines
of \eq{kcm:2}, but modified to account for the diffusion and annihilation
of oscillating pairs of defects, has been presented in
\cite{BuhGar01b}. The resulting FD plots have been shown to display the
characteristic two slope curves of two-timescale systems. However, it is
unclear what is the physical meaning of the piece of the curve with
slope $X<1$ and whether indeed $\Te=T/X$ can be considered as a
thermodynamic temperature. Future research will show what is the true meaning of
these violation factors in KCMs of the coarsening type.

FDT violations in the glassy regime of KCMs are not simply described
within a thermodynamic IS formalism, the Kob-Andersen model perhaps
being an exception. In fact, the blocked states in the 1,1-SFM and the
East model generate identical configurational entropies
\cite{CriRitRocSel00} albeit they show very different dynamics
(diffusive and cooperative respectively). The applicability of
thermodynamic non-equilibrium concepts to models with trivial
equilibrium thermodynamics remains an open question.


\section{QFDT: the experimental evidence} 
\label{exp}

Any valuable physical theory must be successfully challenged by
experiments. Traditionally, the most direct way to experimentally access
FDT violations is through noise measurements. In the frequency domain
the FDT \eq{FDT} corresponds to the Nyquist formula that relates the
power spectrum of an observable to the imaginary part of its
susceptibility. With the same notation we used in
sections  \ref{basic},\ref{ME} we can write define the power spectrum
$S_{A,B}(\omega)$ and the complex susceptibility
$\chi_{A,B}(\omega)$,
\begin{eqnarray}
S_{A,B}(\omega)&=\frac{1}{\pi}\int_{-\infty}^{\infty}C_{A,B}(t)\exp(i\omega
t)dt;
\label{exp1a}
\\
\chi_{A,B}(\omega)&=\frac{\delta
\la\widehat{A}(\omega)\ra}{\delta B(\omega)}
\label{exp1b}
\end{eqnarray}
where $\widehat{A}(\omega)$ (and analogously $B$) is given by,  
\be
\widehat{A}(\omega)=\int_{-\infty}^{\infty}A(t)\exp(i\omega t)dt
\label{exp2}
\ee
If $\chi_{A,B}(\omega)=\chi_{A,B}'(\omega)+i\chi_{A,B}''(\omega)$ the
Nyquist formula \eq{fdr:3} reads \cite{GroMaz},
\be
S_{A,B}(\omega)=\frac{2k_BT}{\pi}\frac{\chi_{A,B}''(\omega) }{\omega}
\label{exp3}
\ee
The power spectrum can be experimentally measured considering the case
where the external perturbation couples to the measured observable
$A=B$. In the experimental protocol, the time evolution of the
observable $A(t)$ and the out-of-phase susceptibility
$\chi_{A,B}''(\omega)$ are recorded. The power spectrum is given by
$S_{A}(\omega)\propto \la |\hat{A}(\omega)|^2\ra$. This allows to verify
the Nyquist relation \eq{exp3}. Typical noise experiments are the
observation of electric voltage fluctuations or the motion of a Brownian
particle. Other measurements included sample to sample fluctuations of
the resistivity in small samples \cite{AleWeiIsr92,Weissman93}.
Observations of magnetic noise were successfully undertaken in
spin-glasses \cite{AlbHamOciRefBou87,RefOci87,Bouchiat90}, more than ten
years ago, and despite of the extremely low noise signal characteristic
of magnetic systems. Globally, these experiments show that, within the
accessible window of frequencies and times, no systematic FDT violations
are observed. The out-of-phase susceptibility $\chi''(\omega)$
associated to $\chi(\omega)$ is practically frequency independent and
the power spectrum shows the characteristic $1/f$ behavior.  The
negative result of these experiments points out one of the most
important difficulties encountered in these type experiments. As they
cover the frequency range $10^{-2},10^2$ Hertz, these frequencies are
much larger than the inverse of the aging time, therefore only the
locally equilibrated regime $\omega t\gg 1$ is explored. Only seldom the
``slow'' regime $\omega t\sim 1$, where FDT violations are expected, is
measured. Very recently, direct measurements of FDT violations in the
time domain have been reported for insulating spin glasses
\cite{HerOci02}. FDT violations have been measured beyond the
quasi-stationary regime and experimental FD plots have been found to be
consistent with results obtained by numerical simulations in spin glass
models (see section \ref{disor:sg}). However the value of the effective
temperature obtained in these experiments is much larger than the
annealing temperature at which the system is equilibrated before the
quenching takes place, casting doubts on the meaning of the effective
temperature as a thermodynamic temperature in these experiments.
Moreover these measurements reveal that, within the experimentally
accessible time window, FD curves $\chi(C)$ \eq{eq7e3} are
time-dependent and quite far from the expected asymptotic curve. This
may explain previously reported discrepancies among FD plots obtained by
analyzing the magnetization data of several spin glass systems
\cite{CugGreKurVin99}. These indirect measurements of FDT violations do
not show any clear evidence of a universal curve $\chi(C)$ and suggest
that experimental measurements are quite far from the asymptotic
regime. Experimental FD plots are shown in figure \ref{exp:fig1}.
\begin{figure}
  \centering
  \includegraphics[scale=0.4]{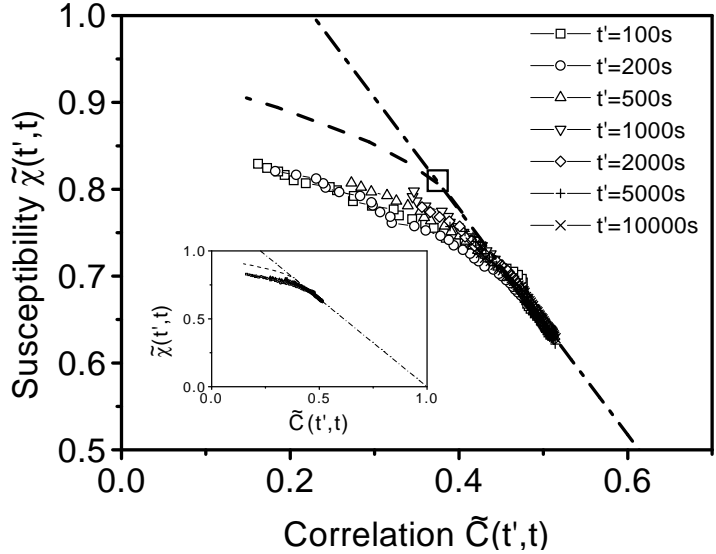}
  \caption{FD plot obtained by measuring voltage autocorrelations and
    relaxation susceptibility in the insulating spin glass
    $CdCr_{1.7}In_{0.3}S_4$ quenched from $T=1.2\Tg$ down to  
    $T=0.8\Tg$ ($\Tg=16.2K$). 
    Effective temperatures are around $30 K$ and much larger than the
    annealing temperature $19.4 K$. From \protect\cite{HerOci02}.
    }
  \label{exp:fig1}
\end{figure}

In structural glasses recent measurements show also the existence of FDT
violations, however the physical interpretation of these experiments is
still unclear. Voltage noise measurements \cite{GriIsr99} in an electric
resonant circuit formed by a capacitor containing glycerol and an
inductance show that the Nyquist formula \eq{exp3} is violated depending
on the waiting time and the frequency. By defining the effective
temperature $\Te(\omega,t_w)$ as in \eq{fdr:3} (i.e. the temperature
that satisfies the Nyquist formula \eq{exp3}) experiments reveal that the
capacitance ages and FDT violations appear also in the range $\omega
t_w\gg 1$. This seems to be in contradiction with previous old experimental results in
spin glasses. The origin of this discrepancy is presently unclear. Other
recent experiments by Ciliberto and
coworkers \cite{BelCilLar01,BelCil02} on Laponite, a synthetic clay of
charged particles that have the shape of a disc, show strong FD
violations. If let evolve inside water, Laponite generates a colloid
glass consisting of a packed irregular structure of disks due to the
complex pattern of quadrupolar interactions.  It is found that the value
of the effective temperature $\Te(\omega,t_w)$, for low enough
frequencies, is up to 3 or 4 orders of magnitude larger than the bath
temperature. Although this has been interpreted as an evidence for
coarsening dynamics (where $\Te\to \infty$) a conclusive explanation
of the origin of these high values is still unknown.  One interesting
aspect of these experiments is the scaling behavior $\omega t^{1/2}$
observed in the power spectrum indicative that FDT violations persist
even in the regime $\omega t_w\gg 1$, in agreement with the previous
results on glycerol. Contrarily to what would be expected, noise measurements in a rheological
experiment for Laponite do not detect significant FDT violations.

Experiments that successfully clearly demonstrate the existence of effective
temperatures related to FDT violations are certainly needed. These
preliminary account of results shows that still much work has to be done
in aging, driven or granular systems to provide a safe ground to many of
these ideas.

\section{Conclusions}
\label{conclusions}

In this review we have presented an overview on what is now an active
area of research, i.e. the study of FDT violations in glassy
systems. Glassy systems are widespread in nature and found in many
different areas covering physics, chemistry or biology. These are
non-equilibrium systems which are either in a non-stationary slowly
relaxing aging state or in a weakly driven stationary state. Equilibrium
systems are often described by a set of intensive parameters such as
temperature, pressure or density. In a similar way, in glassy systems an
important role is played by the waiting time (aging systems) or the
intensity of the driving force and/or its frequency (driven systems). These
parameters describe how far from equilibrium the system is. In
fact, it is becoming steadily clear that a thermodynamic description of
glassy systems can be partially rescued.  In this description the glassy
regime can be rationalized by using some of the concepts of equilibrium
statistical physics such as the existence of a modified version of the
FDT, the so called quasi-FDT (QFDT).

Associated to the existence of the QFDT there is the concept of
effective temperature. Rather than being a useful parameter to
describe the behavior of non-equilibrium systems, the
effective temperature has a deeper physical meaning. It could be the
vestige of the existence of some dynamical measure underlying the
non-equilibrium regime. However we do not know how
to prove the existence of the QFDT from first principles in the same way
we do not know how to prove Boltzmann equal probability hypothesis in
equilibrium theory.  Therefore, establishing the existence of this
dynamical measure is somehow equivalent to assuming the existence of a
QFDT. Then the crucial point is which are the specific predictions, that can
be experimentally tested, one can derive from the existence of a
QFDT. The measurement of the effective temperature itself appears as the
most direct way of challenging the QFDT. However, direct measurements of
this non-equilibrium temperature appear quite difficult, results are
still very preliminary and much progress is yet to be done to reach a
convincing and clear evidence.

The study of several families of models, as described in this review,
appears as a very fruitful source of inspiration for new concepts and
ideas that could be eventually exported to different classes of
problems.  However this path is not free of challenges and
ambiguities. Indeed what are natural concepts for some families of
models appear to be quite artificial in other families. As an example,
key concepts in a thermodynamic formulation of the glassy state are the
existence of an effective temperature associated to the configurational
entropy or complexity, in the same way the bath temperature is
associated to the Boltzmann entropy in equilibrium theory. However, what
appears to be an interesting quantity describing glassy systems with
two-timescales cannot be easily translated into systems with many
timescales. Even more, what appears to be a meaningful quantity for
models having a complex thermodynamics (such as spin-glass models)
appears to be meaningless in models described by a trivial Hamiltonian
(such as kinetically constrained models).

From the point of view of theoretical studies, our understanding of the
existence of a QFDT appears also quite problematic. Most of the
solutions we have described in this review are only valid in the case of
mean-field interactions, however their validity beyond that limit
remains speculative. The use of numerical simulations has aided to
bridge the gap. It is quite interesting that most of the results
predicted in mean-field models are qualitatively also observed in
short-range systems. This tendency to rationalize the behavior of real
systems within a mean-field scenario, i.e. far beyond their natural
domain of applicability, has become quite standard in the study of
glassy systems. The description of the equilibrium properties of
spin-glass systems has for many years followed a similar route. Although
the knowledge we can gain from numerical simulations in glassy systems
is always quite qualitative (either due to the limited range
of timescales or to the inherent simplicity of the simulated model) the
accumulated evidence, as reported along this review, points toward the
emergence of a QFDT in the non-equilibrium regime of glassy systems, reminiscent
of how the Boltzmann measure emerges in equilibrium
systems. The same conclusion holds for the Edwards measure in granular
media.

What will be the future in the research of FDT violations? Although
modeling promises to offer new ideas and will clarify our understanding
we feel that more progress is certainly needed in basic theory and
experiment. In theory we need to understand the origin of the existence
of a QFDT right from the microscopics. Cooperative processes in glasses
involve a few tens of atoms and occur along nanometric length-scales. How
to link the microscopic activated processes to the emergence of
macroscopic properties (such as the effective temperature) is a real
challenge. In this context, it appears quite interesting to pursue the
investigation of the so called fluctuation theorems recently proposed to
quantify transient violations of the second law of
thermodynamics. 
From the experimental side the current accumulated knowledge is still too
poor and more experiments are certainly needed for this field of
research to grow. A future line of progress is
the use of nanotechnology devices to do noise  measurements over
spatially localized regions of nanometric sizes. These devices could
be used as a microscope to measure activated processes occurring in 
small length scales.

Certainly we will see upcoming developments in this exciting area of
research. A continuous exchange of ideas among theory, simulation and
experiments is highly desirable and certainly needed to improve our
current understanding in this field.

{\bf Acknowledgments:} We are grateful to all our collaborators, much of
whose work has been described in this review, including L. L.  Bonilla,
J. P. Bouchaud, S. Ciuchi, B. Coluzzi, L. F. Cugliandolo, K. Dawson,
S. Franz, A.  Garriga, J. Hertz, H. Horner, J. Kurchan, L. Leuzzi,
E. Marinari, U. Marini Bettolo, M. Mezard, Th. M. Nieuwenhuizen,
F. G. Padilla, I. Pagonabarraga, G. Parisi, M. Picco, A. Puglisi,
F. Ricci-Tersenghi, F. Rao, M. Rub\'{\i}, A. Rocco, G. Ruocco, M. Sales,
D. Sherrington, F. Sciortino, M. Sellitto, P. Sollich, H.-J. Sommers,
H. Sompolinsky and P. Tartaglia.  We are also indebted to many other
colleagues for enlightening discussions on many of the topics covered in
this review. We are particularly grateful to P. Sollich for useful
comments and improvement suggestions. FR has been supported by the
Spanish Ministerio de Ciencia y Tecnolog\'{\i}a Grant BFM2001-3525,
Generalitat de Catalunya and Acciones-Integradas Italia-Espa\~na
(HI2000-0087).

\section{List of abbreviations}
\begin{description}
\item[1RSB] One-step Replica Symmetry Breaking
\item[BG] Backgammon
\item[BTM] Bouchaud Trap Model
\item[EA] Edwards-Anderson model
\item[FD] Fluctuation-Dissipation
\item[FDR] Fluctuation-Dissipation Ratio
\item[FDT] Fluctuation-Dissipation Theorem
\item[FILG] Frustrated Ising Lattice Gas
\item[IRF] Integrated Response Function
\item[IS] Inherent Structure
\item[LJ] Lennard-Jones
\item[ME] Master Equation
\item[MF] Mean Field
\item[MCT] Mode-Coupling Theory
\item[OSC] Oscillator
\item[QFDT] Quasi-Fluctuation-Dissipation Theorem
\item[REM] Random Energy Model
\item[RFIM] Random Field Ising Model
\item[ROM] Random Orthogonal Model
\item[RSB] Replica Symmetry Breaking
\item[SK] Sherrington-Kirkpatrick
\item[TAP] Thouless-Anderson-Palmer
\item[$\bi{\Tc}$] Mode-Coupling critical temperature, 
  Dynamical critical temperature
\item[$\bi{\Te}$] Effective temperature
\item[$\bi{\Tg}$] Glass transition temperature
\item[$\bi{\TK}$] Kauzmann temperature
\item[$\bi{\Trsb}$] Static transition temperature in replica calculations
\item[TRM] Thermoremanent Magnetization
\item[TTI] Time Translation Invariance
\item[ZFC] Zero Field Cooled
\end{description}

\section*{References}

\end{document}